\documentclass[a4paper,fleqn,usenatbib]{mnras}
\usepackage{newtxtext,newtxmath}
\usepackage[T1]{fontenc}
\usepackage{ae,aecompl}
\usepackage{graphicx}	
\usepackage{amsmath}	
\usepackage{amssymb}	
\usepackage{booktabs}	
\usepackage{url}

\usepackage{array}
\newcolumntype{H}{>{\setbox0=\hbox\bgroup}c<{\egroup}@{}} 

\title[The $\delta$\,Sct instability strip]{\textit{Gaia}-derived luminosities of \textit{Kepler} A/F stars and the pulsator fraction across the $\delta$\,Scuti instability strip}

\author[Simon J. Murphy et al.]{
Simon J. Murphy$^{1,2}$\thanks{E-mail: simon.murphy@sydney.edu.au (SJM)}, Daniel Hey$^{1,2}$, Timothy Van Reeth$^{1,2}$, Timothy R. Bedding$^{1,2}$
\\
$^{1}$ Sydney Institute for Astronomy (SIfA), School of Physics, University of Sydney, NSW 2006, Australia\\
$^{2}$ Stellar Astrophysics Centre, Department of Physics and Astronomy, Aarhus University, 8000 Aarhus C, Denmark\\
} 

\date{Accepted XXX. Received YYY; in original form ZZZ}
\pubyear{2015}
\begin{document}
\label{firstpage}
\maketitle

\begin{abstract}
We study the fraction of stars in and around the $\delta$\,Scuti instability strip that are pulsating, using \textit{Gaia} DR2 parallaxes to derive precise luminosities. We classify a sample of over 15\,000 \textit{Kepler} A and F stars into $\delta$\,Sct and non-$\delta$\,Sct stars, paying close attention to variability that could have other origins. We find that 18\:per\:cent of the $\delta$\,Sct stars have their dominant frequency above the \textit{Kepler} long-cadence Nyquist frequency (periods $<$ 1\,hr), and 30\:per\:cent have some super-Nyquist variability. We analyse the pulsator fraction as a function of effective temperature and luminosity, finding that many stars in the $\delta$\,Sct instability strip do not pulsate. The pulsator fraction peaks at just over 70\:per\:cent in the middle of the instability strip. The results are insensitive to the amplitude threshold used to identify the pulsators. We define a new empirical instability strip based on the observed pulsator fraction that is systematically hotter than theoretical strips currently in use. The stellar temperatures, luminosities, and pulsation classifications are provided in an online catalogue.
\end{abstract}




\section{Introduction}
\label{sec:intro}

Do all stars in the $\delta$\,Sct instability strip pulsate? This is a long-standing question whose answer requires high-quality light curves and accurate atmospheric parameters for a large number of stars. These have only recently become available. The Kepler Space Telescope \citep{kochetal2010,boruckietal2010} has delivered almost uninterrupted light curves for over 15\,000 A and early-F stars. Catalogues of photometrically-derived stellar properties \citep{brownetal2011,huberetal2014,mathuretal2017} have allowed greater exploitation of these light curves, including placement of the targets on a $T_{\rm eff}$--$\log g$ diagram (e.g. \citealt{uytterhoevenetal2011}). The uncertainties in $\log g$ were large, typically half the height of the main sequence, and there was much discussion over whether the photometric effective temperatures were accurate \citep{lehmannetal2011,pinsonneaultetal2012,tkachenkoetal2012}. Now, precise luminosities are calculable from Gaia parallaxes \citep{gaiacollaboration2016,gaiacollaboration2018a}, and spectroscopic temperatures are available for hundreds of \textit{Kepler} A/F stars \citep[e.g.][]{tkachenkoetal2013b,niemczuraetal2017} to readdress the accuracy of photometric temperatures. At last we possess the data necessary to reliably determine the fraction of stars within the $\delta$\,Sct instability strip that pulsate.

Central to the pulsator-fraction question is the definition of the instability strip. The $\delta$\,Sct pulsators are 1.5--2.3\,M$_{\odot}$ stars that lie at the intersection of the classical instability strip with the main sequence, in the region where the depth of the surface convection zone is a steep function of effective temperature. As such, time-dependent convection (TDC) models are needed to accurately model the pulsations and compute instability strips \citep{dupretetal2004,grigahceneetal2005}.

TDC has an adjustable parameter, which is the mixing length $\alpha_{\rm MLT}$. \citet{dupretetal2004,dupretetal2005a} set this to the solar value, $\alpha_{\rm MLT} = 1.8$, but when known $\delta$\,Sct stars \citep{rodriguezetal2000} were plotted against instability strips with this value, the observed stars had a temperature offset with respect to the instability strip. Better agreement is seen with $\alpha_{\rm MLT}=2.0$ \citep{houdek2000,houdek&dupret2015}, but instability strips with $\alpha_{\rm MLT} = 1.8$ are still commonly plotted and compared against observations. This has led to several observers noting that the instability strip is incorrect for \textit{Kepler} $\delta$\,Sct stars \citep{uytterhoevenetal2011,balona2018c,bowman&kurtz2018}, and has contributed to the discussion of the accuracy of photometric temperatures.

There is a separate class of variables, the $\gamma$\,Dor stars, which pulsate in gravity modes (g\:modes) and lie close to the $\delta$\,Sct stars on the H--R diagram. Indeed, many $\delta$\,Sct--$\gamma$\,Dor hybrids exist (e.g. \citealt{grigahceneetal2010a,balona&dziembowski2011,kurtzetal2014,saioetal2015,schmid&aerts2016}), though even at \textit{Kepler} precision pure examples of both are also known \citep{vanreethetal2015a,bowman2017,lietal2019a}. Originally, these classes were distinguished by their oscillation frequencies, with a division made at $\sim$5\,d$^{-1}$ (e.g. \citealt{handler&shobbrook2002,hareteretal2010}), and the $\gamma$\,Dor stars having lower frequencies than the $\delta$\,Sct stars. However, stellar rotation complicates this picture, allowing rapidly rotating $\gamma$\,Dor stars to have frequencies above 5\,d$^{-1}$ in the observer's frame \citep{bouabidetal2013,saioetal2018b}, and combination frequencies of g\:modes (e.g.\ harmonics) can do the same \citep{kurtzetal2015}. In fact, it is likely that the combination frequencies contribute to the excitation of higher degree g\:modes at these high frequencies \citep{saioetal2018b}. Conversely, combinations of p\:modes can also cause Fourier peaks at low frequency in $\delta$\,Sct stars, and care must be taken not to confuse them with other classes of variability \citep{breger&montgomery2014}. Distinguishing self-excited ($\delta$\,Sct) p\:modes from harmonics, combinations, or high-degree g\:modes is important if we are to understand the pulsational driving and damping mechanisms at play in A and F stars.

It has long been suspected that all stars within the $\delta$\,Sct instability strip might be variable if the detection limit becomes low enough \citep{breger1969b}. Decades of further ground-based observations resulted in the realisation that the vast majority of $\delta$\,Sct stars are low-amplitude radial and non-radial pulsators, but only half the stars in the instability strip appeared to be variable at ground-based ($\sim$1\,mmag) precision \citep{breger2000}. \textit{Kepler} observations confirmed this view -- no more than 60\:per\:cent of stars in the $\delta$\,Sct instability strip actually pulsate, and half of those have amplitudes below 1\,mmag \citep{balona&dziembowski2011,murphy2014}.

In this paper we investigate the location of the $\delta$\,Sct pulsators on the H--R diagram, looking specifically at the pulsator fraction as a function of effective temperature and luminosity. We calculate luminosities for over 15\,000 A and F stars using \textit{Gaia} DR2 parallaxes in Sec.\,\ref{sec:obs}. In Sec.\,\ref{sec:pulsators}, we classify them into $\delta$\,Sct stars and non-$\delta$\,Sct stars, which we then place on H--R diagrams in Sec.\,\ref{sec:HRD}. In Sec.\,\ref{sec:fraction} we look at the pulsator fraction before discussions (Sec.\,\ref{sec:discussion}) and conclusions (Sec.\,\ref{sec:conclusions}).

\section{Observational data}
\label{sec:obs}

\subsection{Target Selection}
\label{ssec:targets}

We selected \textit{Kepler} targets with temperatures between 6500 and 10\,000\,K, according to the `input' temperatures in \citet{mathuretal2017}. This range covers the $\delta$\,Sct instability strip and the region around it where $\delta$\,Sct pulsators are observed. Variable stars hotter than 10\,000\,K are likely to be B-type pulsators ($\beta$\,Cep stars), while stars listed as cooler than 6500\,K may only be $\delta$\,Sct stars if their temperatures are inaccurate. While some genuine $\delta$\,Sct stars might have incorrect temperatures and lie outside this range, the vast majority of $\delta$\,Sct stars will have been captured.

We chose the \citet{mathuretal2017} catalogue because of its homogeneity. The `input' temperatures were chosen to avoid the clustering of stars around isochrones in the $T_{\rm eff}$--$\log g$ plane that occurs in the `output' temperatures. This is further discussed, alongside verification of the effective temperature scale, in Sec.\,\ref{ssec:temp_scale}.

We made no selection based on $\log g$ or [Fe/H] values, which we also gathered from the input values of \citet{mathuretal2017}. We made no cut on \textit{Kepler} magnitude, but we did cut stars with a luminosity 0.4\,dex (i.e.\ 1\,mag) fainter than the zero-age main-sequence (ZAMS) after calculating the luminosities (Sec.\,\ref{ssec:gaia}) because we focus here on stars on the main-sequence and immediate post-main-sequence. We did not include stars for which there are no \textit{Kepler} light curves available. The total sample size in our $T_{\rm eff}$ range and after filtering by luminosity is 15\,229.

\subsection{\textit{Gaia} and supplementary stellar data}
\label{ssec:gaia}

\begin{figure}
\begin{center}
\includegraphics[width=0.5\textwidth]{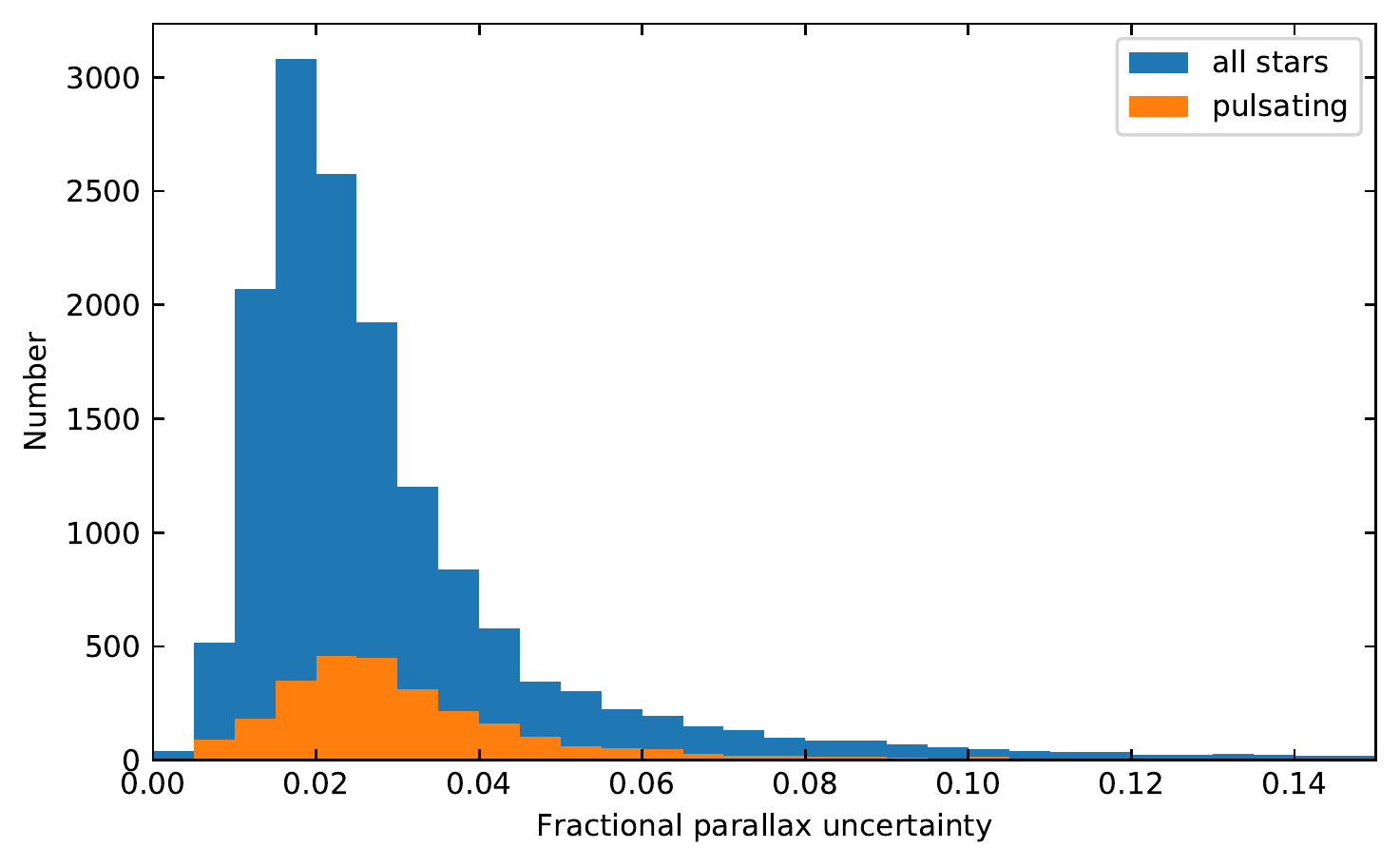}
\caption{Fractional parallax uncertainties for the entire sample. Only 347 of 15\,229 targets have fractional parallax uncertainties beyond the plot limit (i.e. > 0.15).}
\label{fig:frac_pi_err}
\end{center}
\end{figure}

In DR2, stars with \textit{Gaia} temperatures above 8000\,K had no available extinction values in the \textit{Gaia} G band, $A_G$, and thus no luminosities. It also appears that extinctions were not used in calculating the luminosities of stars where $A_G$ was available \citep{andraeetal2018}, hence the {\it Gaia} luminosities as provided in the DR2 catalogue are unreliable for our targets. We therefore calculated luminosities for all targets, taking from {\it Gaia} DR2 only stellar parallaxes, $\pi$, and their uncertainties \citep{gaiacollaboration2018a}. Most of our targets have precise parallaxes (Fig.\,\ref{fig:frac_pi_err}), with only 249 having fractional uncertainties above 0.2. Below, we make stricter selections when analysing H--R diagrams of A/F stars.

To obtain a stellar distance, $d$, from a DR2 parallax, we used the normalized posterior distribution over the distance following equation~2 of \citet{bailer-jonesetal2018} along with the length scale model adopted there. This produces a distribution of distances for each star, from which samples are drawn in the Monte Carlo calculation of the stellar luminosity (Sec.\,\ref{ssec:luminosity}). 

From the analysis of quasars, \citet{lindegrenetal2018} inferred a zero-point offset in \textit{Gaia} parallaxes of $-$0.03\,mas, suggesting that 0.03\,mas should be added to the published values. Similar offsets have been independently confirmed using Gaia DR2 parallaxes and asteroseismology of evolved stars in the \textit{Kepler} field \citep{zinnetal2018}, but the offset is colour and magnitude dependent. \citet{lindegrenetal2018} also noted large-scale variations of a similar size depending on colour, magnitude, and position on the sky. As such, in the \textit{Gaia} DR2 catalogue validation paper \citep{arenouetal2018}, users are discouraged from correcting individual parallaxes to account for the zero-point offset. We experimented with applying the 0.03\,mas offset and found that it led to luminosities that were unrealistically low for many stars (i.e. it pushed a large number of stars below the ZAMS). From this, we infer that the parallax offset is small for bluer stars, and we made no correction for zero-point offset in the luminosities used in the rest of our analysis.

For apparent magnitudes, we used the Kepler Input Catalogue (KIC; \citealt{brownetal2011}) on MAST to obtain $g_{\rm KIC}$ and $r_{\rm KIC}$ magnitudes, and recalibrated the $g$ magnitudes to the SDSS scale using equation 1 of \citet{pinsonneaultetal2012}. Working in the $g$ band is advantageous because this is close to the peak of the spectral energy distribution of A/F stars, hence the bolometric corrections are small, and these are available homogeneously for our sample. The $g$ magnitudes have typical random uncertainties of 0.02\,mag for our targets \citep{brownetal2011}. Although MAST has $V$-band magnitudes (\textit{V\_UBV}) from \citet{everettetal2012} for a large fraction of our targets, the completeness is lower than for $g_{\rm KIC}$ and we found these $V$ magnitudes to be systematically too faint (Appendix\,\ref{app:reddening}).

\begin{figure}
\begin{center}
\includegraphics[width=0.5\textwidth]{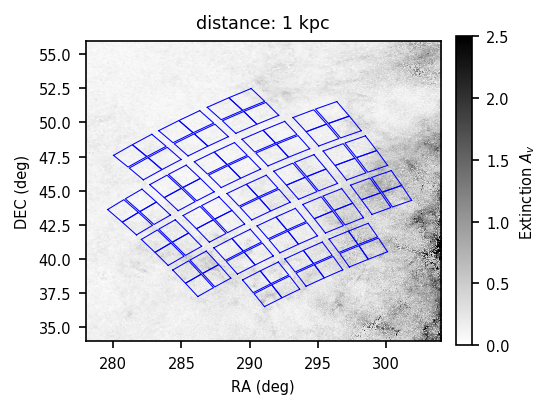}
\caption{$V$-band extinction in the \textit{Kepler} field, using data from the Bayestar\,17 reddening map \citep{greenetal2018}. In the online version of the article, this figure is animated to show different distances. The $g$- and $V$-band extinctions are related as  $A_g = 1.214 A_V$.}
\label{fig:dust}
\end{center}
\end{figure}

We obtained extinctions and their uncertainties with the {\sc dustmaps} python package, which queries the Bayestar\,17 reddening map \citep{greenetal2018}. The dust distribution in the \textit{Kepler} field is shown in Fig.\,\ref{fig:dust}. To convert to extinctions in the appropriate photometric band, we used Table\:1 of \citet{sanders&das2018}. A small grey offset of 0.063\,mag (recommended by \citealt{greenetal2018}) was added to the extinction coefficient. We found that KIC extinctions are strongly overestimated (Appendix\,\ref{app:reddening}), so they were not used. We computed extinctions in the $g$ band, $A_g$.

We calculated bolometric corrections (BCs) with the {\sc isoclassify} python package \citep{huberetal2017}, which takes $T_{\rm eff}$, $\log g$ and [Fe/H] as inputs. The BCs can be computed for any band using the MIST tables \citep{dotter2016,choietal2016}. These, too, were calculated for the SDSS $g$ band.

\subsection{Luminosity calculations}
\label{ssec:luminosity}

Logarithmic bolometric luminosities were calculated via absolute magnitudes using the standard formulae:
\begin{eqnarray}
M_g = m_g - 5 \left( \log d - 1 \right) - A_g,
\label{eq:mag}
\end{eqnarray}
and
\begin{eqnarray}
\log L_{\rm bol}/L_{\odot} = -(M_g + BC - M_{\rm bol,\sun})/2.5.
\label{eq:Lbol}
\end{eqnarray}
For the bolometric magnitude of the Sun, $M_{\rm bol,\sun}$, we adopted the value recommended by the IAU (4.74; \citealt{mamajeketal2015b}). Hereafter, we use ``$\log L$'' to refer to the luminosities obtained via \mbox{eq.\,\ref{eq:Lbol}}. We determined $\log L$ and its uncertainties with a Monte-Carlo process, taking the median value of 200\,000 samples for the $\log L$ value, and the 15.9 and 84.1 percentiles for the 1$\sigma$ uncertainties.

\begin{figure}
\begin{center}
\includegraphics[width=0.5\textwidth]{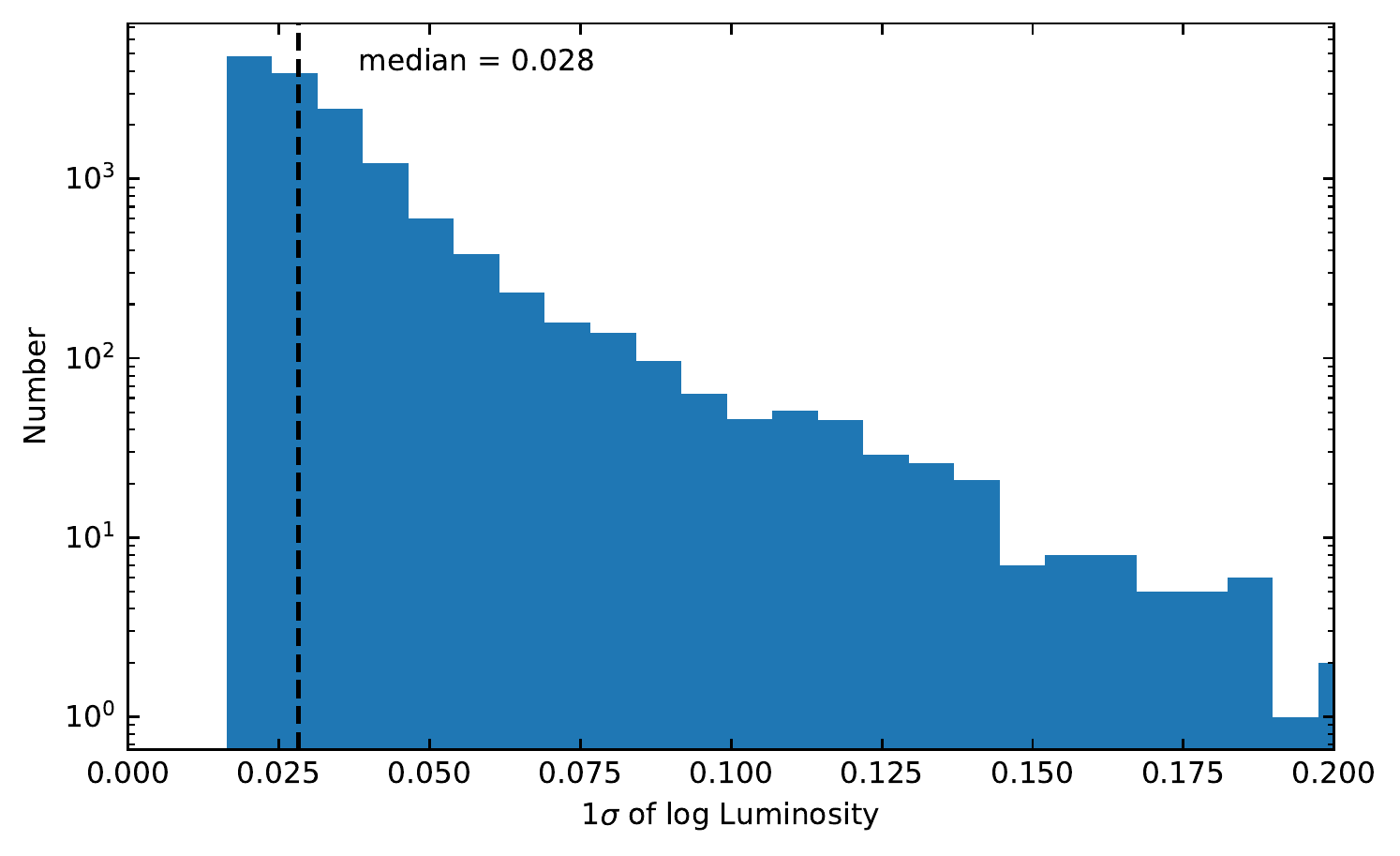}
\caption{The distribution of 1$\sigma$ uncertainties on the stellar luminosities, taken as the standard deviation of the Monte-Carlo samples.}
\label{fig:lum_err}
\end{center}
\end{figure}

The uncertainties in BC, which feed into the luminosity Monte Carlo process, were obtained from uncertainties in other quantities. The typical uncertainties on $T_{\rm eff}$ ($\sim$250\,K) make only a small contribution to the uncertainty in BC ($<0.03$\,mag). This is because BCs for stars in the $\delta$\,Sct instability strip in the SDSS $g$ band are small ($\lvert BC \rvert < 0.15$\,mag) and undergo a turning point at 8500\,K. The other contributions come from the uncertainties in $\log g$, [Fe/H] and $A_V$. We estimate these to be 0.02\,mag for 0.5\,dex in $\log g$, 0.025\,mag for 0.25\,dex in [Fe/H], and an upper limit of 0.002\,mag for extinction. We combined these in quadrature. The resulting distribution of luminosity uncertainties is shown in Fig.\,\ref{fig:lum_err}.

\subsection{Verification of the temperature scale}
\label{ssec:temp_scale}

KIC photometry is very useful for approximating stellar properties for a large number of stars across a broad range of temperatures and surface gravities, but uncertainties for individual stars are typically large, especially at hotter temperatures ($T_{\rm eff} \gtrsim 6500$\,K). A benefit of using a large sample, as we do here, is that individual uncertainties become less important. However, attention must still be paid to any systematic offsets in $T_{\rm eff}$, as we alluded to in Sec.\,\ref{sec:intro}.

There has been some discussion about these systematic offsets. \citet{pinsonneaultetal2012} found that KIC temperatures were systematically 215\,K too low when compared to the infra-red flux method (IRFM), in the $T_{\rm eff}$ range 4000--6500\,K where the IRFM is well defined. On the other hand, \citet{guziketal2015} used atmospheric parameters of A/F stars derived by \citet{molenda-zakowiczetal2013} from LAMOST spectra and found that \citeauthor{pinsonneaultetal2012}'s systematic offset is not evident for these hotter ($T_{\rm eff}$ > 6500\,K) stars. In fact, \citet{guziketal2015} found a small systematic offset between the KIC and LAMOST temperatures in the opposite direction.

Revised stellar properties have also been published for \textit{Kepler} targets \citep{huberetal2014,mathuretal2017}. Those authors conditioned published atmospheric parameters on a grid of stellar isochrones, resulting in catalogues of input and output temperatures. While the output temperatures are the ones most often used, the conditioning on isochrones to generate the output quantities is not especially useful for hotter stars, where the isochrones are relatively further apart in $T_{\rm eff}$--$\log g$ space, resulting in stars heavily concentrated around isochrones with large gaps between them. This is the reason we use the input temperatures from \citet{mathuretal2017} in this paper. Nonetheless, the revised stellar properties catalogues are a valuable resource, and the input temperatures are an improvement on the original KIC, incorporating tens of thousands of new temperature sources from the literature. 

Effective temperatures from high-resolution spectroscopy (e.g.\ \citealt{tkachenkoetal2013a}) and spectral energy distributions (SEDs; \citealt{niemczuraetal2017}) have also been calculated independently of each other for many \textit{Kepler} A/F stars. \citet{niemczuraetal2015} found good agreement between temperatures from SEDs and both the KIC and revised stellar properties catalogues, but the spectroscopic temperatures were systematically 189\,K larger than the KIC. Other spectroscopic analyses have found different levels of agreement with the KIC, even when conducted by the same group (\citealt{tkachenkoetal2012,tkachenkoetal2013a}), suggesting that any offset may depend strongly on the type of stars (e.g. late-B stars versus early-F stars) or that the differences between spectroscopic and photometric temperatures for individual targets can be large.

\begin{figure}
\begin{center}
\includegraphics[width=0.48\textwidth]{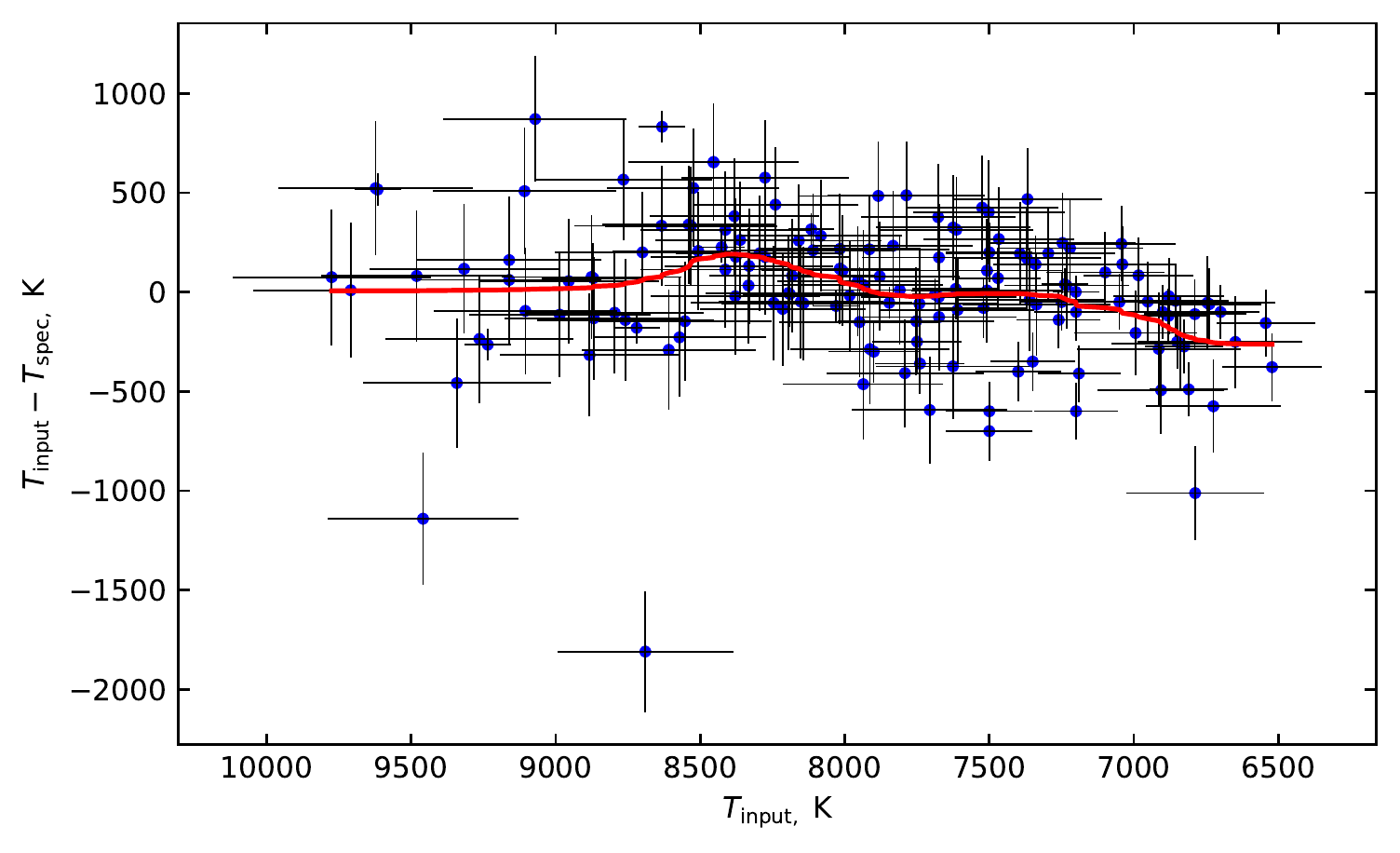}
\caption{The difference between spectroscopic and photometric temperatures for 154 A/F stars. The red line is a 10-point Gaussian smooth of the data. Uncertainties for individual stars are larger than the systematic trend. The extent of disagreement near 6500\,K is probably exaggerated by selection effects.}
\label{fig:teff_scale}
\end{center}
\end{figure}

Here, we have verified the input temperatures in \citet{mathuretal2017}, $T_{\rm input}$, against a homogeneously derived set of spectroscopic temperatures from \citet{niemczuraetal2015,niemczuraetal2017}, $T_{\rm spec}$. Specifically, we used the temperatures determined from Fe lines, in cases where they were distinguished from temperatures obtained from Balmer line profile fitting \citep{niemczuraetal2017}. These two methods generally yield temperatures that agree to 1$\sigma$ and \citet{niemczuraetal2015} made no distinction between the methods in the values they reported. The temperature differences ($T_{\rm spec} - T_{\rm input}$) have a mean of only 4\,K but a standard deviation of 348\,K. However, there is a significant dependence on temperature itself (Fig.\,\ref{fig:teff_scale}). The turnover near 8000\,K is in agreement with \citet{guziketal2015}, and the fact that the input temperatures appear to be too low for stars near 6500\,K is in agreement with the results of \citet{pinsonneaultetal2012} and is a major reason that we kept 6500\,K as our temperature cut-off. 
There may be a selection effect responsible for the discrepancy between 6500 and 7000\,K. Targets in these surveys were not chosen at random or distributed evenly by temperature, but were often selected for their pulsation properties. Thus if a $\delta$\,Sct star happens to have $T_{\rm input}$ near 6500\,K, beyond the red edge of the $\delta$\,Sct instability strip, its spectroscopic temperature is very likely to be higher. Conversely, a randomly chosen star near 6500\,K does not necessarily have an erroneous temperature. For this reason, and for simplicity and reproducibility, we do not apply any temperature correction to the input temperatures from \citet{mathuretal2017}.

\section{Identifying the $\delta$\,Sct pulsators}
\label{sec:pulsators}

Our purpose is to investigate the locations of $\delta$\,Sct and non-$\delta$\,Sct stars with respect to the instability strip, and to calculate the pulsator fraction. To do this efficiently and accurately, we made further selections based on the available data.

We have already removed white dwarfs in our temperature range by means of our luminosity cut. We also removed the 249 targets with fractional parallax uncertainties above 20\:per\:cent. We could have made a stricter selection on parallax uncertainty. For example, there are 400 targets with fractional parallax errors between 10 and 20\:per\:cent whose measured parallaxes still contain valuable information \citep{lurietal2018}. Their removal would make little difference in a sample of 15\,000 stars so we kept the cut at 20\:per\:cent.

We also removed known eclipsing binaries and ellipsoidal variables from the sample because their Fourier transforms often have harmonics of the orbital frequency that complicate the identification of low-amplitude pulsations. For this, we used the binary classifications from \citet{murphyetal2018} and the Villanova Catalogue \citep{kirketal2016}.\footnote{\url{http://keplerebs.villanova.edu/}} We also used the \citet{murphyetal2018} catalogue to remove RR\,Lyr variables from the sample for similar reasons, though we do not expect any RR\,Lyr variable to be a $\delta$\,Sct star. These deselections resulted in a final sample size of 14\,330 stars for analysis.

We did not remove the pulsation-timing binaries from \citet{murphyetal2018} since, unlike for eclipsing binaries, their removal would bias the pulsator fraction calculation by preferentially removing pulsating stars. The impact of binary stars on the analysis is discussed in Sec.\,\ref{sec:discussion}.

\subsection{\textit{Kepler} data}
\label{ssec:kepler}

We used \textit{Kepler} long-cadence (LC; 29.45-min sampling) light curves from Q0--Q17, processed with the msMAP pipeline \citep{stumpeetal2014}, and calculated discrete Fourier transforms of these light curves. Since the aim was to investigate the $\delta$\,Sct pulsator fraction, we did not calculate the Fourier transform below 5\,d$^{-1}$, where $\gamma$\,Dor and other variability is often seen. The sampling properties of the \textit{Kepler} spacecraft allow the real oscillation frequencies to be distinguished from Nyquist aliases, even when those frequencies are above the Nyquist frequency \citep{murphyetal2012b}. We therefore calculated our Fourier transforms between 5 and 43.9\,d$^{-1}$, where the latter value is the sampling frequency minus 5\,d$^{-1}$. We found that 17.9\:per\:cent of \textit{Kepler} $\delta$\,Sct stars have their strongest oscillation mode above the LC Nyquist frequency, and the number of $\delta$\,Sct stars with {\it any} oscillation above the LC Nyquist frequency is approximately 30\:per\:cent.

\subsection{Pulsation classifications}
\label{ssec:pulsation_classification}

A major component of the analysis was to decide which \textit{Kepler} A/F stars are $\delta$\,Sct pulsators. We began with the manual pulsation classifications from \citet{murphyetal2018}, with an extension to stars with $T_{\rm eff}$ in the range 6500--6600\,K, classified in the same manner. We refer to these as the `manual classifications'. The purpose of those classifications was different from the purpose here. \citet{murphyetal2018} were using pulsation timing to look for binary stars, for which any stable frequency with a high enough signal-to-noise ratio (SNR) is sufficient \citep{murphyetal2014}. This has two important consequences. The first is that harmonics of low-frequency variability were viable for pulsation timing but these are not $\delta$\,Sct stars (unless there are also independent p\:modes). In this work, we reinspected the Fourier transforms of all pulsators under the manual classifications, including at low frequencies ($<$5\,d$^{-1}$), to verify the presence of independent p\:modes. The p\:modes did not have to be dominant, but they did have to be independent. We refer to pulsation classifications following this reinspection as `revised' classifications. 

\begin{figure}
\begin{center}
\includegraphics[width=0.48\textwidth]{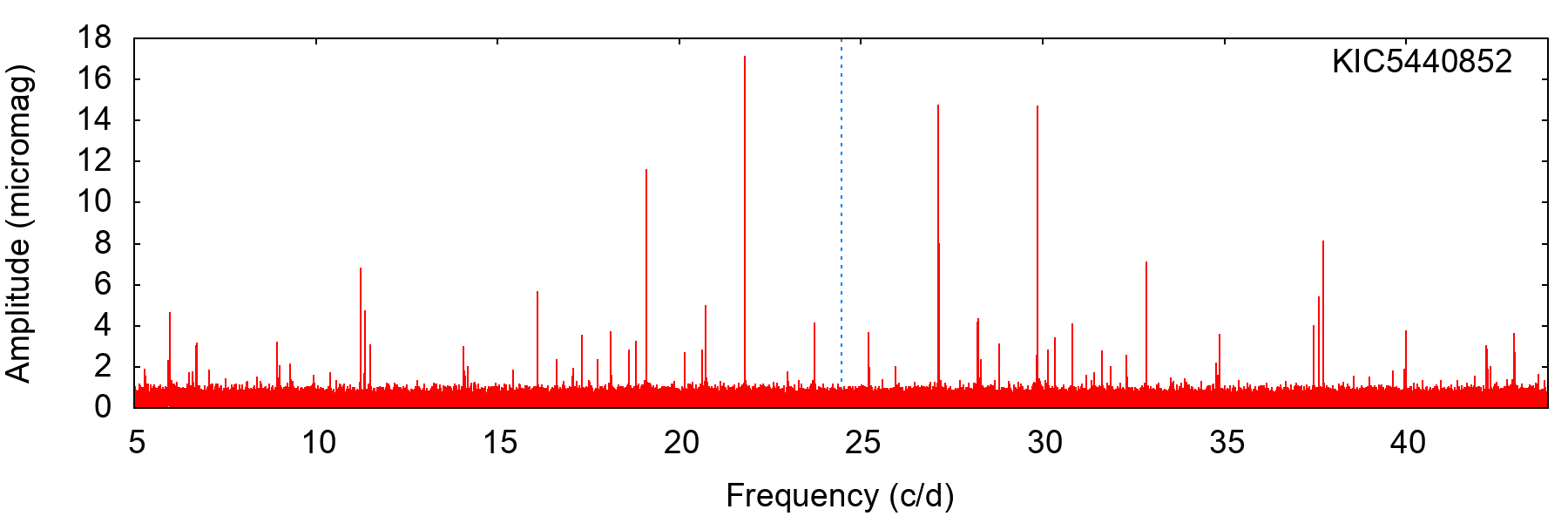}\\
\includegraphics[width=0.48\textwidth]{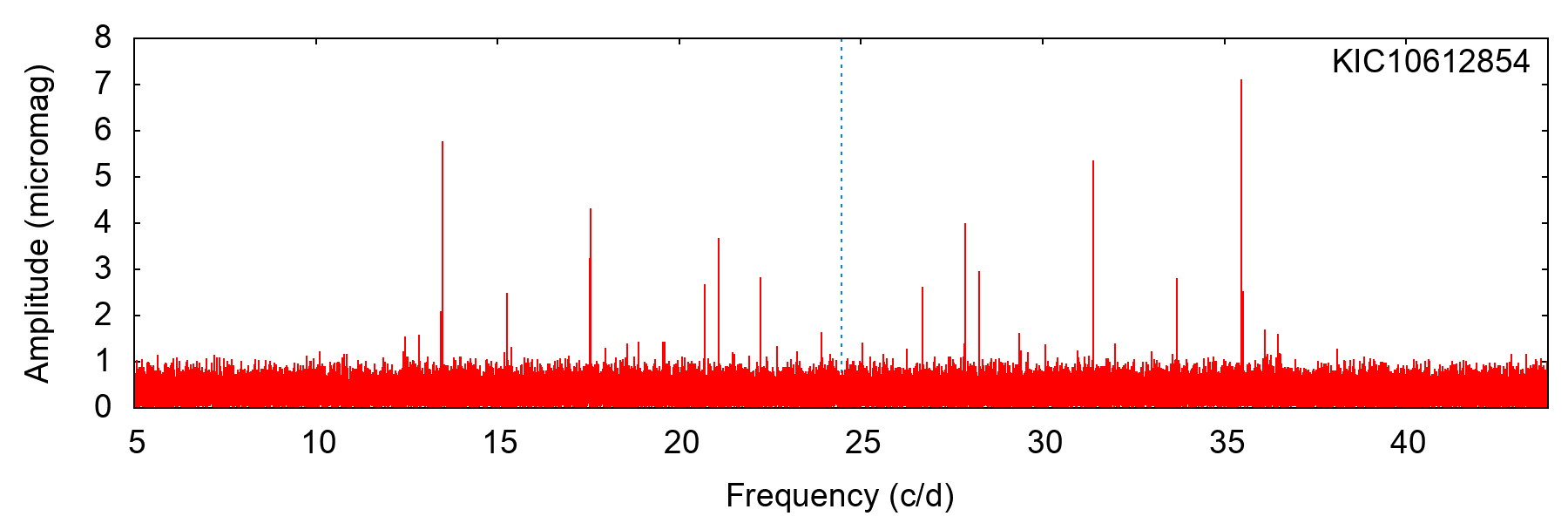}
\caption{Multiperiodic $\delta$\,Sct pulsators with pulsation amplitudes below 20\,$\upmu$mag in  \textit{Kepler} data. Both happen to have real oscillation frequencies above the Nyquist frequency (24.48\,d$^{-1}$, dashed blue line).}
\label{fig:low_amp}
\end{center}
\end{figure}

The second consequence is that \citet{murphyetal2018} automatically classified stars as non-pulsators if their strongest Fourier peak was below 20\,$\upmu$mag, because these peaks had insufficient SNR to detect binaries. Tens of examples of $\delta$\,Sct stars with amplitudes below 20\,$\upmu$mag exist in the \textit{Kepler} data; two are shown in Fig.\,\ref{fig:low_amp}. We reclassified these as pulsators, while noting that this becomes subjective at some level.

\begin{figure*}
\begin{center}
\includegraphics[width=0.32\textwidth]{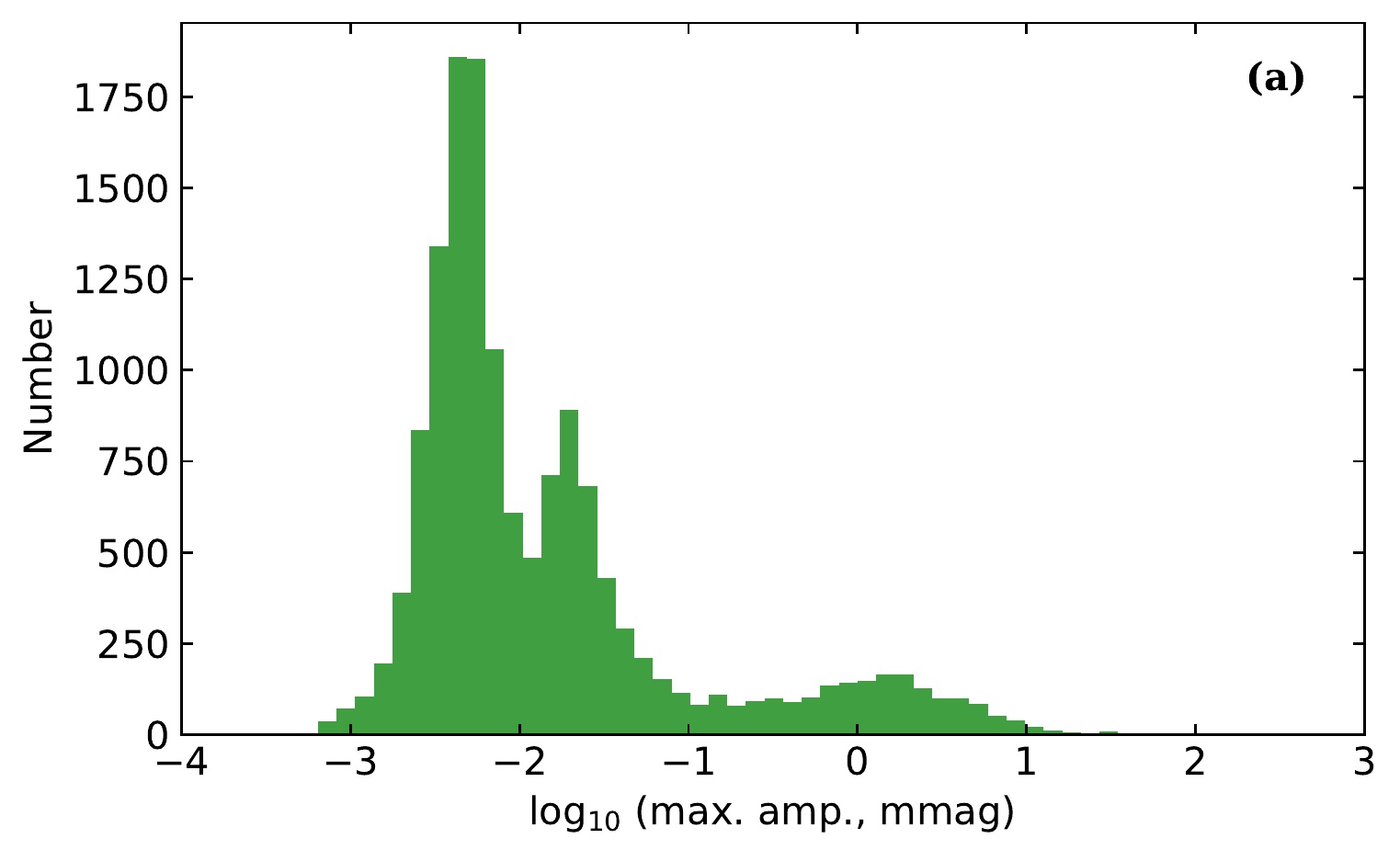} \includegraphics[width=0.32\textwidth]{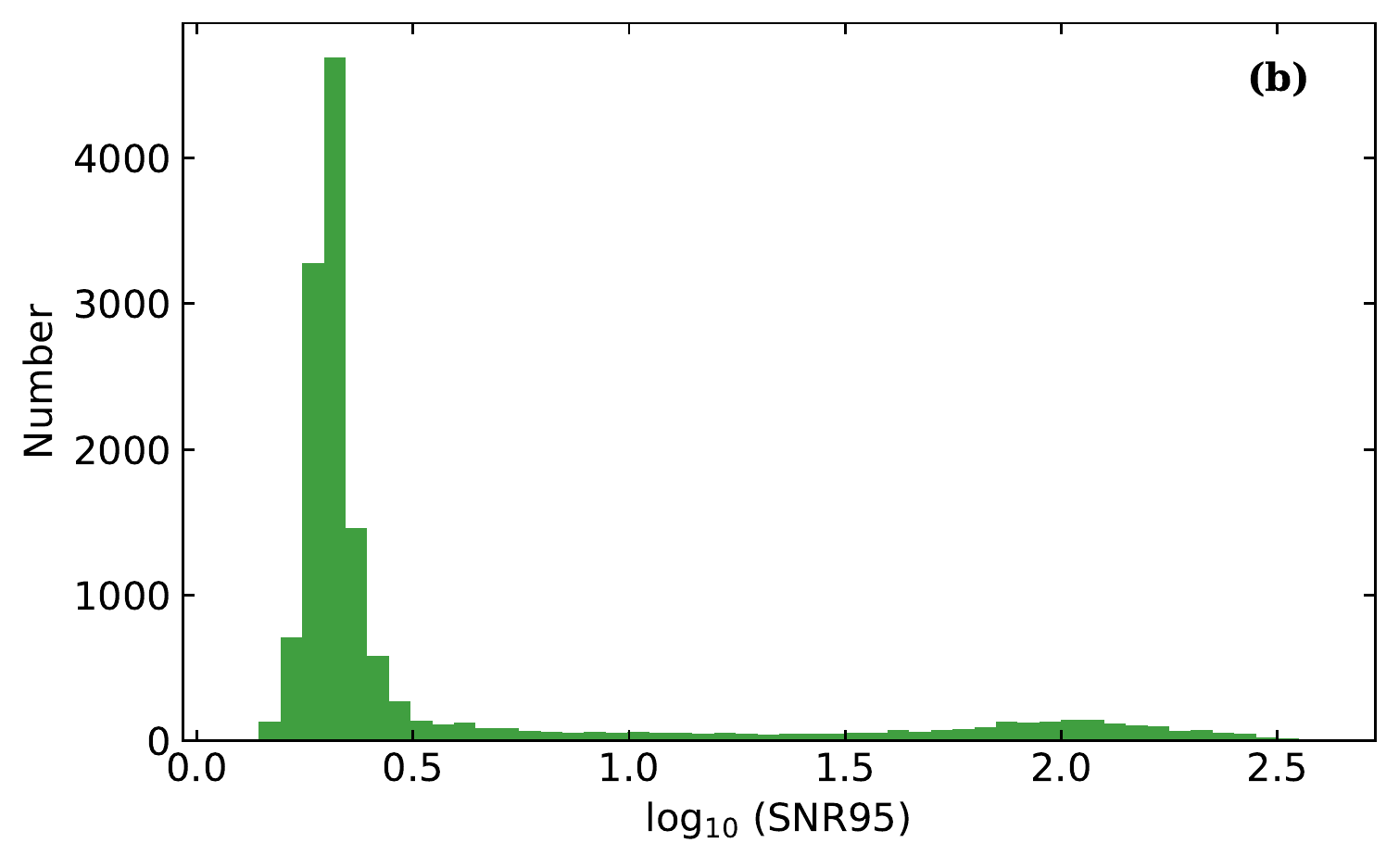} \includegraphics[width=0.32\textwidth]{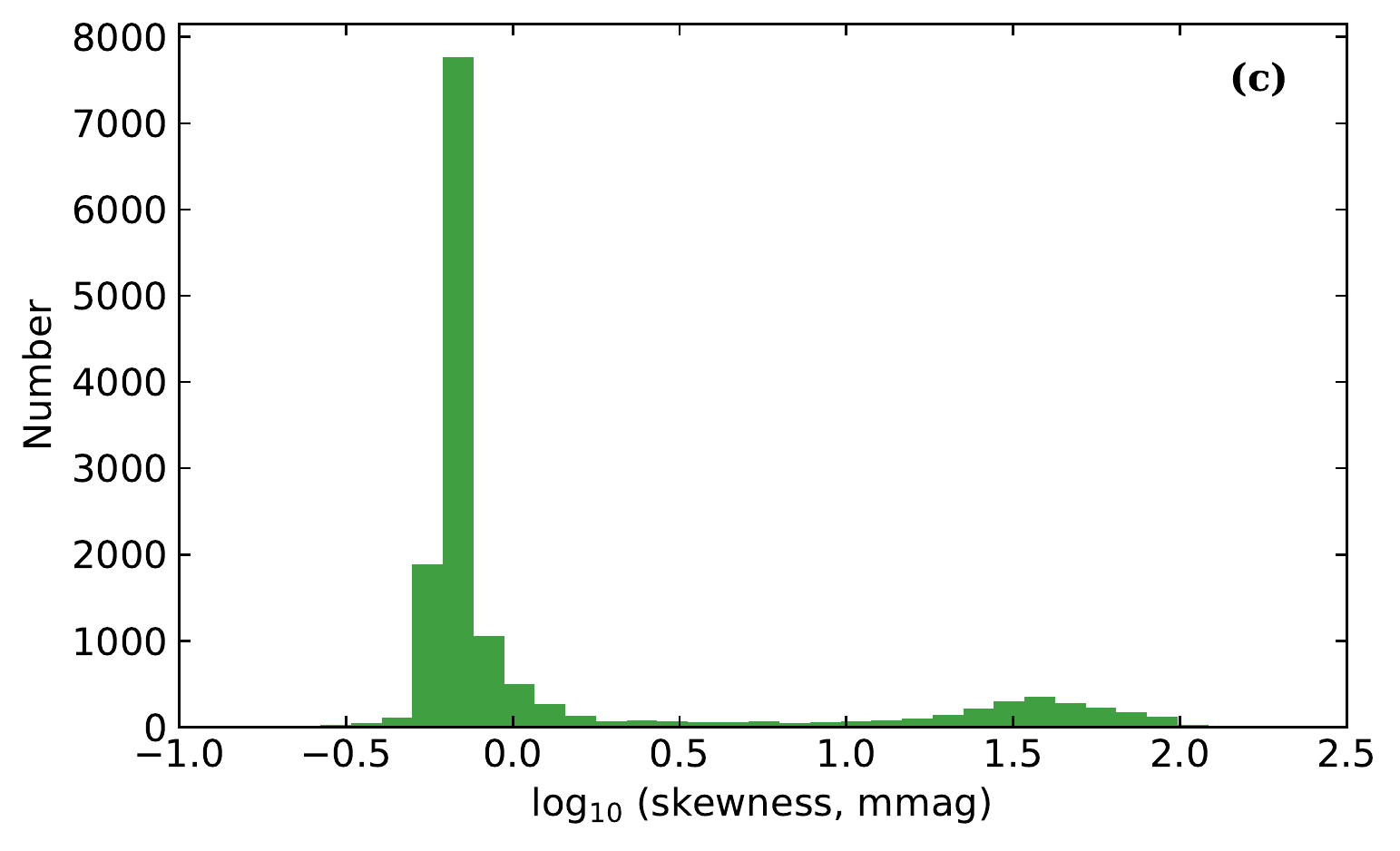}
\caption{{\bf(a)} The amplitude distribution of the strongest Fourier peak from each star. {\bf(b)} The distribution in signal-to-noise ratio (SNR), using the 95th percentile of the Fourier amplitudes to estimate the noise in a given star. {\bf(c)} The distribution of the skewness of each star's Fourier data. In each plot, higher-amplitude pulsators are found to the right and non-pulsators to the left.}
\label{fig:fou_statistics}
\end{center}
\end{figure*}

An obvious progression is to use an algorithmic amplitude threshold or SNR metric to classify pulsators. A simple amplitude threshold, say 10\,$\upmu$mag, does not work because it would classify thousands of noisy non-pulsators as $\delta$\,Sct stars. Fig.\,\ref{fig:fou_statistics}a shows the multi-modal distribution of the maximum Fourier amplitude of the 14\,330 stars. While the strong pulsators with maximum amplitudes of $\sim$1\,mmag stand out, there is also a large number of targets whose strongest peak is a few tens of $\upmu$mag, and then several thousand non-pulsators whose strongest peak is less than 10\,$\upmu$mag. Importantly, the latter two groups have some overlap, suggesting that a simple amplitude cut is not appropriate for selecting pulsators.

A SNR threshold using the strongest peak is therefore more appropriate, since the non-pulsators have very low SNR and are better distinguished from low-amplitude pulsators. Fig.\,\ref{fig:fou_statistics}b shows the distribution of SNR. For the noise estimate we used the 95th percentile of Fourier amplitudes, following \citet{murphy2014}. 
Traditionally in analyses of classical pulsators, peaks are considered significant if their SNR exceeds 4 \citep{bregeretal1993}, i.e. log SNR > 0.6, but this is usually applied to oscillations of borderline significance in multiperiodic pulsators. A more appropriate threshold for the strongest peak in the Fourier transform might be SNR$\sim$10. Fig.\,\ref{fig:fou_statistics}b shows the sample is well divided into `pulsators' and `non-pulsators' by such a definition, though a small tail does connect the two groups, so the exact threshold chosen does have some effect.

The pulsators and non-pulsators are even better distinguished by the skewness of their Fourier amplitude distributions. For this calculation, we took the Fourier transform of each stellar light curve and calculated the skewness of the amplitudes. High-amplitude pulsators have very high skewness whereas pure white noise has zero skewness. Under this metric, pulsators and non-pulsators are well separated with an even smaller tail connecting the two populations (Fig.\,\ref{fig:fou_statistics}c).

\begin{figure}
\begin{center}
\includegraphics[width=0.5\textwidth]{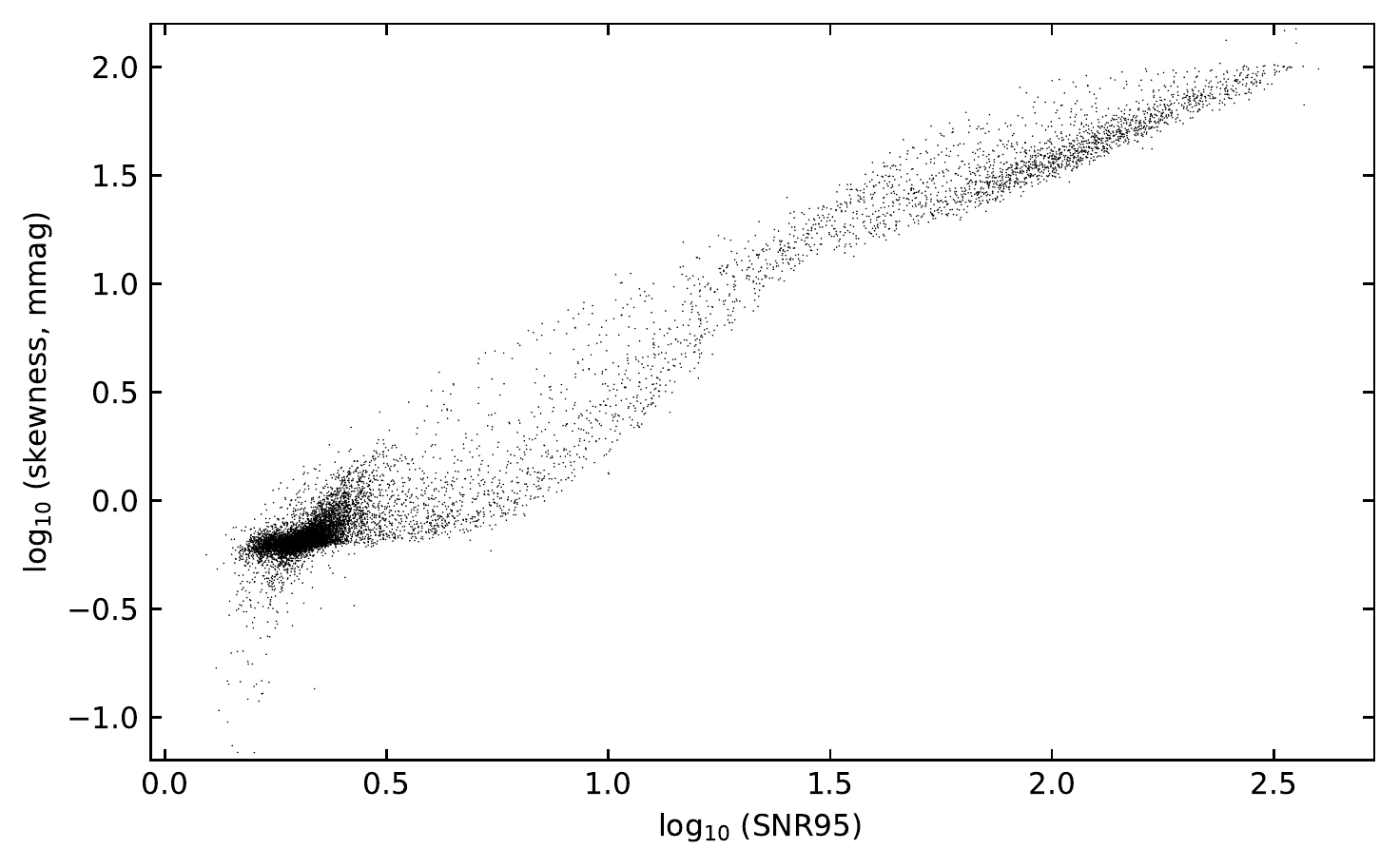}
\caption{Skewness versus SNR for the Fourier data of 14\,982 stars. Pulsators (upper-right) and non-pulsators (lower-left) are separated into two islands connected by a small stream of points.}
\label{fig:skewsnr}
\end{center}
\end{figure}

The skewness and SNR measurements can be combined to make a two-dimensional selection of pulsators (Fig.\,\ref{fig:skewsnr}). Pulsators and non-pulsators are mostly separated into two dense groups, but there is a continuous stream of points connecting them. Both the groups and the stream have structure. This can be used to distinguish mono-periodic pulsators from multiperiodic ones, to identify poor quality light curves, to find high-amplitude $\delta$\,Sct stars (HADS), or even to find new classes of variable stars that pulsate at similar frequencies. This is discussed in more detail in Appendix\,\ref{app:skewplot}. Here, we simply note that several different divisions could be made to isolate the pulsators based on these statistical properties, without making a specific recommendation.

The relationship between the skewness and SNR is quite linear, implying that either statistic alone has almost as much diagnostic power as the two used together. There is future potential in calculating different statistics on the light curve or Fourier transform and combining them to have greater utility, not necessarily limited to two dimensions or to just $\delta$\,Sct stars. This is left as future work.

\begin{figure}
\begin{center}
\includegraphics[width=0.5\textwidth]{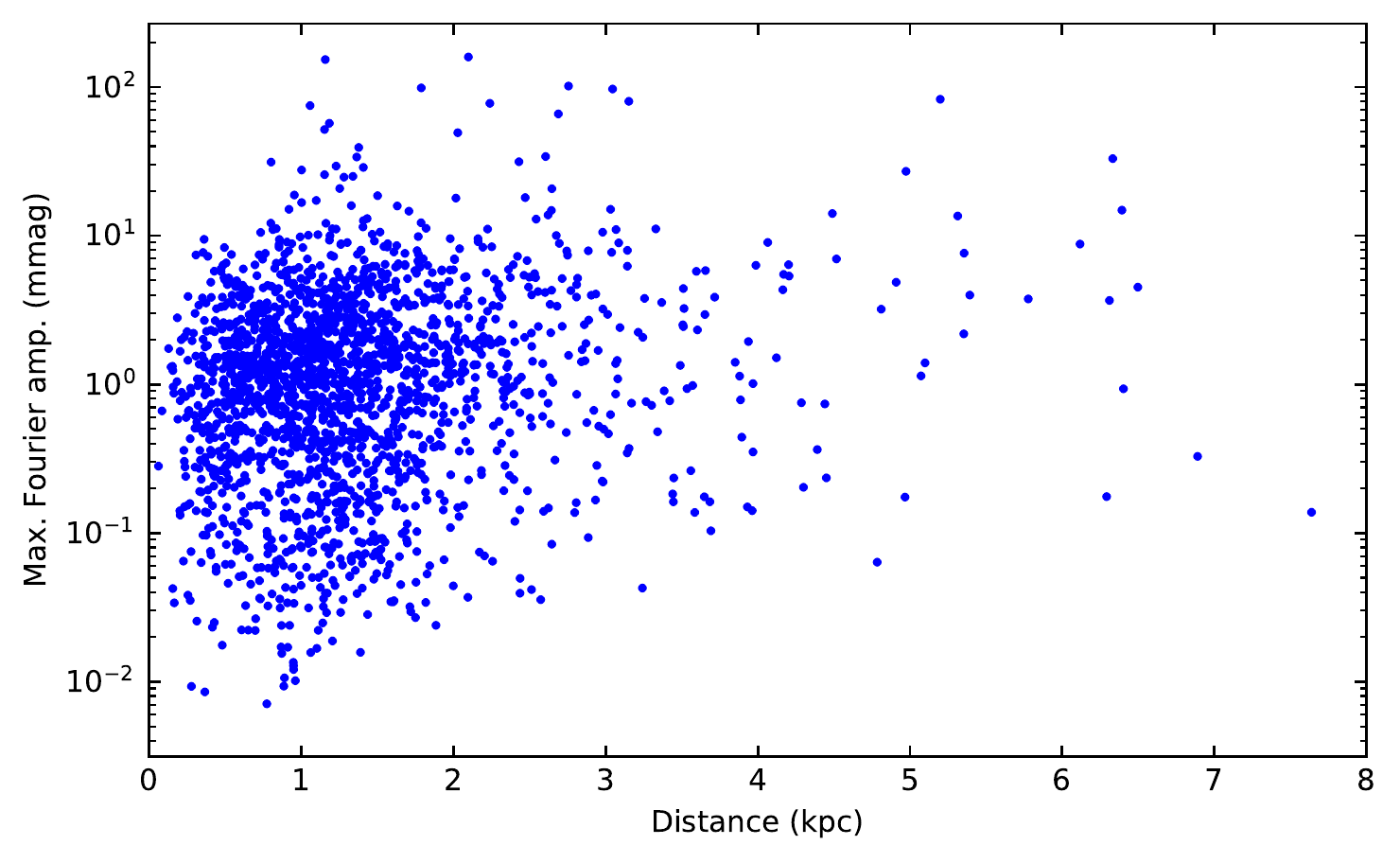}
\caption{Pulsation amplitude as a function of distance for the 1988 $\delta$\,Sct stars in the sample.}
\label{fig:distance_amplitude}
\end{center}
\end{figure}

We use the `revised' classifications in the rest of this paper, rather than an algorithmic classification, and we show in Sec.\,\ref{sec:frac_temp} that the choice is unimportant. This is because the pulsators and non-pulsators are densely grouped in Fig.\,\ref{fig:skewsnr} and the stream between them is sparsely populated, meaning that objects in the stream make up only a small fraction of the total, while the difference between the `manual' and `revised' classifications has a much bigger effect. There were 1988 $\delta$\,Sct stars in the `revised' classifications and 207 stars with other variability at $>5$\,d$^{-1}$ (mostly $\gamma$\,Dor stars with harmonics), which were added to the 12\,135 non-pulsators to form our `non-dSct' sample. The distribution of pulsation amplitude as a function of distance is shown in Fig.\,\ref{fig:distance_amplitude}.

\section{The H--R diagram for A/F stars}
\label{sec:HRD}

\subsection{Distribution of the stars}

\begin{figure}
\begin{center}
\includegraphics[width=0.5\textwidth]{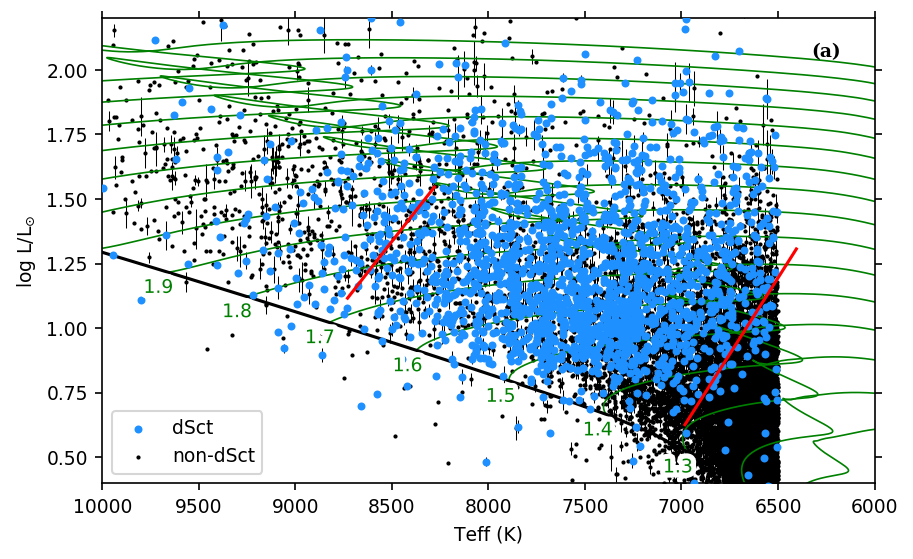}
\includegraphics[width=0.5\textwidth]{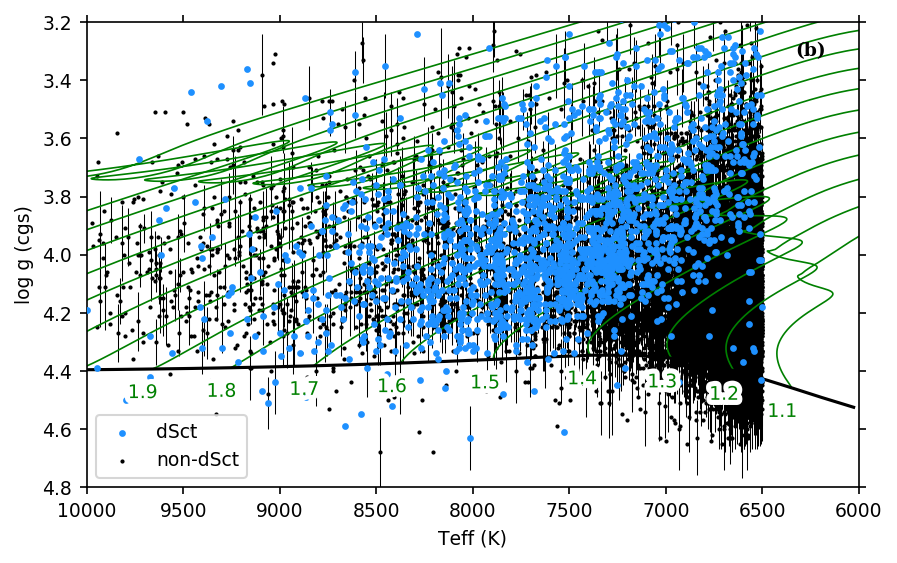}
\includegraphics[width=0.5\textwidth]{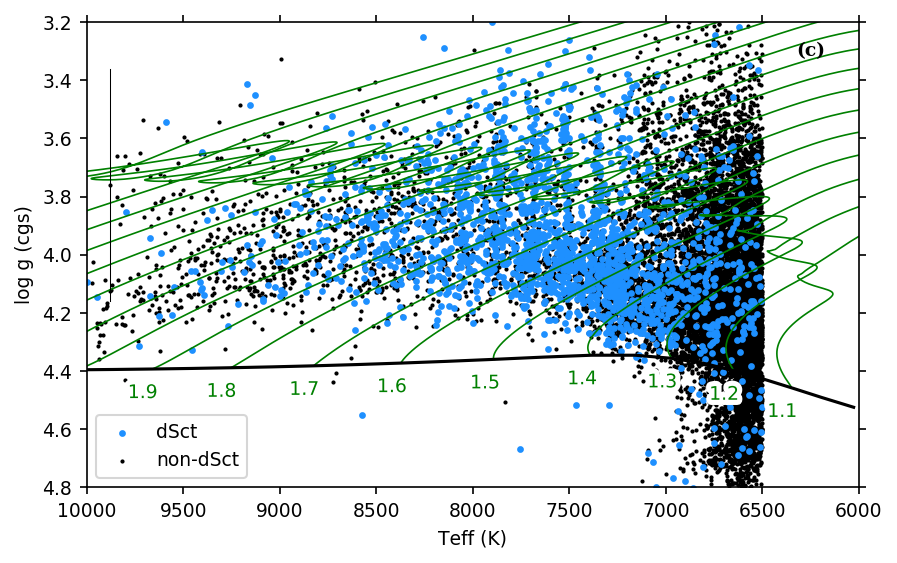}
\caption{The $\delta$\,Sct pulsators and non-pulsators across the H--R diagram {\bf (a)}, and the $T_{\rm eff}$--$\log g$ diagram {\bf (b,c)} using our recomputed $\log g$ and the $\log g$ from the revised stellar properties catalogue, respectively. Uncertainties on $\log L$ and $\log g$ are shown for every 5th point in {\bf (a)} and {\bf (b)}, respectively, but for clarity in {\bf (c)} the typical 0.4\,dex uncertainty in $\log g$ is shown for just one point. Evolutionary tracks are shown in green at intervals of 0.1\,M$_{\odot}$, with their corresponding masses written beneath the ZAMS (black line). The $\delta$\,Sct instability strip for radial modes with $n \leq 7$ and $\alpha_{\rm MLT} = 1.8$ from \citet{dupretetal2005b} is shown as solid red lines.}
\label{fig:hrd}
\end{center}
\end{figure}

Our sample spans a wide range of $T_{\rm eff}$ and $L$, with pulsators found throughout this range. Fig.\,\ref{fig:hrd} shows that pulsators are found well outside the theoretical instability strip. The instability strip shown is from \citet{dupretetal2005b}, for radial modes with radial orders between $n=1$ and $7$, giving the broadest instability strip depicted there. Their study used evolutionary models with time-dependent convection with $\alpha_{\rm MLT} = 1.8$. As we wrote in Sec.\,\ref{sec:intro}, the boundaries depend strongly on the value of $\alpha_{\rm MLT}$, and 1.8 is the value used for the Sun \citep{dupretetal2004}. Space-based observations clearly show that, despite the blue-edge being theoretically well-determined \citep{balona&dziembowski2011}, no clear observational blue edge is apparent. This led \citet{bowman&kurtz2018} to suggest that $\alpha_{\rm MLT}$ should not be treated as a fixed quantity, but should vary across the instability strip. A new investigation of appropriate values of $\alpha_{\rm MLT}$ is clearly warranted.

\subsection{Calculation of evolutionary tracks}

In order to meaningfully place our objects on an H--R diagram and study the pulsator fraction with reference against the stellar masses and the ZAMS/TAMS boundaries, we computed some stellar models. Our evolutionary tracks were computed in {\sc MESA} \citep{paxtonetal2011}, v10108. Our `standard model' at each mass had $X = 0.71$, $Z=0.010$ (corresponding to [M/H] $=-0.11$).\footnote{We calculated [M/H] from Z using the photospheric solar abundance, $Z_{\odot} = 0.0134$ with $X_{\odot} = 0.7381$.}  We used a metallicity below solar because the median [Fe/H] of our sample is $-0.139$\,dex, according to the \citet{mathuretal2017} stellar properties catalogue. This is corroborated by spectroscopy of over 100 chemically-normal \textit{Kepler} A stars \citep{niemczuraetal2015,niemczuraetal2017}, where a mean [Fe/H] of $-0.1$\,dex was found. The other parameters of our standard model are $\alpha_{\rm MLT} = 1.8$, exponential core overshooting of 0.015\,Hp, exponential H-burning shell over- and under-shooting of 0.015\,Hp, exponential envelope overshooting of 0.025\,Hp, diffusive mixing $\log D_{\rm mix} = 0$ (in cm$^{2}$s$^{-1}$), OPAL opacities, and the \citet{asplundetal2009} solar abundance mixture. These parameters were chosen as `reasonable' for our targets; a full investigation of the influence of each parameter is beyond the scope of this paper. Interested readers are directed to \citet{aertsetal2018} and references therein. The tracks were computed for the mass range 1.0--3.0\,M$_{\odot}$, at intervals of 0.05\,M$_{\odot}$, and include $T_{\rm eff}$, $\log L$, $\log g$ and age at 600 steps. We make them available as supplementary online material, along with tracks computed with the same physics and a solar metallicity ($Z = 0.014$, corresponding to [M/H] = 0.036). We also provide an example {\sc MESA} inlist at \url{mesastar.org}.

\begin{figure}
\begin{center}
\includegraphics[width=0.5\textwidth]{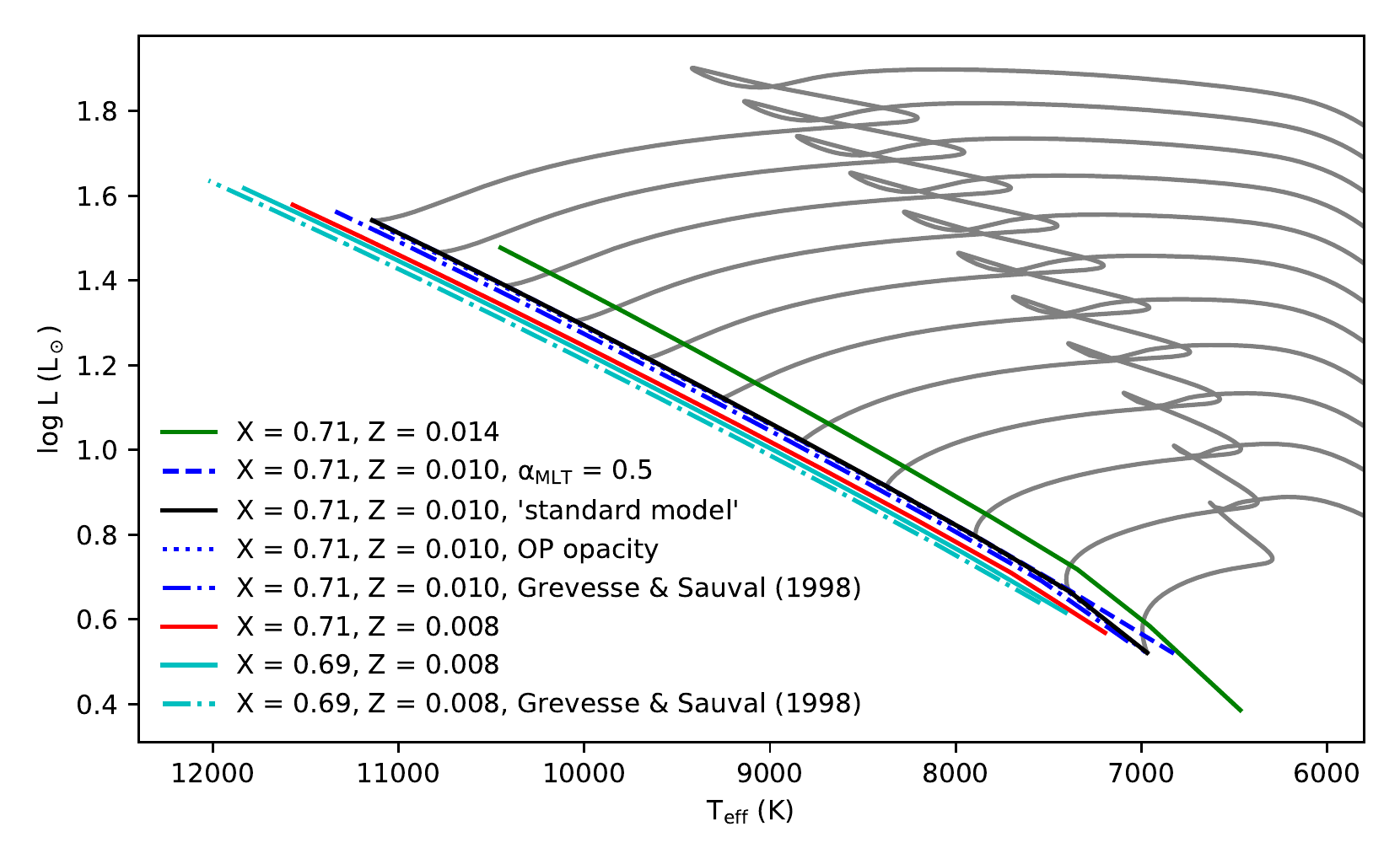}
\caption{Evolutionary tracks (grey) in the mass range 1.3--2.3\,M$_{\odot}$ for our `standard model' with $X = 0.71$, $Z=0.010$, $\alpha_{\rm MLT} = 1.8$, OPAL opacities, and the \citet{asplundetal2009} solar abundance mixture, resulting in the ZAMS shown in black. 
The effect on the ZAMS of changing the input physics is illustrated, with the other ZAMS lines all spanning the same mass range.}
\label{fig:zams}
\end{center}
\end{figure}

The choice of input parameters influences the position of the ZAMS and the duration of the main-sequence phase, and thus affects whether our targets are on the main sequence. Since the \textit{Kepler} field spans 5--22$^{\circ}$ in galactic latitude \citep{vancleve&caldwell2016}, there should be few young (i.e. zero-age) massive stars in our sample, except those rejuvenated by binary mass transfer (e.g. \citealt{stepienetal2017,streameretal2018}). In Fig.\,\ref{fig:hrd} this deficit is seen at a low level in $\log L$ and our derived $\log g$ (described below), but is exaggerated in the $\log g$ values determined from KIC photometry. Such comparisons highlight the importance of using stellar models that accurately place the ZAMS. The location of the ZAMS is mostly sensitive to the $X, Y, Z$ fractions, as shown in Fig.\,\ref{fig:zams}.

Our tracks are in good agreement with the MIST tracks computed with no rotation and similar metallicities ([Fe/H] $= -0.1, 0.0$) from \citet{dotter2016} and \citet{choietal2016}, except that we find a longer main-sequence phase. It is important to acknowledge that rotation affects the observed temperature and luminosity of a star, depending on the rotation rate and the inclination angle \citep{gray&corbally2009}, neither of which are known for our stars. Rotation also induces extra mixing, with similar effects to convective core overshooting \citep{jermynetal2018}, but modelling this is beyond the scope of this work.

\subsection{Estimation of stellar parameters}

\begin{table*}
\centering
\caption{Stellar data. The full table is available online.}
\begin{tabular}{rHcr@{\,$\pm$\,}lr@{\,$\pm$\,}lr@{\,$\pm$\,}lr@{\,$\pm$\,}lr@{\,$\pm$\,}lr@{\,$\pm$\,}lr@{\,$\pm$\,}lcr@{\,}lc}
\toprule
\multicolumn{1}{c}{KIC} & DR2 & \hspace{-1mm}$g_{\rm SDSS}$\hspace{-1mm} & \multicolumn{2}{c}{\hspace{-1mm}$\pi$\hspace{-1mm}} & \multicolumn{2}{c}{\hspace{-1mm}$A_g$\hspace{-1mm}} & \multicolumn{2}{c}{\hspace{-1mm}$T_{\rm eff}$\hspace{-1mm}} & \multicolumn{2}{c}{[Fe/H]\hspace{-1mm}} & \multicolumn{2}{c}{\hspace{-1mm}$\log g$\hspace{-1mm}} & \multicolumn{2}{c}{\hspace{-1mm}Radius\hspace{-1mm}} & \multicolumn{2}{c}{\hspace{-1mm}Mass\hspace{-1mm}} & \hspace{-2mm}qflag\hspace{-2mm} & \multicolumn{2}{c}{\hspace{-1mm}$\log L/ L_{\odot}$\hspace{-1mm}} & \hspace{-2mm}$\delta$\,Sct\hspace{-2mm} \\
 &  & mag & \multicolumn{2}{c}{mas} & \multicolumn{2}{c}{mag} & \multicolumn{2}{c}{K} &  \multicolumn{2}{c}{} & \multicolumn{2}{c}{} & \multicolumn{2}{c}{R$_{\odot}$} & \multicolumn{2}{c}{M$_{\odot}$} &  &  &  & \\
\midrule
\vspace{1.5mm}
10000009 & 2130602090064518272 & 11.810 & 1.654 & 0.023 & 0.11 & 0.03 & 6850 & 197 & $-$0.15 & 0.3 & 4.06 & 0.10 & 1.81 & 0.11 & 1.38 & 0.25 & 0 & 0.812 & $^{+0.020}_{-0.020}$ & 0 \\
\vspace{1.5mm}
10000135 & 2130601780826873216 & 12.581 & 1.953 & 0.021 & 0.10 & 0.02 & 6758 & 216 & $-$0.10 & 0.3 & 4.42 & 0.11 & 1.11 & 0.07 & 1.18 & 0.25 & 0 &0.365 & $^{+0.019}_{-0.019}$ & 0 \\
\vspace{1.5mm}
10000823 & 2130644940957601152 & 15.708 & 0.687 & 0.081 & 0.19 & 0.02 & 6578 & 303 & $-$0.17 & 0.3 & 4.56 & 0.19 & 0.87 & 0.15 & 1.03 & 0.25 & 5 & 0.110 & $^{+0.115}_{-0.103}$ & 0 \\
\vspace{1.5mm}
10001045 & 2130633327366078592 & 13.943 & 0.798 & 0.014 & 0.17 & 0.01 & 6593 & 200 & $-$0.44 & 0.3 & 4.11 & 0.11 & 1.62 & 0.11 & 1.26 & 0.25 & 0 & 0.649 & $^{+0.022}_{-0.022}$ & 0 \\
\vspace{1.5mm}
10001145 & 2130638305228782464 & 11.675 & 0.967 & 0.022 & 0.15 & 0.01 & 7864 & 275 & $-$0.33 & 0.3 & 3.92 & 0.09 & 2.44 & 0.19 & 1.82 & 0.25 & 0 &1.312 & $^{+0.026}_{-0.026}$ & 1 \\
\bottomrule
\end{tabular}
\label{tab:stellar}
\end{table*}

We derived mass estimates for the stars in our sample by evaluating their positions in the H--R diagram against our tracks. First, the theoretical $T_{\rm eff,th}$ and $\log L_{\rm th}$ values of the evolution tracks were interpolated onto a refined grid of stellar mass and age. The interpolated mass values $\mathrm{M}_*$ varied from 1 to 3\,M$_{\odot}$ with a step of $0.005\,\mathrm{M}_\odot$. The stellar ages along each track were linearly mapped onto a common age scale $a_r$, whereby the ZAMS and the TAMS correspond to $a_r$ values of 1 and 2, respectively. Subsequently, we estimated the mass $\mathrm{M}_*$ for each star with observed $T_{\rm eff,obs}\pm\sigma_{T_{\rm eff}}$ and $\log\,L_{\rm obs}\pm\sigma_{\log\,L}$ values by taking the mode of the probability distribution
\begin{eqnarray}
 \mathcal{P}\left(\mathrm{M}_*\right) = \sum_{a_{r,i}} W\left(\mathrm{M}_*,a_{r,i}\right) 
\times \exp\left[-\frac{1}{2}\left(\left(\frac{T_{\rm eff,th} - T_{\rm eff,obs}}{\sigma_{T_{\rm eff}}}\right)^2 \right.\right. \quad\quad \notag\\
+ \left.\left. \left(\frac{\log\,L_{\rm th} - \log\,L_{\rm obs}}{\sigma_{\log\,L}}\right)^2\right)\right],
\end{eqnarray}
where $W\left(\mathrm{M}_*,a_{r,i}\right)$ is a normalising weight factor that scales with the relative duration of the stellar evolution at each time step. Stars between 0 and 1\,mag (0 and 0.4\,dex in $\log L$) below the ZAMS were assigned a different quality flag, as their masses are more uncertain. We did not take metallicities of individual stars into account when computing masses, hence the metallicity spread of our sample dominates the uncertainties. We therefore emphasize that these masses are valid only in a statistical sense. An accurate stellar mass for any individual target should first precisely measure and then incorporate the metallicity into the model, alongside asteroseismic information if available. The 1$\sigma$ uncertainties in [Fe/H], $T_{\rm eff}$ and $\log L$ contribute 0.2, 0.1, and 0.05\,M$_{\odot}$ to the uncertainty in mass. To this, we added another 0.1\,M$_{\odot}$ for uncertain model physics, accounting for the unknown overshooting, mixing length and similar parameters \citep[e.g.][]{claret&torres2018}. Of these, overshooting dominates and the value 0.1\,M$_{\odot}$ was determined empirically by studying the change in position of the TAMS resulting from a change in overshooting of 0.015\,Hp. The four contributions were combined in quadrature, giving $\sigma M = 0.25$\,M$_{\odot}$ for each star. We also calculated stellar radii
\begin{eqnarray}
R / R_{\odot}  = \sqrt{\frac{L/L_{\odot}}{(T_{\rm eff}/T_{\rm eff,\odot})^4}}
\end{eqnarray}
and surface gravities
\begin{eqnarray}
\log g~{\rm (cgs)}   = 2 + \log \frac{G M}{R^2},
\end{eqnarray}
where the stellar masses and radii are given in SI units. We used the IAU-recommended nominal solar conversion constants \citep{mamajeketal2015c}. Uncertainties on these parameters were determined by a Monte-Carlo process, and are given in Table\:\ref{tab:stellar}.


\section{The $\delta$\,Sct pulsator fraction across the HR Diagram}
\label{sec:fraction}

When comparing theoretical calculations of the instability strip with the observed population of $\delta$\,Sct stars, it is important to consider the sample from which the latter are drawn. Looking at the distribution of $\delta$\,Sct stars alone (Fig.\,\ref{fig:hrd}a), it appears that the instability strips from \citet{dupretetal2005b} are a good fit to the observed pulsators. However, at the lower right the pulsators are drawn from a sample of thousands of stars, whereas in the upper left they are drawn from a population of tens. Hence, it is necessary to consider the {\it pulsator fraction} as the observable quantity when attempting to model the instability strip. We investigate this quantity here.

\subsection{As a function of effective temperature}
\label{sec:frac_temp}

\begin{figure}
\begin{center}
\includegraphics[width=0.5\textwidth]{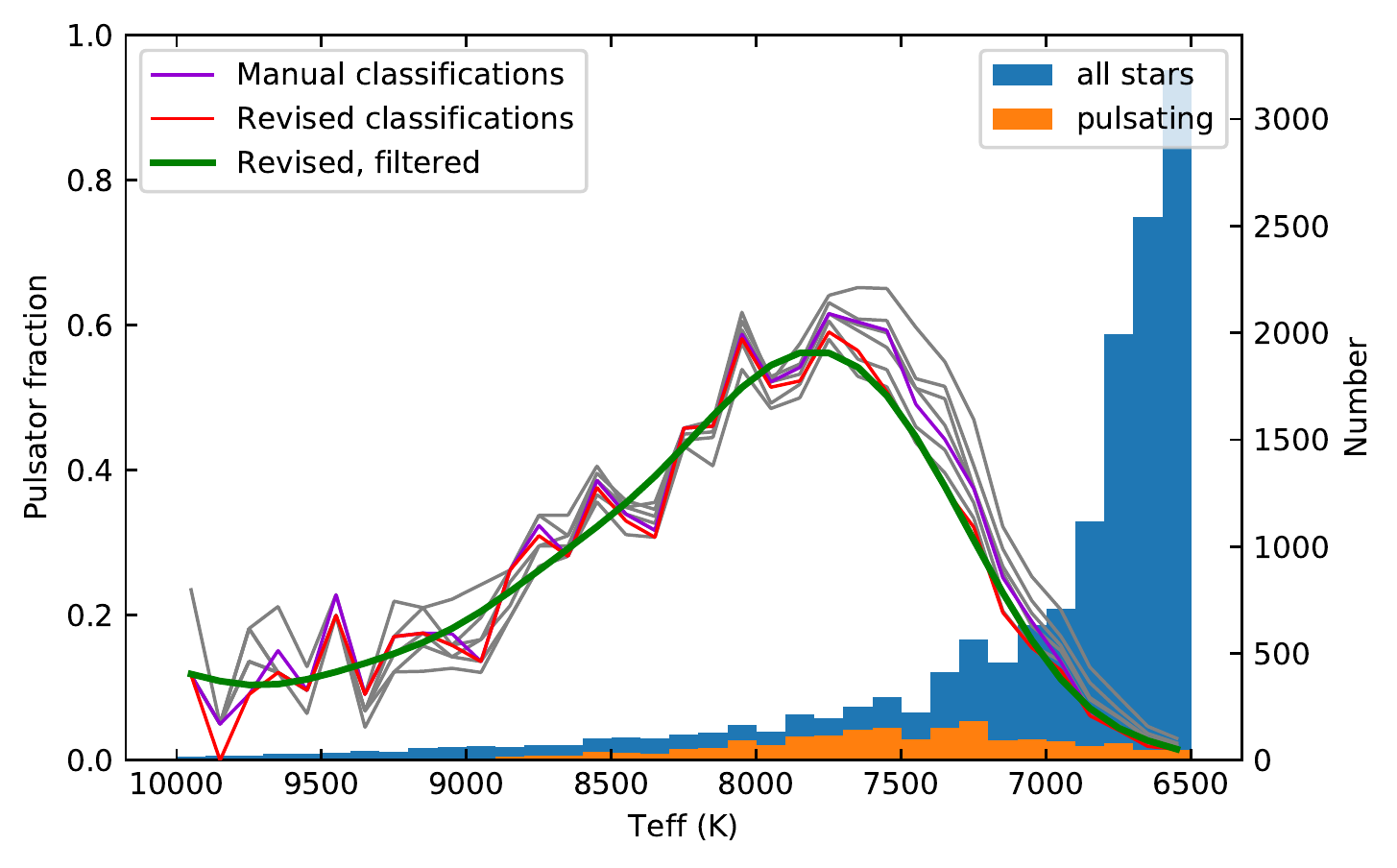}
\caption{The $\delta$\,Sct pulsator fraction (left axis) as a function of $T_{\rm eff}$, with the number of targets in each histogram bin on the right y-axis. Grey lines show the pulsator fraction according to various algorithmic divisions of pulsators and non-pulsators, for which details are given in the text, while the violet line shows the pulsator fraction using the `manual' classifications. It is important that the grey and violet lines are in agreement. However, not all of those stars are actually $\delta$\,Sct stars, since some $\gamma$\,Dor stars have harmonics that spill into the $>$5\,d$^{-1}$ region (Sec.\,\ref{ssec:pulsation_classification}). The red curve uses the `revised' classifications, obtained by reclassifying those stars as non-$\delta$\,Sct stars. The green curve is a filtered version of the red one, describing the $\delta$\,Sct pulsator fraction as a function of temperature.}
\label{fig:frac_teff}
\end{center}
\end{figure}

It is evident from Fig.\,\ref{fig:hrd} that not all stars inside the instability strip are pulsating, especially towards the red edge. A histogram of the pulsator fraction as a function of temperature (Fig.\,\ref{fig:frac_teff}) shows that it only reaches 60\:per\:cent, even in the middle of the instability strip. At 7200\,K, still inside the instability strip and 200--500\,K from the cool edge, the pulsator fraction is as low as 25\:per\:cent. Clearly, some mechanism is inhibiting the oscillations in many of these A stars. Meanwhile, the pulsator fraction drops only a little to $\sim$40\:per\:cent at the blue edge, and is still at 20\:per\:cent at temperatures several hundred kelvin beyond the blue edge. The observations confirm that our understanding of the driving and damping mechanisms is incomplete.

These results do not depend on which method we used to identify the $\delta$\,Sct stars (see Sec.\,\ref{ssec:pulsation_classification}). In addition to the manual classifications, pulsator fractions from a further six classification algorithms are presented in Fig.\,\ref{fig:frac_teff} as the grey lines, all giving similar results. These are based empirically on the skewness--SNR diagram (Fig.\,\ref{fig:skewsnr}) as follows:
\begin{itemize}
\item log (skewness) > 1.1
\item log (skewness) > 0.6
\item log (skewness) > 0.3
\item log (SNR95) > 1.0
\item log (SNRmed) > 1.0
\item log (skewness) > $-$3 log (SNR95) + 4.5.
\end{itemize}
The SNRmed measurement uses the median Fourier amplitude instead of the 95th percentile as an alternative noise characterisation. The sixth division is based on both skewness and SNR. These automated classifications do not distinguish between $\delta$\,Sct stars and $\gamma$\,Dor stars with harmonics (see Sec.\,\ref{ssec:pulsation_classification}). Selecting only the $\delta$\,Sct stars (i.e. using the `revised' classifications), we arrived at the pulsator fraction in red in Fig.\,\ref{fig:frac_teff}. This curve is slightly noisy because of the fine $T_{\rm eff}$ resolution used, so a low-pass filter is applied to give a smoother representation (green).

How can we use these observations to redefine the observational instability strip as a function of effective temperature? We suggest adopting the region in which $>$20\:per\:cent of the stars pulsate. In that case, the instability strip is located between 7100 and 9000\,K, but remains a function of luminosity. This definition shifts the instability strip to hotter values than those currently accepted, but the observations clearly warrant such a shift. A major effort to improve our understanding of driving and damping mechanisms is required to understand the mismatch between the observed and theoretical instability strips.

\subsection{In two dimensions}
\label{ssec:twod}

The instability strip boundaries are primarily a function of $T_{\rm eff}$ but there is also strong dependence on the $y$-coordinate \citep{xiongetal2016}. This has previously seen little observational investigation due to the difficulty in gathering accurate $\log g$ or $\log L$ values for a large sample. While the KIC contains $\log g$ values for approximately 85\:per\:cent of our sample, the uncertainties are typically $\pm$0.4\,dex. Subsequent revised stellar properties catalogues \citep{huberetal2014,mathuretal2017} provide $\log g$ for more stars but still only 29\:per\:cent of the sample have uncertainties $<0.4$\,dex. For reference, the main sequence extends from 4.3 to 3.5 in $\log g$, so the uncertainties are between one half and one quarter of the main-sequence range in each direction.

\begin{figure*}
\begin{center}
\includegraphics[width=0.5\textwidth]{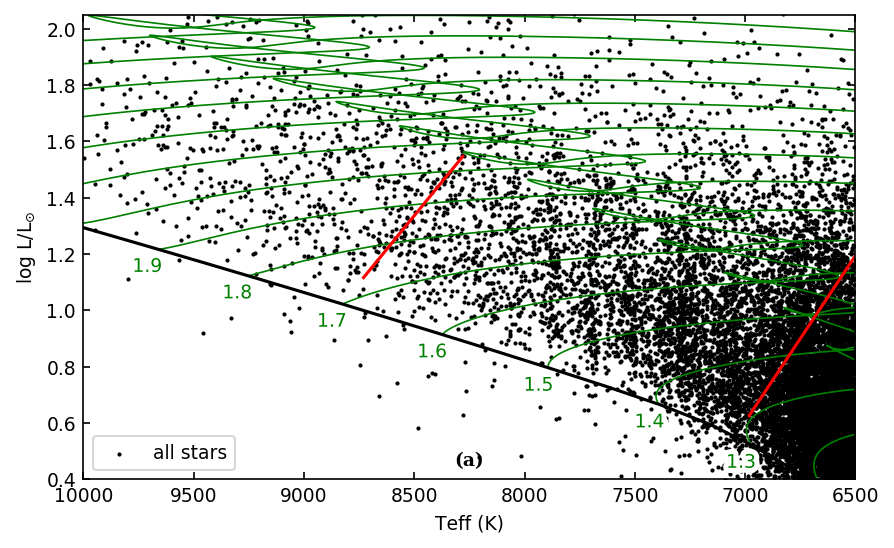}\includegraphics[width=0.5\textwidth]{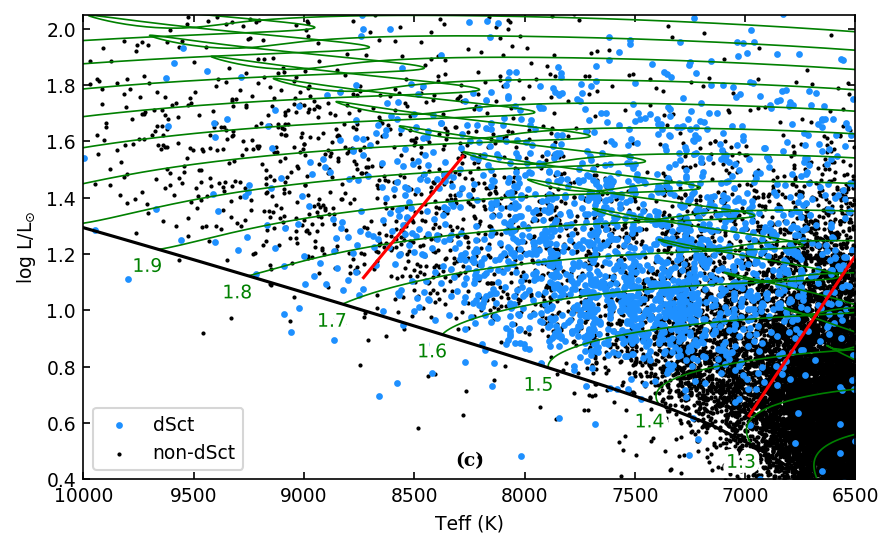}\\
\includegraphics[width=0.5\textwidth]{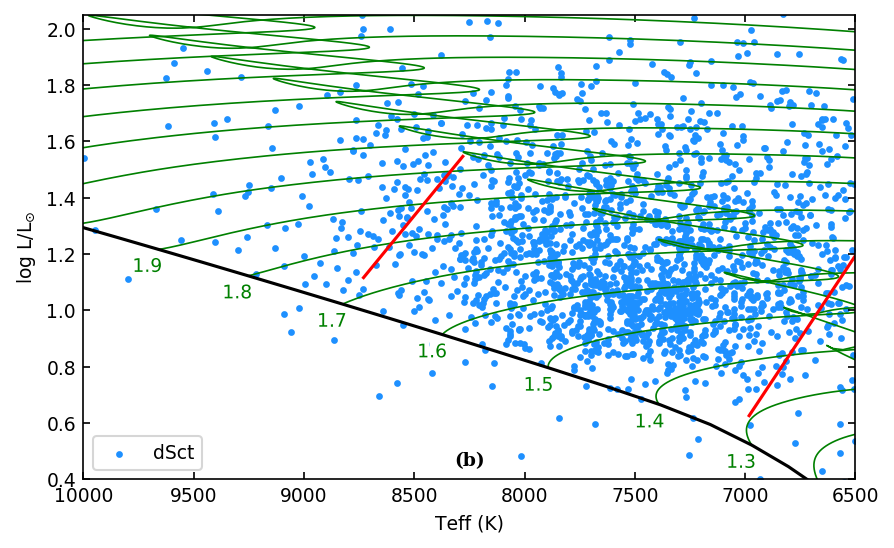}\includegraphics[width=0.5\textwidth]{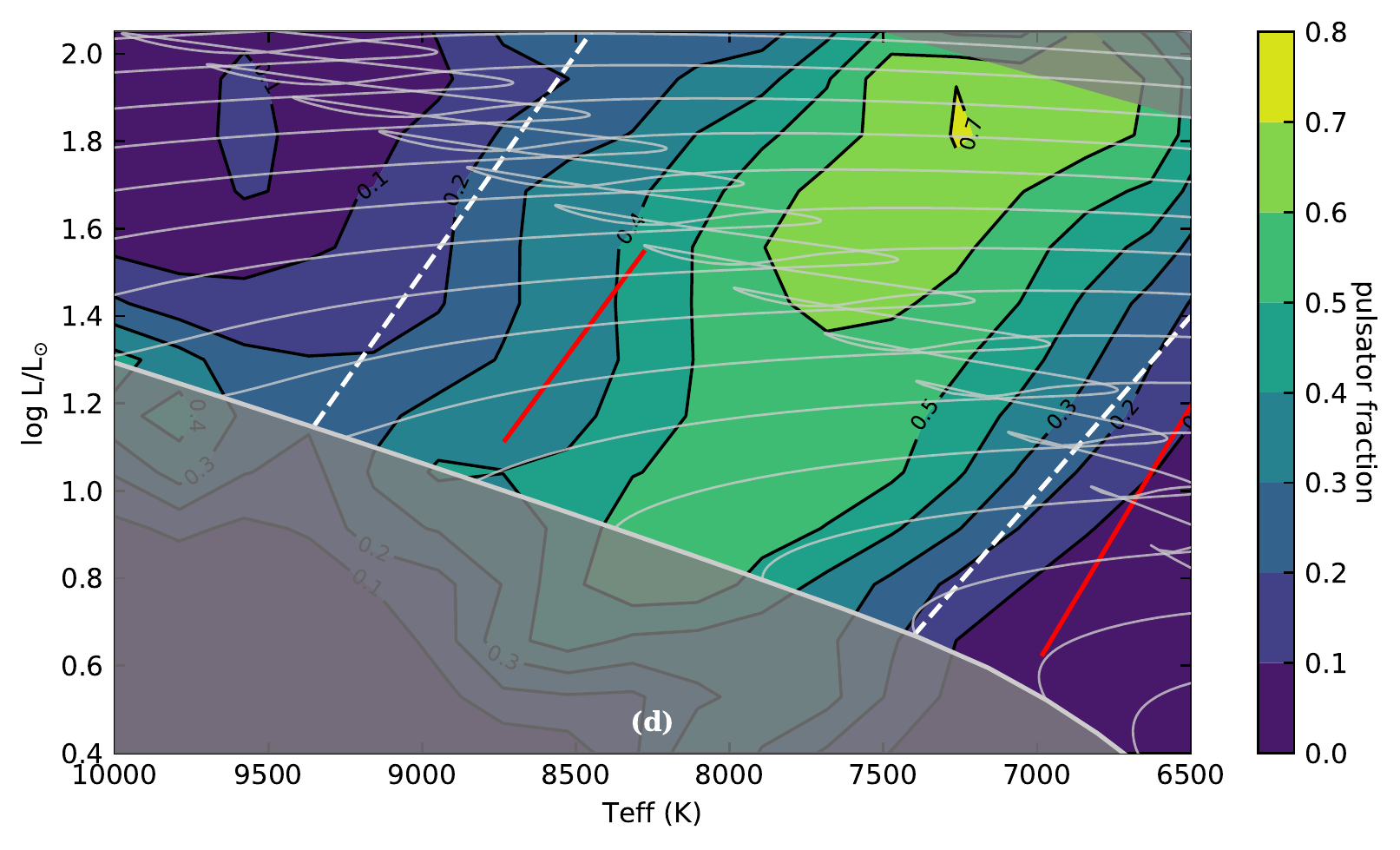}
\caption{The target stars on the H--R diagram, with evolutionary tracks (green) and the instability strip (red) as in Fig.\,\ref{fig:hrd}. Panels show: {\bf (a)} All targets. {\bf (b)} Pulsators only. {\bf (c)} All targets classified into $\delta$\,Sct and non-$\delta$\,Sct stars. {\bf (d)} Contour plot illustrating the $\delta$\,Sct pulsator fraction. Theoretical instability strip boundaries (red) from \citet{dupretetal2005b} clearly do not represent the observed pulsator fraction and we suggest new observed boundaries as dashed white lines, bracketing the region where $>$20\:per\:cent of stars pulsate. Greyed-out areas have too few stars for the fraction to be computed reliably.}
\label{fig:contour}
\end{center}
\end{figure*}

Luminosities derived from {\it Gaia} DR2 parallaxes offer significant improvement. The main sequence spans a range of $\sim$0.9 in $\log L$ and the uncertainties are $\sim$0.03 (refer back to Fig.\,\ref{fig:lum_err}), leading to 10--15 times better precision. The pulsator fraction can now be explored in two dimensions, as shown in Fig.\,\ref{fig:contour}. We computed a contour plot of the pulsator fraction by dividing the A/F star region into a grid with cells of width 200\,K and height 0.12\,dex ($\log L$). We calculated the pulsator fraction in each cell and applied a Gaussian smoothing to the results. The difference between the observed and theoretical instability strips is striking. Although $\delta$\,Sct stars can exist at temperatures cooler than the red edge (Fig.\,\ref{fig:contour}b), the number of stars at these temperatures is large (Fig.\,\ref{fig:contour}a) and so the pulsator fraction is low (Fig.\,\ref{fig:contour}d). On the other hand, the fraction of $\delta$\,Sct pulsators beyond the theoretical blue edge remains high over a range of luminosities.
\clearpage

We can measure instability strip boundaries based on these observations of the pulsator fraction. Specifically, we define straight lines that follow the 20\:per\:cent threshold for pulsator fraction. Although the sample size is much larger than any previously available sample, the contours near the blue edge do suffer from small-number statistics, whereas the red edge is much better defined. The resulting instability strip edges are:
\begin{description}
\item red edge: $\phantom{l}\log L = -0.000811~T_{\rm eff}/{\rm K} + 6.672$
\item blue edge: $\log L = -0.001000~T_{\rm eff}/{\rm K} + 10.500$
\end{description}
or equivalently, in log form:
\begin{description}
\item red edge: $\phantom{l}\log L = -12.962 \log T_{\rm eff}/{\rm K}  + 50.823$
\item blue edge: $\log L = -20.476 \log T_{\rm eff}/{\rm K} + 82.454$
\end{description}

We stress that the validity is for the main-sequence and immediate post-main-sequence evolutionary phases. At high luminosities and low effective temperatures (i.e. in the Hertzsprung gap), there are too few stars for the pulsator fraction to be determined reliably, hence this area is greyed out in Fig.\,\ref{fig:contour}d. We do not necessarily expect that an extrapolation of our results to the RR\,Lyr or $\delta$\,Cep instability strips would be compatible with observations (e.g. \citealt{marconietal2015,muravevaetal2018,anderson2018}).

\section{Discussion}
\label{sec:discussion}

\subsection{Instability strip boundaries}

It is clear that the theoretical blue edge does not match the observed pulsator fraction. A major reason for this may be the need to include the contribution of turbulent pressure to the pulsational driving \citep{antocietal2014}, which is not typically accounted for when computing instability strips for the standard $\kappa$\,mechanism.
Another reason may be the chosen value of $\alpha_{\rm MLT}$. Following previous suggestions \citep{houdek2000,houdek&dupret2015}, a value of $\alpha_{\rm MLT} = 2.0$ is more appropriate than the commonly adopted value of 1.8, but a more thorough investigation, including a mass-dependent value of $\alpha_{\rm MLT}$, is warranted \citep{bowman&kurtz2018}. That investigation might also consider a different $\alpha_{\rm MLT}$ for the core and envelope.

At the red edge, additional pulsation damping by convection may be needed to bring models into agreement with our observed pulsator fraction. The red edge is particularly sensitive to the chosen value of $\alpha_{\rm MLT}$: a change from 1.8 to 2.0 shifts the red edge towards hotter temperatures by 200\,K \citep{houdek2000}.

Comparing the distribution of observed $\delta$\,Sct stars against the evolutionary tracks in Fig.\,\ref{fig:contour}, the mass range of main-sequence $\delta$\,Sct stars is 1.5--2.3\,M$_{\odot}$, with some exceptions. It should be noted that these stars are not necessarily pulsators for their whole main-sequence lifetimes. Stars above 2.0\,M$_{\odot}$ are not expected to be pulsators until the end of their main-sequence evolution, while stars below 1.7\,M$_{\odot}$ will pass out of the instability strip before the end of theirs.

The observed temperature distribution of $\gamma$\,Dor stars does not appear to completely overlap with the $\delta$\,Sct stars, contrary to the suggestion by \citet{balona2018c}. The 58 $\gamma$\,Dor stars with spectroscopic temperatures in \citet{vanreethetal2015b} form the largest sample available. The range of temperatures is 6835--7365\,K, putting many of them cooler than the red edge of our observed $\delta$\,Sct instability strip. Only 16\:per\:cent of their targets were hybrids, and the majority of those were binaries, so the $\delta$\,Sct and $\gamma$\,Dor pulsations may be coming from different stars. It is therefore clear that not all $\delta$\,Sct stars are also $\gamma$\,Dor stars. We reiterate that while many genuine $\delta$\,Sct--$\gamma$\,Dor hybrids exist (e.g. \citealt{grigahceneetal2010a,kurtzetal2014,saioetal2015,schmid&aerts2016}), there are also pure examples of both classes (\citealt{bowman2017, lietal2019a}).

\subsection{Observational uncertainties}

\begin{figure*}
\begin{center}
\includegraphics[width=0.32\textwidth]{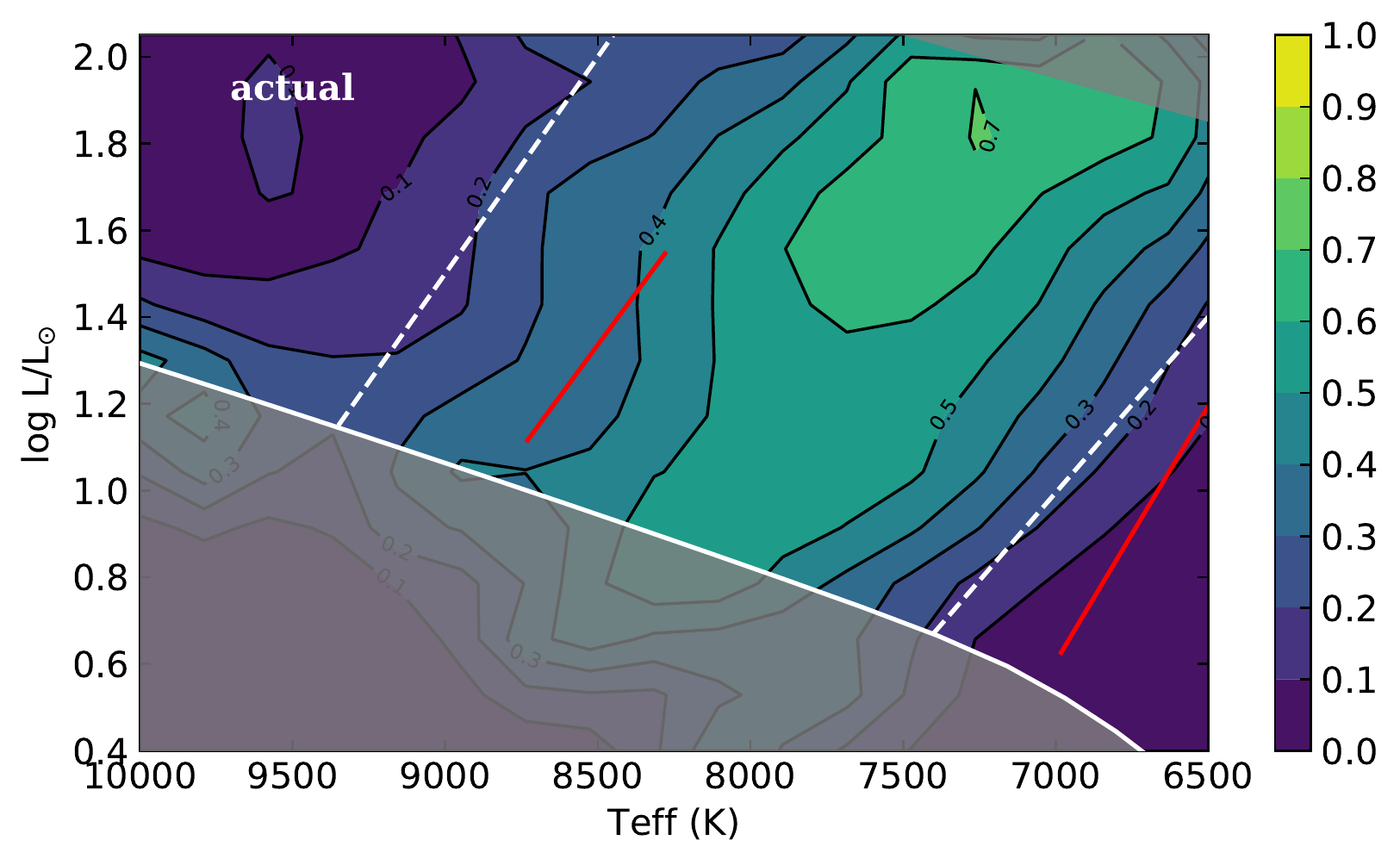} \includegraphics[width=0.32\textwidth]{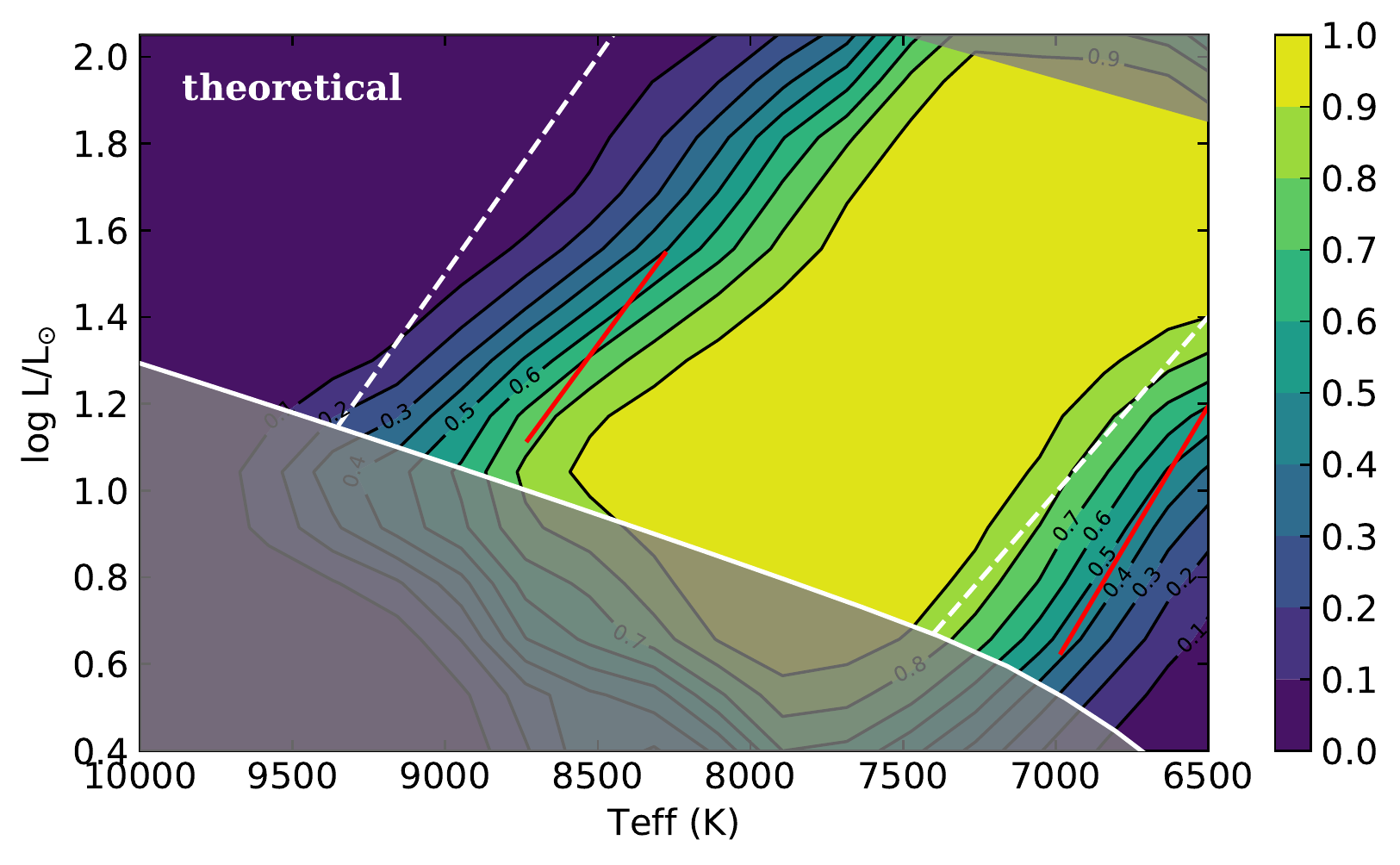} \includegraphics[width=0.32\textwidth]{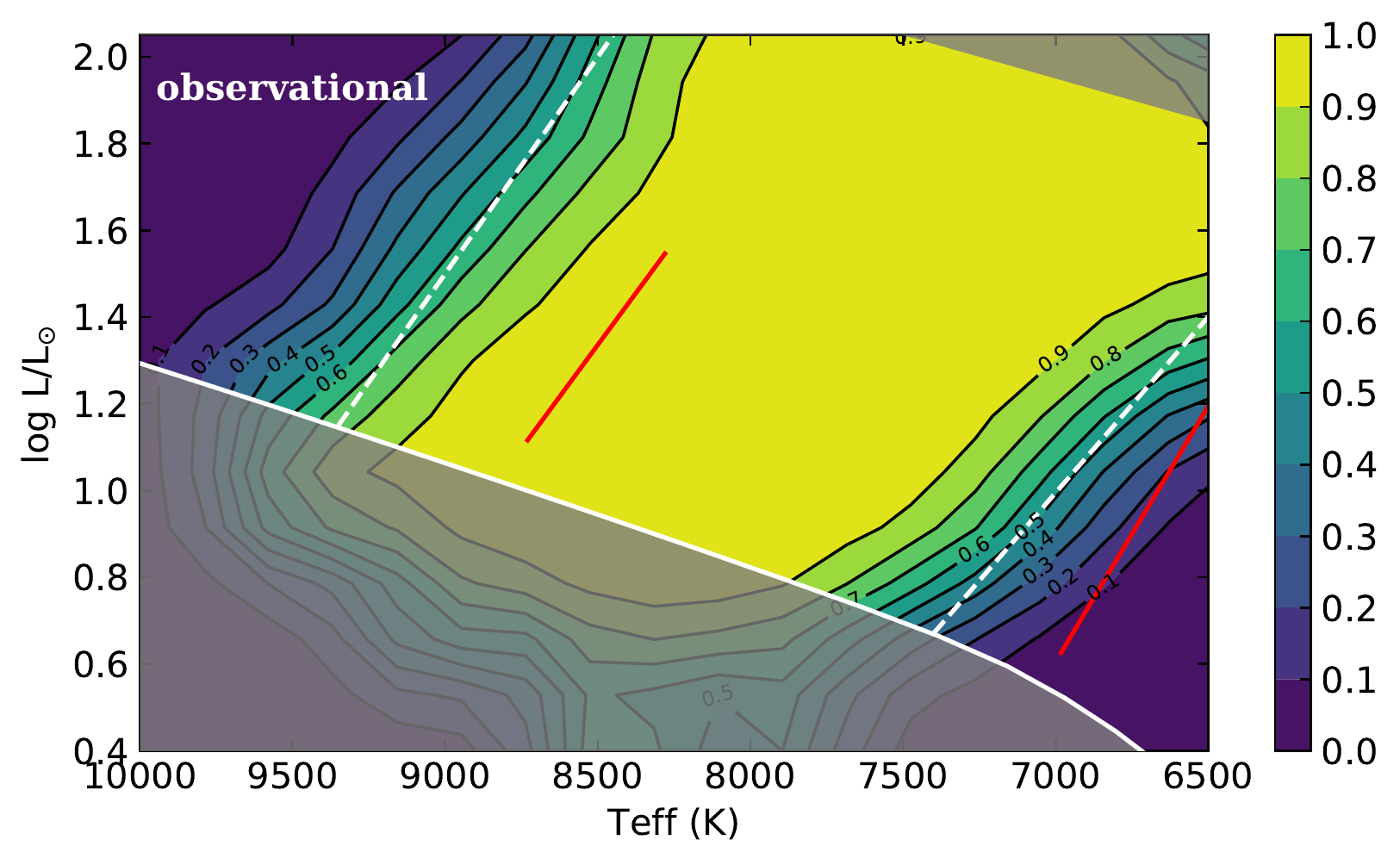}
\caption{Simulation of a pure instability strip observed with typical observational uncertainties. The solid red lines are the `theoretical' instability strip, and dashed white lines are the new `observational' instability strip based on the pulsator fraction (see Fig.\,\ref{fig:contour}). The leftmost panel is the actual distribution of pulsator fraction from Fig.\,\ref{fig:contour}d, with the contour range extended to 1.0 to match the other panels here. The simulation in the middle panel began with all stars within the red instability strip being pulsators, while the right panel used the white instability strip. This simulation allows us to reject the hypothesis that the instability strip is pure.}
\label{fig:simulation}
\end{center}
\end{figure*}

Here, we consider the possibility put forward by \citet{murphyetal2015} that the $\delta$\,Sct instability strip is actually pure (pulsator fraction 100\:per\:cent), but the observational uncertainties (particularly in temperature) cause some contamination of the instability strip with non-pulsators, and also cause some pulsators to fall outside of the instability strip. To evaluate this hypothesis, we took the same set of targets used from Sec.\,\ref{sec:pulsators} onwards, with their observed temperatures and luminosities, and created a simulation in which all stars inside the instability strip pulsate and all those outside do not. We then perturbed the observed temperatures and luminosities by a Gaussian of width equal to their 1$\sigma$ uncertainties, and recomputed the pulsator fraction. The simulation was performed twice: once using the theoretical instability strip computed with TDC models \citep{dupretetal2005b}, and once using our new observational instability strip, based on the pulsator fraction. In Fig.\,\ref{fig:simulation} they are compared to the actual distribution from Fig.\,\ref{fig:contour}. The simulation seems incompatible with the actual distribution, regardless of which instability strip is used, so we are able to reject the hypothesis that the $\delta$\,Sct instability strip is actually pure. The conclusion in \citet{murphyetal2015} that all stars in the instability strip pulsate is now seen to be incorrect, presumably due to the small sample size used there.

\subsection{Chemically peculiar stars}
\label{ssec:peculiar}

There are two classes of chemically peculiar stars among which the pulsator fraction is expected to be lower than for normal stars. Firstly, the strong magnetic fields of Ap stars suppress the p\:modes of low radial order that are normally excited in $\delta$\,Sct stars \citep{saio2005}. Secondly, metallic-lined A (Am) stars are slowly rotating A stars where mixing processes associated with rotation are inefficient \citep{baglinetal1973}, so diffusive processes become dominant. Helium gravitationally settles out of the partial ionization zone where p\:modes are driven, hence Am stars are expected (and observed) to have a low pulsator fraction \citep{breger1970,kurtz1989,smalleyetal2011,smalleyetal2017}. Suggestions that the non-pulsators in the instability strip are actually Ap/Am stars have naturally followed \citep{murphy2014,murphyetal2015}.

The fraction of stars that are Ap reaches a peak of $\sim$13\:per\:cent near A0 and is lower elsewhere \citep{wolff1983}, so they are not numerous enough to explain the non-pulsators. The Am stars represent nearly 50\:per\:cent of the population at $\sim$7400\,K, but this rapidly falls to 20\:per\:cent at 7200\,K and 5\:per\:cent at 7100\,K \citep{smith1973,smalleyetal2017}. Hence the Am stars also cannot explain the lack of pulsators.

There remains no convincing explanation why tens of percent of stars in the middle of the $\delta$\,Sct instability strip are non-pulsators.

\subsection{Low-amplitude pulsators}

It is interesting to speculate on the nature of the extremely-low amplitude pulsators from Fig.\,\ref{fig:low_amp}. Are the pulsation modes in these stars really intrinsically limited to an amplitude of 10\,$\upmu$mag? If so, what mechanism limits this amplitude growth but allows other $\delta$\,Sct stars to reach amplitudes of tenths of a magnitude \citep{mcnamara2000,templetonetal2002}? One possibility is that some are distant, and hence faint, $\delta$\,Sct stars contaminating the light curves of foreground objects. This might give the illusion of a single A star if the targets had weighted spectral energy distributions similar to an A star, or if both stars were indeed A stars. Contamination such as this might also be the explanation for pulsating stars beyond the blue edge. We investigated this in two ways. Firstly, in Fig.\,\ref{fig:distance_amplitude} we showed the distribution of pulsation amplitude as a function of distance for the pulsating stars of the sample. There was no indication there that the low-amplitude pulsators are more distant stars. Indeed, detecting such low amplitude pulsation requires a low noise level, so the low-amplitude pulsators tend to be bright (nearby) stars for which photon noise is smaller. Secondly, we examined the KIC `contamination' parameter, which is a floating point value between 0 and 1 that estimates the fraction of flux in the aperture not attributable to the target star. We found no significant difference in the contaminations of pulsators and non-pulsators, and we found that the `hot pulsators' ($T_{\rm eff} > 8500$\,K) beyond the blue edge had lower contaminations, on average, than the cool pulsators ($T_{\rm eff} \leq 8500$\,K). Hence we reject the hypothesis that contamination or dilution of light curves is responsible for low-amplitude pulsators.

We note that the low amplitudes are not the result of attenuation due to undersampling \citep{chaplinetal2011,shibahashi&murphy2018}, since many other $\delta$\,Sct stars are seen with higher amplitudes at similar frequencies. It is also worth noting that $\delta$\,Sct stars will not be missed because of our choice of frequency range (5.0--43.9\,d$^{-1}$), since stars oscillating well above the sampling frequency will have Nyquist aliases detectable in the observed frequency range (KIC\,10977859, with oscillations near 60\,d$^{-1}$, is an excellent example; \citealt{murphy2012a}).

\subsection{The effect of binaries}
It is well known that a large fraction of A stars have companions. That fraction approaches unity when integrated over all orbital periods \citep{moe&distefano2017,guszejnovetal2017}. For binaries, the `observed' luminosity is higher than the true luminosity because the companion contributes some flux. The extreme case is for two equal stars (`twins'), where luminosity is overestimated by a factor of two. Fortunately, the extreme case is uncommon: the mass-ratio distribution of binaries with A-star primaries is well-constrained at intermediate periods \citep{murphyetal2018}, and peaks around 0.2. This means that for $\delta$\,Sct stars, whose mass distribution has a mean of $\sim$1.8\,M$_{\odot}$, the most common companions are K/M dwarfs. Luminosity is a very steep function of mass on the main sequence, with $L \sim M^{3.5}$, so the contribution of these companions to the observed luminosity is tiny, at $\sim$0.2$^{3.5}$ = 0.0036, with considerable spread. Short-period binaries generally show more equal masses (\citealt{kratter2011} and references therein) and the fraction of twins among A stars is a strong function of orbital period \citep{moe&distefano2017}, but the removal of eclipsing binaries from our sample preferentially removed twins and hence strongly mitigated the impact of binaries on our results.

\section{Conclusions}
\label{sec:conclusions}

We have calculated the luminosities of over 15\,000 \textit{Kepler} A and F stars using Gaia DR2 parallaxes, and carefully considered various extinction maps and photometric data to give the most precise results. We found the Bayestar17 extinction maps and the KIC $g$ photometry (rescaled to the SDSS system) to be the most accurate.

We inspected the Fourier transforms of the \textit{Kepler} light curves of our targets and classified them into $\delta$\,Sct and non-$\delta$\,Sct stars. Algorithmic classifications of variability were in good agreement with the manual classifications. After removing other types of variables (primarily eclipsing binaries and $\gamma$\,Dor stars with harmonics or high-frequency modes) we obtained a sample of 1988 genuine $\delta$\,Sct stars. We found the location of that sample on the H--R diagram to be broadly consistent with theoretical instability strips. However, when we consider the sample from which the pulsators are drawn (i.e. all A and early F stars observed by \textit{Kepler}), it is clear that the theoretical instability strip is underpopulated with pulsators at the red edge, and overpopulated beyond the blue edge. 

We calculated new instability strips based on the observed pulsator fraction on the H--R diagram. The observational strip for main-sequence stars lies between 7100 and 9000\,K, corresponding roughly to spectral types A3--F0. While $\delta$\,Sct stars can be found outside this range, they comprise only a small fraction of the population there. A better description also takes the luminosity dependence into account, leading to boundaries at
\begin{description}
\item red edge: $\phantom{l}\log L/L_{\odot} = -0.000811~T_{\rm eff}/{\rm K} + 6.672$
\item blue edge: $\log L/L_{\odot} = -0.001000~T_{\rm eff}/{\rm K} + 10.500$.
\end{description}

Inside the instability strip, the pulsator fraction is well below 100\:per\:cent. We dismissed chemical peculiarity as the origin of the non-pulsators. The large sample used in this study also argues strongly against the possibility that the non-pulsators actually lie outside the instability strip, even after the uncertainties on their temperatures and luminosities are considered. The distribution of amplitudes in the Fourier transforms of \textit{Kepler} light curves suggests that the non-pulsators do not have low-amplitude pulsation below the micro-magnitude detection threshold. This observational work, focussed on the pulsator fraction, provides a new viewpoint and new results upon which a renewed modelling effort of the $\delta$\,Sct instability strip can be built.

\section*{Acknowledgements}

We thank Daniel~Huber, Dan~Foreman-Mackey, Andrew~Casey, Coryn Bailer-Jones, and Matteo~Cantiello for useful discussions, and we thank the anonymous referee for comments that improved the quality of this manuscript. This work was supported by the Australian Research Council, an Australian Government Research Training Program (RTP) scholarship, and the Danish National Research Foundation (Grant DNRF106) through its funding for the Stellar Astrophysics Centre (SAC). This work has made use of data from the European Space Agency (ESA) mission {\it Gaia} (\url{https://www.cosmos.esa.int/gaia}), processed by the {\it Gaia} Data Processing and Analysis Consortium (DPAC, \url{https://www.cosmos.esa.int/web/gaia/dpac/consortium}). Funding for the DPAC has been provided by national institutions, in particular the institutions participating in the {\it Gaia} Multilateral Agreement.

\appendix

\section{Reddening Maps in the \textit{Kepler} Field}
\label{app:reddening}

\begin{figure}
	\centering
	\includegraphics[width=\linewidth]{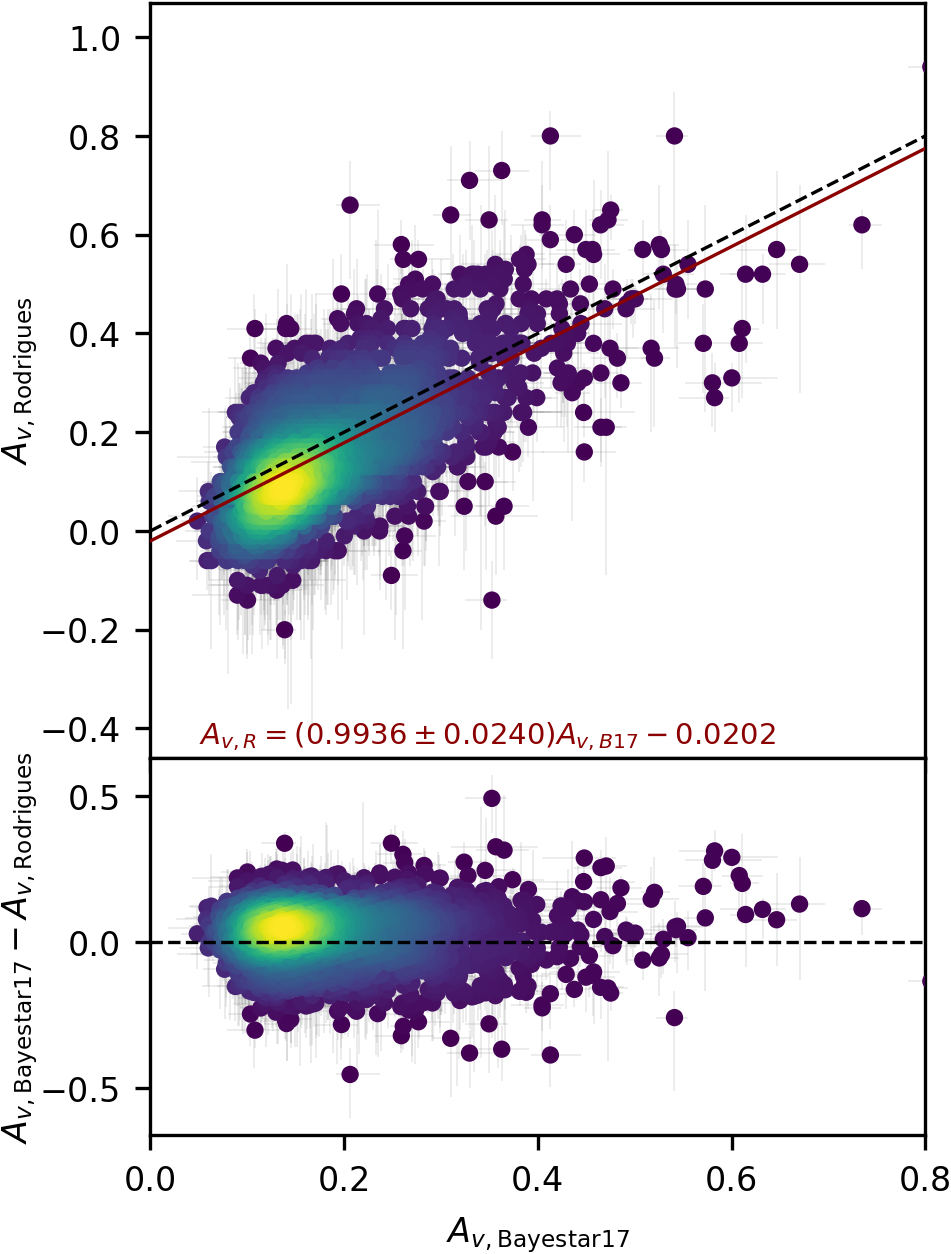}
    \caption{Comparison of a selection of stars in the \textit{Kepler} field from extinctions in the Rodrigues dataset \citep{rodriguesetal2014} against those of Bayestar\:17 \citep{greenetal2018}. The line of unity is dashed black, while a linear fit is provided in red. The bottom panel depicts the absolute difference between both maps.}
    \label{fig:rodrigues_comparison}
\end{figure}

\begin{figure*}
\centering
	\includegraphics[width=\linewidth]{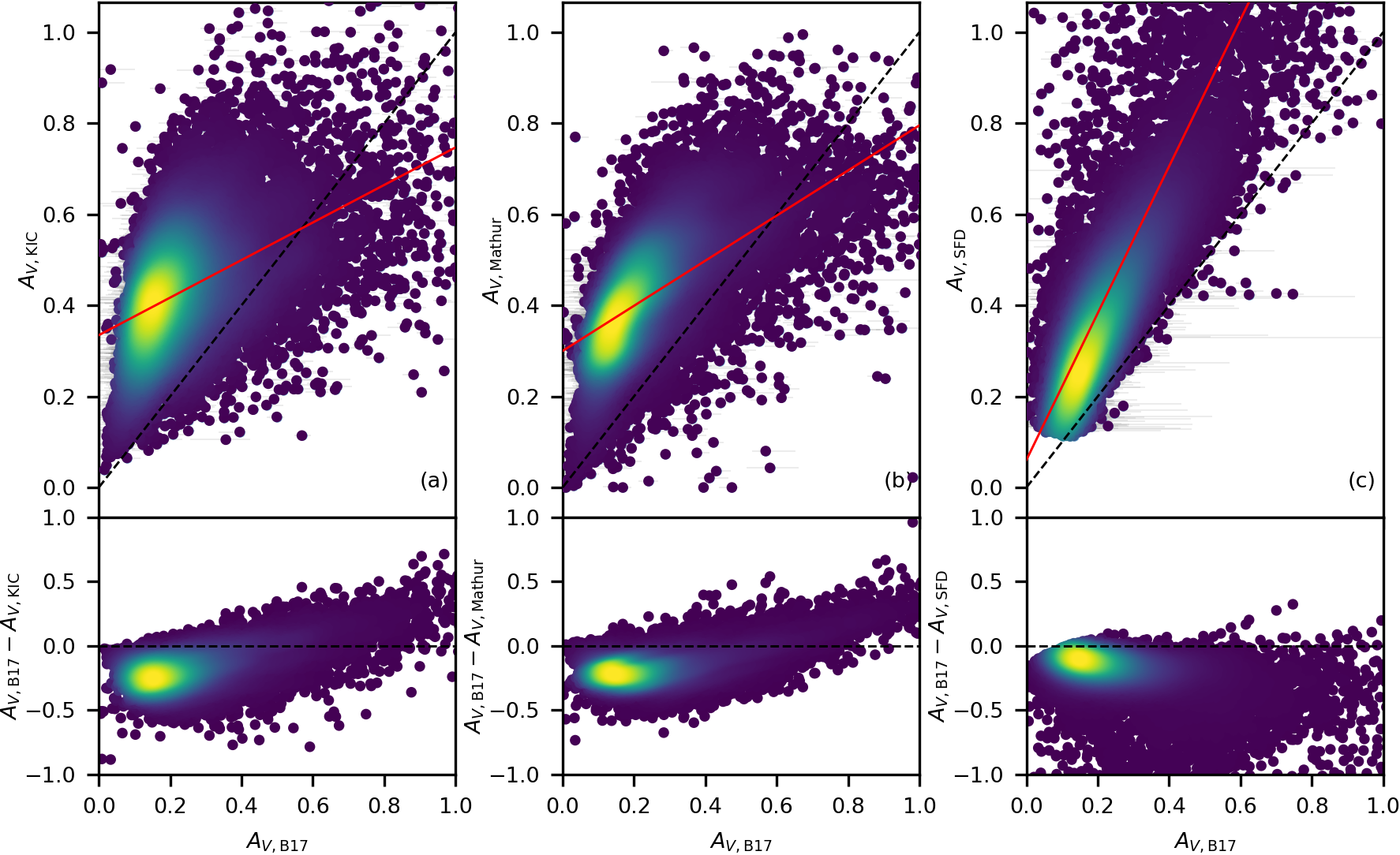}
    \caption{Density plots of extinctions from Bayestar 17 compared against those of the KIC \citep{brownetal2011}, of the \citet{mathuretal2017} stellar properties catalogue, and the SFD \citep{schlafly&finkbeiner2011} reddening maps respectively. Error bars are provided in grey where available. The lines of unity and best linear fit are shown as the black dashed and red lines, respectively. The lower panels depict the residuals of the linear fits.}
    \label{fig:comparison}
\end{figure*}

As described in Sec.\,\ref{ssec:gaia}, extinction corrections $A_V$ were obtained from the {\sc dustmaps} Python package \citep{green2018} provided for interfacing with the Bayestar\:17 (hereafter B17) reddening map \citep{greenetal2018}. B17 is a 3D map of interstellar reddening covering declinations above $\delta \geq -30^{\circ}$ out to a distance of several kiloparsecs. Reddening in this map depends on stellar photometry obtained through PAN-STARSS\,1 and 2MASS.

Despite B17 superseding the previous iteration of the map (Bayestar\:15; B15; \citealt{greenetal2015}), units between the two maps are not identical. B17 has been normalised such that one unit of reddening predicts the same $E(g-r)$  as one unit of the original Schlegel, Finkbeiner \& Davis map (hereafter SFD; \citealt{schlegeletal1998}, recalibrated by \citealt{schlafly&finkbeiner2011}). B15 uses units of colour excess $E(B-V)$ on the SFD scale, so care must be taken when applying extinction corrections from either map. Although \cite{greenetal2018} provided extinction coefficients to convert B17 into either 2MASS or PAN-STARRS\,1 passbands, a reddening law must be assumed prior to obtaining extinctions in other bands. \cite{sanders&das2018} provided coefficients to be used with B17 that were derived following the extinction curve of \cite{schlaflyetal2016}. An appropriate \textit{SDSS} \textit{g}-band extinction coefficient of $R(\lambda) = 3.613$ has been chosen for our purposes.

Fig. \ref{fig:rodrigues_comparison} compares the \textit{V}-band extinction corrections from the APOKASC catalogue \citep{rodriguesetal2014} and B17 for the sample of stars in APOKASC. Although the sample in the APOKASC catalogue differs from the sample used in this work, it is still worthwhile to point out the excellent agreement between both sets of extinctions. Stars in the Rodrigues catalogue are spread over several high- and low-density regions of dust between \mbox{$0.5$ and $5$\,kpc}. Such a large sample over different dust conditions leads to an excellent test case for the B17 map in the \textit{\textit{Kepler}} field. Indeed, assuming both a minor offset and linear relationship between the maps we find that
\begin{equation}
A_{V,\rm{APOKASC}} = (0.9936 \pm 0.0240) A_{V, \rm{Bayestar}} - 0.0202,
\end{equation}
yielding a gradient of unity within 1$\sigma$ errors. 

In the APOKASC catalogue, extinctions were derived via Bayesian inferences of stellar parameters from joint observations by \textit{Kepler} and APOGEE. Extinctions and distances that fit the overall spectrum of the star were simultaneously estimated alongside constraints imposed by asteroseismic and spectroscopic data. Since the \citet{rodriguesetal2014} analysis predates even the earliest iteration of the Bayestar maps, we provide this comparison here. For a further comparison of the APOKASC extinction corrections against those of other maps we refer the reader to \citet{rodriguesetal2014}.

Figure~\ref{fig:comparison} compares the extinctions obtained from B17 with those available from \cite{mathuretal2017}, the KIC \citep{brownetal2011}, and the SFD map \citep{schlafly&finkbeiner2011}. Extinctions in the KIC were derived from a simple reddening model for the distribution of dust, based on galactic latitude and distance \citep{brownetal2011}. The KIC model assumed that dust is distributed in a smooth disk aligned with the plane of the Milky Way, having an exponential decay of density with height above the plane. The final values for reddening provided by the KIC were derived simultaneously with other stellar parameters, meaning that the reddening is implicitly tied to parameters such as temperature and metallicity whose values are potentially imprecise. Figure~\ref{fig:comparison} indicates a significant overestimation of extinctions in the KIC with respect to B17, similar to the overestimates found by \citet{rodriguesetal2014}. This significant discrepancy probably reflects the fact that dust distribution is much more patchy along the Galactic disc than assumed in the relatively simple model adopted for the KIC.

Based on Fig.\,\ref{fig:comparison}, extinctions in the \citeauthor{mathuretal2017} catalogue also appear somewhat overestimated with respect to B17. At high extinctions there is relatively good agreement between the two maps, yet the largest concentration of extinctions is similar to those of the KIC. This is expected because the \citeauthor{mathuretal2017} catalogue largely inherits from the KIC.

The SFD extinction corrections tend to be much higher than the B17 catalogue (right panel of Fig.\,\ref{fig:comparison}). This is because SFD is a two dimensional map that provides extinctions at infinity, rather than along the line of sight to each star. In general, the linear gradient between two maps depends closely on the band in which the comparison is made. However, both the \citeauthor{mathuretal2017} and KIC values are significantly offset from the zero-point, regardless of any discrepancies in gradient. On the other hand, the SFD reddening map in Fig \ref{fig:comparison} has only a minor offset.

Good agreement between extinctions in the KIC and the \citet{gontcharov2017} 3D maps has been noted \citep{balona2018c}, even though these are in stark disagreement with B17 values. Given the independent spectroscopic and asteroseismic confirmation that KIC extinctions are inaccurate \citep{rodriguesetal2014}, we consider the B17 maps to be the best available for the \textit{Kepler} field. Using KIC or Gontcharov extinctions would result in our log luminosities being overestimated by 0.12\,dex, on average.

\begin{figure}
\centering
\includegraphics[width=\linewidth]{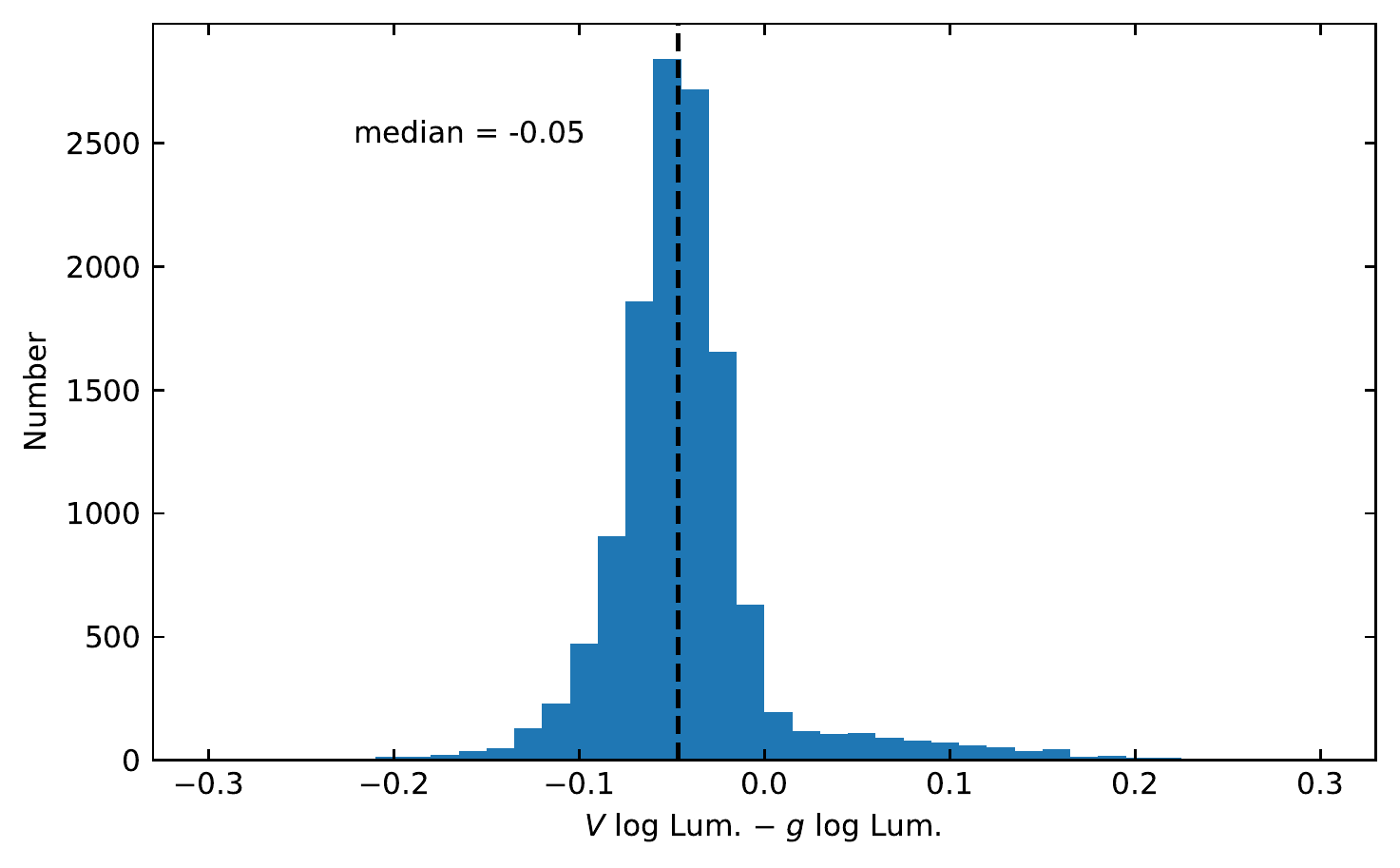}
\caption{Difference between log luminosities derived via the {\it V\_UBV} data and the KIC $g$ data on MAST. KIC $g$ was first converted to the SDSS scale according to the relations in \citet{pinsonneaultetal2012}, see Sec.\,\ref{ssec:gaia}, here.}
\label{fig:Vg}
\end{figure}

In Sec.\,\ref{ssec:gaia} we noted that luminosities derived using {\it V\_UBV} data from MAST were systematically lower, by 0.05\,dex, than those derived via the $g$ band (Fig.\,\ref{fig:Vg}). This is probably because the {\it UBV} data collected by \citet{everettetal2012} were not calibrated against standard stars, but instead tied to the KIC. While the offset between the KIC scale and the SDSS scale has been addressed \citep{pinsonneaultetal2012}, the same is not true for these {\it UBV} data and the Johnson-Cousins system.

\bibliographystyle{mnras}
\bibliography{sjm_bibliography} 

\begin{thebibliography}{}
\makeatletter
\relax
\def\mn@urlcharsother{\let\do\@makeother \do\$\do\&\do\#\do\^\do\_\do\%\do\~}
\def\mn@doi{\begingroup\mn@urlcharsother \@ifnextchar [ {\mn@doi@}
  {\mn@doi@[]}}
\def\mn@doi@[#1]#2{\def\@tempa{#1}\ifx\@tempa\@empty \href
  {http://dx.doi.org/#2} {doi:#2}\else \href {http://dx.doi.org/#2} {#1}\fi
  \endgroup}
\def\mn@eprint#1#2{\mn@eprint@#1:#2::\@nil}
\def\mn@eprint@arXiv#1{\href {http://arxiv.org/abs/#1} {{\tt arXiv:#1}}}
\def\mn@eprint@dblp#1{\href {http://dblp.uni-trier.de/rec/bibtex/#1.xml}
  {dblp:#1}}
\def\mn@eprint@#1:#2:#3:#4\@nil{\def\@tempa {#1}\def\@tempb {#2}\def\@tempc
  {#3}\ifx \@tempc \@empty \let \@tempc \@tempb \let \@tempb \@tempa \fi \ifx
  \@tempb \@empty \def\@tempb {arXiv}\fi \@ifundefined
  {mn@eprint@\@tempb}{\@tempb:\@tempc}{\expandafter \expandafter \csname
  mn@eprint@\@tempb\endcsname \expandafter{\@tempc}}}

\bibitem[\protect\citeauthoryear{{Aerts} et~al.,}{{Aerts}
  et~al.}{2018}]{aertsetal2018}
{Aerts} C.,  et~al., 2018, \mn@doi [\apjs] {10.3847/1538-4365/aaccfb}, \href
  {http://adsabs.harvard.edu/abs/2018ApJS..237...15A} {237, 15}

\bibitem[\protect\citeauthoryear{{Anderson}}{{Anderson}}{2018}]{anderson2018}
{Anderson} R.~I.,  2018, \mn@doi [\aap] {10.1051/0004-6361/201832585}, \href
  {http://adsabs.harvard.edu/abs/2018A%26A...611L...7A} {611, L7}

\bibitem[\protect\citeauthoryear{{Andrae} et~al.,}{{Andrae}
  et~al.}{2018}]{andraeetal2018}
{Andrae} R.,  et~al., 2018, \mn@doi [\aap] {10.1051/0004-6361/201732516}, \href
  {http://adsabs.harvard.edu/abs/2018A%26A...616A...8A} {616, A8}

\bibitem[\protect\citeauthoryear{{Antoci} et~al.,}{{Antoci}
  et~al.}{2014}]{antocietal2014}
{Antoci} V.,  et~al., 2014, \mn@doi [\apj] {10.1088/0004-637X/796/2/118}, \href
  {http://adsabs.harvard.edu/abs/2014ApJ...796..118A} {796, 118}

\bibitem[\protect\citeauthoryear{{Arenou} et~al.,}{{Arenou}
  et~al.}{2018}]{arenouetal2018}
{Arenou} F.,  et~al., 2018, \mn@doi [\aap] {10.1051/0004-6361/201833234}, \href
  {http://adsabs.harvard.edu/abs/2018A%26A...616A..17A} {616, A17}

\bibitem[\protect\citeauthoryear{{Asplund}, {Grevesse}, {Sauval}  \&
  {Scott}}{{Asplund} et~al.}{2009}]{asplundetal2009}
{Asplund} M.,  {Grevesse} N.,  {Sauval} A.~J.,   {Scott} P.,  2009, \mn@doi
  [\araa] {10.1146/annurev.astro.46.060407.145222}, \href
  {http://adsabs.harvard.edu/abs/2009ARA%26A..47..481A} {47, 481}

\bibitem[\protect\citeauthoryear{{Baglin}, {Breger}, {Chevalier}, {Hauck}, {Le
  Contel}, {Sareyan}  \& {Valtier}}{{Baglin} et~al.}{1973}]{baglinetal1973}
{Baglin} A.,  {Breger} M.,  {Chevalier} C.,  {Hauck} B.,  {Le Contel} J.~M.,
  {Sareyan} J.~P.,   {Valtier} J.~C.,  1973, \aap, \href
  {http://adsabs.harvard.edu/abs/1973A%26A....23..221B} {23, 221}

\bibitem[\protect\citeauthoryear{{Bailer-Jones}, {Rybizki}, {Fouesneau},
  {Mantelet}  \& {Andrae}}{{Bailer-Jones} et~al.}{2018}]{bailer-jonesetal2018}
{Bailer-Jones} C.~A.~L.,  {Rybizki} J.,  {Fouesneau} M.,  {Mantelet} G.,
  {Andrae} R.,  2018, \mn@doi [\aj] {10.3847/1538-3881/aacb21}, \href
  {http://adsabs.harvard.edu/abs/2018AJ....156...58B} {156, 58}

\bibitem[\protect\citeauthoryear{{Balona}}{{Balona}}{2018}]{balona2018c}
{Balona} L.~A.,  2018, \mn@doi [\mnras] {10.1093/mnras/sty1511}, \href
  {http://adsabs.harvard.edu/abs/2018MNRAS.tmp.1437B} {}

\bibitem[\protect\citeauthoryear{{Balona} \& {Dziembowski}}{{Balona} \&
  {Dziembowski}}{2011}]{balona&dziembowski2011}
{Balona} L.~A.,  {Dziembowski} W.~A.,  2011, \mn@doi [\mnras]
  {10.1111/j.1365-2966.2011.19301.x}, \href
  {http://adsabs.harvard.edu/abs/2011MNRAS.417..591B} {417, 591}

\bibitem[\protect\citeauthoryear{{Borucki} et~al.,}{{Borucki}
  et~al.}{2010}]{boruckietal2010}
{Borucki} W.~J.,  et~al., 2010, \mn@doi [Science] {10.1126/science.1185402},
  \href {http://adsabs.harvard.edu/abs/2010Sci...327..977B} {327, 977}

\bibitem[\protect\citeauthoryear{{Bouabid}, {Dupret}, {Salmon},
  {Montalb{\'a}n}, {Miglio}  \& {Noels}}{{Bouabid}
  et~al.}{2013}]{bouabidetal2013}
{Bouabid} M.-P.,  {Dupret} M.-A.,  {Salmon} S.,  {Montalb{\'a}n} J.,  {Miglio}
  A.,   {Noels} A.,  2013, \mn@doi [\mnras] {10.1093/mnras/sts517}, \href
  {http://adsabs.harvard.edu/abs/2013MNRAS.429.2500B} {429, 2500}

\bibitem[\protect\citeauthoryear{{Bowman}}{{Bowman}}{2017}]{bowman2017}
{Bowman} D.~M.,  2017, {Amplitude Modulation of Pulsation Modes in Delta Scuti
  Stars}.
{Springer}, \mn@doi{10.1007/978-3-319-66649-5}

\bibitem[\protect\citeauthoryear{{Bowman} \& {Kurtz}}{{Bowman} \&
  {Kurtz}}{2018}]{bowman&kurtz2018}
{Bowman} D.~M.,  {Kurtz} D.~W.,  2018, \mn@doi [\mnras] {10.1093/mnras/sty449},
  \href {http://adsabs.harvard.edu/abs/2018MNRAS.476.3169B} {476, 3169}

\bibitem[\protect\citeauthoryear{{Breger}}{{Breger}}{1969}]{breger1969b}
{Breger} M.,  1969, \mn@doi [\apjs] {10.1086/190199}, \href
  {http://adsabs.harvard.edu/abs/1969ApJS...19...79B} {19, 79}

\bibitem[\protect\citeauthoryear{{Breger}}{{Breger}}{1970}]{breger1970}
{Breger} M.,  1970, \mn@doi [\apj] {10.1086/150691}, \href
  {http://adsabs.harvard.edu/abs/1970ApJ...162..597B} {162, 597}

\bibitem[\protect\citeauthoryear{{Breger}}{{Breger}}{2000}]{breger2000}
{Breger} M.,  2000, in {M.~Breger \& M.~Montgomery} ed.,  Astronomical Society
  of the Pacific Conference Series Vol. 210, Delta Scuti and Related Stars.
  pp~3--+

\bibitem[\protect\citeauthoryear{{Breger} \& {Montgomery}}{{Breger} \&
  {Montgomery}}{2014}]{breger&montgomery2014}
{Breger} M.,  {Montgomery} M.~H.,  2014, \mn@doi [\apj]
  {10.1088/0004-637X/783/2/89}, \href
  {http://adsabs.harvard.edu/abs/2014ApJ...783...89B} {783, 89}

\bibitem[\protect\citeauthoryear{{Breger} et~al.,}{{Breger}
  et~al.}{1993}]{bregeretal1993}
{Breger} M.,  et~al., 1993, \aap, \href
  {http://adsabs.harvard.edu/abs/1993A%26A...271..482B} {271, 482}

\bibitem[\protect\citeauthoryear{{Brown}, {Latham}, {Everett}  \&
  {Esquerdo}}{{Brown} et~al.}{2011}]{brownetal2011}
{Brown} T.~M.,  {Latham} D.~W.,  {Everett} M.~E.,   {Esquerdo} G.~A.,  2011,
  \mn@doi [\aj] {10.1088/0004-6256/142/4/112}, \href
  {http://adsabs.harvard.edu/abs/2011AJ....142..112B} {142, 112}

\bibitem[\protect\citeauthoryear{{Chaplin} et~al.,}{{Chaplin}
  et~al.}{2011}]{chaplinetal2011}
{Chaplin} W.~J.,  et~al., 2011, \mn@doi [\apj] {10.1088/0004-637X/732/1/54},
  \href {http://adsabs.harvard.edu/abs/2011ApJ...732...54C} {732, 54}

\bibitem[\protect\citeauthoryear{{Choi}, {Dotter}, {Conroy}, {Cantiello},
  {Paxton}  \& {Johnson}}{{Choi} et~al.}{2016}]{choietal2016}
{Choi} J.,  {Dotter} A.,  {Conroy} C.,  {Cantiello} M.,  {Paxton} B.,
  {Johnson} B.~D.,  2016, \mn@doi [\apj] {10.3847/0004-637X/823/2/102}, \href
  {http://adsabs.harvard.edu/abs/2016ApJ...823..102C} {823, 102}

\bibitem[\protect\citeauthoryear{{Claret} \& {Torres}}{{Claret} \&
  {Torres}}{2018}]{claret&torres2018}
{Claret} A.,  {Torres} G.,  2018, \mn@doi [\apj] {10.3847/1538-4357/aabd35},
  \href {http://adsabs.harvard.edu/abs/2018ApJ...859..100C} {859, 100}

\bibitem[\protect\citeauthoryear{{Dotter}}{{Dotter}}{2016}]{dotter2016}
{Dotter} A.,  2016, \mn@doi [\apjs] {10.3847/0067-0049/222/1/8}, \href
  {http://adsabs.harvard.edu/abs/2016ApJS..222....8D} {222, 8}

\bibitem[\protect\citeauthoryear{{Dupret}, {Grigahc{\`e}ne}, {Garrido},
  {Gabriel}  \& {Scuflaire}}{{Dupret} et~al.}{2004}]{dupretetal2004}
{Dupret} M.-A.,  {Grigahc{\`e}ne} A.,  {Garrido} R.,  {Gabriel} M.,
  {Scuflaire} R.,  2004, \mn@doi [\aap] {10.1051/0004-6361:20031740}, \href
  {http://adsabs.harvard.edu/abs/2004A%26A...414L..17D} {414, L17}

\bibitem[\protect\citeauthoryear{{Dupret}, {Grigahc{\`e}ne}, {Garrido}, {De
  Ridder}, {Scuflaire}  \& {Gabriel}}{{Dupret} et~al.}{2005a}]{dupretetal2005a}
{Dupret} M.~A.,  {Grigahc{\`e}ne} A.,  {Garrido} R.,  {De Ridder} J.,
  {Scuflaire} R.,   {Gabriel} M.,  2005a, \mn@doi [\mnras]
  {10.1111/j.1365-2966.2005.09187.x}, \href
  {http://adsabs.harvard.edu/abs/2005MNRAS.361..476D} {361, 476}

\bibitem[\protect\citeauthoryear{{Dupret}, {Grigahc{\`e}ne}, {Garrido},
  {Gabriel}  \& {Scuflaire}}{{Dupret} et~al.}{2005b}]{dupretetal2005b}
{Dupret} M.,  {Grigahc{\`e}ne} A.,  {Garrido} R.,  {Gabriel} M.,   {Scuflaire}
  R.,  2005b, \mn@doi [\aap] {10.1051/0004-6361:20041817}, \href
  {http://adsabs.harvard.edu/abs/2005A%26A...435..927D} {435, 927}

\bibitem[\protect\citeauthoryear{{Everett}, {Howell}  \& {Kinemuchi}}{{Everett}
  et~al.}{2012}]{everettetal2012}
{Everett} M.~E.,  {Howell} S.~B.,   {Kinemuchi} K.,  2012, \mn@doi [\pasp]
  {10.1086/665529}, \href {http://adsabs.harvard.edu/abs/2012PASP..124..316E}
  {124, 316}

\bibitem[\protect\citeauthoryear{{Gaia Collaboration} et~al.,}{{Gaia
  Collaboration} et~al.}{2016}]{gaiacollaboration2016}
{Gaia Collaboration} et~al., 2016, \mn@doi [\aap]
  {10.1051/0004-6361/201629512}, \href
  {http://adsabs.harvard.edu/abs/2016A%26A...595A...2G} {595, A2}

\bibitem[\protect\citeauthoryear{{Gaia Collaboration}, {Brown}, {Vallenari},
  {Prusti}, {de Bruijne}, {Babusiaux}  \& {Bailer-Jones}}{{Gaia Collaboration}
  et~al.}{2018}]{gaiacollaboration2018a}
{Gaia Collaboration} {Brown} A.~G.~A.,  {Vallenari} A.,  {Prusti} T.,  {de
  Bruijne} J.~H.~J.,  {Babusiaux} C.,   {Bailer-Jones} C.~A.~L.,  2018,
  preprint, \href {http://adsabs.harvard.edu/abs/2018arXiv180409365G} {}
  (\mn@eprint {arXiv} {1804.09365})

\bibitem[\protect\citeauthoryear{{Gontcharov}}{{Gontcharov}}{2017}]{gontcharov2017}
{Gontcharov} G.~A.,  2017, \mn@doi [Astronomy Letters]
  {10.1134/S1063773717070039}, \href
  {http://adsabs.harvard.edu/abs/2017AstL...43..472G} {43, 472}

\bibitem[\protect\citeauthoryear{{Gray} \& {Corbally}}{{Gray} \&
  {Corbally}}{2009}]{gray&corbally2009}
{Gray} R.~O.,  {Corbally} J. C.,  2009, {Stellar Spectral Classification}.
{Princeton University Press}

\bibitem[\protect\citeauthoryear{Green}{Green}{2018}]{green2018}
Green G.,  2018, \mn@doi [Journal of Open Source Software]
  {10.21105/joss.00695}, 3, 695

\bibitem[\protect\citeauthoryear{{Green} et~al.,}{{Green}
  et~al.}{2015}]{greenetal2015}
{Green} G.~M.,  et~al., 2015, \mn@doi [\apj] {10.1088/0004-637X/810/1/25},
  \href {http://adsabs.harvard.edu/abs/2015ApJ...810...25G} {810, 25}

\bibitem[\protect\citeauthoryear{{Green} et~al.,}{{Green}
  et~al.}{2018}]{greenetal2018}
{Green} G.~M.,  et~al., 2018, \mn@doi [\mnras] {10.1093/mnras/sty1008}, \href
  {http://adsabs.harvard.edu/abs/2018MNRAS.478..651G} {478, 651}

\bibitem[\protect\citeauthoryear{{Grigahc{\`e}ne}, {Dupret}, {Gabriel},
  {Garrido}  \& {Scuflaire}}{{Grigahc{\`e}ne}
  et~al.}{2005}]{grigahceneetal2005}
{Grigahc{\`e}ne} A.,  {Dupret} M.-A.,  {Gabriel} M.,  {Garrido} R.,
  {Scuflaire} R.,  2005, \mn@doi [\aap] {10.1051/0004-6361:20041816}, \href
  {http://adsabs.harvard.edu/abs/2005A%26A...434.1055G} {434, 1055}

\bibitem[\protect\citeauthoryear{{Grigahc{\`e}ne} et~al.,}{{Grigahc{\`e}ne}
  et~al.}{2010}]{grigahceneetal2010a}
{Grigahc{\`e}ne} A.,  et~al., 2010, \mn@doi [\apjl]
  {10.1088/2041-8205/713/2/L192}, \href
  {http://adsabs.harvard.edu/abs/2010ApJ...713L.192G} {713, L192}

\bibitem[\protect\citeauthoryear{{Guszejnov}, {Hopkins}  \&
  {Krumholz}}{{Guszejnov} et~al.}{2017}]{guszejnovetal2017}
{Guszejnov} D.,  {Hopkins} P.~F.,   {Krumholz} M.~R.,  2017, \mn@doi [\mnras]
  {10.1093/mnras/stx725}, \href
  {http://adsabs.harvard.edu/abs/2017MNRAS.468.4093G} {468, 4093}

\bibitem[\protect\citeauthoryear{{Guzik}, {Bradley}, {Jackiewicz},
  {Molenda-Zakowicz}, {Uytterhoeven}  \& {Kinemuchi}}{{Guzik}
  et~al.}{2015}]{guziketal2015}
{Guzik} J.~A.,  {Bradley} P.~A.,  {Jackiewicz} J.,  {Molenda-Zakowicz} J.,
  {Uytterhoeven} K.,   {Kinemuchi} K.,  2015, preprint, \href
  {http://adsabs.harvard.edu/abs/2015arXiv150200175G} {} (\mn@eprint {arXiv}
  {1502.00175})

\bibitem[\protect\citeauthoryear{{Handler} \& {Shobbrook}}{{Handler} \&
  {Shobbrook}}{2002}]{handler&shobbrook2002}
{Handler} G.,  {Shobbrook} R.~R.,  2002, \mn@doi [\mnras]
  {10.1046/j.1365-8711.2002.05401.x}, \href
  {http://adsabs.harvard.edu/abs/2002MNRAS.333..251H} {333, 251}

\bibitem[\protect\citeauthoryear{{Hareter} et~al.,}{{Hareter}
  et~al.}{2010}]{hareteretal2010}
{Hareter} M.,  et~al., 2010, preprint, \href
  {http://adsabs.harvard.edu/abs/2010arXiv1007.3176H} {} (\mn@eprint {arXiv}
  {1007.3176})

\bibitem[\protect\citeauthoryear{{Houdek}}{{Houdek}}{2000}]{houdek2000}
{Houdek} G.,  2000, in {Breger} M.,  {Montgomery} M.,  eds,  Astronomical
  Society of the Pacific Conference Series Vol. 210, Delta Scuti and Related
  Stars. p.~454

\bibitem[\protect\citeauthoryear{{Houdek} \& {Dupret}}{{Houdek} \&
  {Dupret}}{2015}]{houdek&dupret2015}
{Houdek} G.,  {Dupret} M.-A.,  2015, \mn@doi [Living Reviews in Solar Physics]
  {10.1007/lrsp-2015-8}, \href
  {http://adsabs.harvard.edu/abs/2015LRSP...12....8H} {12, 8}

\bibitem[\protect\citeauthoryear{{Huber} et~al.,}{{Huber}
  et~al.}{2014}]{huberetal2014}
{Huber} D.,  et~al., 2014, \mn@doi [\apjs] {10.1088/0067-0049/211/1/2}, \href
  {http://adsabs.harvard.edu/abs/2014ApJS..211....2H} {211, 2}

\bibitem[\protect\citeauthoryear{{Huber} et~al.,}{{Huber}
  et~al.}{2017}]{huberetal2017}
{Huber} D.,  et~al., 2017, \mn@doi [\apj] {10.3847/1538-4357/aa75ca}, \href
  {http://adsabs.harvard.edu/abs/2017ApJ...844..102H} {844, 102}

\bibitem[\protect\citeauthoryear{{Jermyn}, {Tout}  \& {Chitre}}{{Jermyn}
  et~al.}{2018}]{jermynetal2018}
{Jermyn} A.~S.,  {Tout} C.~A.,   {Chitre} S.~M.,  2018, \mn@doi [\mnras]
  {10.1093/mnras/sty1831}, \href
  {http://adsabs.harvard.edu/abs/2018MNRAS.480.5427J} {480, 5427}

\bibitem[\protect\citeauthoryear{{Kirk} et~al.,}{{Kirk}
  et~al.}{2016}]{kirketal2016}
{Kirk} B.,  et~al., 2016, \mn@doi [\aj] {10.3847/0004-6256/151/3/68}, \href
  {http://adsabs.harvard.edu/abs/2016AJ....151...68K} {151, 68}

\bibitem[\protect\citeauthoryear{{Koch} et~al.,}{{Koch}
  et~al.}{2010}]{kochetal2010}
{Koch} D.~G.,  et~al., 2010, \mn@doi [\apjl] {10.1088/2041-8205/713/2/L79},
  \href {http://adsabs.harvard.edu/abs/2010ApJ...713L..79K} {713, L79}

\bibitem[\protect\citeauthoryear{{Kratter}}{{Kratter}}{2011}]{kratter2011}
{Kratter} K.~M.,  2011, in {Schmidtobreick} L.,  {Schreiber} M.~R.,   {Tappert}
  C.,  eds,  Astronomical Society of the Pacific Conference Series Vol. 447,
  Evolution of Compact Binaries. p.~47 (\mn@eprint {arXiv} {1109.3740})

\bibitem[\protect\citeauthoryear{{Kurtz}}{{Kurtz}}{1989}]{kurtz1989}
{Kurtz} D.~W.,  1989, \mnras, \href
  {http://adsabs.harvard.edu/abs/1989MNRAS.238.1077K} {238, 1077}

\bibitem[\protect\citeauthoryear{{Kurtz}, {Saio}, {Takata}, {Shibahashi},
  {Murphy}  \& {Sekii}}{{Kurtz} et~al.}{2014}]{kurtzetal2014}
{Kurtz} D.~W.,  {Saio} H.,  {Takata} M.,  {Shibahashi} H.,  {Murphy} S.~J.,
  {Sekii} T.,  2014, \mn@doi [\mnras] {10.1093/mnras/stu1329}, \href
  {http://adsabs.harvard.edu/abs/2014MNRAS.444..102K} {444, 102}

\bibitem[\protect\citeauthoryear{{Kurtz}, {Shibahashi}, {Murphy}, {Bedding}  \&
  {Bowman}}{{Kurtz} et~al.}{2015}]{kurtzetal2015}
{Kurtz} D.~W.,  {Shibahashi} H.,  {Murphy} S.~J.,  {Bedding} T.~R.,   {Bowman}
  D.~M.,  2015, \mn@doi [\mnras] {10.1093/mnras/stv868}, \href
  {http://adsabs.harvard.edu/abs/2015MNRAS.450.3015K} {450, 3015}

\bibitem[\protect\citeauthoryear{{Lehmann} et~al.,}{{Lehmann}
  et~al.}{2011}]{lehmannetal2011}
{Lehmann} H.,  et~al., 2011, \mn@doi [\aap] {10.1051/0004-6361/201015769},
  \href {http://adsabs.harvard.edu/abs/2011A%26A...526A.124L} {526, A124}

\bibitem[\protect\citeauthoryear{{Li}, {Bedding}, {Murphy}, {Van Reeth},
  {Antoci}  \& {Ouazzani}}{{Li} et~al.}{2018}]{lietal2019a}
{Li} G.,  {Bedding} T.~R.,  {Murphy} S.~J.,  {Van Reeth} T.,  {Antoci} V.,
  {Ouazzani} R.-M.,  2018, preprint, \href
  {http://adsabs.harvard.edu/abs/2018arXiv181003362L} {} (\mn@eprint {arXiv}
  {1810.03362})

\bibitem[\protect\citeauthoryear{{Lindegren} et~al.,}{{Lindegren}
  et~al.}{2018}]{lindegrenetal2018}
{Lindegren} L.,  et~al., 2018, preprint, \href
  {http://adsabs.harvard.edu/abs/2018arXiv180409366L} {} (\mn@eprint {arXiv}
  {1804.09366})

\bibitem[\protect\citeauthoryear{{Luri} et~al.,}{{Luri}
  et~al.}{2018}]{lurietal2018}
{Luri} X.,  et~al., 2018, \mn@doi [\aap] {10.1051/0004-6361/201832964}, \href
  {http://adsabs.harvard.edu/abs/2018A%26A...616A...9L} {616, A9}

\bibitem[\protect\citeauthoryear{{Mamajek} et~al.,}{{Mamajek}
  et~al.}{2015a}]{mamajeketal2015b}
{Mamajek} E.~E.,  et~al., 2015a, preprint, \href
  {http://adsabs.harvard.edu/abs/2015arXiv151006262M} {} (\mn@eprint {arXiv}
  {1510.06262})

\bibitem[\protect\citeauthoryear{{Mamajek} et~al.,}{{Mamajek}
  et~al.}{2015b}]{mamajeketal2015c}
{Mamajek} E.~E.,  et~al., 2015b, preprint, \href
  {http://adsabs.harvard.edu/abs/2015arXiv151007674M} {} (\mn@eprint {arXiv}
  {1510.07674})

\bibitem[\protect\citeauthoryear{{Marconi} et~al.,}{{Marconi}
  et~al.}{2015}]{marconietal2015}
{Marconi} M.,  et~al., 2015, \mn@doi [\apj] {10.1088/0004-637X/808/1/50}, \href
  {http://adsabs.harvard.edu/abs/2015ApJ...808...50M} {808, 50}

\bibitem[\protect\citeauthoryear{{Mathur} et~al.,}{{Mathur}
  et~al.}{2017}]{mathuretal2017}
{Mathur} S.,  et~al., 2017, \mn@doi [\apjs] {10.3847/1538-4365/229/2/30}, \href
  {http://adsabs.harvard.edu/abs/2017ApJS..229...30M} {229, 30}

\bibitem[\protect\citeauthoryear{{McNamara}}{{McNamara}}{2000}]{mcnamara2000}
{McNamara} D.~H.,  2000, in {M.~Breger \& M.~Montgomery} ed.,  Astronomical
  Society of the Pacific Conference Series Vol. 210, Delta Scuti and Related
  Stars. pp 373--+

\bibitem[\protect\citeauthoryear{{Moe} \& {Di Stefano}}{{Moe} \& {Di
  Stefano}}{2017}]{moe&distefano2017}
{Moe} M.,  {Di Stefano} R.,  2017, \mn@doi [\apjs] {10.3847/1538-4365/aa6fb6},
  \href {http://adsabs.harvard.edu/abs/2017ApJS..230...15M} {230, 15}

\bibitem[\protect\citeauthoryear{{Molenda-{\.Z}akowicz}
  et~al.,}{{Molenda-{\.Z}akowicz} et~al.}{2013}]{molenda-zakowiczetal2013}
{Molenda-{\.Z}akowicz} J.,  et~al., 2013, \mn@doi [\mnras]
  {10.1093/mnras/stt1095}, \href
  {http://adsabs.harvard.edu/abs/2013MNRAS.434.1422M} {434, 1422}

\bibitem[\protect\citeauthoryear{{Muraveva}, {Delgado}, {Clementini}, {Sarro}
  \& {Garofalo}}{{Muraveva} et~al.}{2018}]{muravevaetal2018}
{Muraveva} T.,  {Delgado} H.~E.,  {Clementini} G.,  {Sarro} L.~M.,   {Garofalo}
  A.,  2018, \mn@doi [\mnras] {10.1093/mnras/sty2241}, \href
  {http://adsabs.harvard.edu/abs/2018MNRAS.481.1195M} {481, 1195}

\bibitem[\protect\citeauthoryear{{Murphy}}{{Murphy}}{2012}]{murphy2012a}
{Murphy} S.~J.,  2012, \mn@doi [\mnras] {10.1111/j.1365-2966.2012.20644.x},
  \href {http://adsabs.harvard.edu/abs/2012MNRAS.422..665M} {422, 665}

\bibitem[\protect\citeauthoryear{{Murphy}}{{Murphy}}{2014}]{murphy2014}
{Murphy} S.~J.,  2014, PhD thesis, Univ. Central Lancashire,
  \mn@doi{10.1007/978-3-319-09417-5}

\bibitem[\protect\citeauthoryear{{Murphy}, {Shibahashi}  \& {Kurtz}}{{Murphy}
  et~al.}{2013}]{murphyetal2012b}
{Murphy} S.~J.,  {Shibahashi} H.,   {Kurtz} D.~W.,  2013, \mn@doi [\mnras]
  {10.1093/mnras/stt105}, \href
  {http://adsabs.harvard.edu/abs/2013MNRAS.430.2986M} {430, 2986}

\bibitem[\protect\citeauthoryear{{Murphy}, {Bedding}, {Shibahashi}, {Kurtz}  \&
  {Kjeldsen}}{{Murphy} et~al.}{2014}]{murphyetal2014}
{Murphy} S.~J.,  {Bedding} T.~R.,  {Shibahashi} H.,  {Kurtz} D.~W.,
  {Kjeldsen} H.,  2014, \mn@doi [\mnras] {10.1093/mnras/stu765}, \href
  {http://adsabs.harvard.edu/abs/2014MNRAS.441.2515M} {441, 2515}

\bibitem[\protect\citeauthoryear{{Murphy}, {Bedding}, {Niemczura}, {Kurtz}  \&
  {Smalley}}{{Murphy} et~al.}{2015}]{murphyetal2015}
{Murphy} S.~J.,  {Bedding} T.~R.,  {Niemczura} E.,  {Kurtz} D.~W.,   {Smalley}
  B.,  2015, \mn@doi [\mnras] {10.1093/mnras/stu2749}, \href
  {http://adsabs.harvard.edu/abs/2015MNRAS.447.3948M} {447, 3948}

\bibitem[\protect\citeauthoryear{{Murphy}, {Moe}, {Kurtz}, {Bedding},
  {Shibahashi}  \& {Boffin}}{{Murphy} et~al.}{2018}]{murphyetal2018}
{Murphy} S.~J.,  {Moe} M.,  {Kurtz} D.~W.,  {Bedding} T.~R.,  {Shibahashi} H.,
   {Boffin} H.~M.~J.,  2018, \mn@doi [\mnras] {10.1093/mnras/stx3049}, \href
  {http://adsabs.harvard.edu/abs/2018MNRAS.474.4322M} {474, 4322}

\bibitem[\protect\citeauthoryear{{Niemczura} et~al.,}{{Niemczura}
  et~al.}{2015}]{niemczuraetal2015}
{Niemczura} E.,  et~al., 2015, \mn@doi [\mnras] {10.1093/mnras/stv528}, \href
  {http://adsabs.harvard.edu/abs/2015MNRAS.450.2764N} {450, 2764}

\bibitem[\protect\citeauthoryear{{Niemczura} et~al.,}{{Niemczura}
  et~al.}{2017}]{niemczuraetal2017}
{Niemczura} E.,  et~al., 2017, \mn@doi [\mnras] {10.1093/mnras/stx1256}, \href
  {http://adsabs.harvard.edu/abs/2017MNRAS.470.2870N} {470, 2870}

\bibitem[\protect\citeauthoryear{{Paxton}, {Bildsten}, {Dotter}, {Herwig},
  {Lesaffre}  \& {Timmes}}{{Paxton} et~al.}{2011}]{paxtonetal2011}
{Paxton} B.,  {Bildsten} L.,  {Dotter} A.,  {Herwig} F.,  {Lesaffre} P.,
  {Timmes} F.,  2011, \mn@doi [\apjs] {10.1088/0067-0049/192/1/3}, \href
  {http://adsabs.harvard.edu/abs/2011ApJS..192....3P} {192, 3}

\bibitem[\protect\citeauthoryear{{Pinsonneault}, {An}, {Molenda-{\.Z}akowicz},
  {Chaplin}, {Metcalfe}  \& {Bruntt}}{{Pinsonneault}
  et~al.}{2012}]{pinsonneaultetal2012}
{Pinsonneault} M.~H.,  {An} D.,  {Molenda-{\.Z}akowicz} J.,  {Chaplin} W.~J.,
  {Metcalfe} T.~S.,   {Bruntt} H.,  2012, \mn@doi [\apjs]
  {10.1088/0067-0049/199/2/30}, \href
  {http://adsabs.harvard.edu/abs/2012ApJS..199...30P} {199, 30}

\bibitem[\protect\citeauthoryear{{Rodrigues} et~al.,}{{Rodrigues}
  et~al.}{2014}]{rodriguesetal2014}
{Rodrigues} T.~S.,  et~al., 2014, \mn@doi [\mnras] {10.1093/mnras/stu1907},
  \href {http://adsabs.harvard.edu/abs/2014MNRAS.445.2758R} {445, 2758}

\bibitem[\protect\citeauthoryear{{Rodr{\'{\i}}guez}, {L{\'o}pez-Gonz{\'a}lez}
  \& {L{\'o}pez de Coca}}{{Rodr{\'{\i}}guez} et~al.}{2000}]{rodriguezetal2000}
{Rodr{\'{\i}}guez} E.,  {L{\'o}pez-Gonz{\'a}lez} M.~J.,   {L{\'o}pez de Coca}
  P.,  2000, \mn@doi [\aaps] {10.1051/aas:2000221}, \href
  {http://adsabs.harvard.edu/abs/2000A%26AS..144..469R} {144, 469}

\bibitem[\protect\citeauthoryear{{Saio}}{{Saio}}{2005}]{saio2005}
{Saio} H.,  2005, \mn@doi [\mnras] {10.1111/j.1365-2966.2005.09091.x}, \href
  {http://adsabs.harvard.edu/abs/2005MNRAS.360.1022S} {360, 1022}

\bibitem[\protect\citeauthoryear{{Saio}, {Kurtz}, {Takata}, {Shibahashi},
  {Murphy}, {Sekii}  \& {Bedding}}{{Saio} et~al.}{2015}]{saioetal2015}
{Saio} H.,  {Kurtz} D.~W.,  {Takata} M.,  {Shibahashi} H.,  {Murphy} S.~J.,
  {Sekii} T.,   {Bedding} T.~R.,  2015, \mn@doi [\mnras]
  {10.1093/mnras/stu2696}, \href
  {http://adsabs.harvard.edu/abs/2015MNRAS.447.3264S} {447, 3264}

\bibitem[\protect\citeauthoryear{{Saio}, {Bedding}, {Kurtz}, {Murphy},
  {Antoci}, {Shibahashi}, {Li}  \& {Takata}}{{Saio}
  et~al.}{2018}]{saioetal2018b}
{Saio} H.,  {Bedding} T.~R.,  {Kurtz} D.~W.,  {Murphy} S.~J.,  {Antoci} V.,
  {Shibahashi} H.,  {Li} G.,   {Takata} M.,  2018, \mn@doi [\mnras]
  {10.1093/mnras/sty784}, \href
  {http://adsabs.harvard.edu/abs/2018MNRAS.477.2183S} {477, 2183}

\bibitem[\protect\citeauthoryear{{Sanders} \& {Das}}{{Sanders} \&
  {Das}}{2018}]{sanders&das2018}
{Sanders} J.~L.,  {Das} P.,  2018, \mn@doi [\mnras] {10.1093/mnras/sty2490},
  \href {http://adsabs.harvard.edu/abs/2018MNRAS.tmp.2388S} {}

\bibitem[\protect\citeauthoryear{{Schlafly} \& {Finkbeiner}}{{Schlafly} \&
  {Finkbeiner}}{2011}]{schlafly&finkbeiner2011}
{Schlafly} E.~F.,  {Finkbeiner} D.~P.,  2011, \mn@doi [\apj]
  {10.1088/0004-637X/737/2/103}, \href
  {http://adsabs.harvard.edu/abs/2011ApJ...737..103S} {737, 103}

\bibitem[\protect\citeauthoryear{{Schlafly} et~al.,}{{Schlafly}
  et~al.}{2016}]{schlaflyetal2016}
{Schlafly} E.~F.,  et~al., 2016, \mn@doi [\apj] {10.3847/0004-637X/821/2/78},
  \href {http://adsabs.harvard.edu/abs/2016ApJ...821...78S} {821, 78}

\bibitem[\protect\citeauthoryear{{Schlegel}, {Finkbeiner}  \&
  {Davis}}{{Schlegel} et~al.}{1998}]{schlegeletal1998}
{Schlegel} D.~J.,  {Finkbeiner} D.~P.,   {Davis} M.,  1998, \mn@doi [\apj]
  {10.1086/305772}, \href {http://adsabs.harvard.edu/abs/1998ApJ...500..525S}
  {500, 525}

\bibitem[\protect\citeauthoryear{{Schmid} \& {Aerts}}{{Schmid} \&
  {Aerts}}{2016}]{schmid&aerts2016}
{Schmid} V.~S.,  {Aerts} C.,  2016, \mn@doi [\aap]
  {10.1051/0004-6361/201628617}, \href
  {http://adsabs.harvard.edu/abs/2016A%26A...592A.116S} {592, A116}

\bibitem[\protect\citeauthoryear{{Shibahashi} \& {Murphy}}{{Shibahashi} \&
  {Murphy}}{2018}]{shibahashi&murphy2018}
{Shibahashi} H.,  {Murphy} S.~J.,  2018, preprint, \href
  {http://adsabs.harvard.edu/abs/2018arXiv181110205S} {} (\mn@eprint {arXiv}
  {1811.10205})

\bibitem[\protect\citeauthoryear{{Smalley} et~al.,}{{Smalley}
  et~al.}{2011}]{smalleyetal2011}
{Smalley} B.,  et~al., 2011, \mn@doi [\aap] {10.1051/0004-6361/201117230},
  \href {http://adsabs.harvard.edu/abs/2011A%26A...535A...3S} {535, A3+}

\bibitem[\protect\citeauthoryear{{Smalley} et~al.,}{{Smalley}
  et~al.}{2017}]{smalleyetal2017}
{Smalley} B.,  et~al., 2017, \mn@doi [\mnras] {10.1093/mnras/stw2903}, \href
  {http://adsabs.harvard.edu/abs/2017MNRAS.465.2662S} {465, 2662}

\bibitem[\protect\citeauthoryear{{Smith}}{{Smith}}{1973}]{smith1973}
{Smith} M.~A.,  1973, \mn@doi [\apjs] {10.1086/190270}, \href
  {http://adsabs.harvard.edu/abs/1973ApJS...25..277S} {25, 277}

\bibitem[\protect\citeauthoryear{{St{\c e}pie{\'n}}, {Pamyatnykh}  \&
  {Rozyczka}}{{St{\c e}pie{\'n}} et~al.}{2017}]{stepienetal2017}
{St{\c e}pie{\'n}} K.,  {Pamyatnykh} A.~A.,   {Rozyczka} M.,  2017, \mn@doi
  [\aap] {10.1051/0004-6361/201629511}, \href
  {http://adsabs.harvard.edu/abs/2017A%26A...597A..87S} {597, A87}

\bibitem[\protect\citeauthoryear{{Streamer}, {Ireland}, {Murphy}  \&
  {Bento}}{{Streamer} et~al.}{2018}]{streameretal2018}
{Streamer} M.,  {Ireland} M.~J.,  {Murphy} S.~J.,   {Bento} J.,  2018, \mn@doi
  [\mnras] {10.1093/mnras/sty1881}, \href
  {http://adsabs.harvard.edu/abs/2018MNRAS.480.1372S} {480, 1372}

\bibitem[\protect\citeauthoryear{{Stumpe}, {Smith}, {Catanzarite}, {Van Cleve},
  {Jenkins}, {Twicken}  \& {Girouard}}{{Stumpe} et~al.}{2014}]{stumpeetal2014}
{Stumpe} M.~C.,  {Smith} J.~C.,  {Catanzarite} J.~H.,  {Van Cleve} J.~E.,
  {Jenkins} J.~M.,  {Twicken} J.~D.,   {Girouard} F.~R.,  2014, \mn@doi [\pasp]
  {10.1086/674989}, \href {http://adsabs.harvard.edu/abs/2014PASP..126..100S}
  {126, 100}

\bibitem[\protect\citeauthoryear{{Templeton}, {Basu}  \&
  {Demarque}}{{Templeton} et~al.}{2002}]{templetonetal2002}
{Templeton} M.,  {Basu} S.,   {Demarque} P.,  2002, \mn@doi [\apj]
  {10.1086/341805}, \href {http://adsabs.harvard.edu/abs/2002ApJ...576..963T}
  {576, 963}

\bibitem[\protect\citeauthoryear{{Tkachenko}, {Lehmann}, {Smalley},
  {Debosscher}  \& {Aerts}}{{Tkachenko} et~al.}{2012}]{tkachenkoetal2012}
{Tkachenko} A.,  {Lehmann} H.,  {Smalley} B.,  {Debosscher} J.,   {Aerts} C.,
  2012, \mn@doi [\mnras] {10.1111/j.1365-2966.2012.20687.x}, \href
  {http://adsabs.harvard.edu/abs/2012MNRAS.422.2960T} {422, 2960}

\bibitem[\protect\citeauthoryear{{Tkachenko}, {Lehmann}, {Smalley}  \&
  {Uytterhoeven}}{{Tkachenko} et~al.}{2013a}]{tkachenkoetal2013a}
{Tkachenko} A.,  {Lehmann} H.,  {Smalley} B.,   {Uytterhoeven} K.,  2013a,
  \mn@doi [\mnras] {10.1093/mnras/stt453}, \href
  {http://adsabs.harvard.edu/abs/2013MNRAS.431.3685T} {431, 3685}

\bibitem[\protect\citeauthoryear{{Tkachenko} et~al.,}{{Tkachenko}
  et~al.}{2013b}]{tkachenkoetal2013b}
{Tkachenko} A.,  et~al., 2013b, \mn@doi [\aap] {10.1051/0004-6361/201220978},
  \href {http://adsabs.harvard.edu/abs/2013A%26A...556A..52T} {556, A52}

\bibitem[\protect\citeauthoryear{{Uytterhoeven} et~al.,}{{Uytterhoeven}
  et~al.}{2011}]{uytterhoevenetal2011}
{Uytterhoeven} K.,  et~al., 2011, \mn@doi [\aap] {10.1051/0004-6361/201117368},
  \href {http://adsabs.harvard.edu/abs/2011A%26A...534A.125U} {534, A125}

\bibitem[\protect\citeauthoryear{{Van Cleve} \& {Caldwell}}{{Van Cleve} \&
  {Caldwell}}{2016}]{vancleve&caldwell2016}
{Van Cleve} J.~E.,  {Caldwell} D.~A.,  2016, {Kepler Instrument Handbook}

\bibitem[\protect\citeauthoryear{{Van Reeth} et~al.,}{{Van Reeth}
  et~al.}{2015a}]{vanreethetal2015b}
{Van Reeth} T.,  et~al., 2015a, \mn@doi [\apjs] {10.1088/0067-0049/218/2/27},
  \href {http://adsabs.harvard.edu/abs/2015ApJS..218...27V} {218, 27}

\bibitem[\protect\citeauthoryear{{Van Reeth} et~al.,}{{Van Reeth}
  et~al.}{2015b}]{vanreethetal2015a}
{Van Reeth} T.,  et~al., 2015b, \mn@doi [\aap] {10.1051/0004-6361/201424585},
  \href {http://adsabs.harvard.edu/abs/2015A%26A...574A..17V} {574, A17}

\bibitem[\protect\citeauthoryear{{Wolff}}{{Wolff}}{1983}]{wolff1983}
{Wolff} S.~C.,  1983, {The A-type stars: problems and perspectives.}.
{NASA SP-463, Washington D.C.}

\bibitem[\protect\citeauthoryear{{Xiong}, {Deng}, {Zhang}  \& {Wang}}{{Xiong}
  et~al.}{2016}]{xiongetal2016}
{Xiong} D.~R.,  {Deng} L.,  {Zhang} C.,   {Wang} K.,  2016, \mn@doi [\mnras]
  {10.1093/mnras/stw047}, \href
  {http://adsabs.harvard.edu/abs/2016MNRAS.457.3163X} {457, 3163}

\bibitem[\protect\citeauthoryear{{Zinn}, {Pinsonneault}, {Huber}  \&
  {Stello}}{{Zinn} et~al.}{2018}]{zinnetal2018}
{Zinn} J.~C.,  {Pinsonneault} M.~H.,  {Huber} D.,   {Stello} D.,  2018,
  preprint, \href {http://adsabs.harvard.edu/abs/2018arXiv180502650Z} {}
  (\mn@eprint {arXiv} {1805.02650})

\makeatother
\end{thebibliography}


\section{Analysis of groups in the Skewness -- SNR plot}
\label{app:skewplot}

\begin{figure}
\begin{center}
\includegraphics[width=0.48\textwidth]{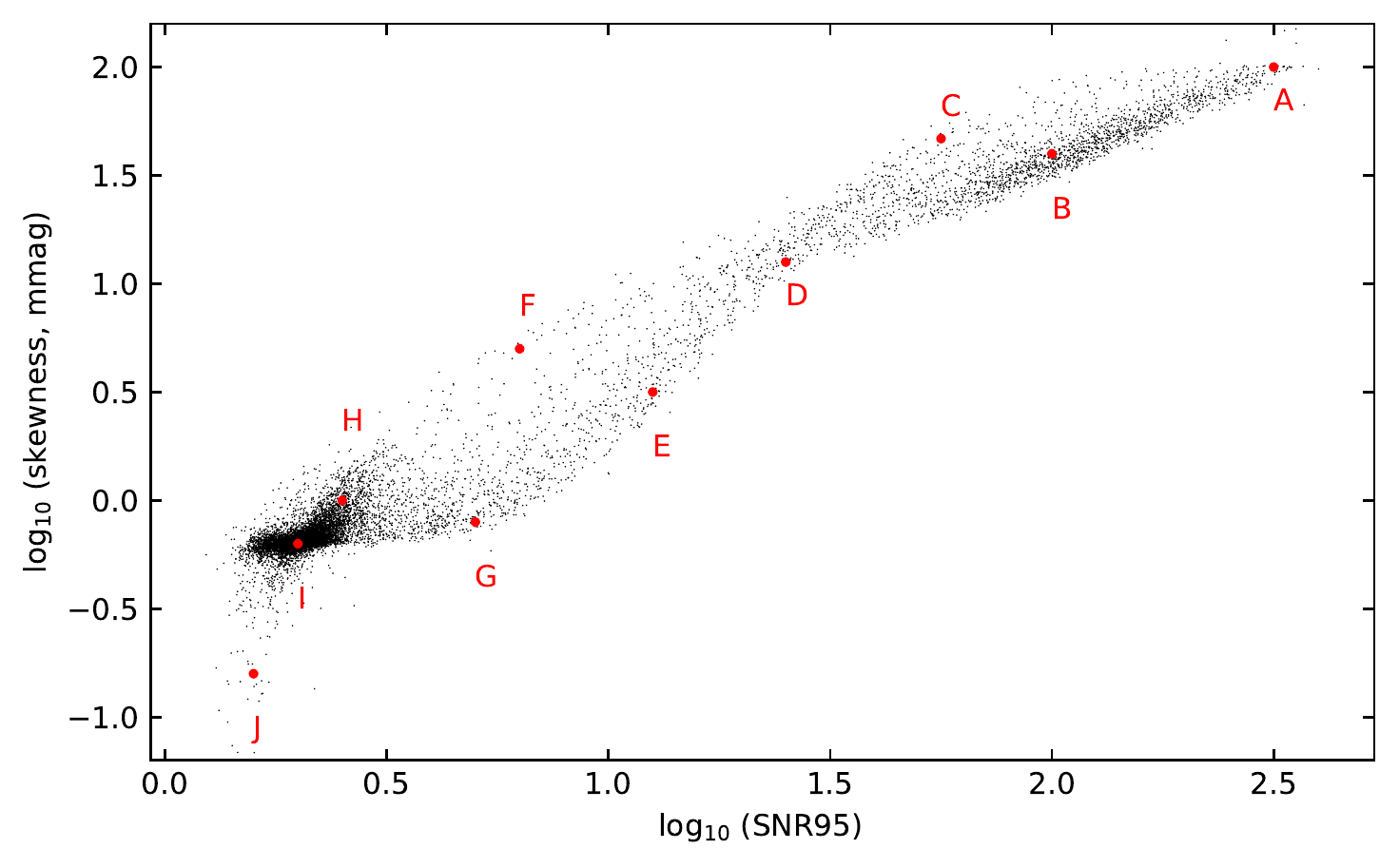}
\caption{Plotting the skewness of the Fourier transform against the signal-to-noise ratio (SNR) reveals groupings between stars with similar pulsation properties. The lettered groups are discussed below.}
\label{fig:skew}
\vspace{-3mm}
\end{center}
\end{figure}

\begin{figure}
\begin{center}
\includegraphics[width=0.48\textwidth]{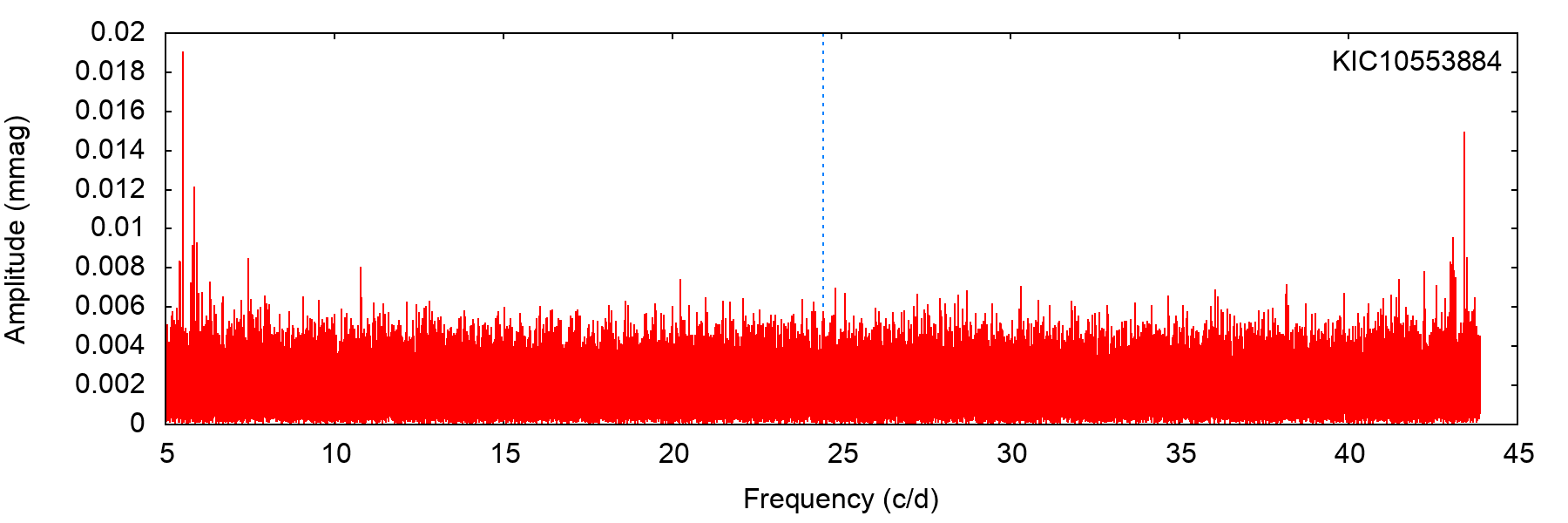}
\includegraphics[width=0.48\textwidth]{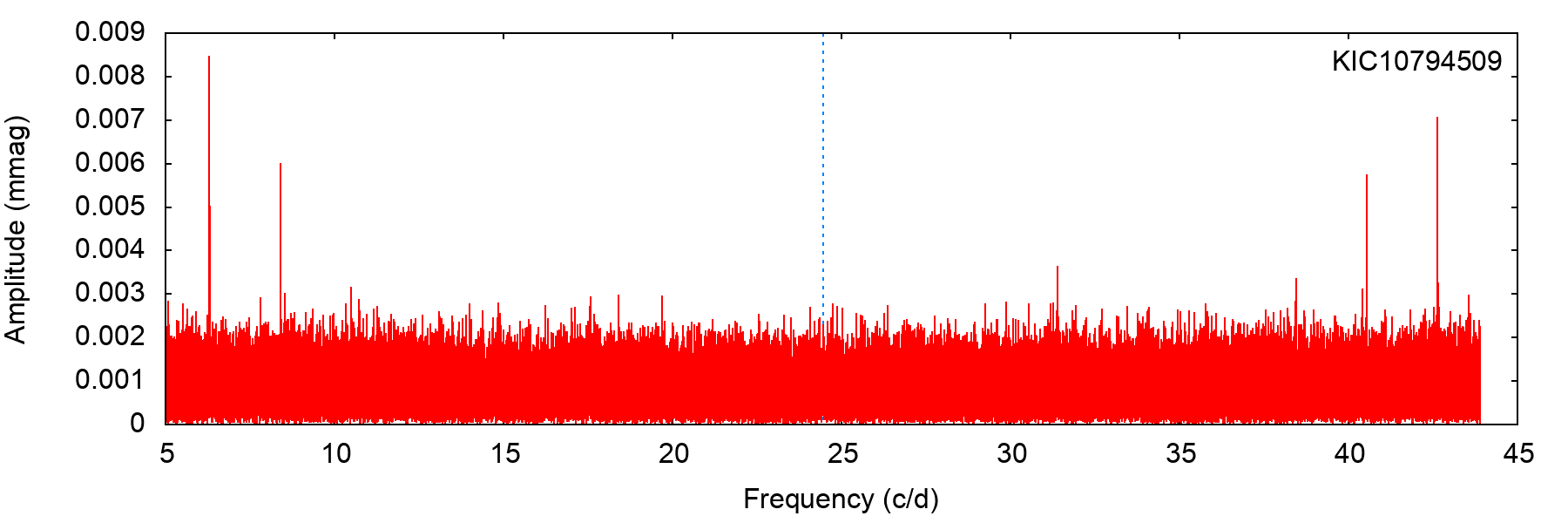}
\includegraphics[width=0.48\textwidth]{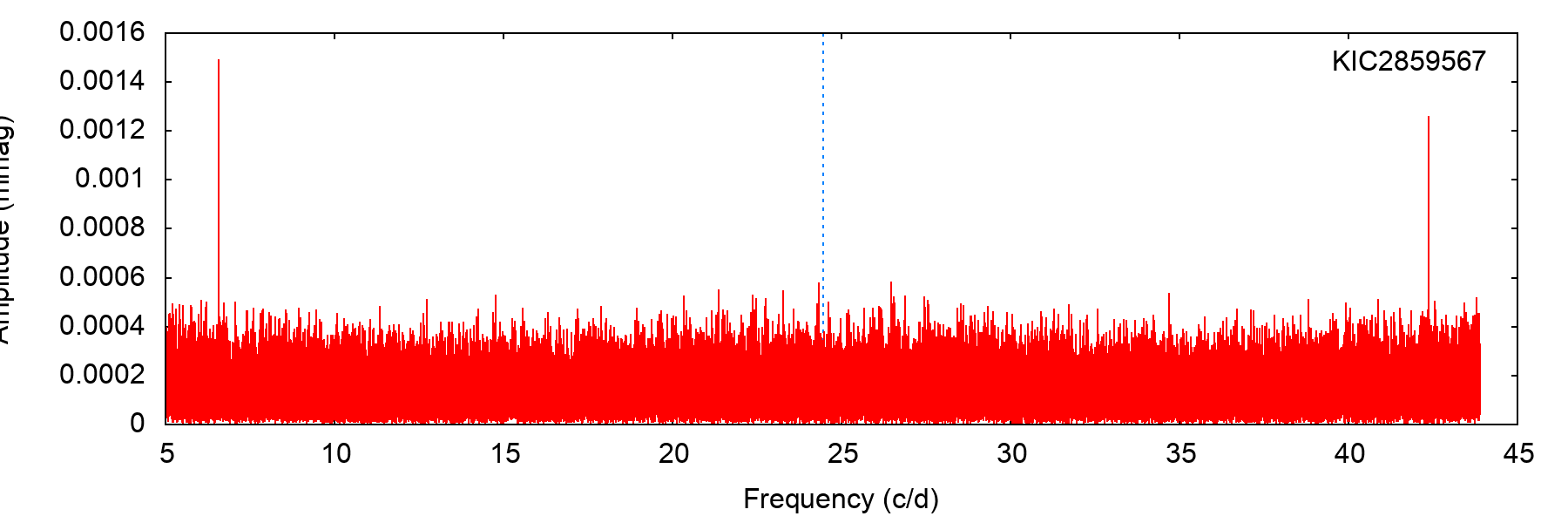}
\includegraphics[width=0.48\textwidth]{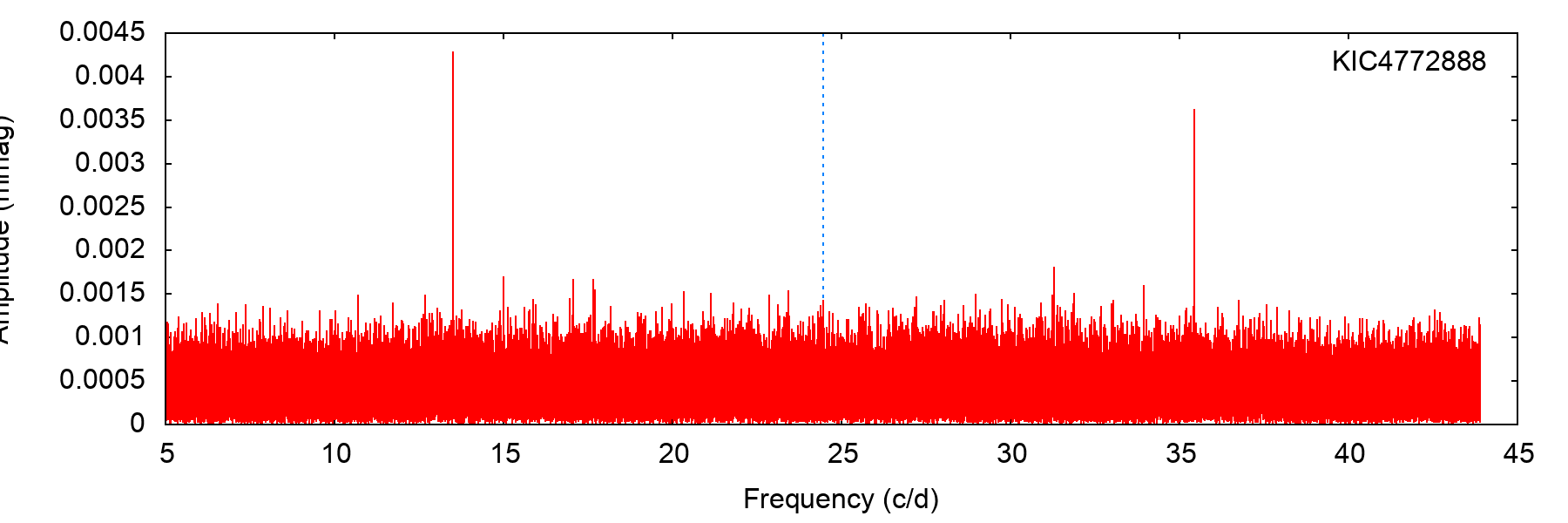}
\includegraphics[width=0.48\textwidth]{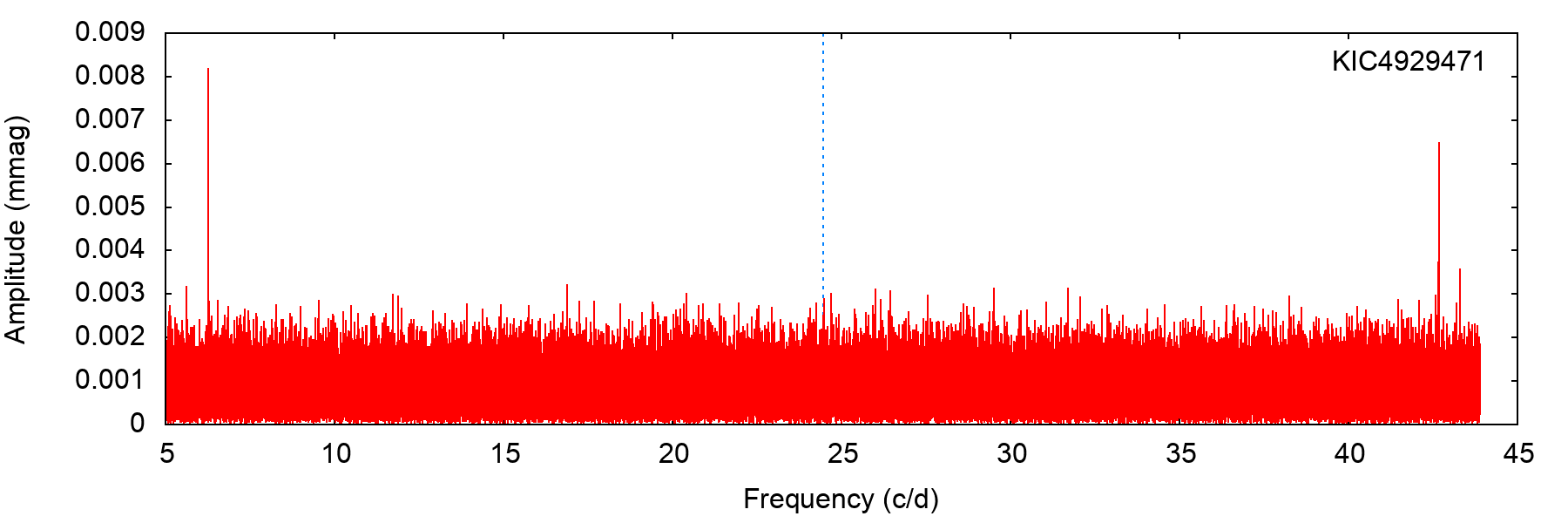}
\includegraphics[width=0.48\textwidth]{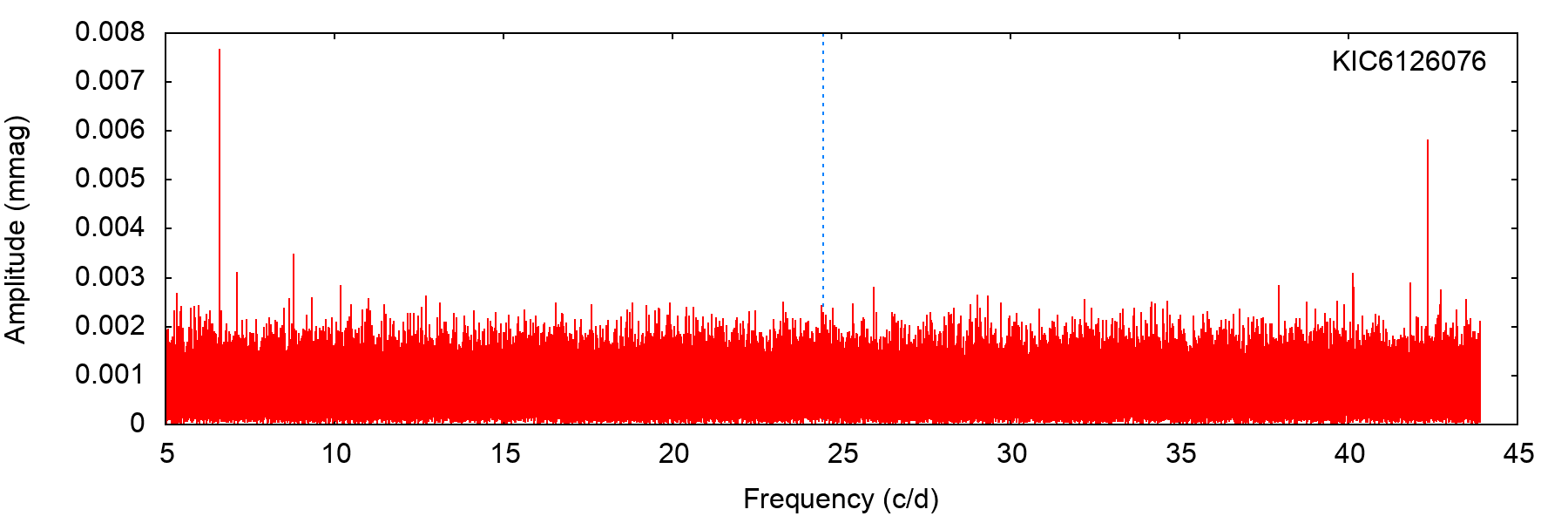}
\includegraphics[width=0.48\textwidth]{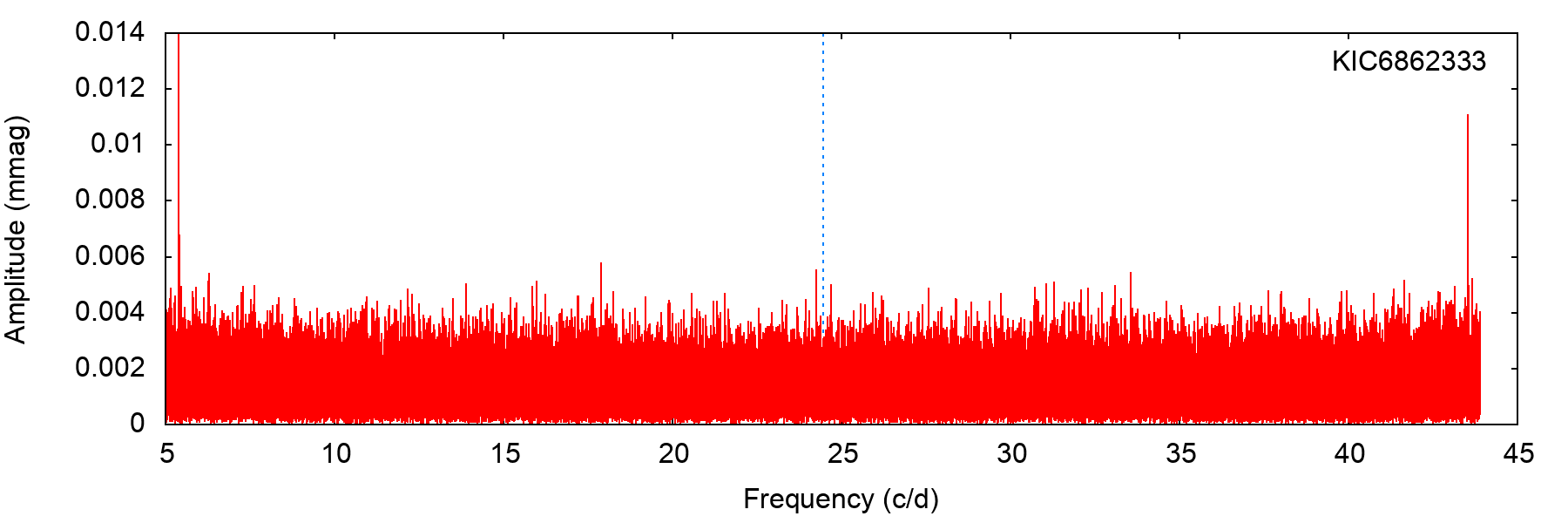}
\includegraphics[width=0.48\textwidth]{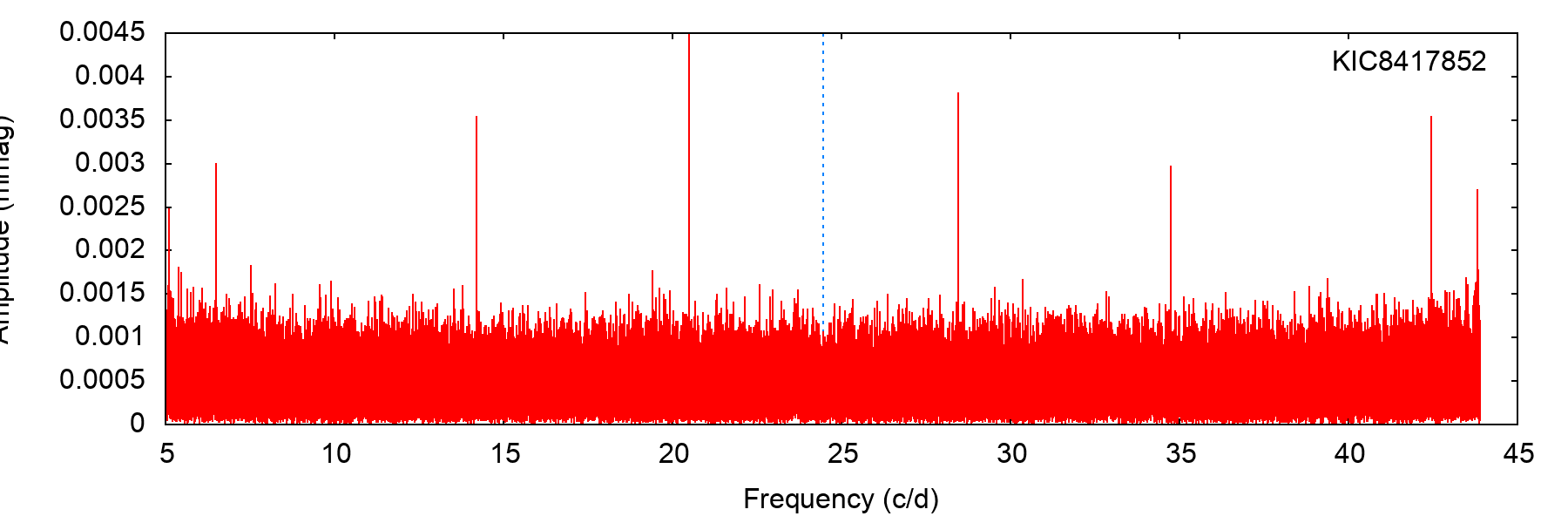}
\caption{Some typical examples of stars in Group G (see Sec.\,\ref{ssec:GroupG}).}
\label{fig:GroupG}
\vspace{-3mm}
\end{center}
\end{figure}

\begin{figure}
\begin{center}
\includegraphics[width=0.48\textwidth]{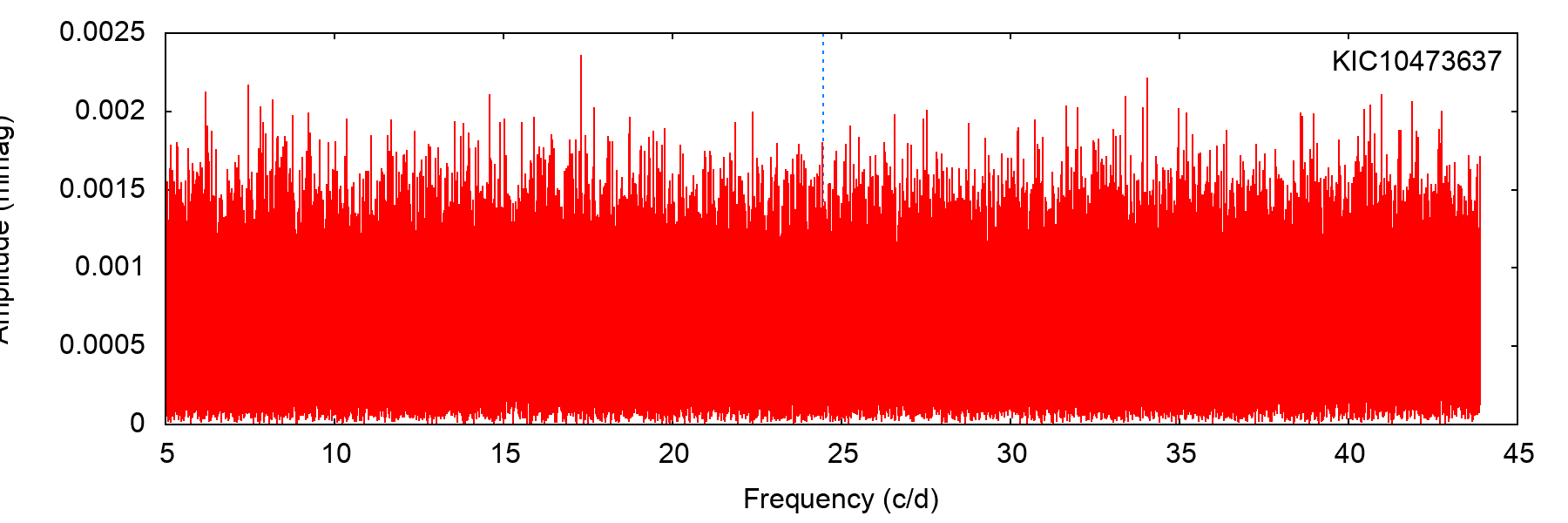}
\includegraphics[width=0.48\textwidth]{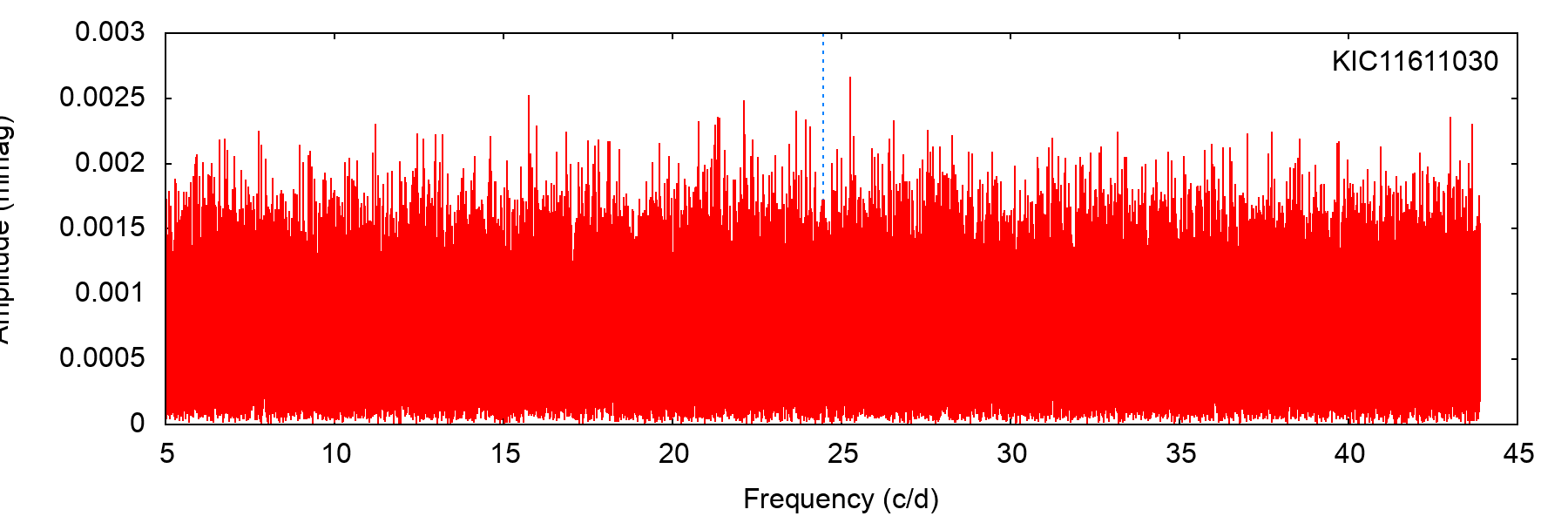}
\includegraphics[width=0.48\textwidth]{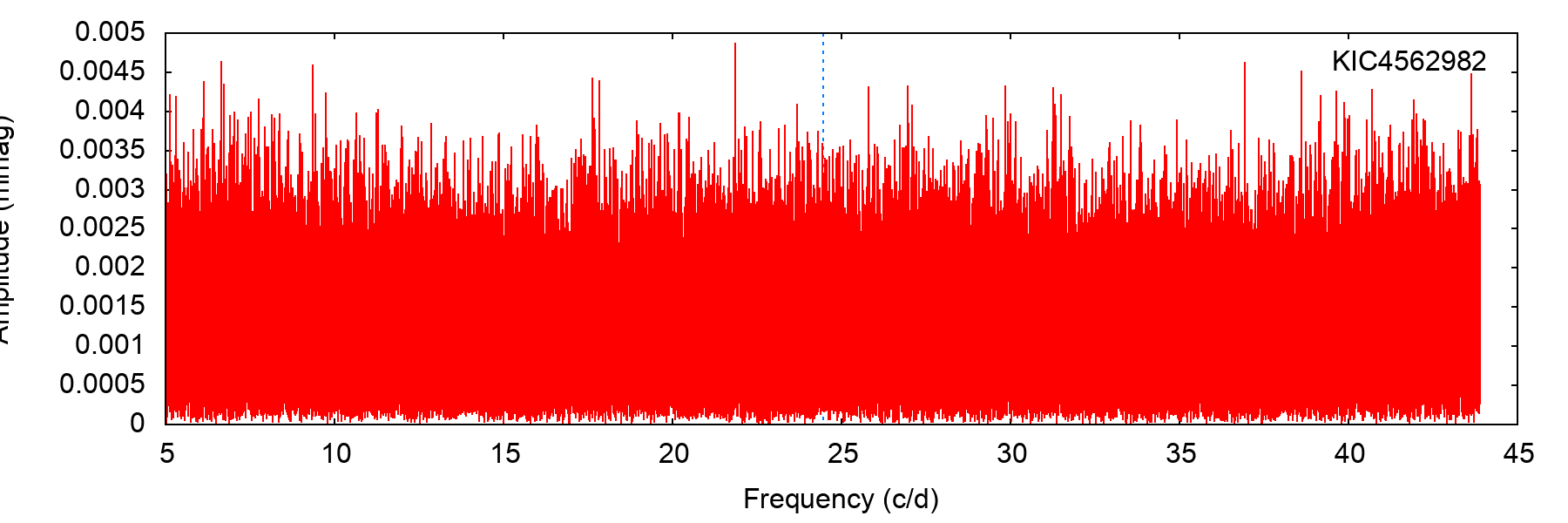}
\includegraphics[width=0.48\textwidth]{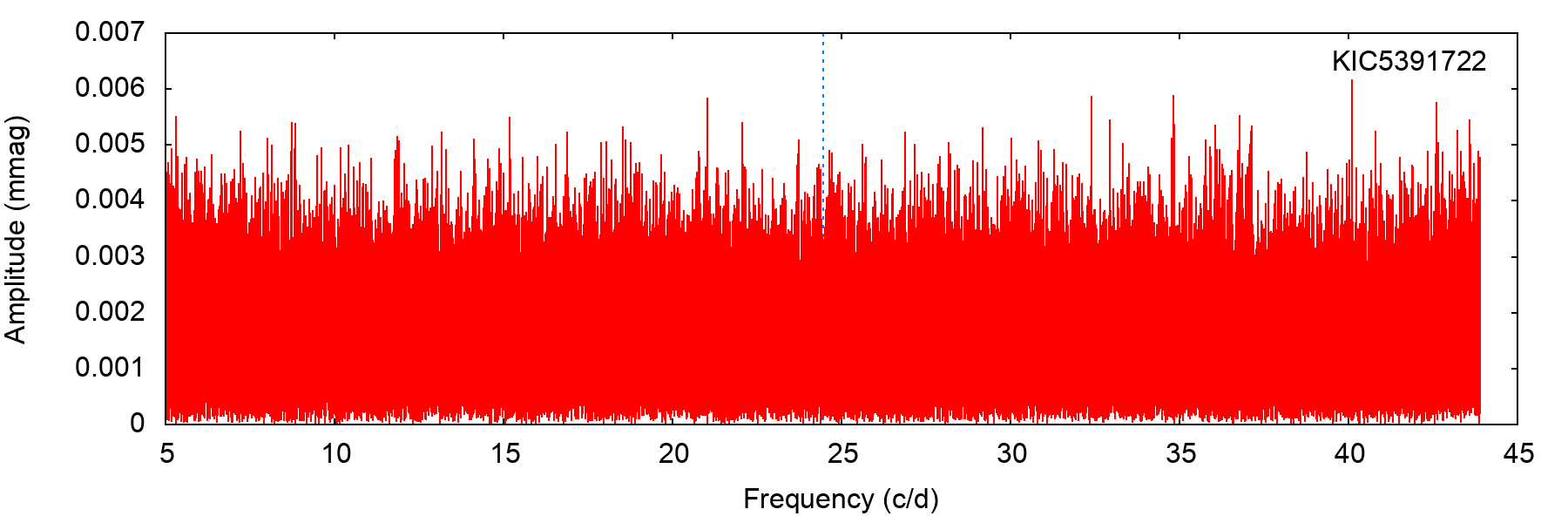}
\includegraphics[width=0.48\textwidth]{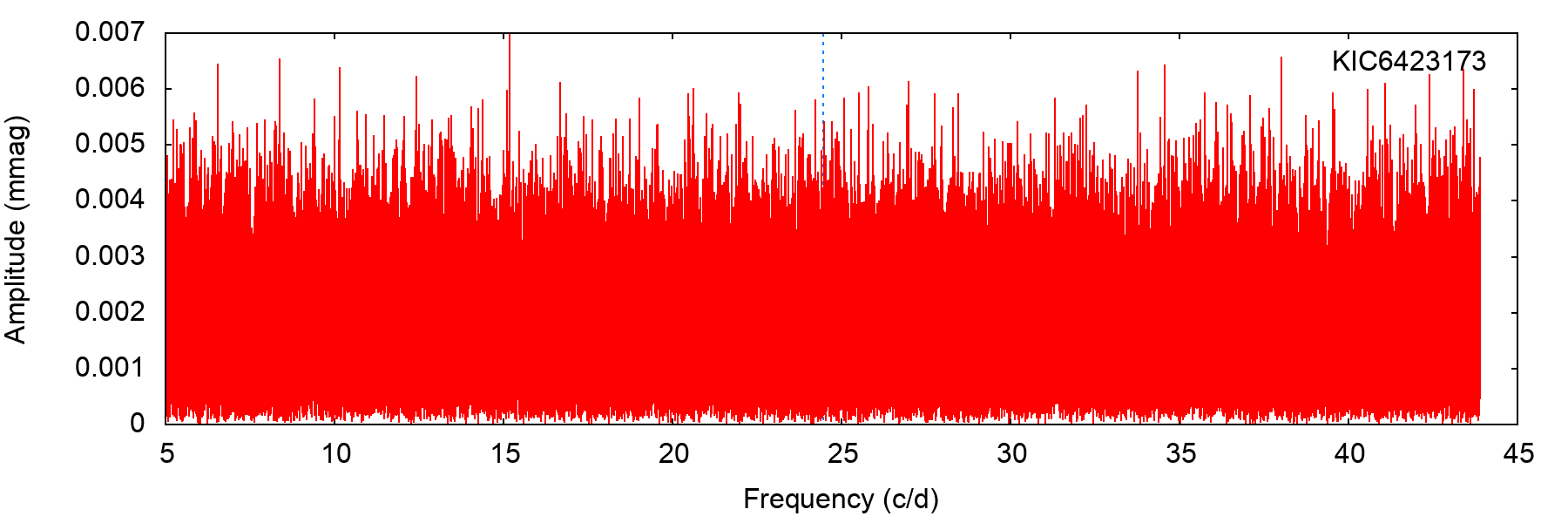}
\includegraphics[width=0.48\textwidth]{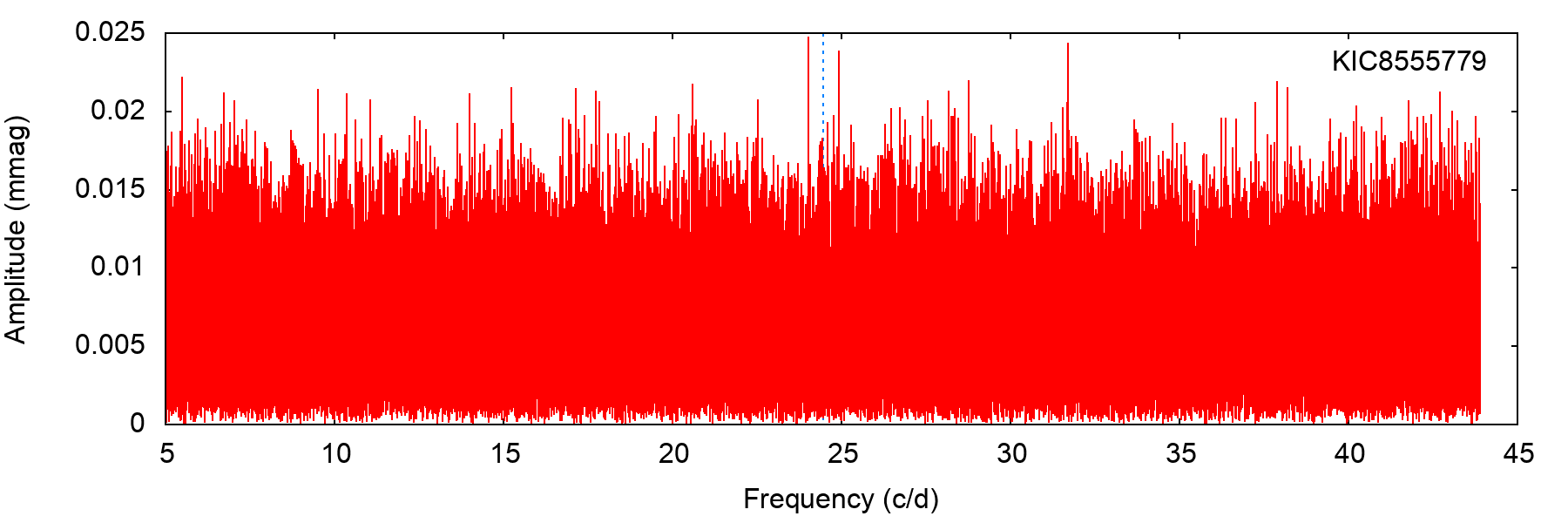}
\includegraphics[width=0.48\textwidth]{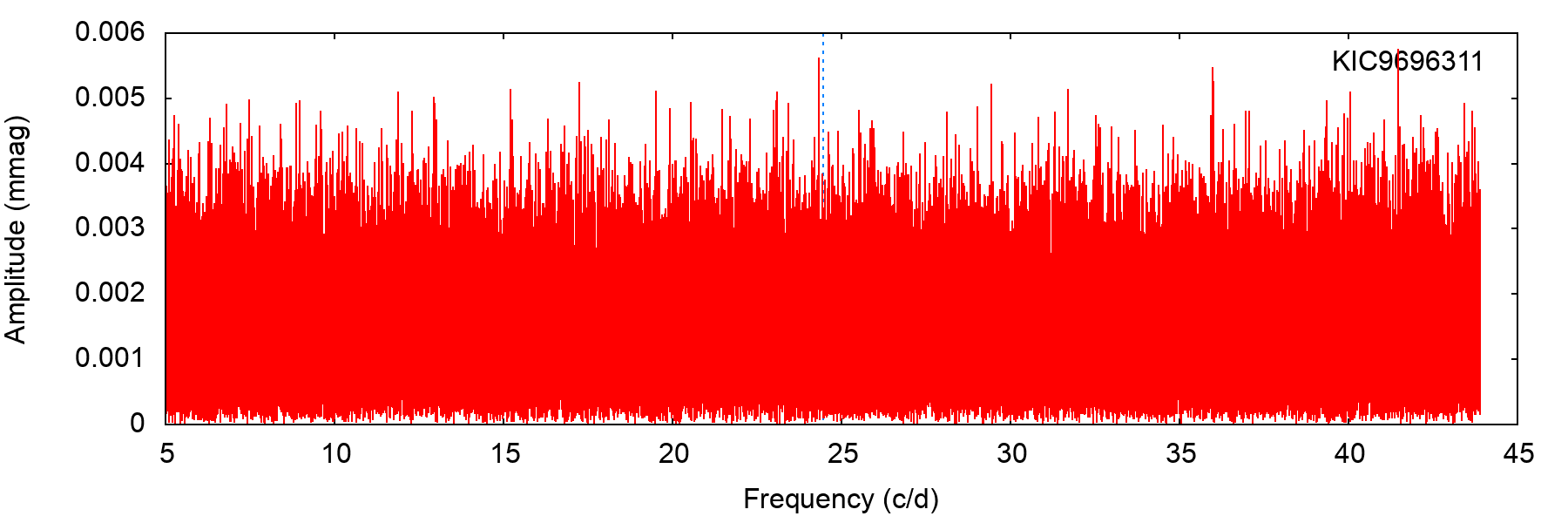}
\includegraphics[width=0.48\textwidth]{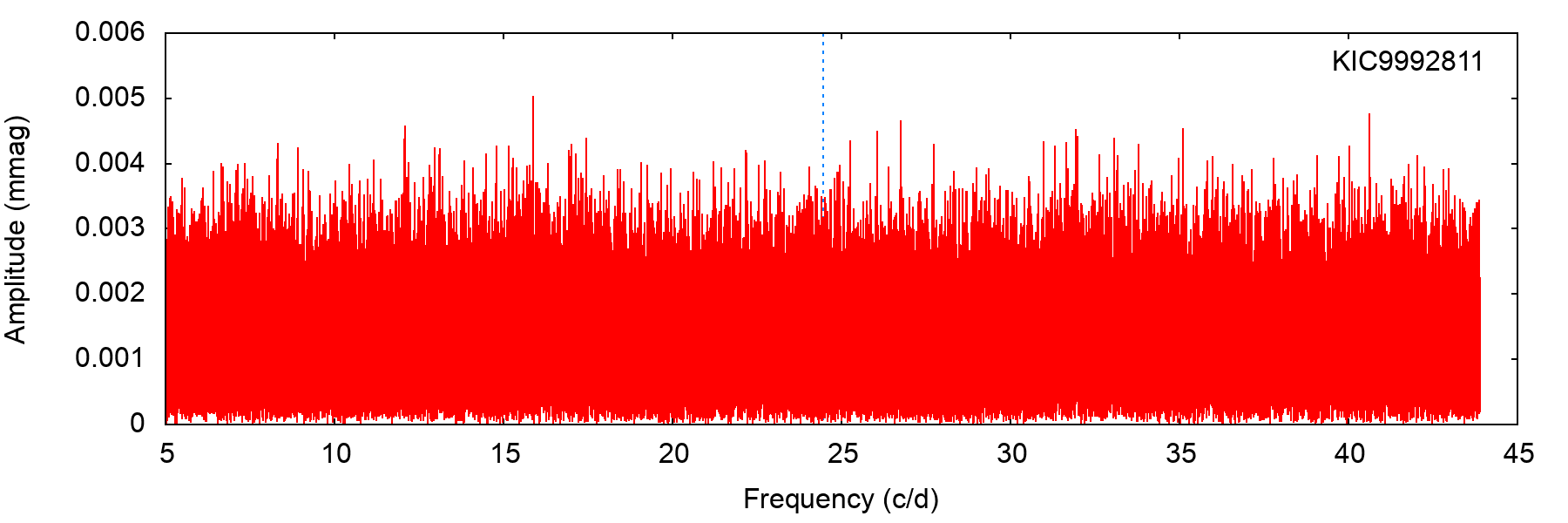}
\caption{Some typical examples of stars in Group I (see Sec.\,\ref{ssec:GroupI}).}
\label{fig:GroupI}
\vspace{-3mm}
\end{center}
\end{figure}

\begin{figure}
\begin{center}
\includegraphics[width=0.48\textwidth]{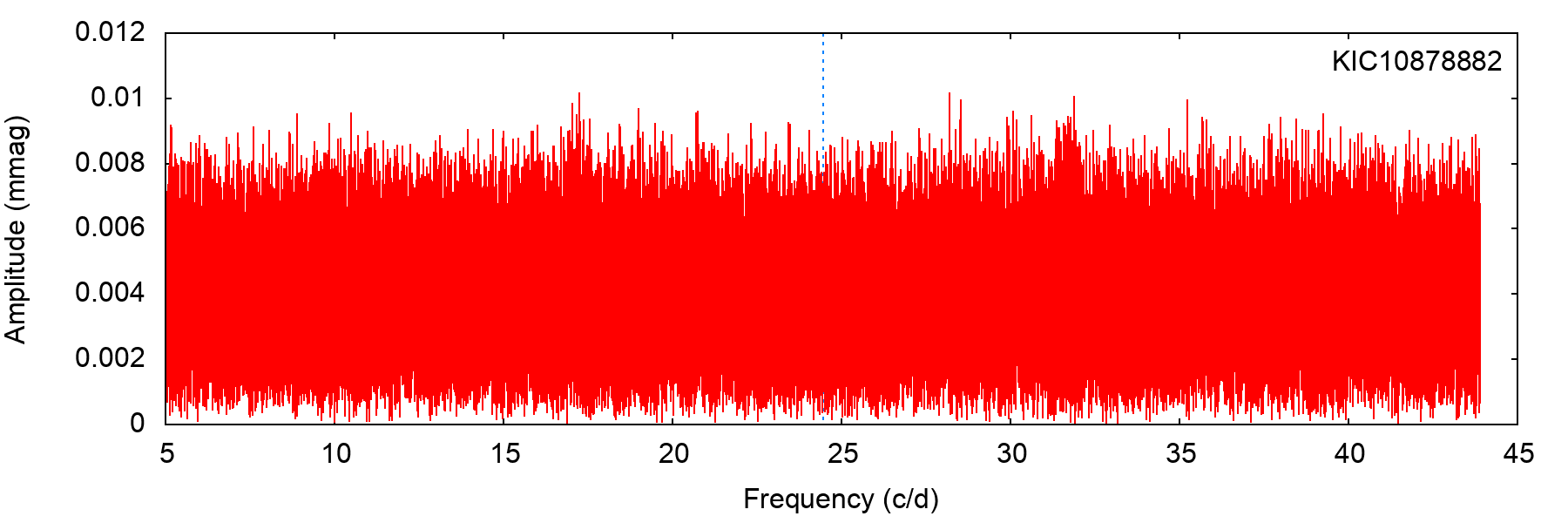}
\includegraphics[width=0.48\textwidth]{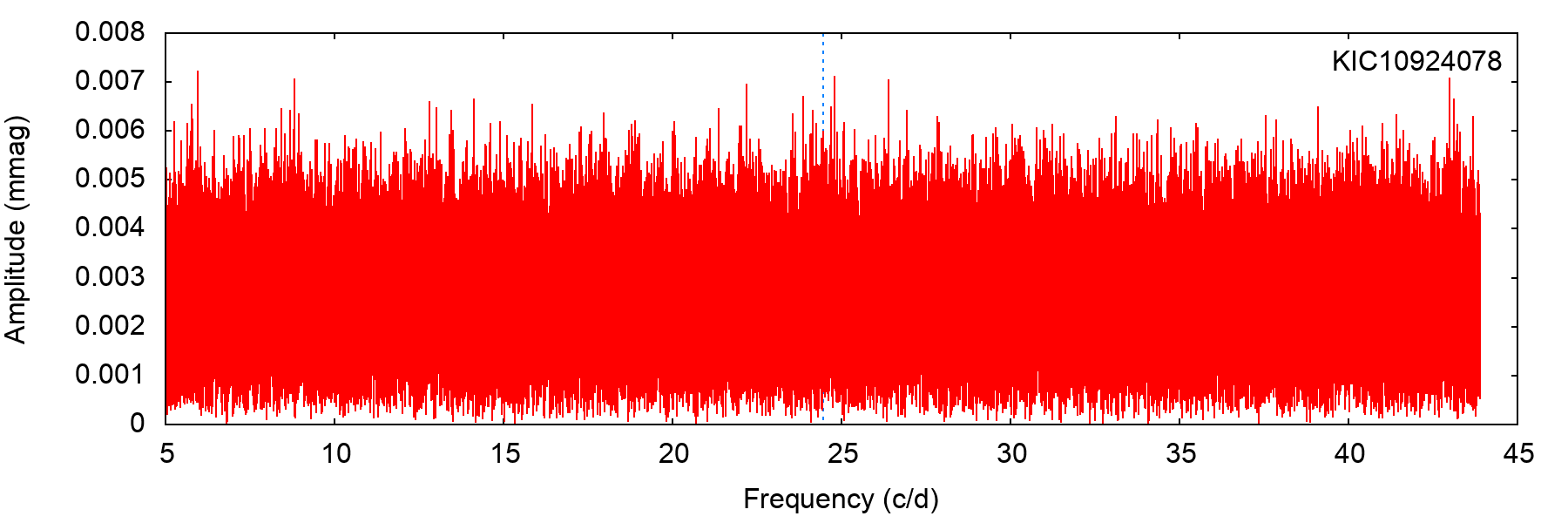}
\includegraphics[width=0.48\textwidth]{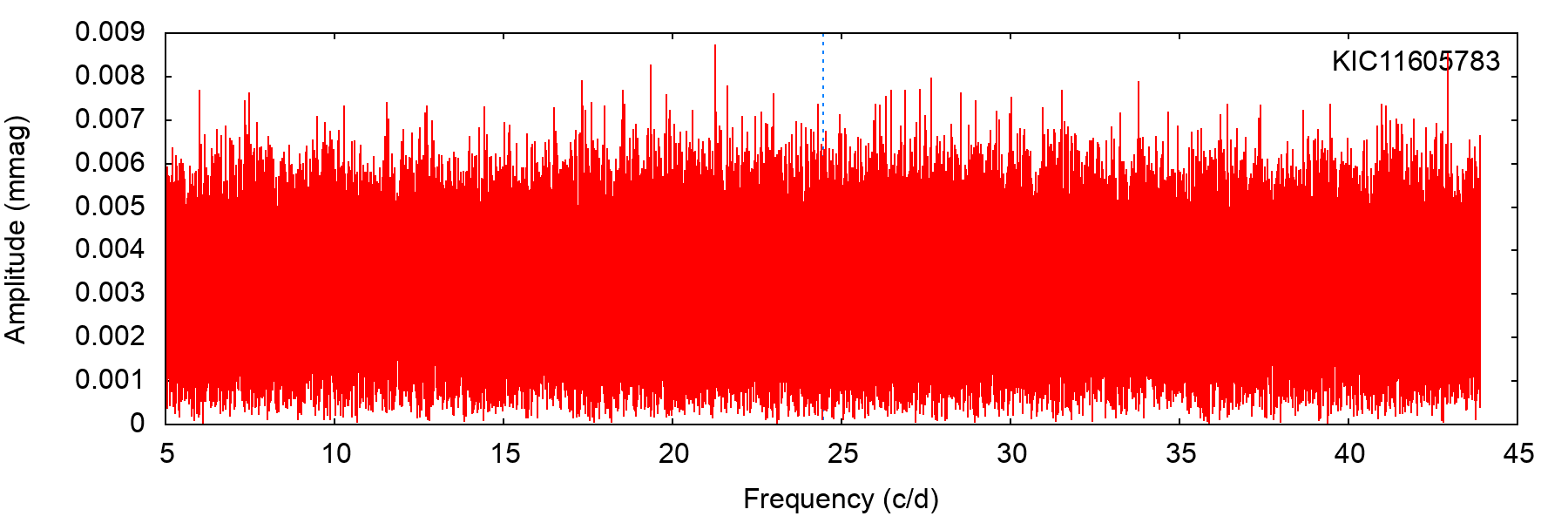}
\includegraphics[width=0.48\textwidth]{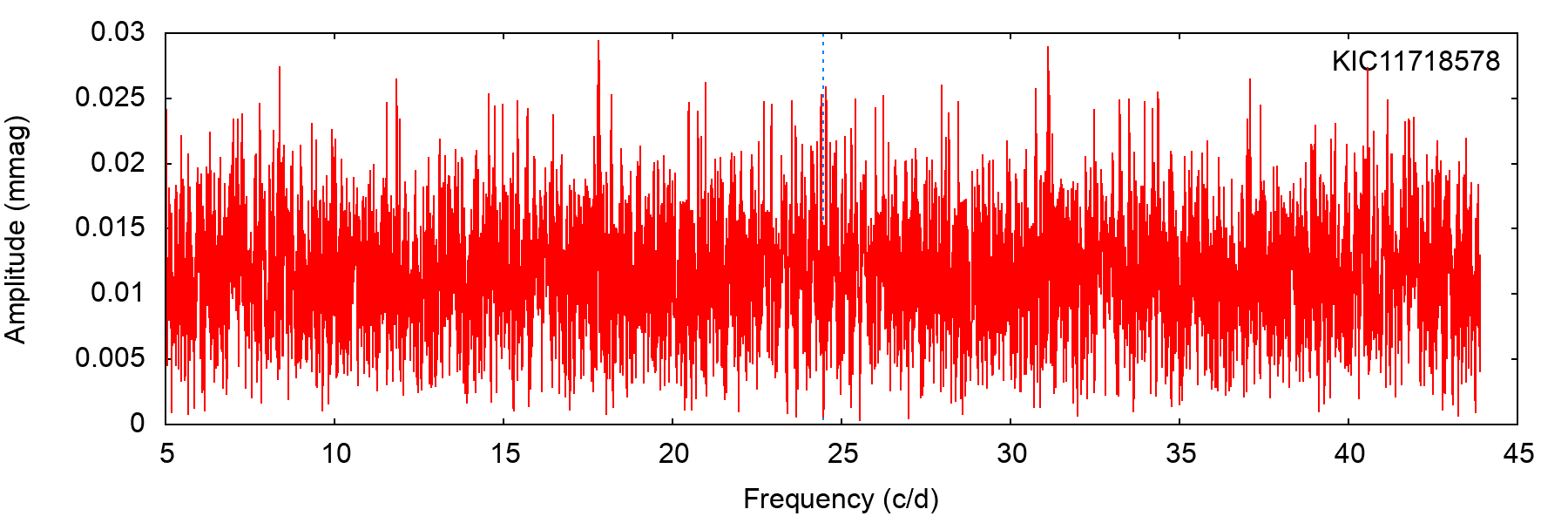}
\includegraphics[width=0.48\textwidth]{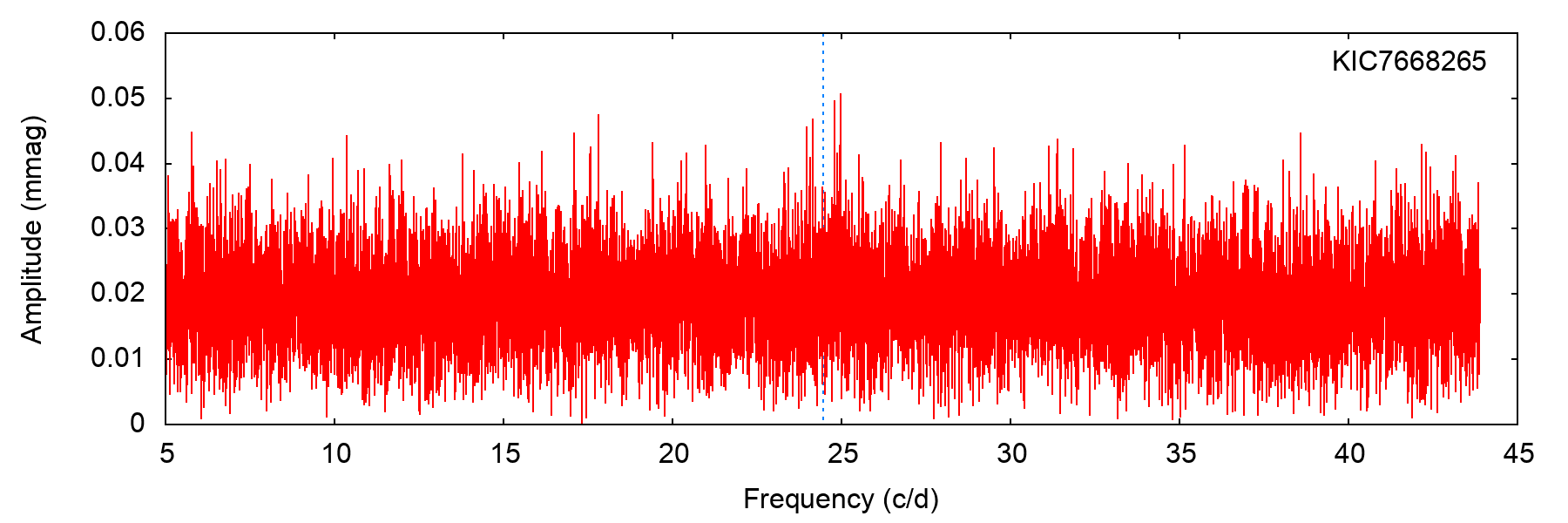}
\includegraphics[width=0.48\textwidth]{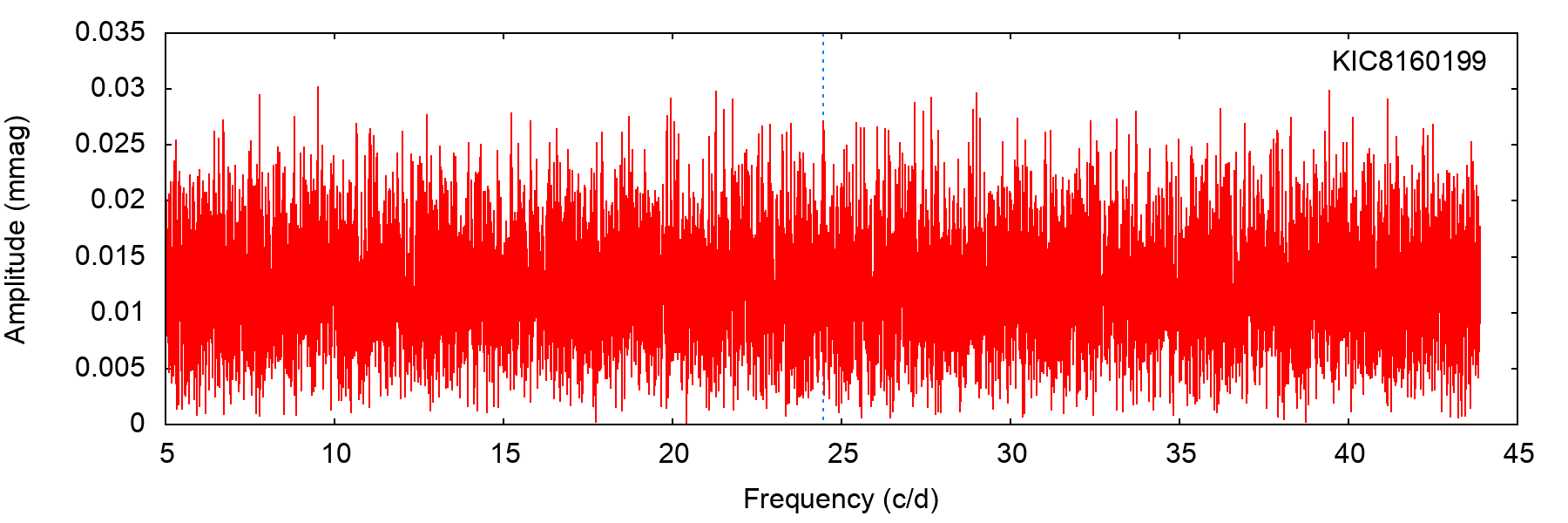}
\includegraphics[width=0.48\textwidth]{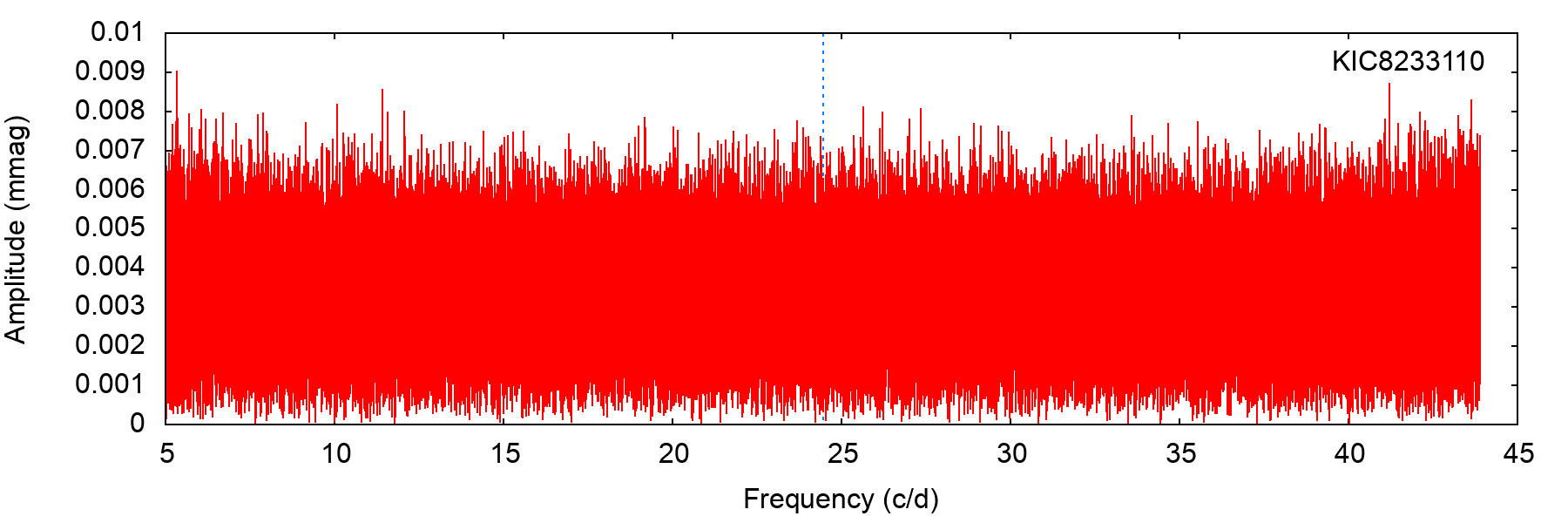}
\includegraphics[width=0.48\textwidth]{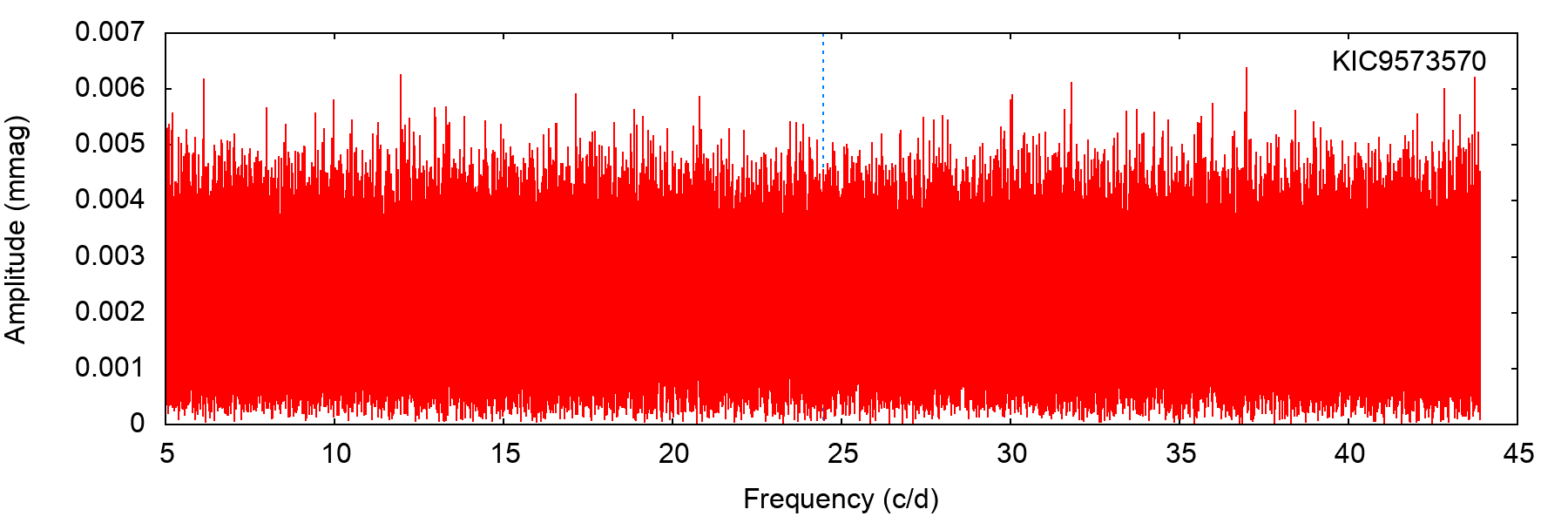}
\caption{Some typical examples of stars in Group J (see Sec.\,\ref{ssec:GroupJ}).}
\label{fig:GroupJ}
\vspace{-3mm}
\end{center}
\end{figure}

A description of the typical pulsation properties in each group of Fig.\,\ref{fig:skew} is given below. Fourier transforms of typical objects in each group are given in Figs\,\ref{fig:GroupA}-\ref{fig:GroupJ}. Although the parameters of skewness and SNR serve well at separating pulsators and non-pulsators, as well as some intermediate groups, there are probably other parameter combinations that better separate these groups, and a third dimension may assist in this considerably.

\subsection{Group A}
\label{ssec:GroupA}
\begin{figure}
\begin{center}
\includegraphics[width=0.48\textwidth]{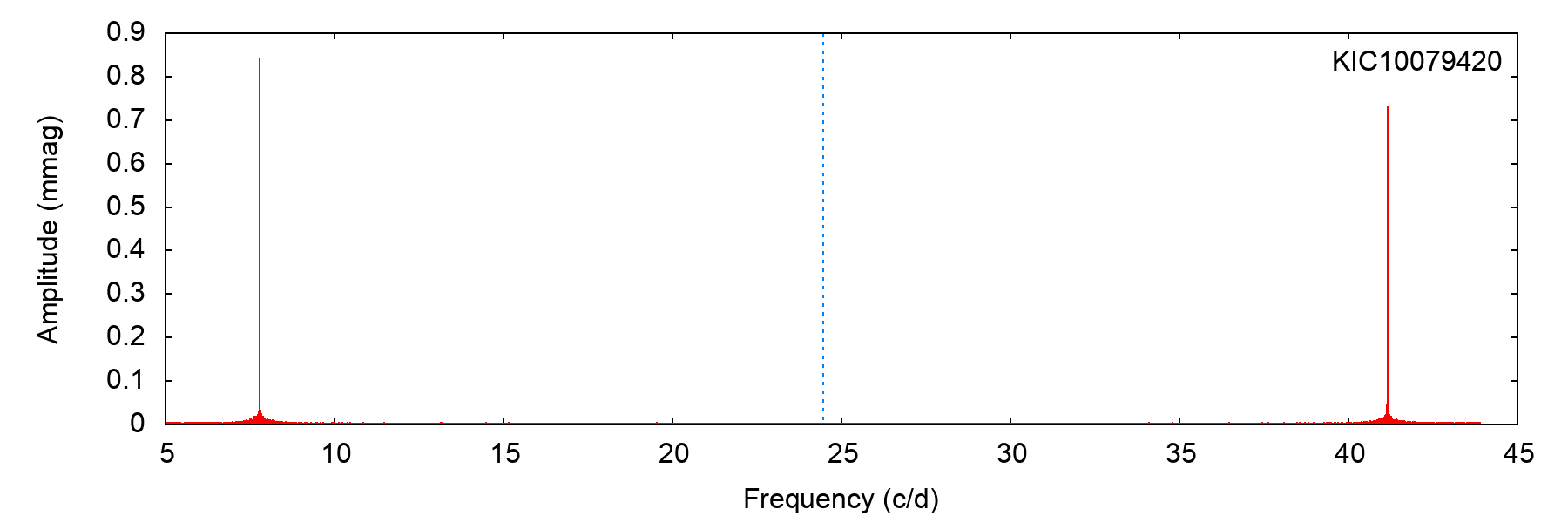}\\
\includegraphics[width=0.48\textwidth]{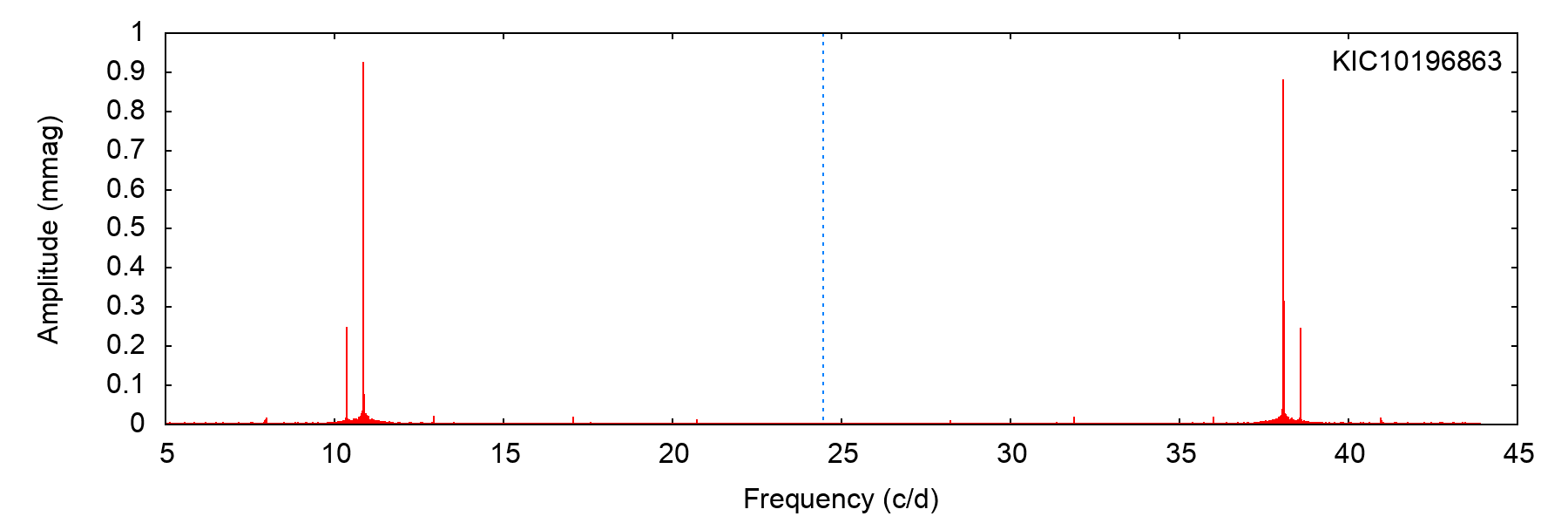}\\
\includegraphics[width=0.48\textwidth]{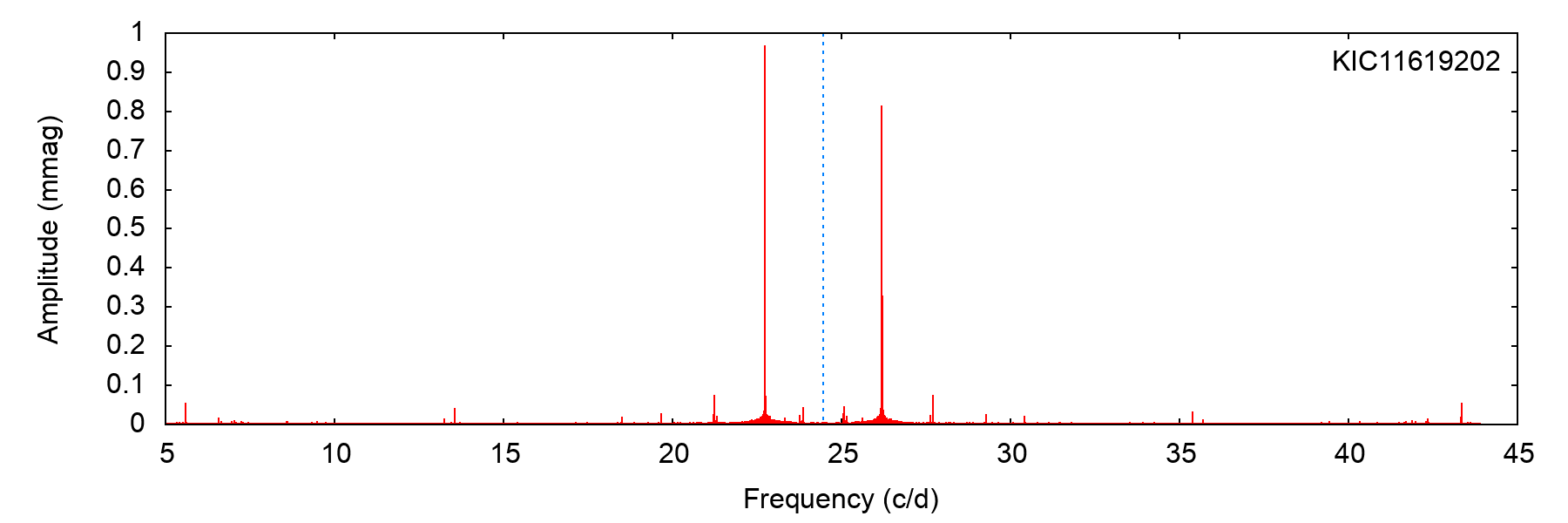}\\
\includegraphics[width=0.48\textwidth]{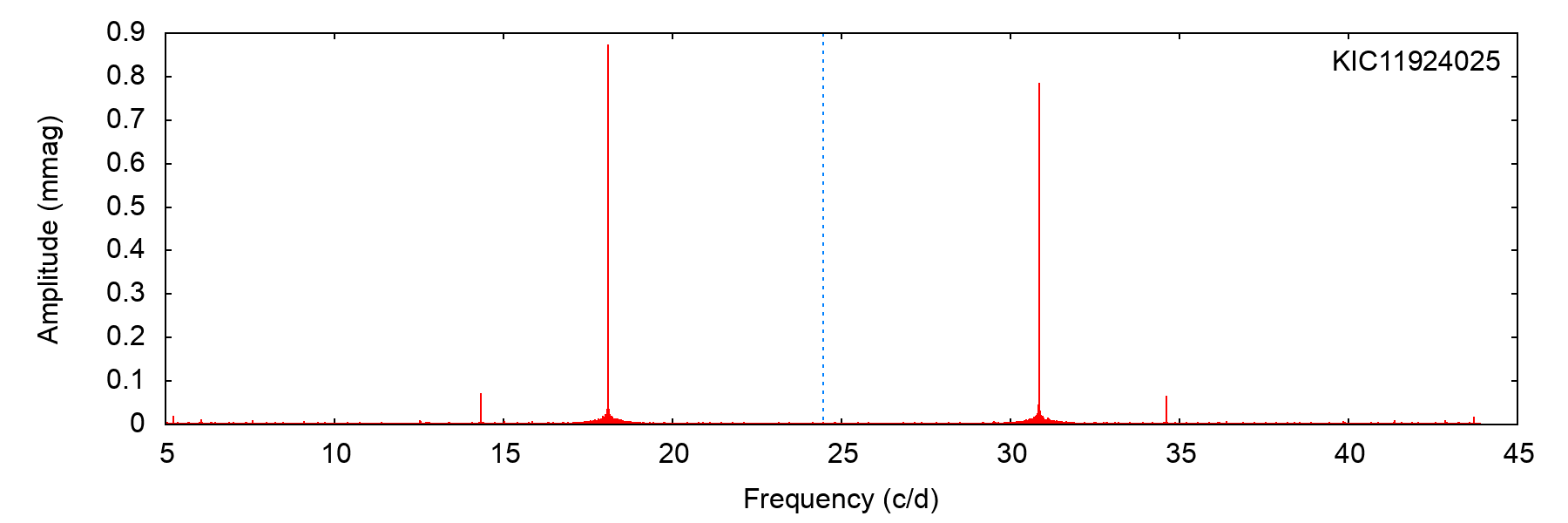}\\
\includegraphics[width=0.48\textwidth]{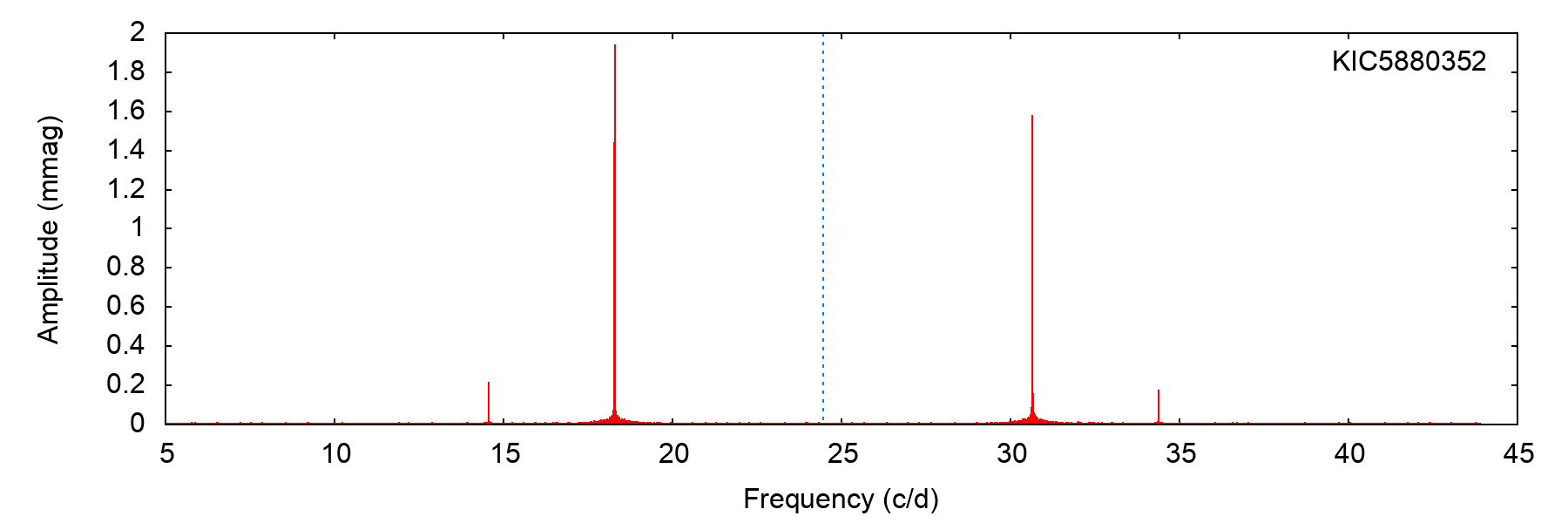}\\
\includegraphics[width=0.48\textwidth]{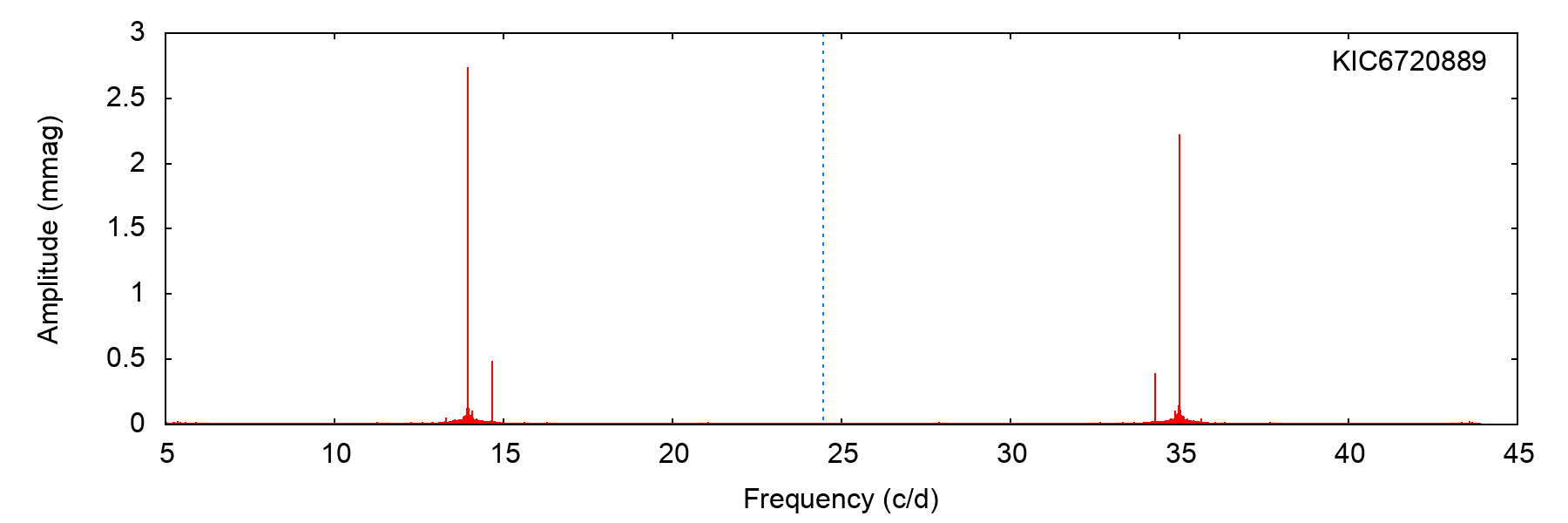}\\
\includegraphics[width=0.48\textwidth]{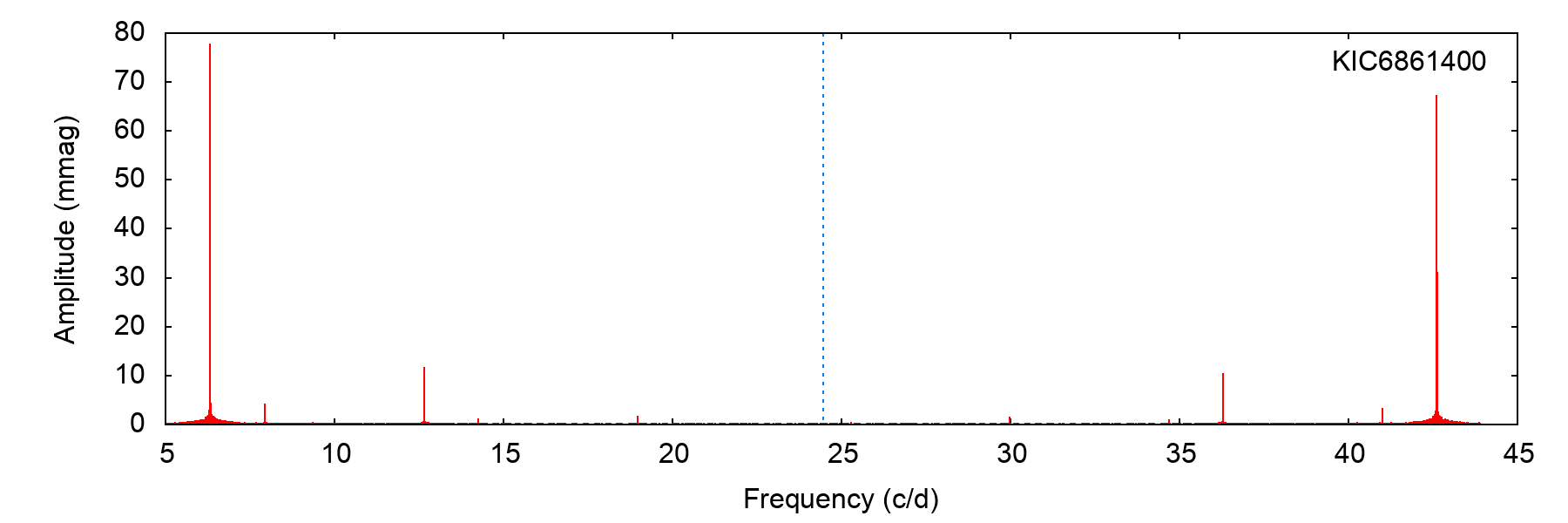}\\
\includegraphics[width=0.48\textwidth]{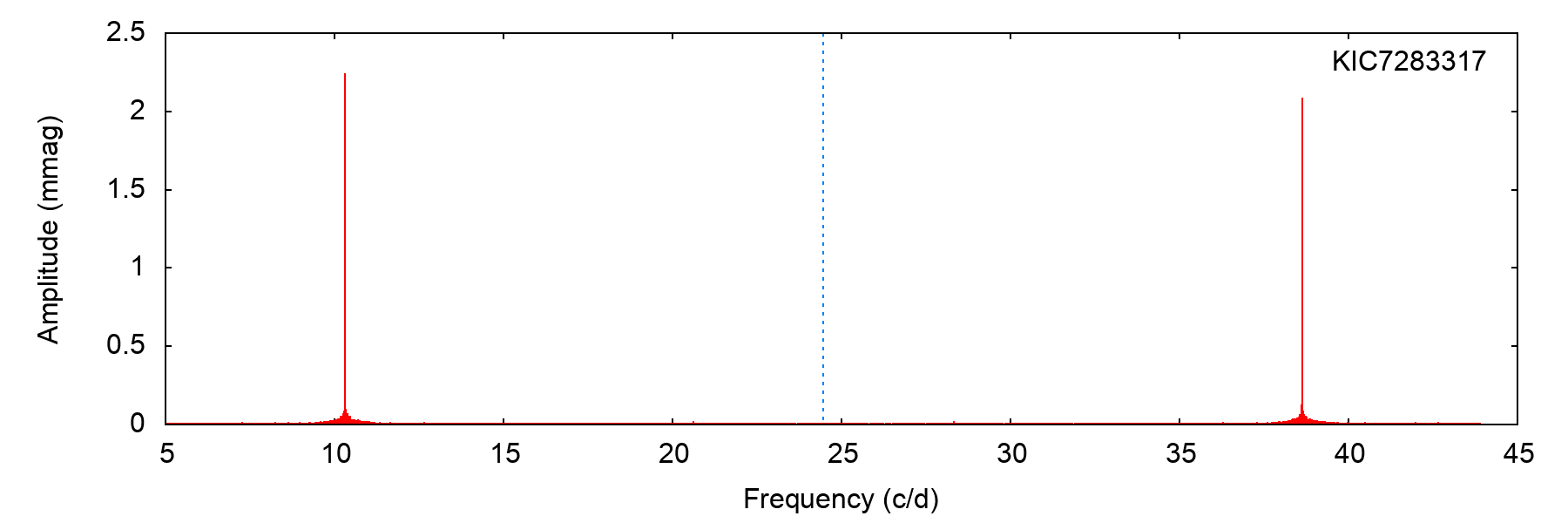}\\
\caption{Some typical examples of stars in Group A (see Sec.\,\ref{ssec:GroupA}).}
\label{fig:GroupA}
\vspace{-3mm}
\end{center}
\end{figure}

Group A contains the Fourier transforms with the highest S/N and also the highest skewness. They are often mono-periodic, or exhibit few pulsation modes (Fig.\,\ref{fig:GroupA}). Their relatively simple Fourier spectra may make them good targets for modelling, on the basis that mode identification should be straight-forward under the assumption that the modes are radial or of low spherical degree, i.e. $\ell = 0, 1$. Two objects in this group, KIC\,5880352 and KIC\,11924025, are remarkably similar.

\subsection{Group B}
\label{ssec:GroupB}
\begin{figure}
\begin{center}
\includegraphics[width=0.48\textwidth]{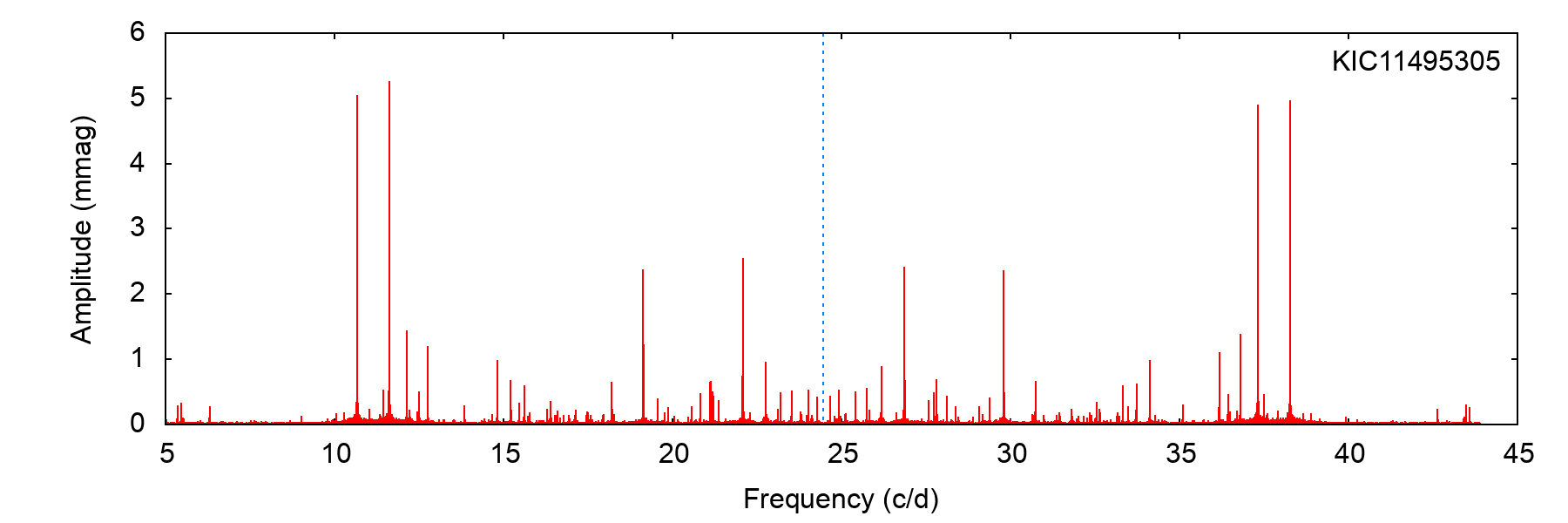}\\
\includegraphics[width=0.48\textwidth]{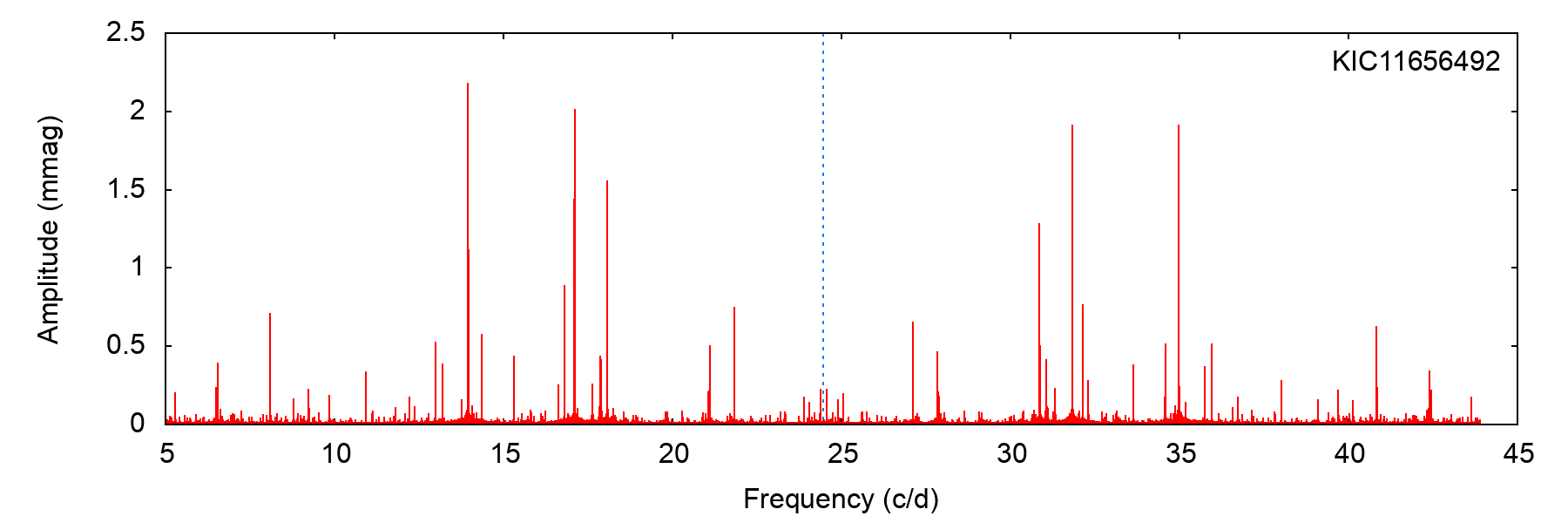}\\
\includegraphics[width=0.48\textwidth]{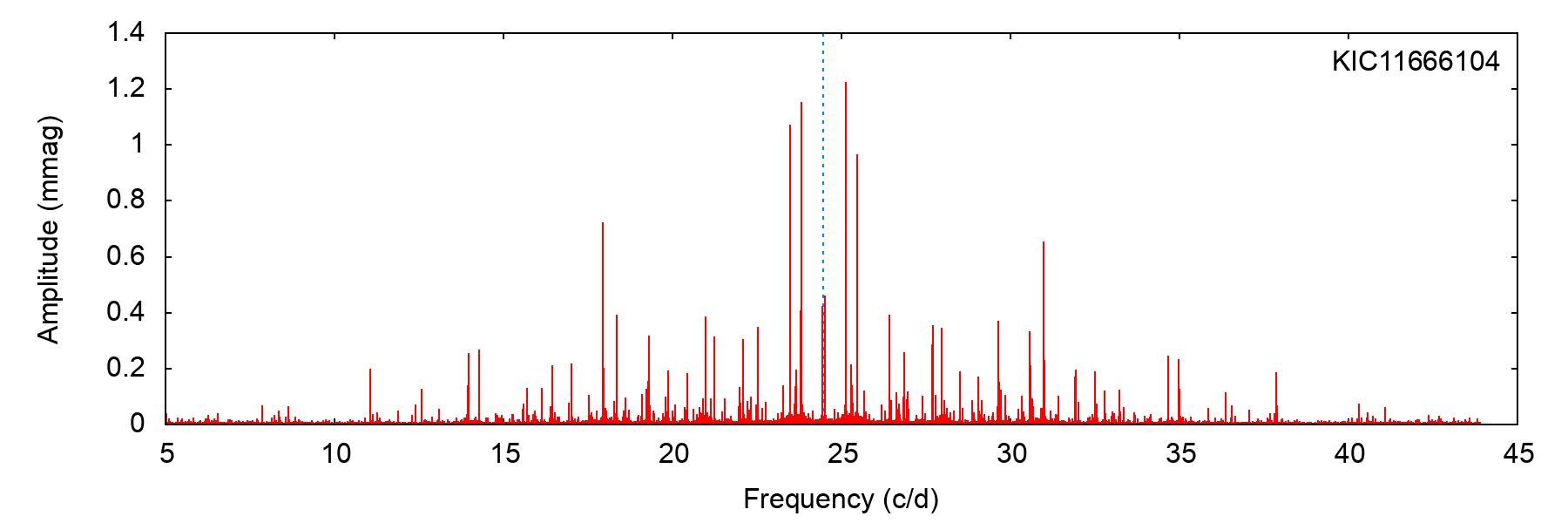}\\
\includegraphics[width=0.48\textwidth]{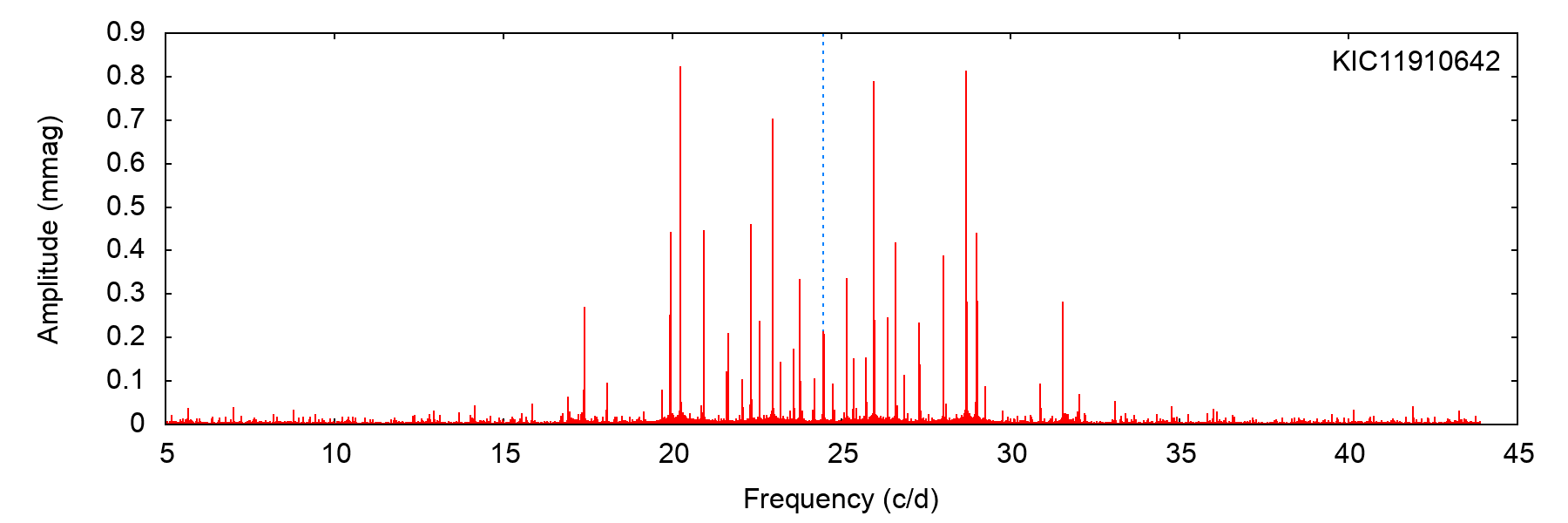}\\
\includegraphics[width=0.48\textwidth]{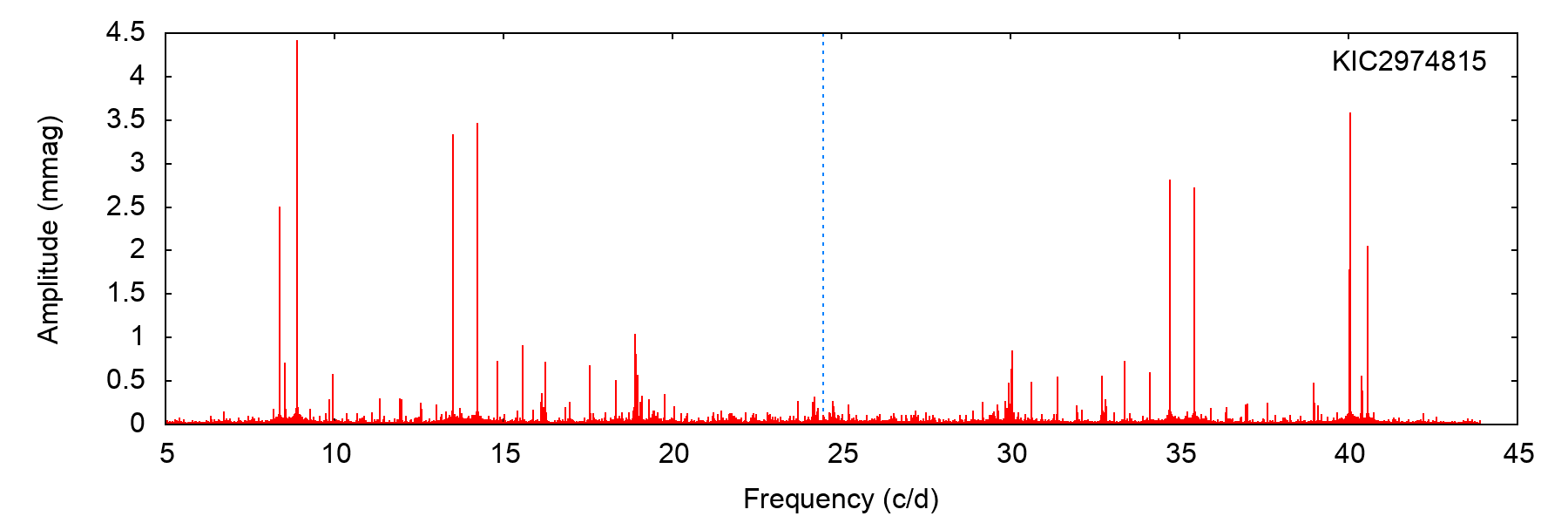}\\
\includegraphics[width=0.48\textwidth]{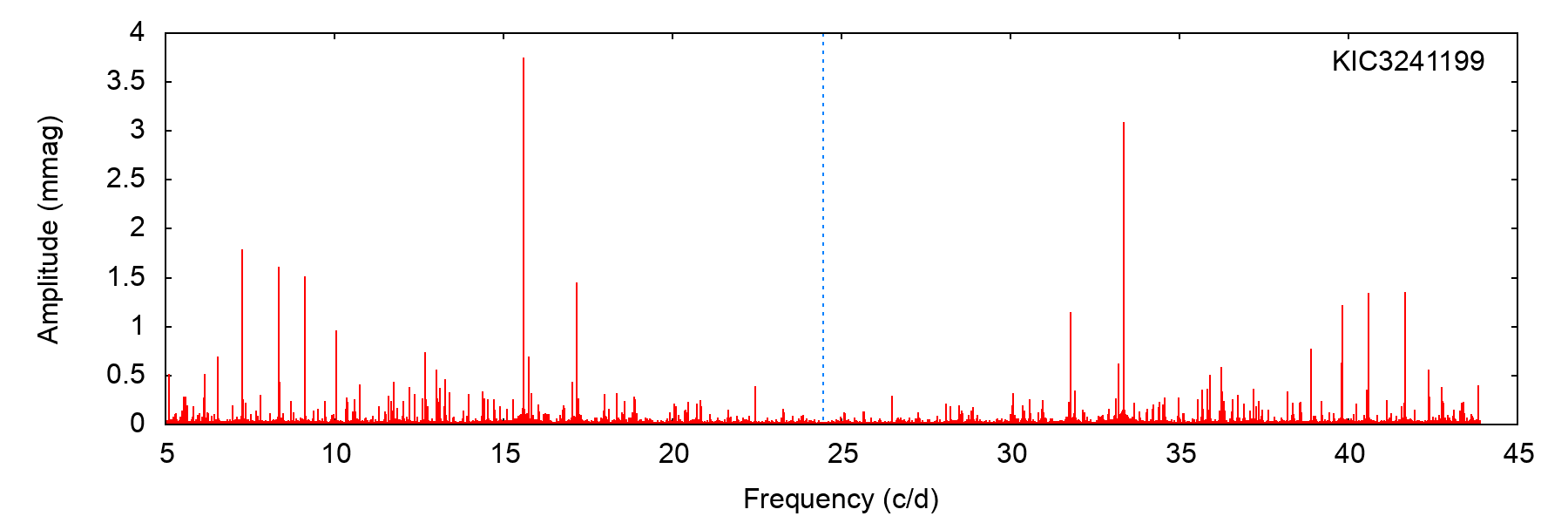}\\
\includegraphics[width=0.48\textwidth]{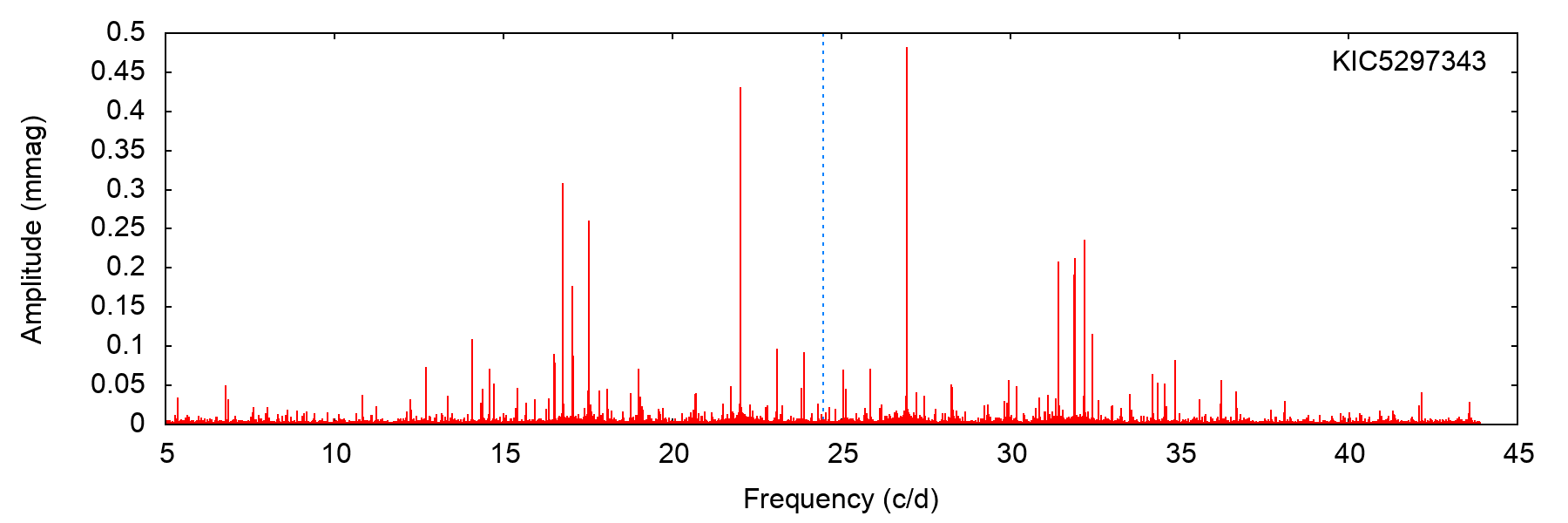}\\ \includegraphics[width=0.48\textwidth]{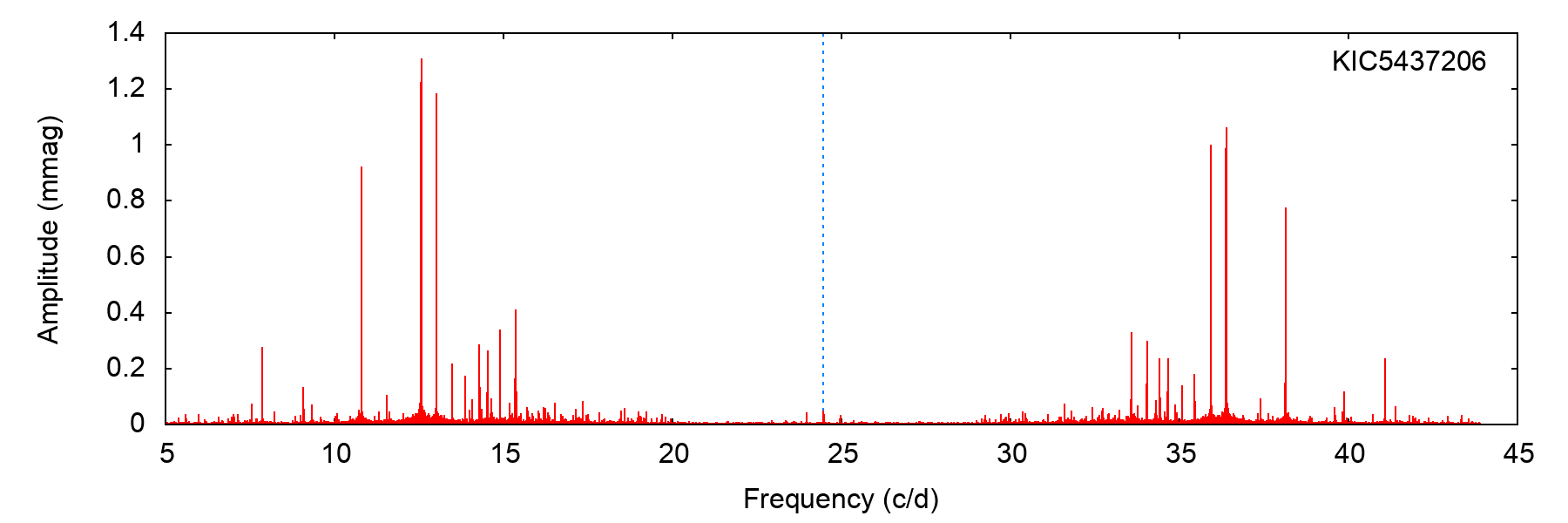}\\
\caption{Some typical examples of stars in Group B (see Sec.\,\ref{ssec:GroupB}).}
\label{fig:GroupB}
\vspace{-3mm}
\end{center}
\end{figure}

Group B lies in the centre of the pulsators cluster, and is characterised by highly multi-modal pulsation (Fig.\,\ref{fig:GroupB}). Oscillation frequencies commonly exceed the \textit{Kepler} LC Nyquist frequency. The high mode-density would make mode identification difficult.

\subsection{Group C}
\label{ssec:GroupC}
\begin{figure}
\begin{center}
\includegraphics[width=0.48\textwidth]{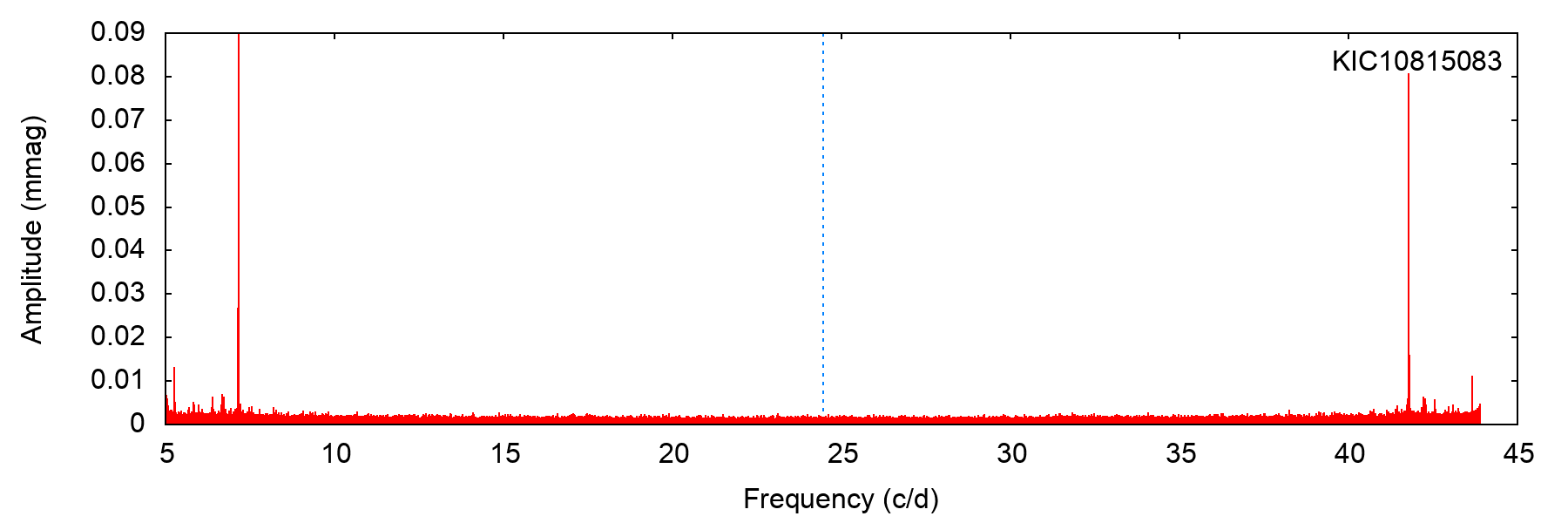}\\
\includegraphics[width=0.48\textwidth]{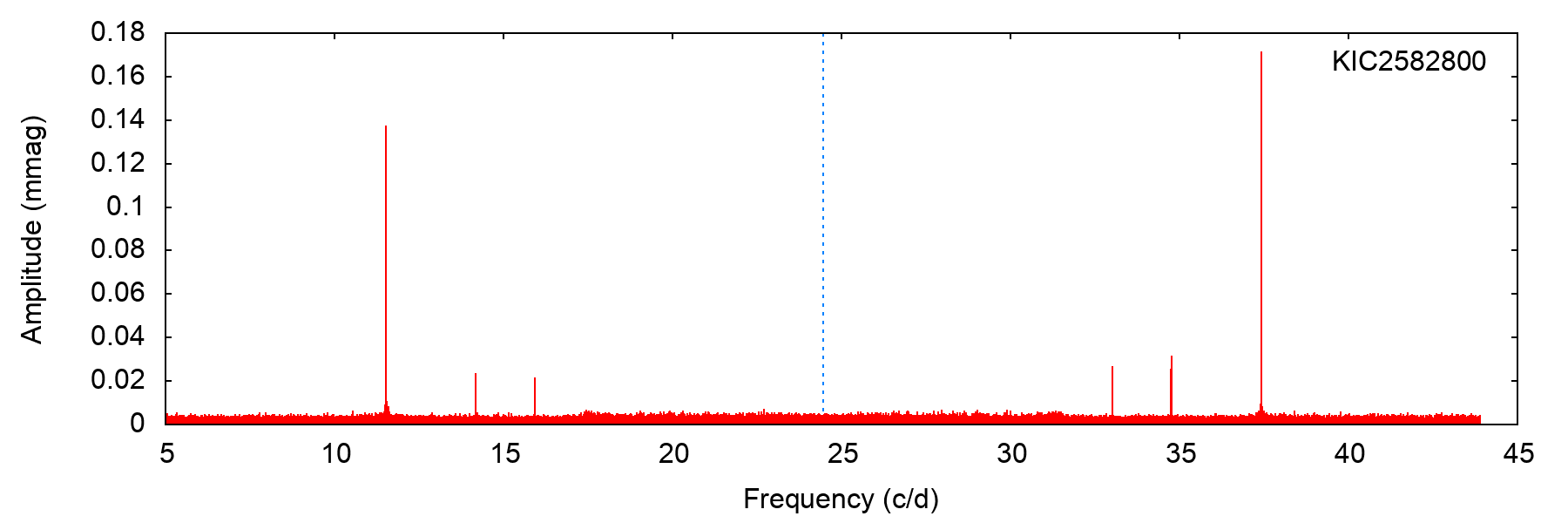}\\
\includegraphics[width=0.48\textwidth]{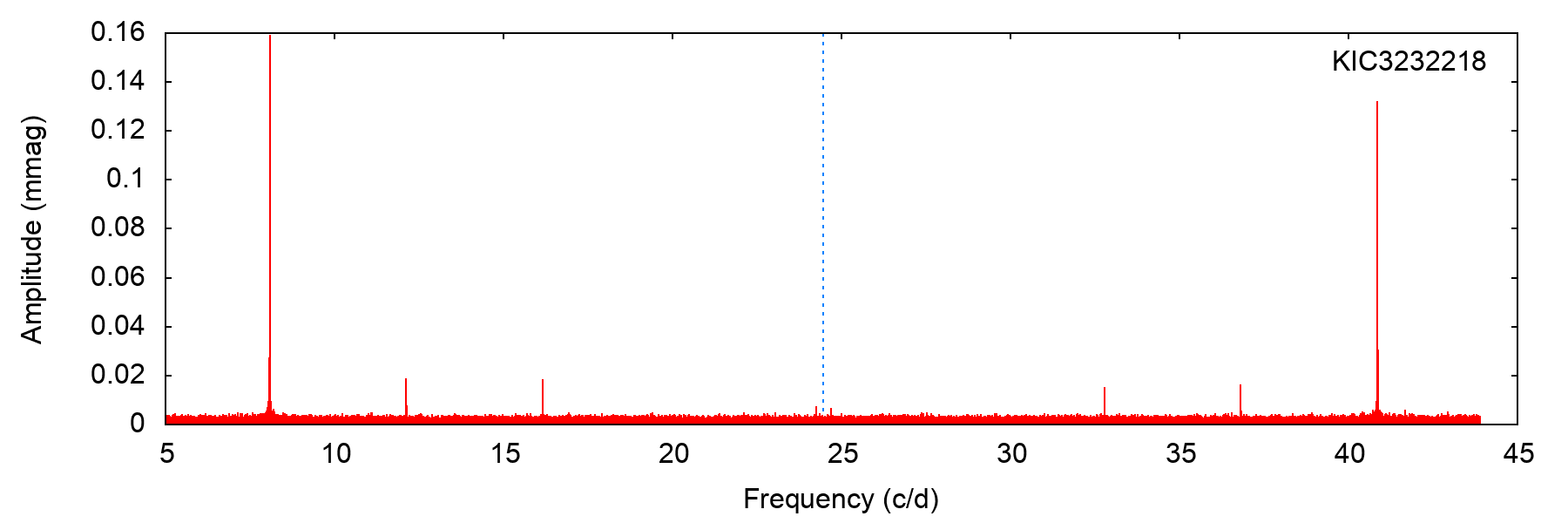}\\ 
\includegraphics[width=0.48\textwidth]{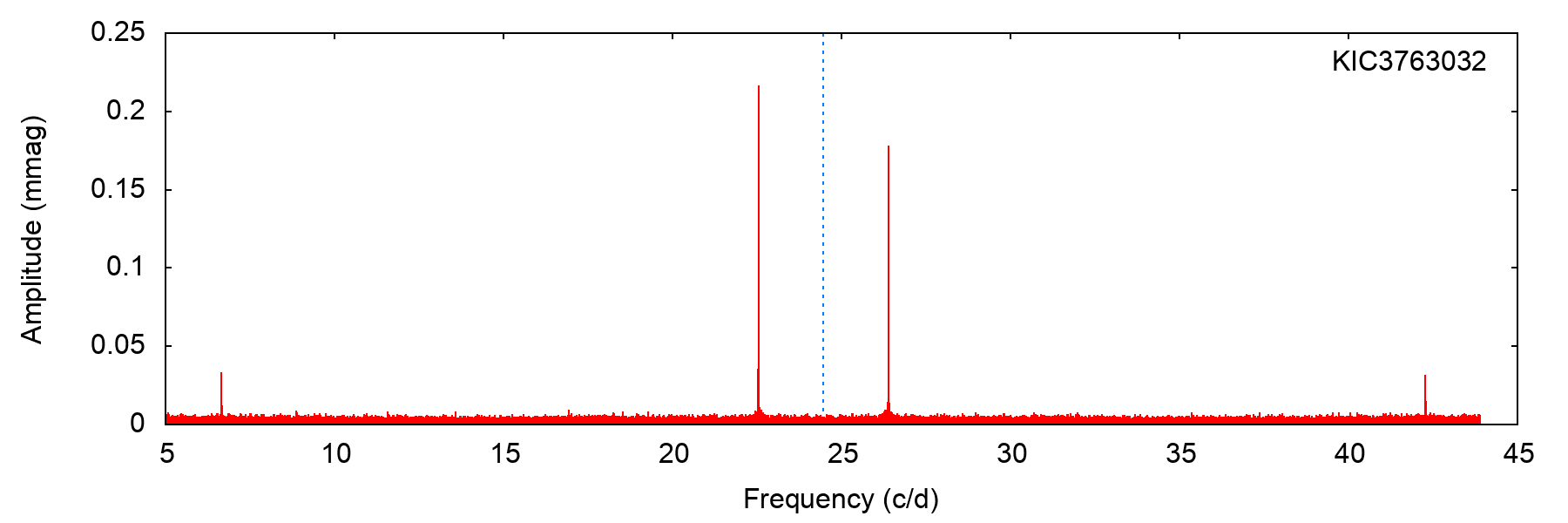}\\
\includegraphics[width=0.48\textwidth]{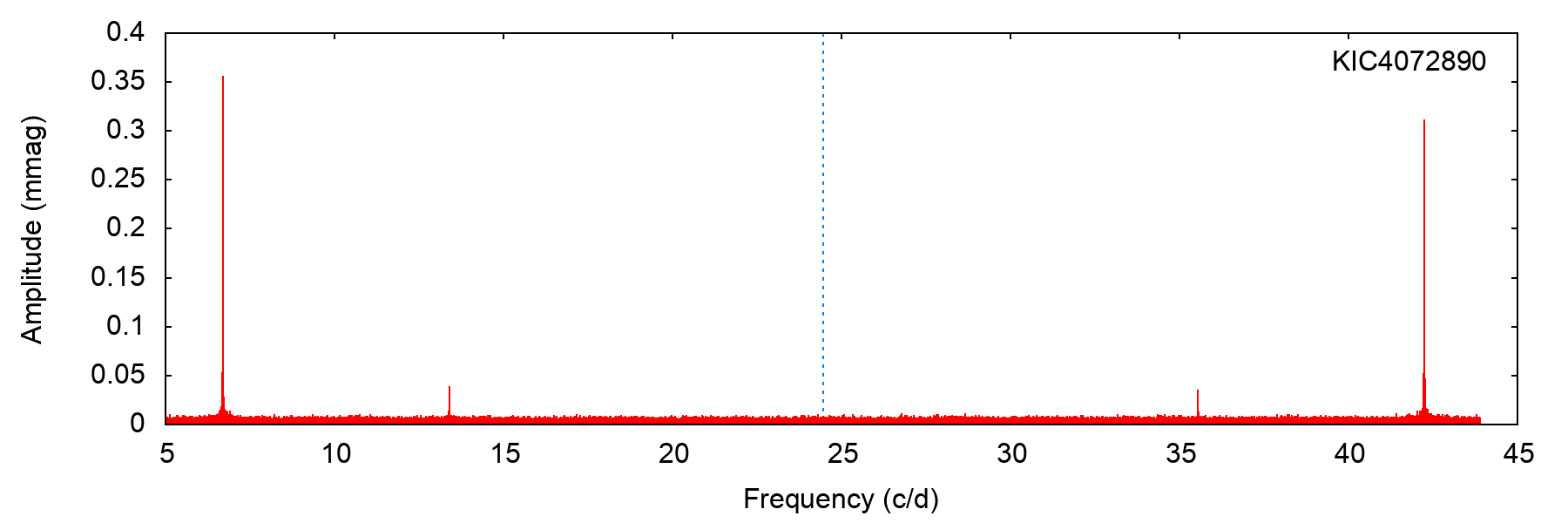}\\
\includegraphics[width=0.48\textwidth]{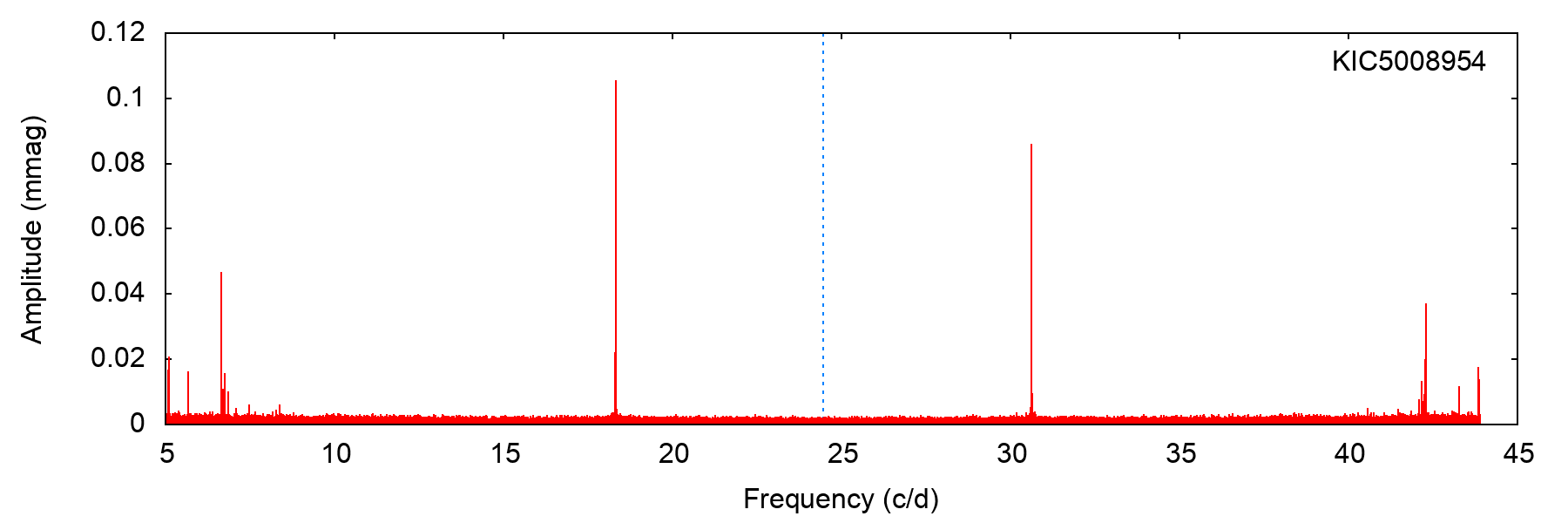}\\
\includegraphics[width=0.48\textwidth]{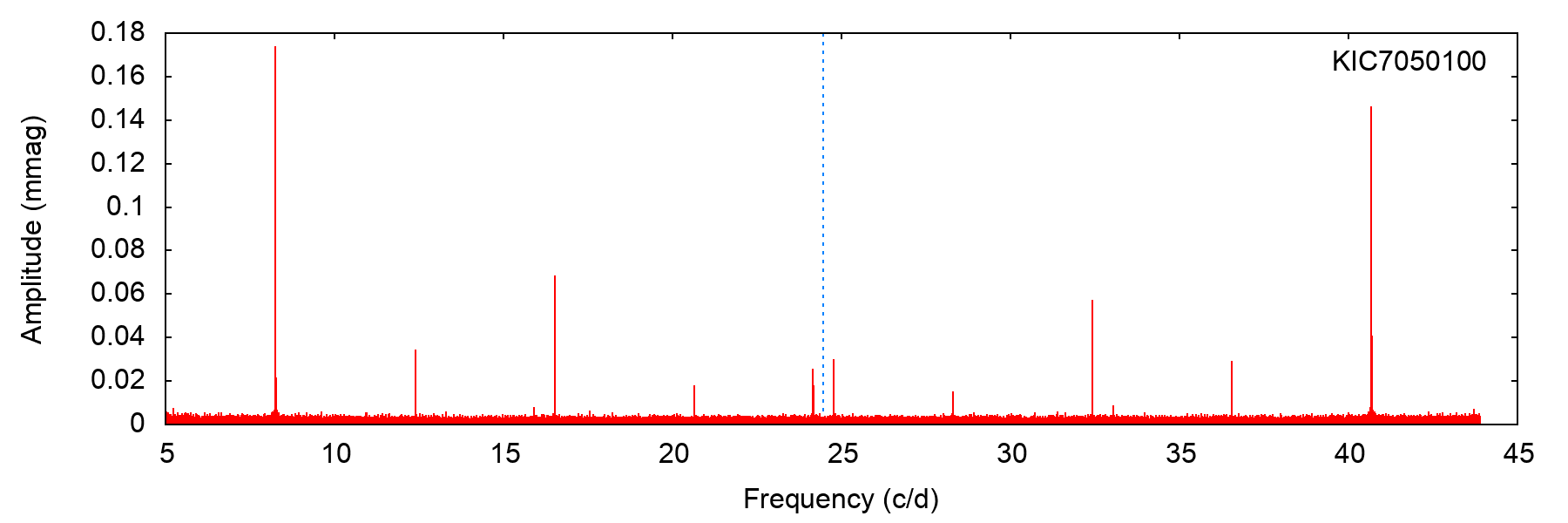}\\
\includegraphics[width=0.48\textwidth]{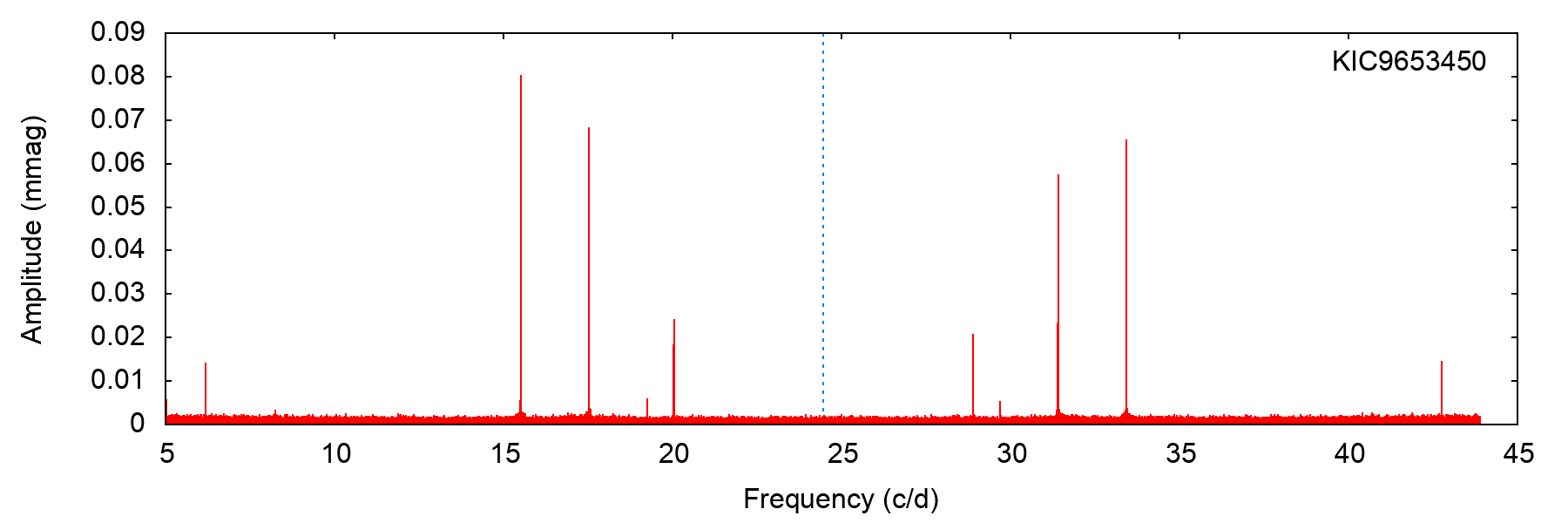}\\
\caption{Some typical examples of stars in Group C (see Sec.\,\ref{ssec:GroupC}).}
\label{fig:GroupC}
\vspace{-3mm}
\end{center}
\end{figure}

This group has high skewness but lower S/N than Group B. Here, stars exhibit only a few significant peaks, but of much lower amplitude than objects in Group A (Fig.\,\ref{fig:GroupC}). While most are $\delta$\,Sct stars, some objects here only have Fourier peaks near the 5\,d$^{-1}$ limit and are $\gamma$\,Dor stars with harmonics.

\subsection{Group D}
\label{ssec:GroupD}
\begin{figure}
\begin{center}
\includegraphics[width=0.48\textwidth]{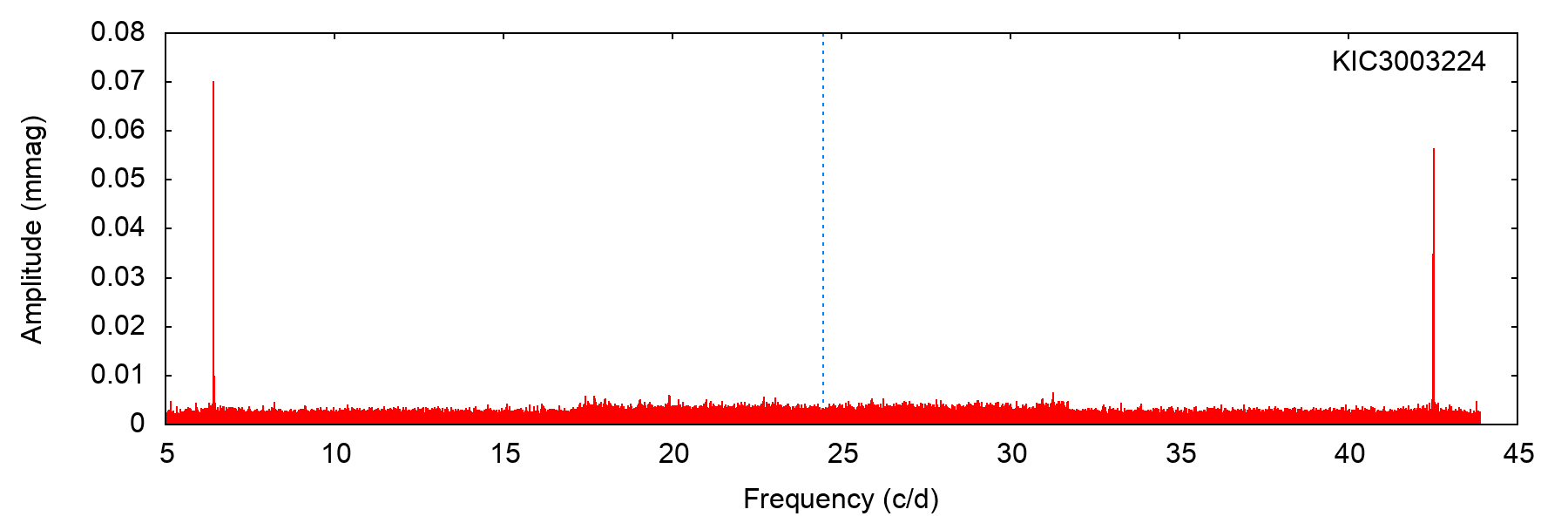}\\
\includegraphics[width=0.48\textwidth]{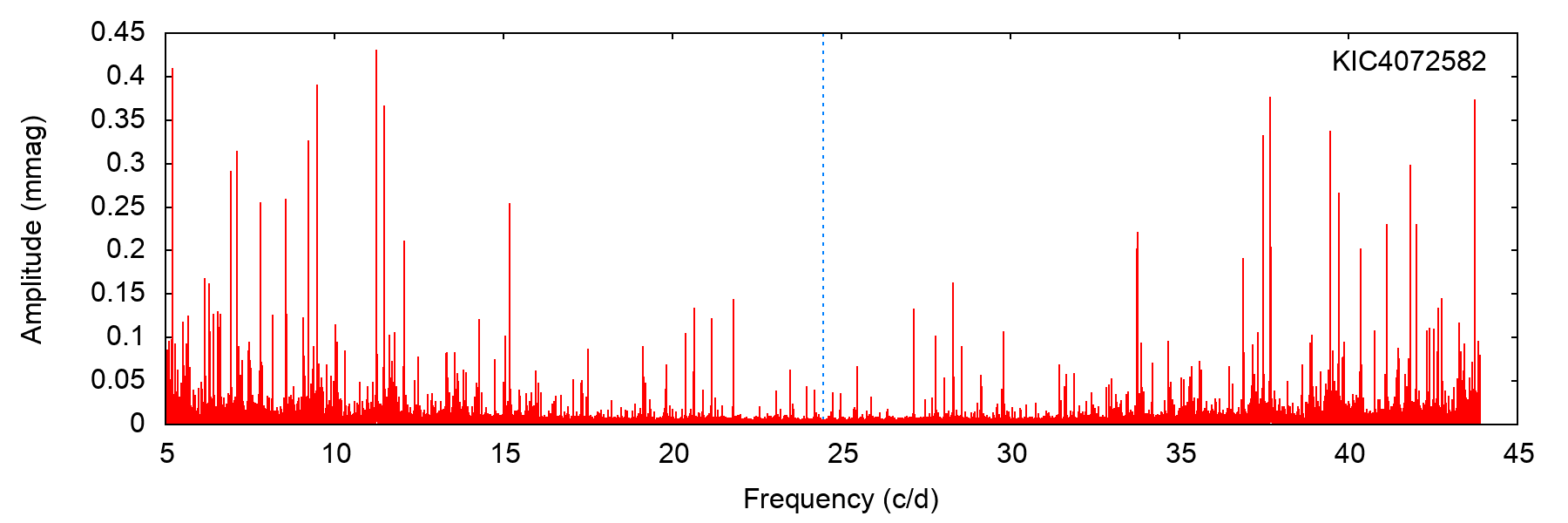}\\
\includegraphics[width=0.48\textwidth]{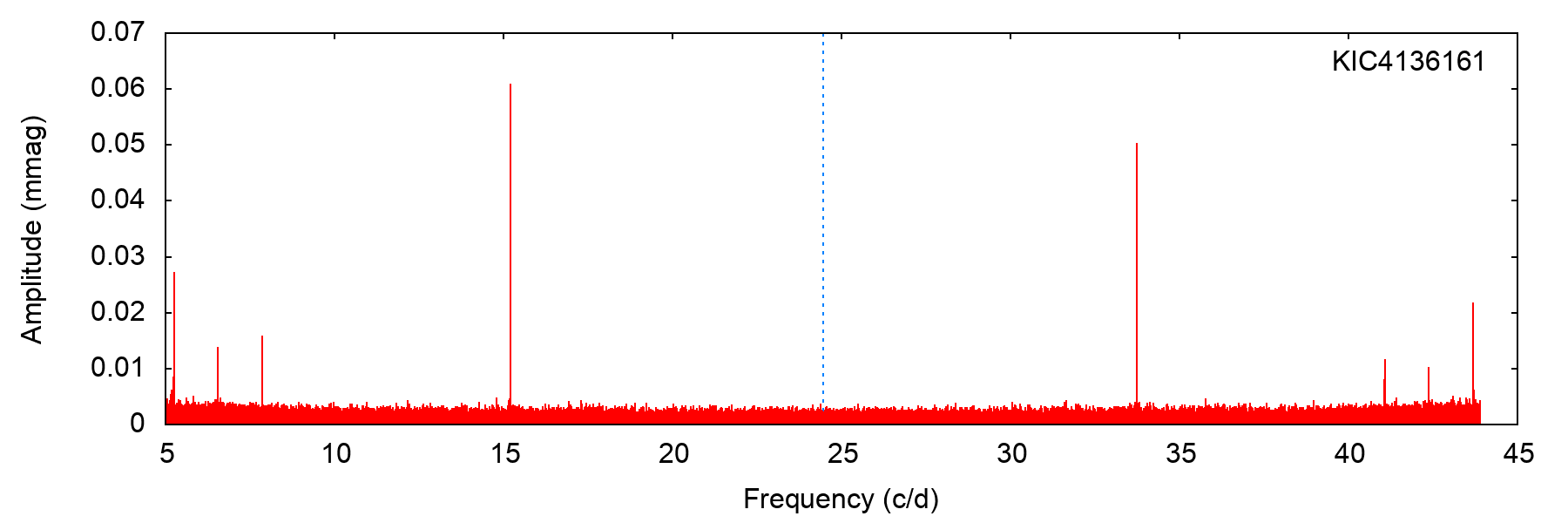}\\
\includegraphics[width=0.48\textwidth]{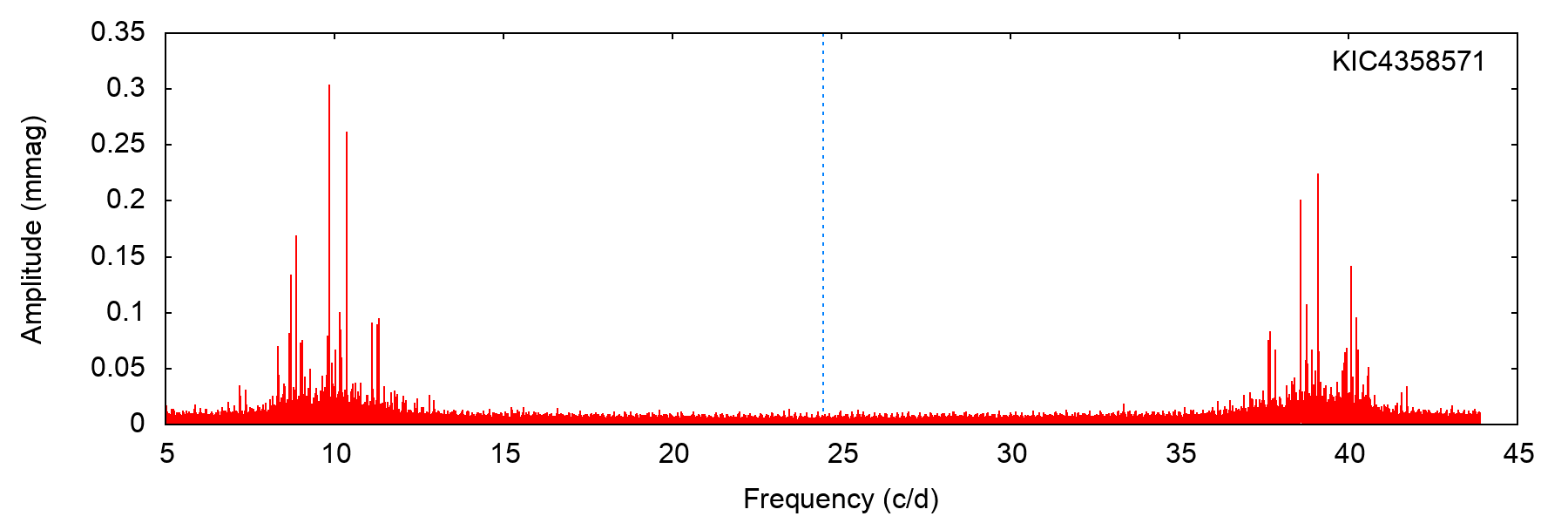}\\
\includegraphics[width=0.48\textwidth]{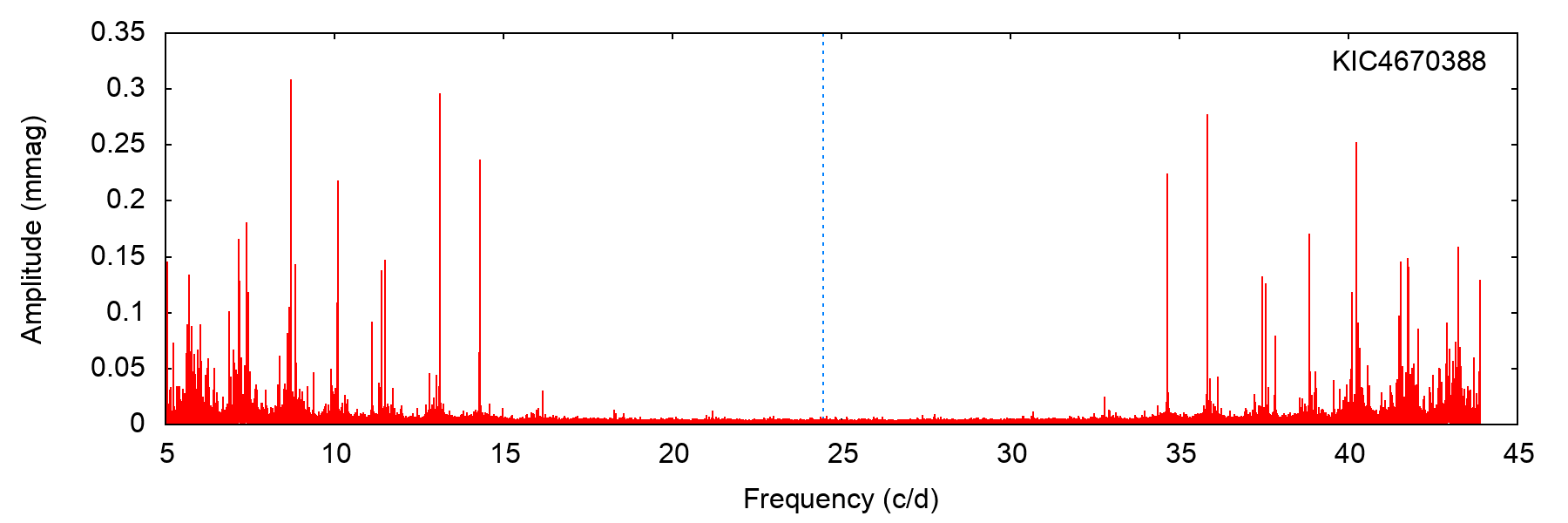}\\
\includegraphics[width=0.48\textwidth]{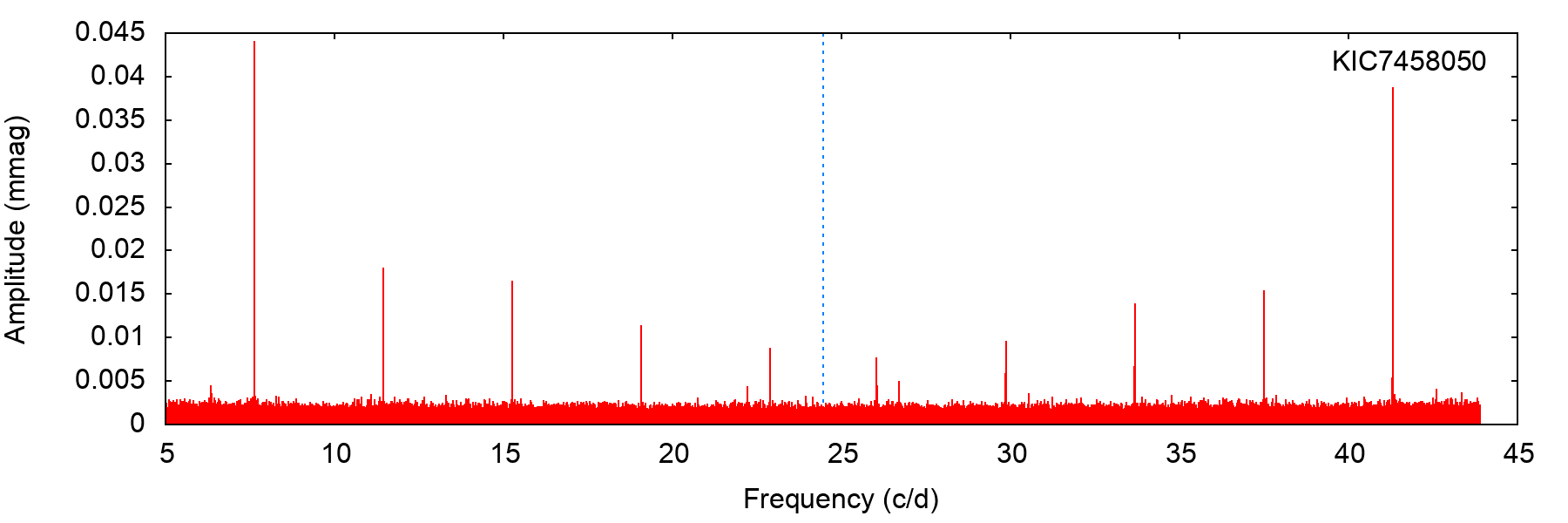}\\
\includegraphics[width=0.48\textwidth]{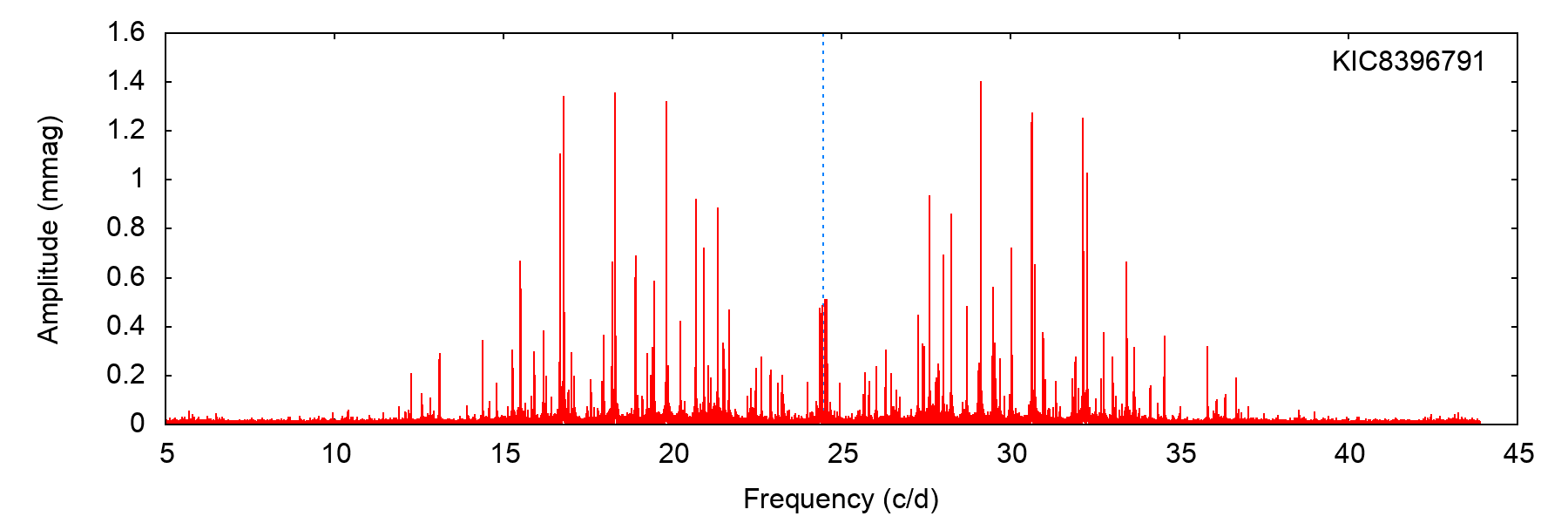}\\
\includegraphics[width=0.48\textwidth]{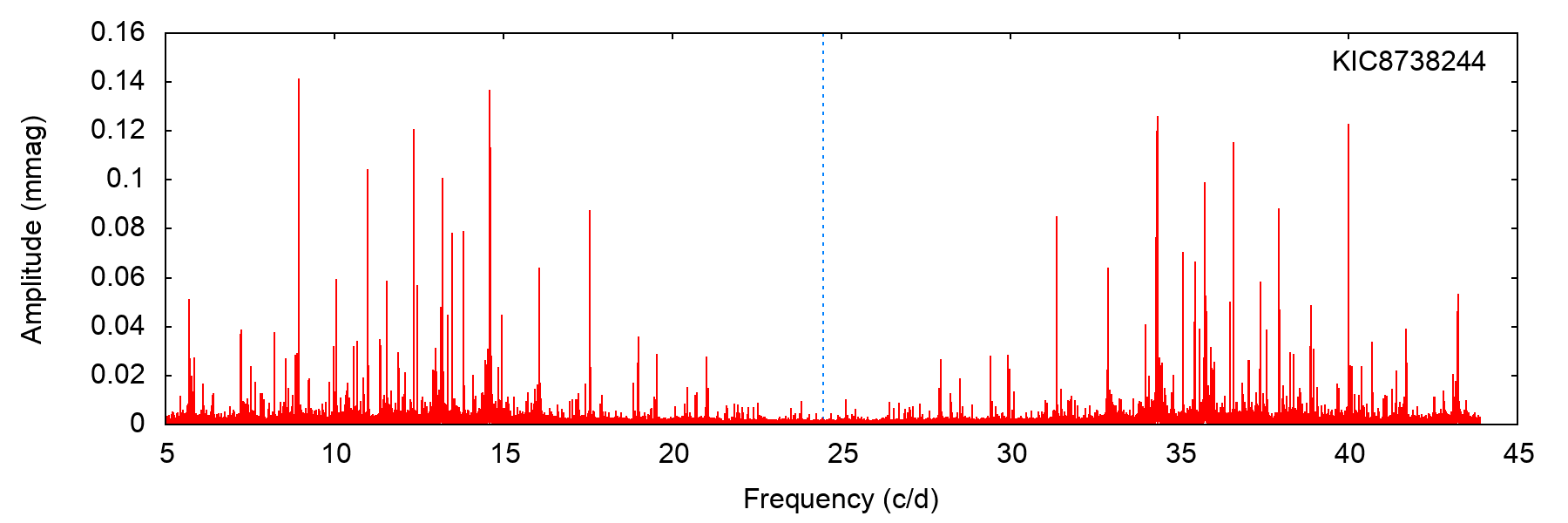}\\
\caption{Some typical examples of stars in Group D (see Sec.\,\ref{ssec:GroupD}).}
\label{fig:GroupD}
\vspace{-3mm}
\end{center}
\end{figure}

Group D lies at the narrow intersection of the pulsators and the non-pulsators, in terms of the S/N and skewness properties. It contains a broad array of pulsational properties, not all of which are consistent with the p-mode pulsation usually seen in $\delta$\,Sct stars (Fig.\,\ref{fig:GroupD}). There are highly multiperiodic pulsators (KIC\,4072582, KIC\,8396791, KIC\,8738244), almost mono-periodic pulsators (KIC\,3003224, KIC\,4136161), objects whose variability is not of the $\delta$\,Sct type despite overlapping in frequency (KIC\,4670388, KIC\,7458050), and one example (KIC\,4358571) of what are perhaps a new class of pulsators that have Fourier spectra distinct from known variable star classes (i.e.\ not a $\delta$\,Sct or $\gamma$\,Dor star) but that overlap partially in frequency. They show very high mode density in a small frequency range, typically between 2 and 6\,d$^{-1}$ across (Fig.\,\ref{fig:newpulsators}).

\subsection{Group E}
\label{ssec:GroupE}
\begin{figure}
\begin{center}
\includegraphics[width=0.48\textwidth]{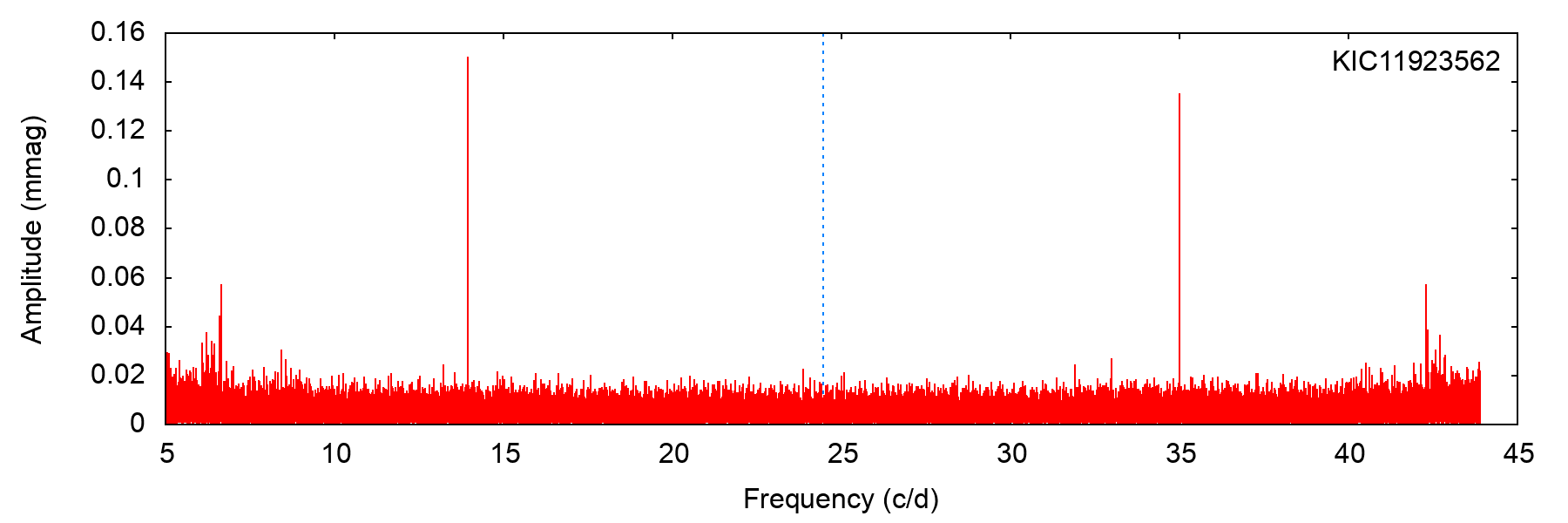}
\includegraphics[width=0.48\textwidth]{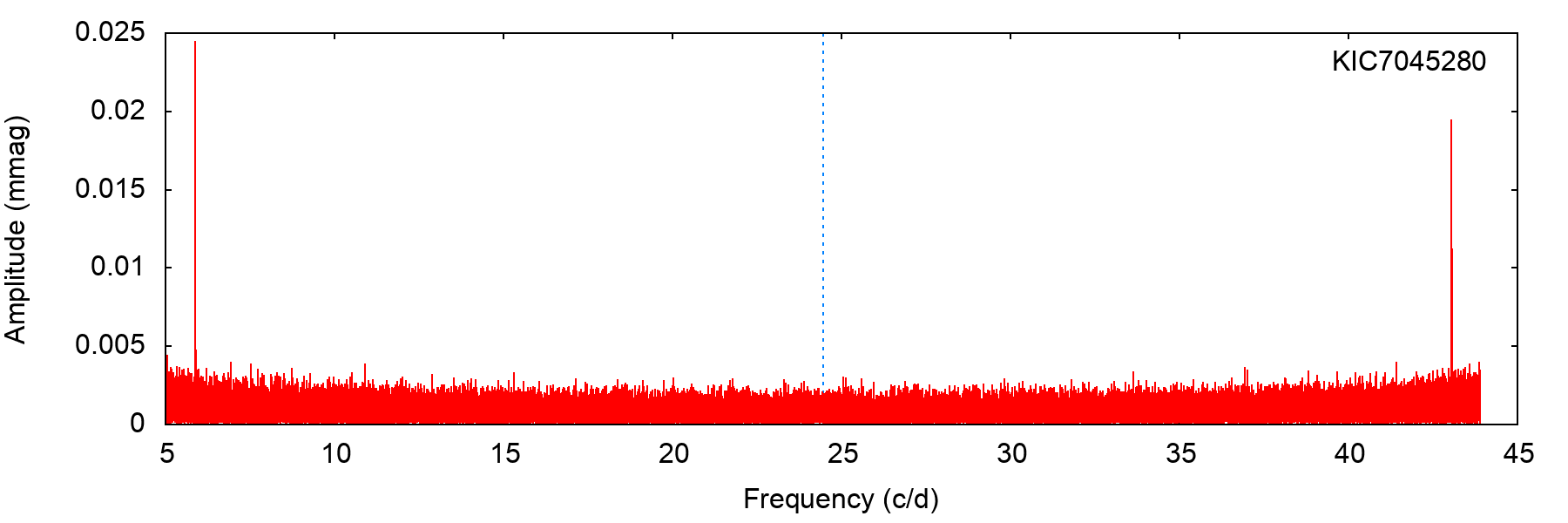}\\
\includegraphics[width=0.48\textwidth]{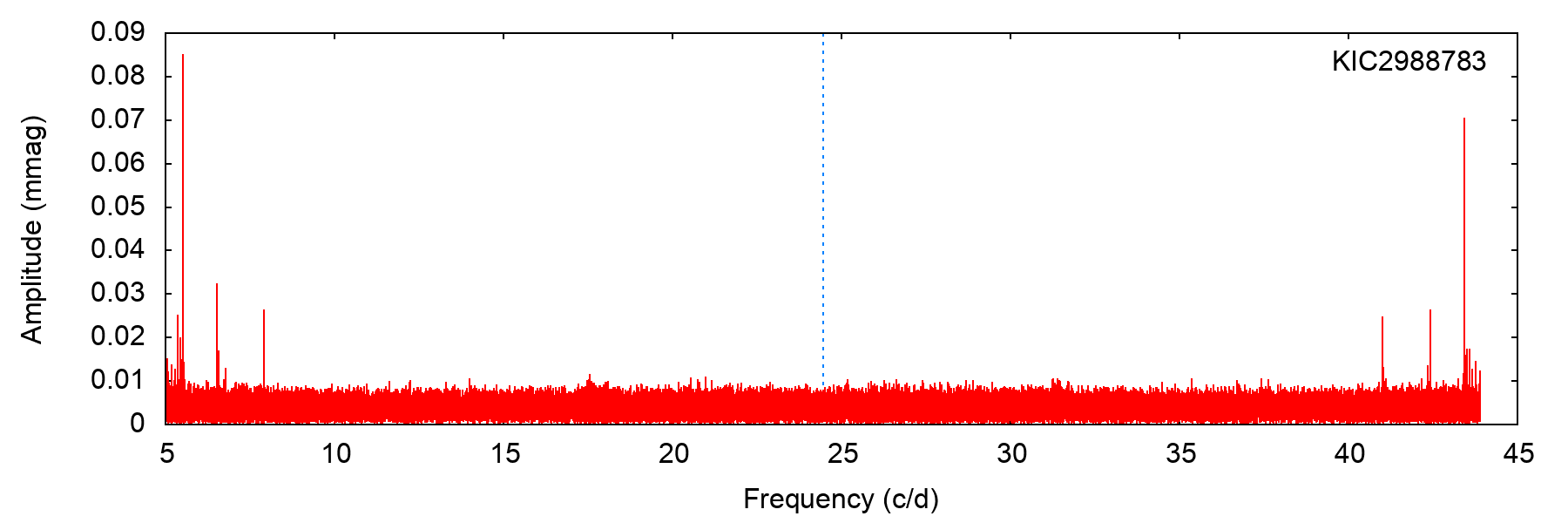}
\includegraphics[width=0.48\textwidth]{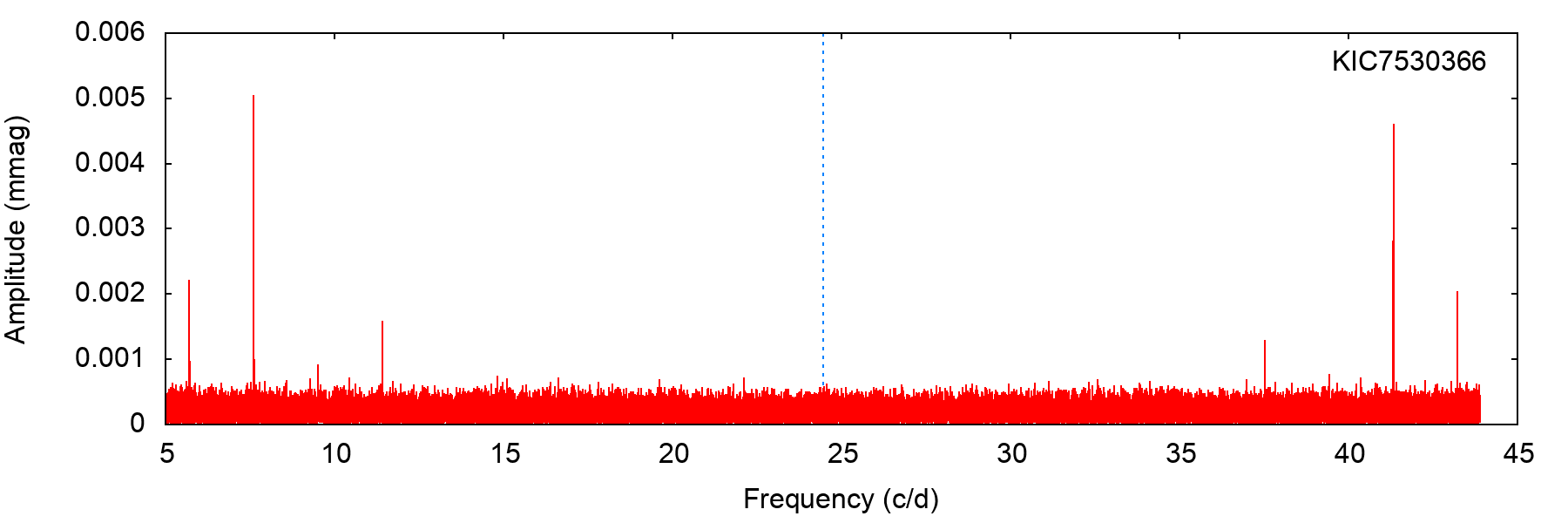}\\
\includegraphics[width=0.48\textwidth]{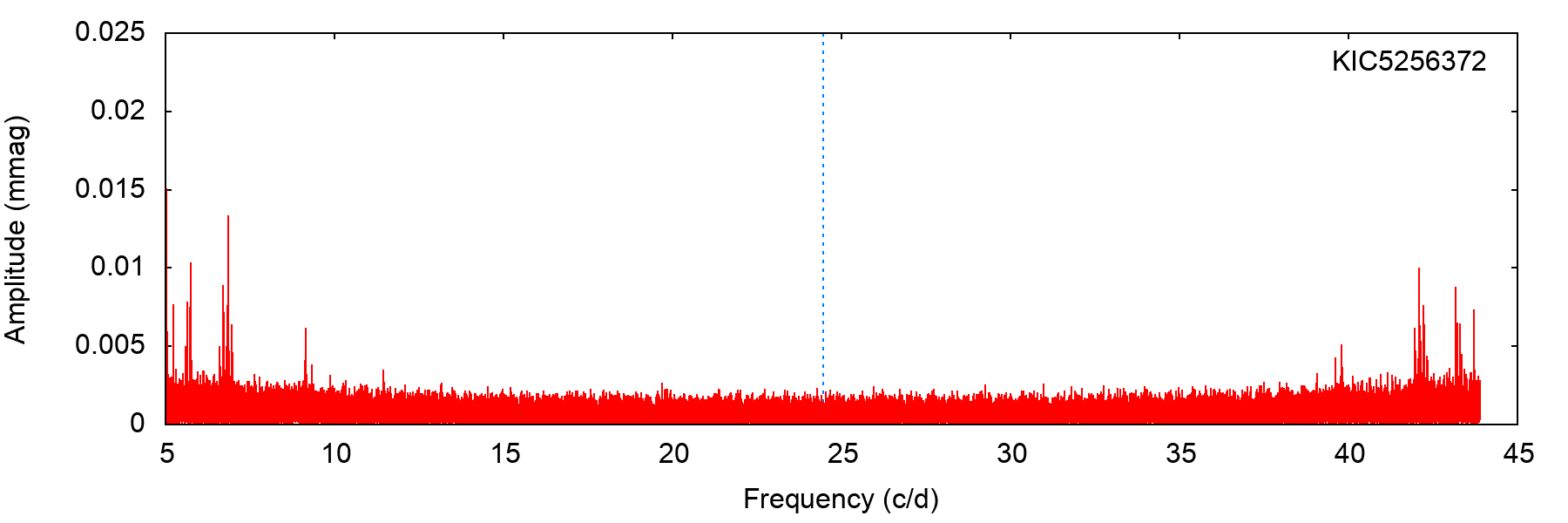}
\includegraphics[width=0.48\textwidth]{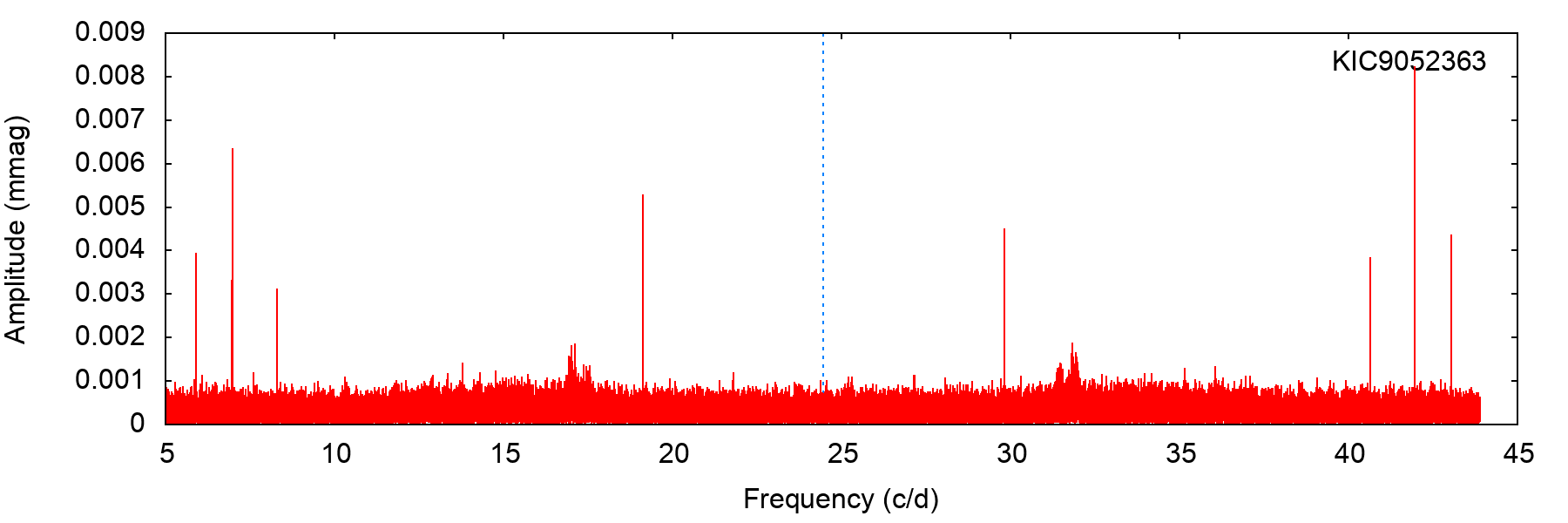}\\
\includegraphics[width=0.48\textwidth]{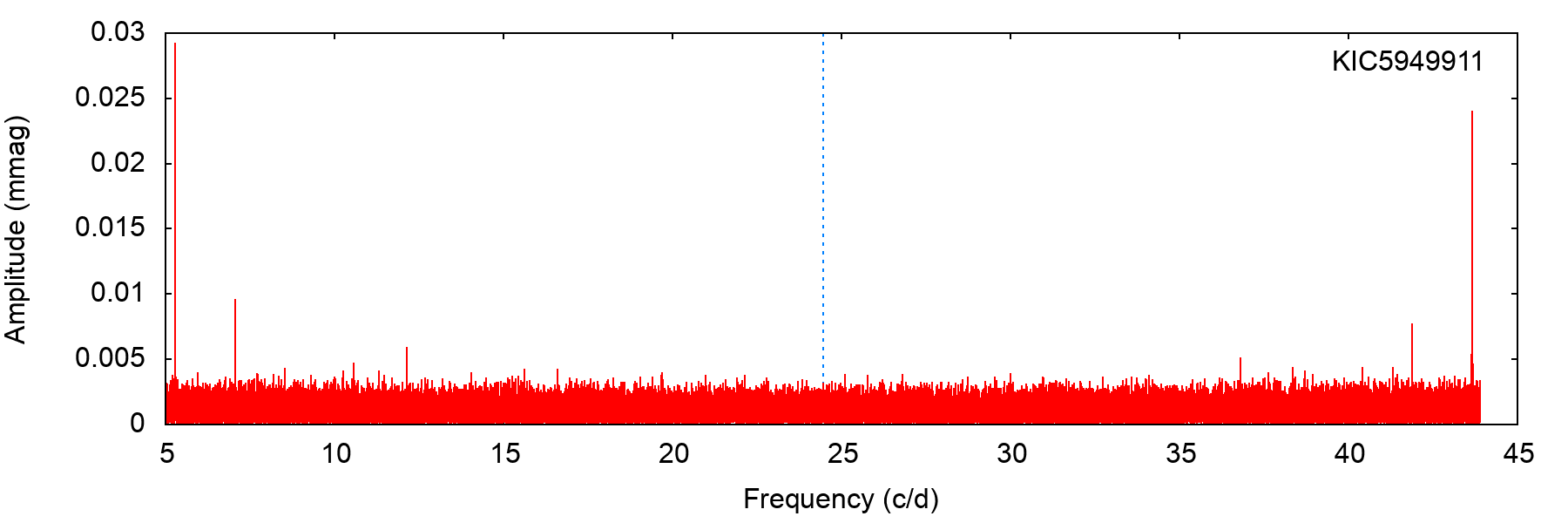}
\includegraphics[width=0.48\textwidth]{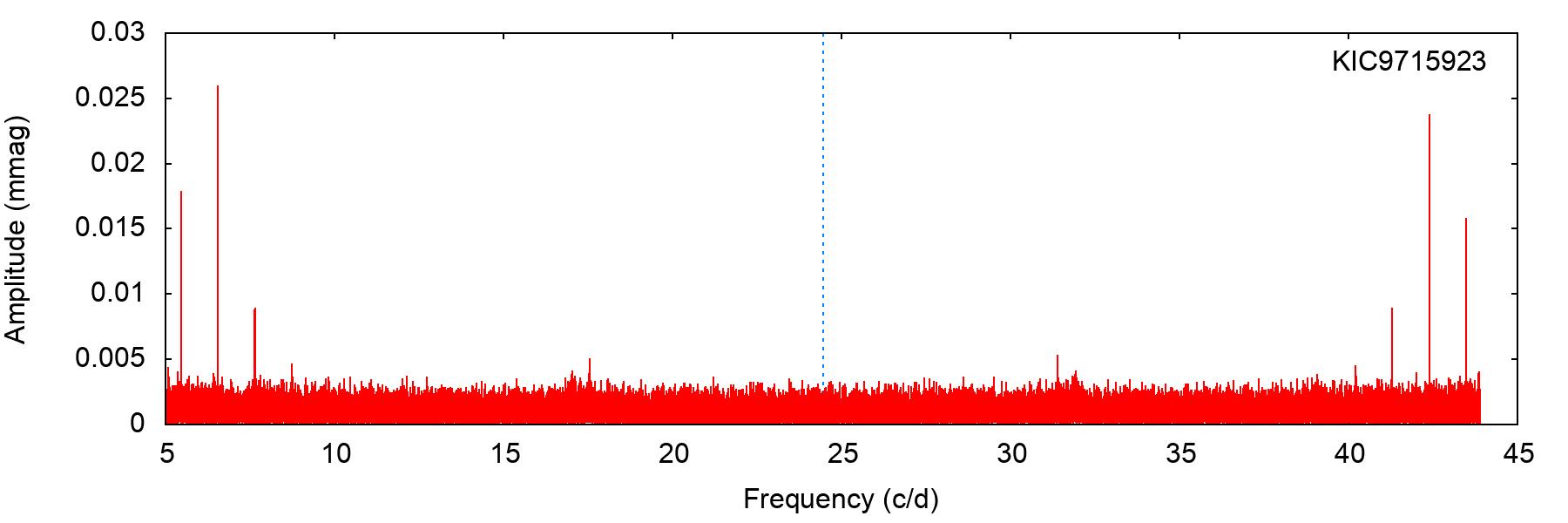}\\
\caption{Some typical examples of stars in Group E (see Sec.\,\ref{ssec:GroupE}).}
\label{fig:GroupE}
\vspace{-3mm}
\end{center}
\end{figure}

Stars in this group have lower S/N and lower skewness than Group D, representing the lower-right of the stream connecting the pulsators and the non-pulsators. While they do have Fourier peaks that are statistically significant, in almost all cases they are not $\delta$\,Sct stars (Fig.\,\ref{fig:GroupE}). Most of them show equidistant peaks near the lower-limit of 5\,d$^{-1}$, which are harmonics of variability at lower frequencies, perhaps originating from low-amplitude ellipsoidal variables.

\subsection{Group F}
\label{ssec:GroupF}
\begin{figure}
\begin{center}
\includegraphics[width=0.48\textwidth]{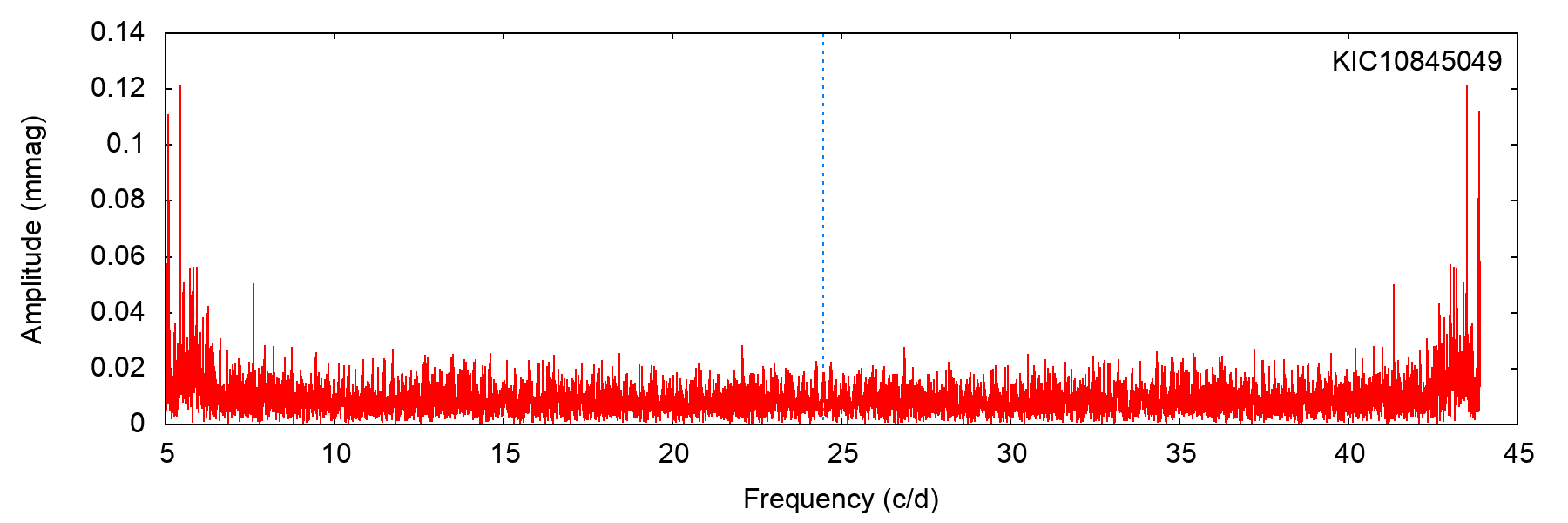}
\includegraphics[width=0.48\textwidth]{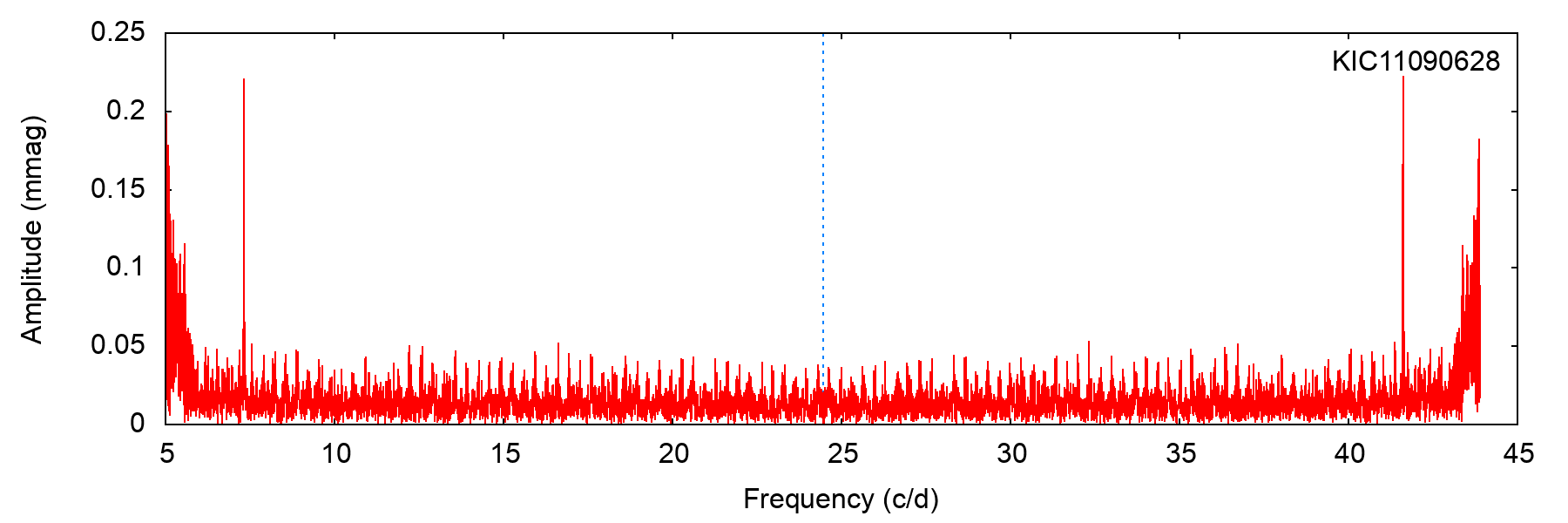}
\includegraphics[width=0.48\textwidth]{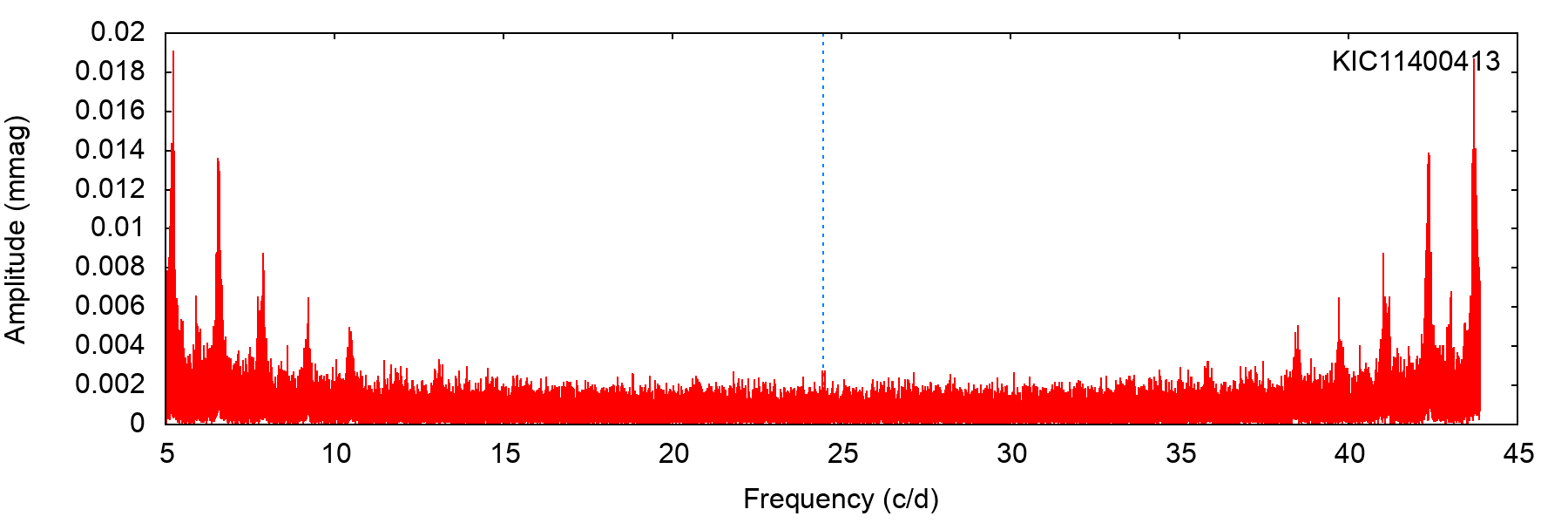}
\includegraphics[width=0.48\textwidth]{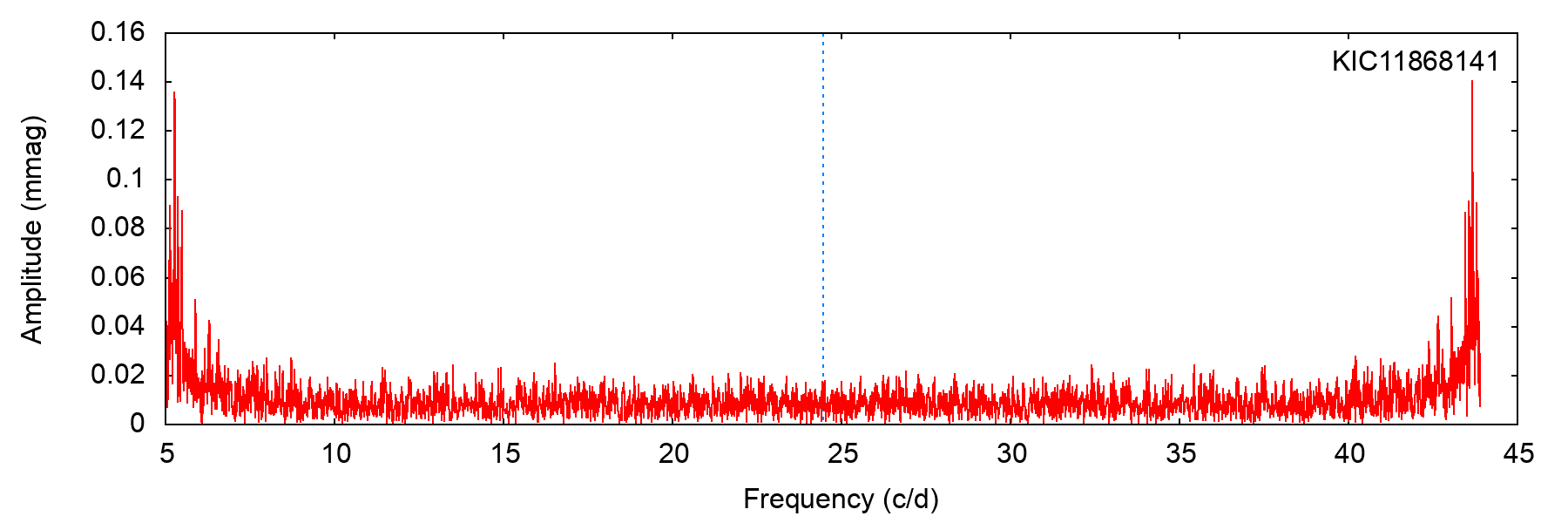}
\includegraphics[width=0.48\textwidth]{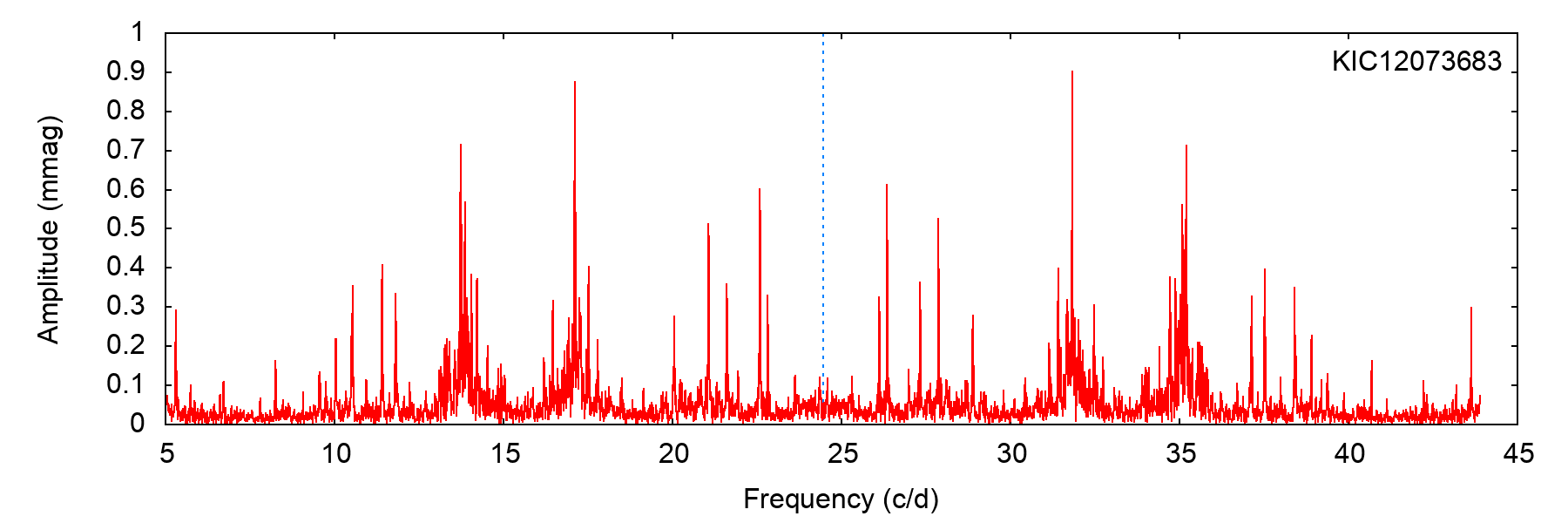}
\includegraphics[width=0.48\textwidth]{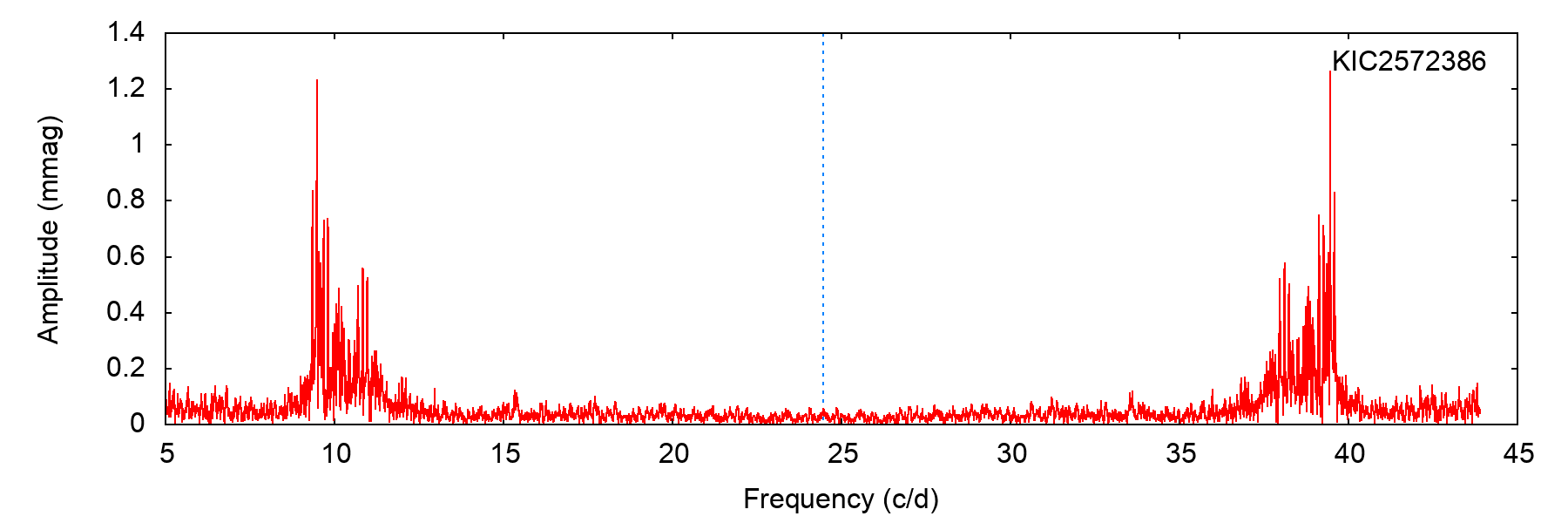}
\includegraphics[width=0.48\textwidth]{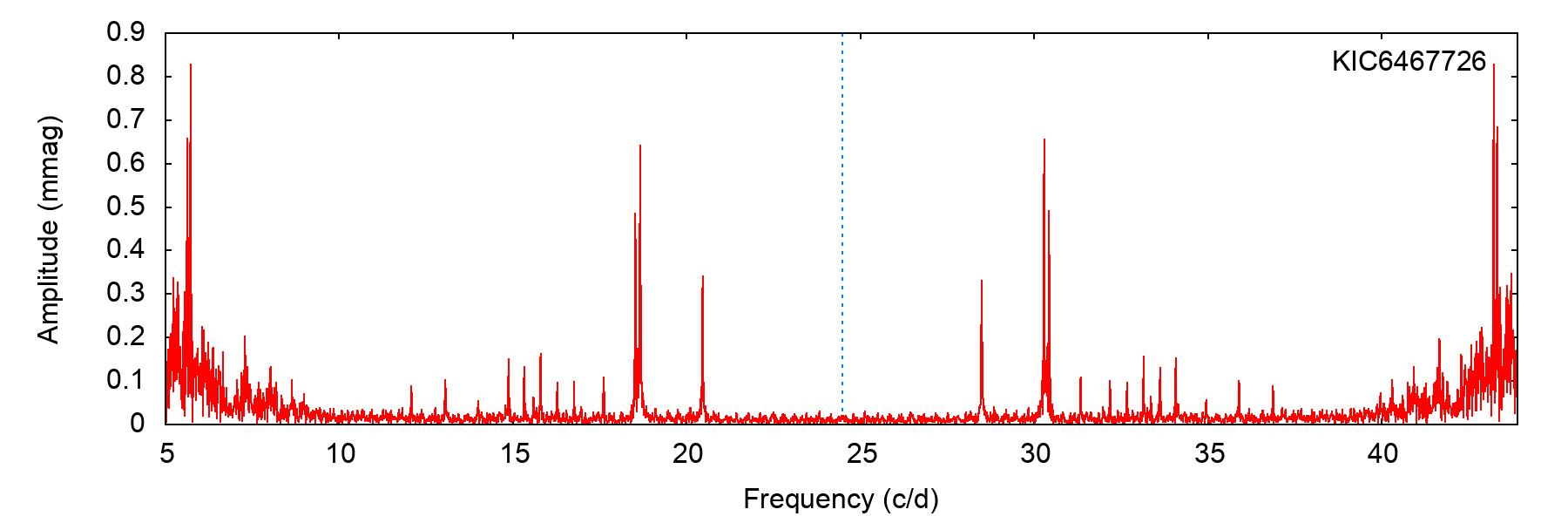}
\includegraphics[width=0.48\textwidth]{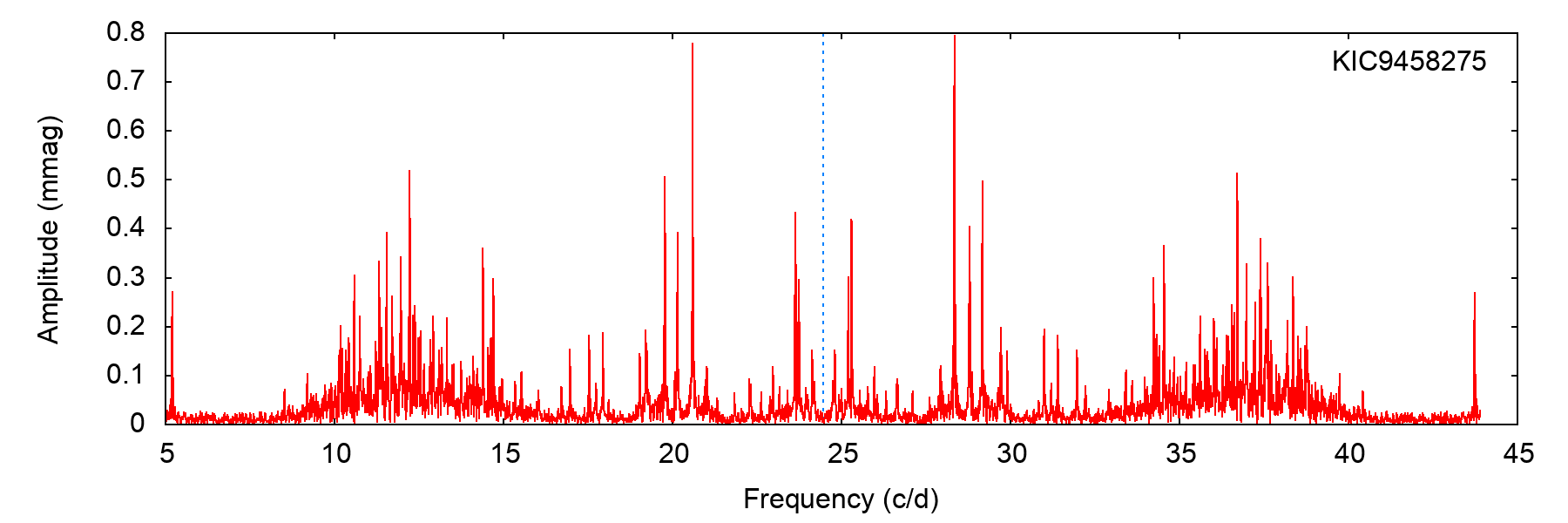}
\caption{Some typical examples of stars in Group F (see Sec.\,\ref{ssec:GroupF}).}
\label{fig:GroupF}
\vspace{-3mm}
\end{center}
\end{figure}

Group F encompasses stars at the upper-left of the stream connecting the pulsators and the non-pulsators. This group is heterogeneous (Fig.\,\ref{fig:GroupF}). Some objects in Group F (e.g. KIC\,6467726) appear to be $\delta$\,Sct stars that were only briefly observed by \textit{Kepler}, and therefore have much poorer frequency resolution and higher noise than if they were observed for the full mission. Their intrinsic properties would be similar to Group B. Many objects in this group (KIC\,10845049, KIC\,11090628, KIC\,11400413, KIC\,11868141, KIC\,7908851) are not $\delta$\,Sct pulsators, but do show some form of multi-periodic variability very close to the 5\,d$^{-1}$ limit. Group F also contains examples of the new class of pulsators (KIC\,12073683, KIC\,2572386, KIC\,9458275), with additional properties: all three examples appear to be super-Nyquist in origin, as evidenced by their higher amplitudes beyond the Nyquist frequency; KIC\,12073683 has two of the power excesses, and its Fourier transform also appears to contain ordinary p\:modes, indicating that it is a hybrid or in a binary with a normal $\delta$\,Sct star; KIC\,2572386 has a single power excess; and finally KIC\,9458275 has a very broad excess spanning 6\,d$^{-1}$, accompanied by normal $\delta$\,Sct pulsation.

\subsection{Group G}
\label{ssec:GroupG}
Objects in this group have only one or two low-amplitude Fourier peaks, usually near the 5-d$^{-1}$ limit. Two cases are exceptional: KIC\,4772888 and KIC\,8417852, having higher oscillation frequencies. Low amplitude pulsators such as these are discussed in Sec.\,\ref{sec:discussion}. Where the peaks are harmonics of lower frequencies, the stars are classified as non-$\delta$\,Sct stars in the `revised' classifications.

\subsection{Group H}
\label{ssec:GroupH}
\begin{figure}
\begin{center}
\includegraphics[width=0.48\textwidth]{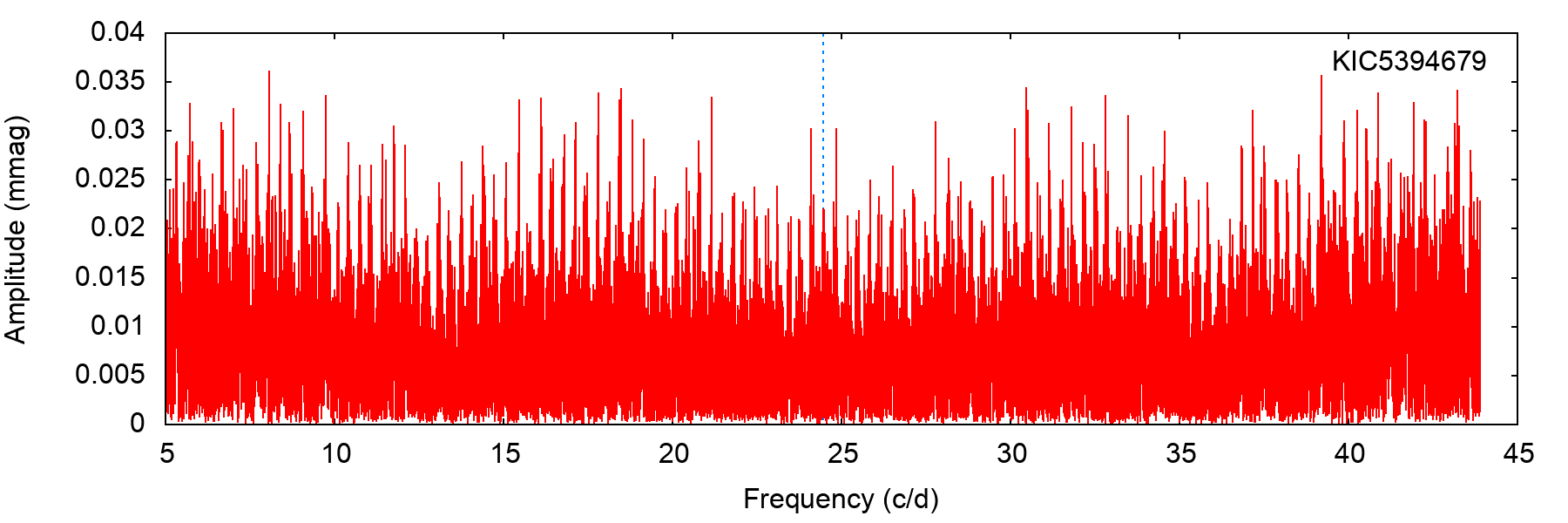}
\includegraphics[width=0.48\textwidth]{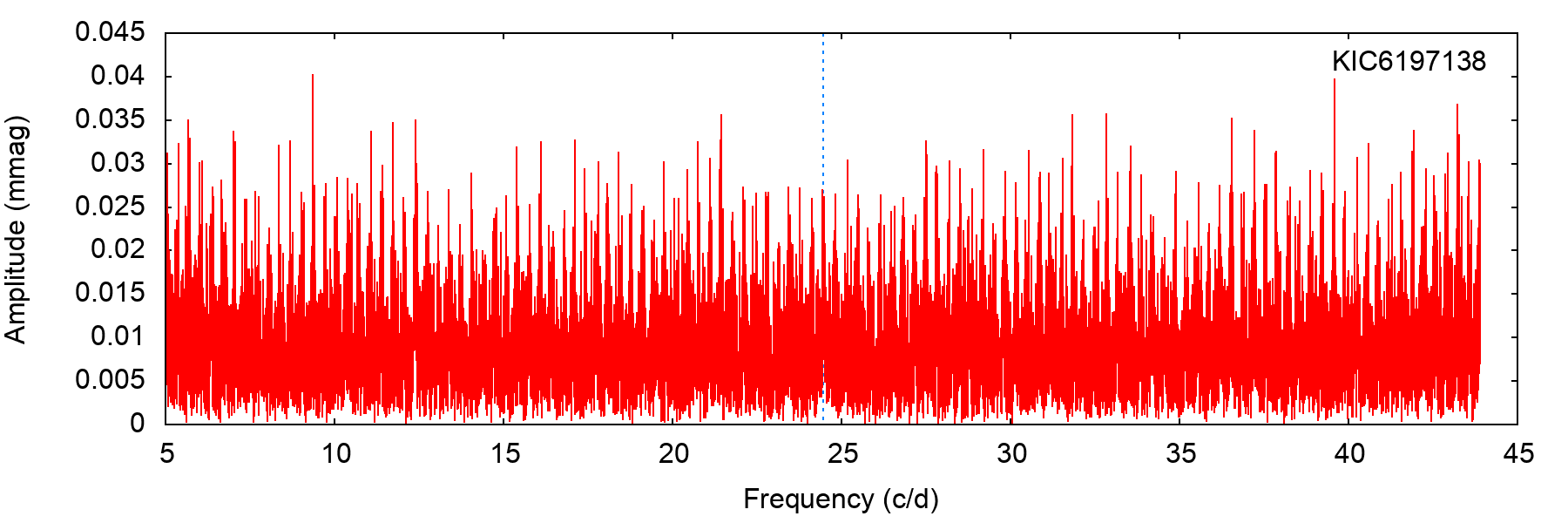}
\includegraphics[width=0.48\textwidth]{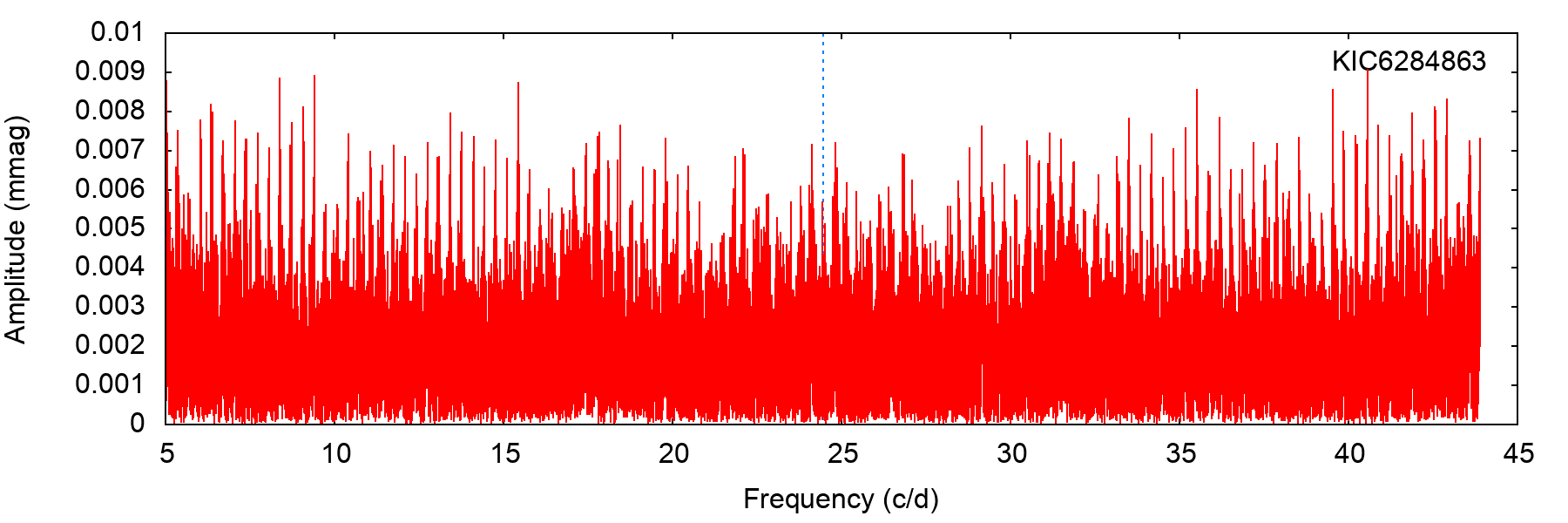}
\includegraphics[width=0.48\textwidth]{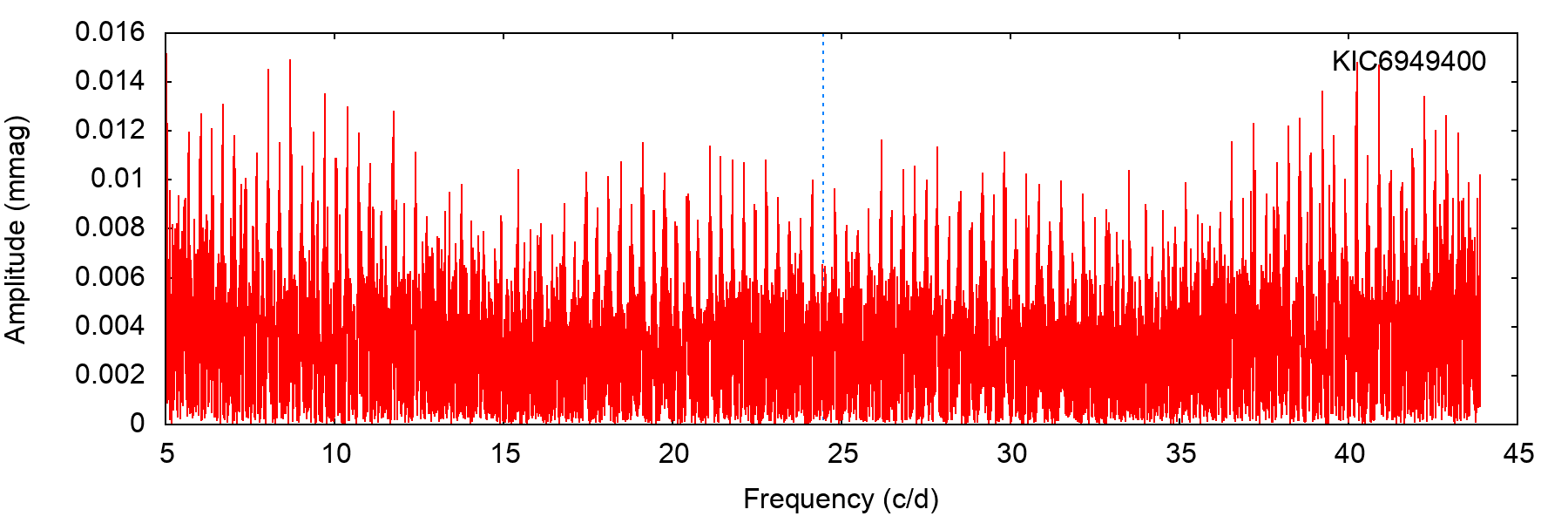}
\includegraphics[width=0.48\textwidth]{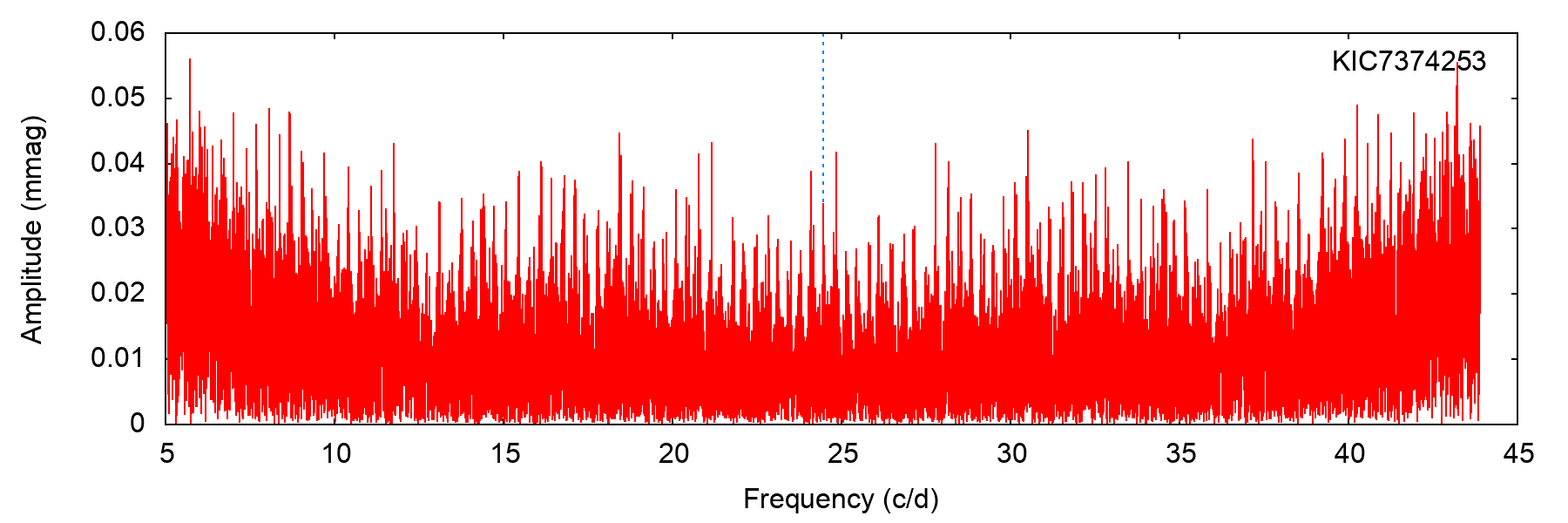}
\includegraphics[width=0.48\textwidth]{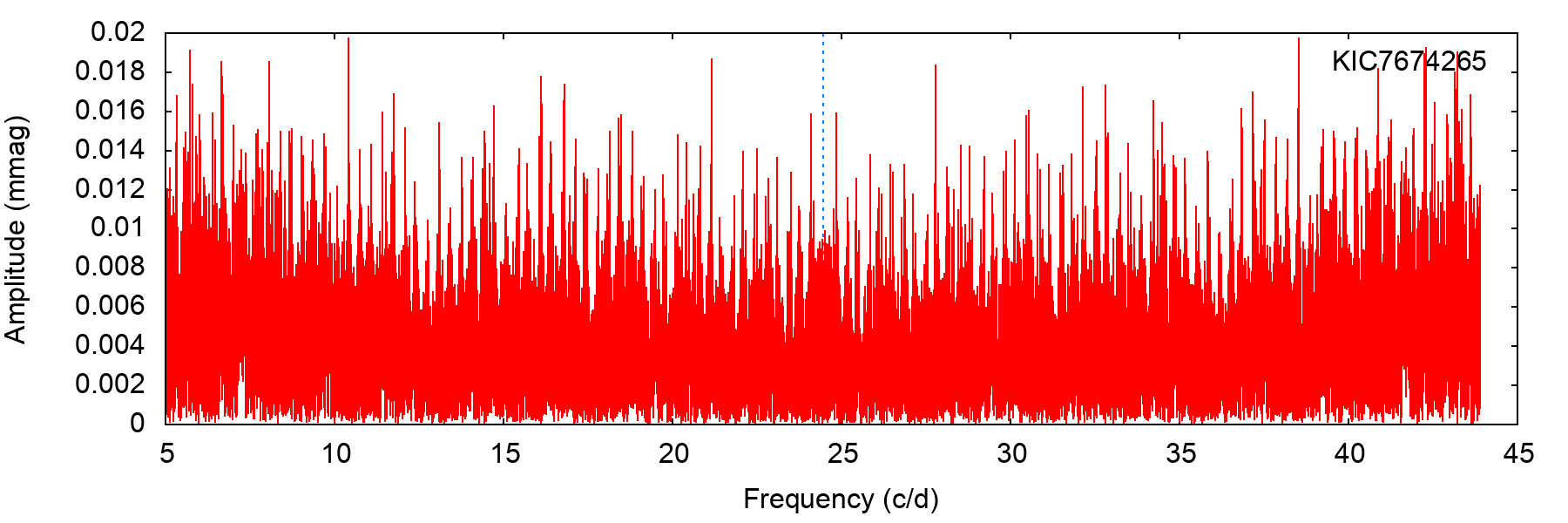}
\includegraphics[width=0.48\textwidth]{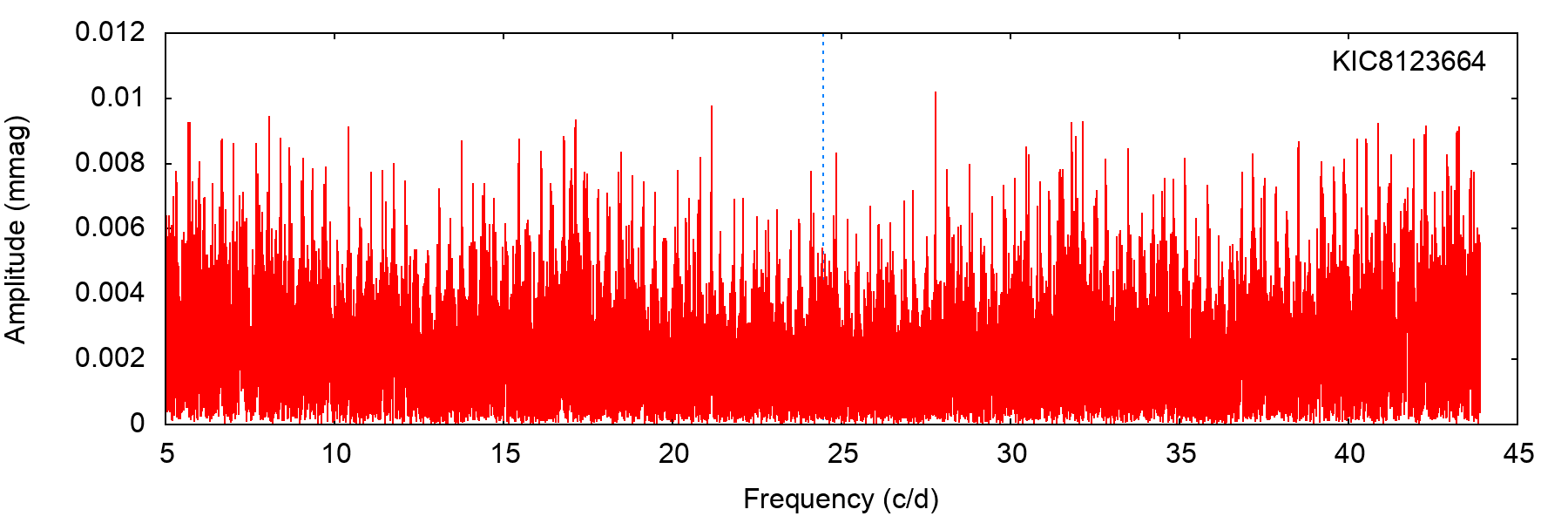}
\includegraphics[width=0.48\textwidth]{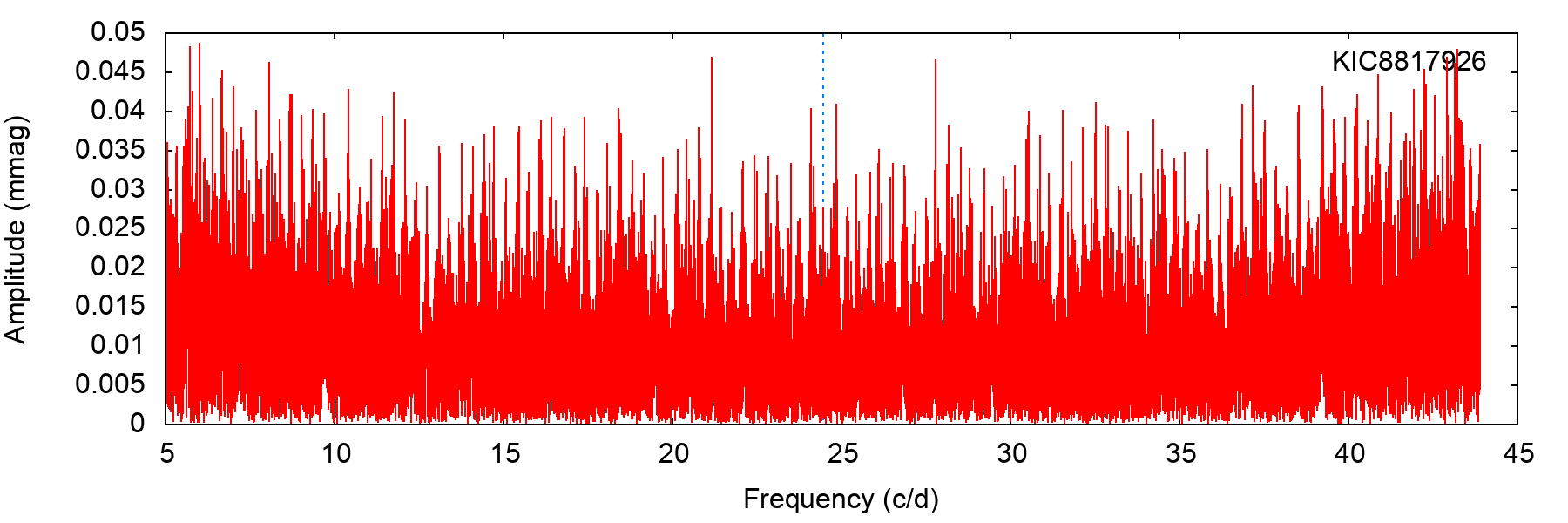}
\caption{Some typical examples of stars in Group H (see Sec.\,\ref{ssec:GroupH}).}
\label{fig:GroupH}
\vspace{-3mm}
\end{center}
\end{figure}

Group H lies just to the upper-right of the main group of non-pulsators. None of these stars are $\delta$\,Sct stars, but their noise is structured. Noise spikes appear at regular intervals, and often in the same place from star to star (Fig.\,\ref{fig:GroupH}). In many cases this is caused by peculiar Q2 data, where there are flux excursions of tens of ppt (tens of mmag). These peculiar data remain the current version available via KASOC. In other cases the cause of the structured noise is short datasets of only one or two Quarters. This group highlights the utility of the SNR-skewness plane in identifying problematic data.

\subsection{Group I}
\label{ssec:GroupI}
These stars are the non-$\delta$\,Sct stars, typically with noise-levels in the Fourier transform of only 1--2\,$\upmu$mag. A cross-examination of their low-frequency (<5\,d$^{-1}$) variability would identify whether they are truly non-variable. If so, their lack of strong surface convection and strong stellar winds must place them among the least variable objects on the H--R diagram. Otherwise, they may serve well as targets to study pure g-mode or r-mode pulsation in isolation.

\subsection{Group J}
\label{ssec:GroupJ}
These objects are also non-pulsators, where extreme outliers in the light curve have strongly skewed the Fourier amplitudes. Without the outliers, these stars would fall in group I.

\subsection{New pulsators}
\label{ssec:GroupNew}
\begin{figure}
\begin{center}
\includegraphics[width=0.48\textwidth]{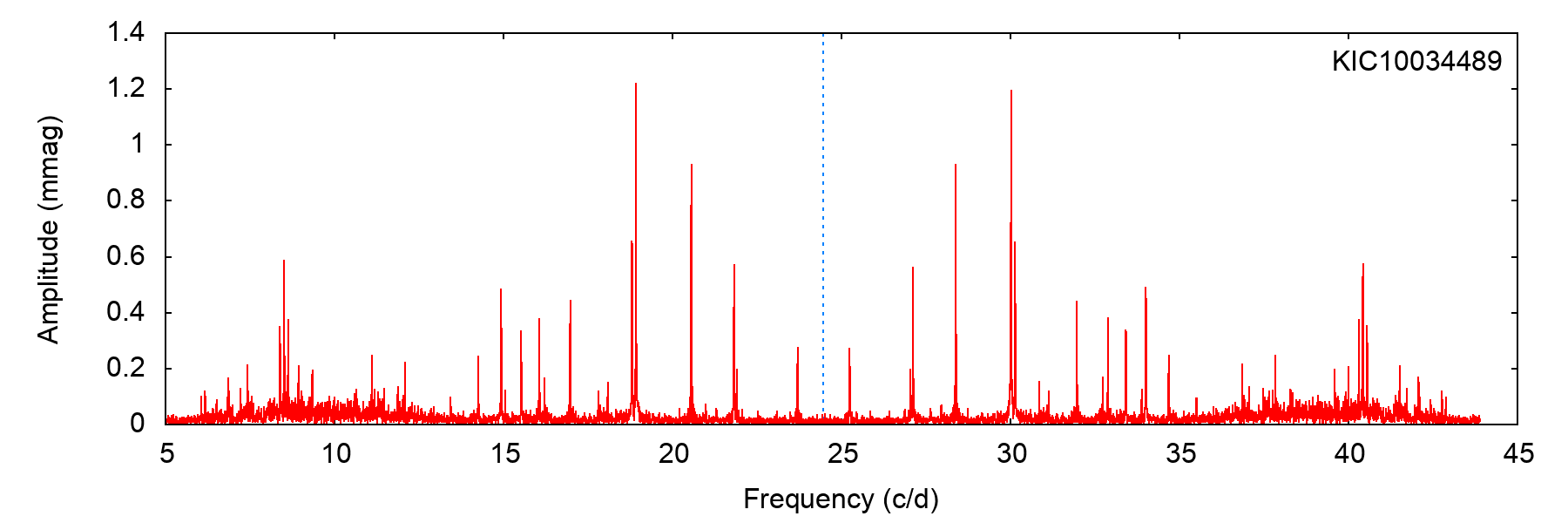}
\includegraphics[width=0.48\textwidth]{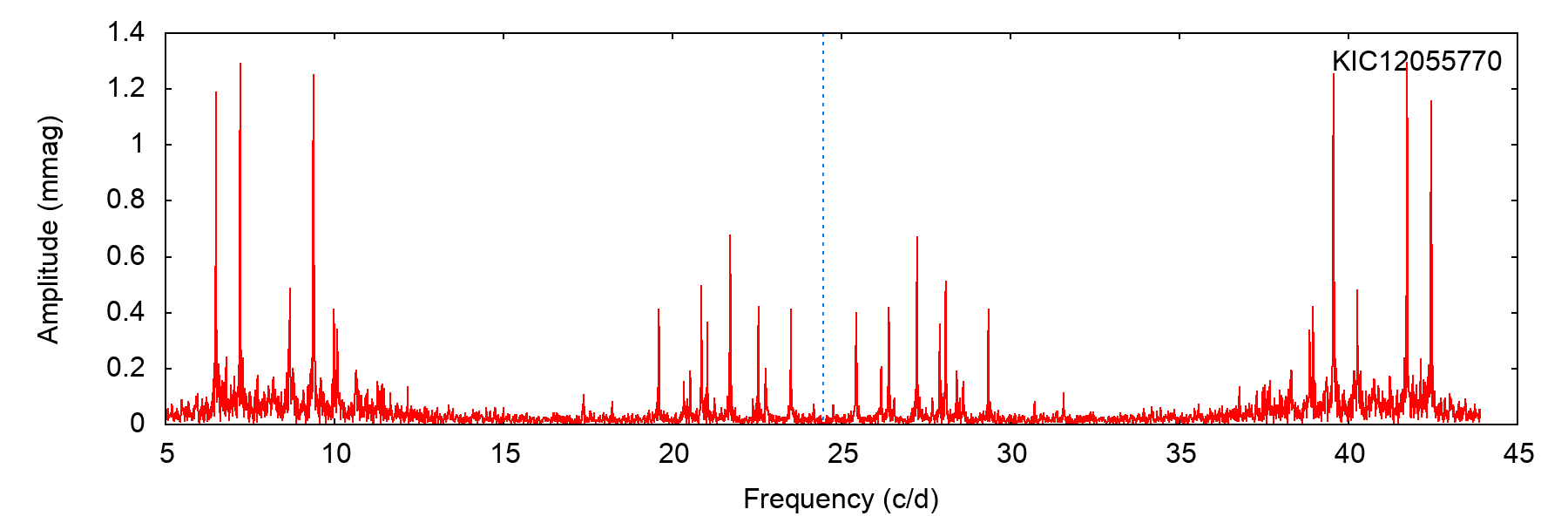}
\includegraphics[width=0.48\textwidth]{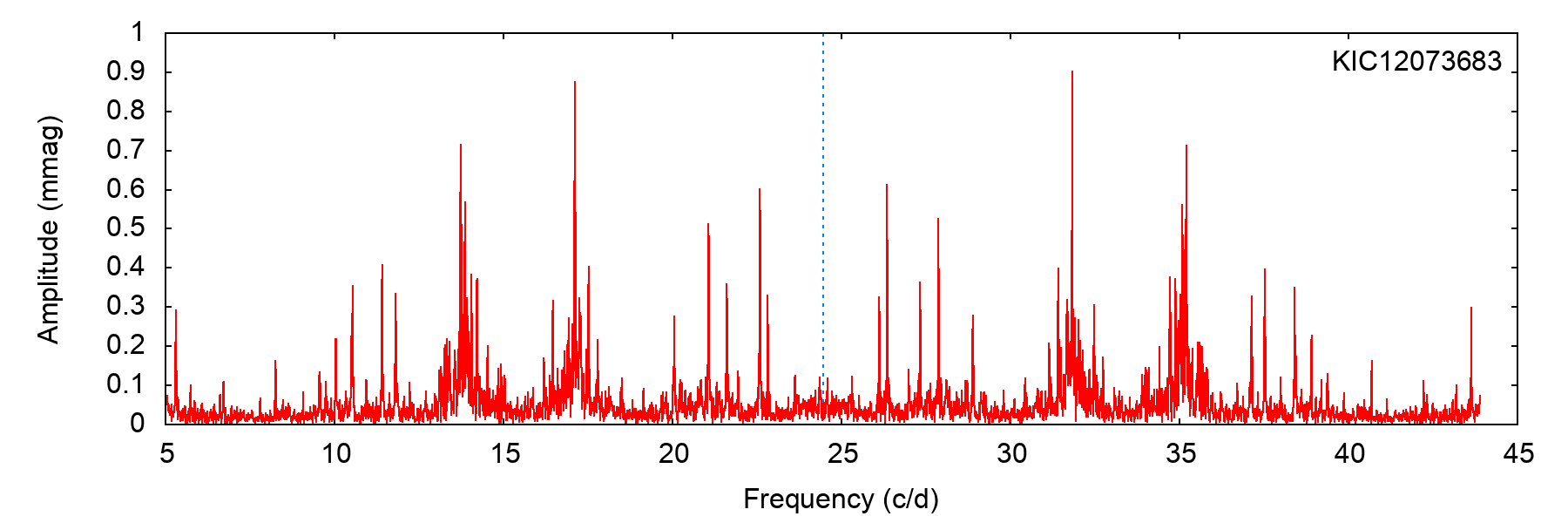}
\includegraphics[width=0.48\textwidth]{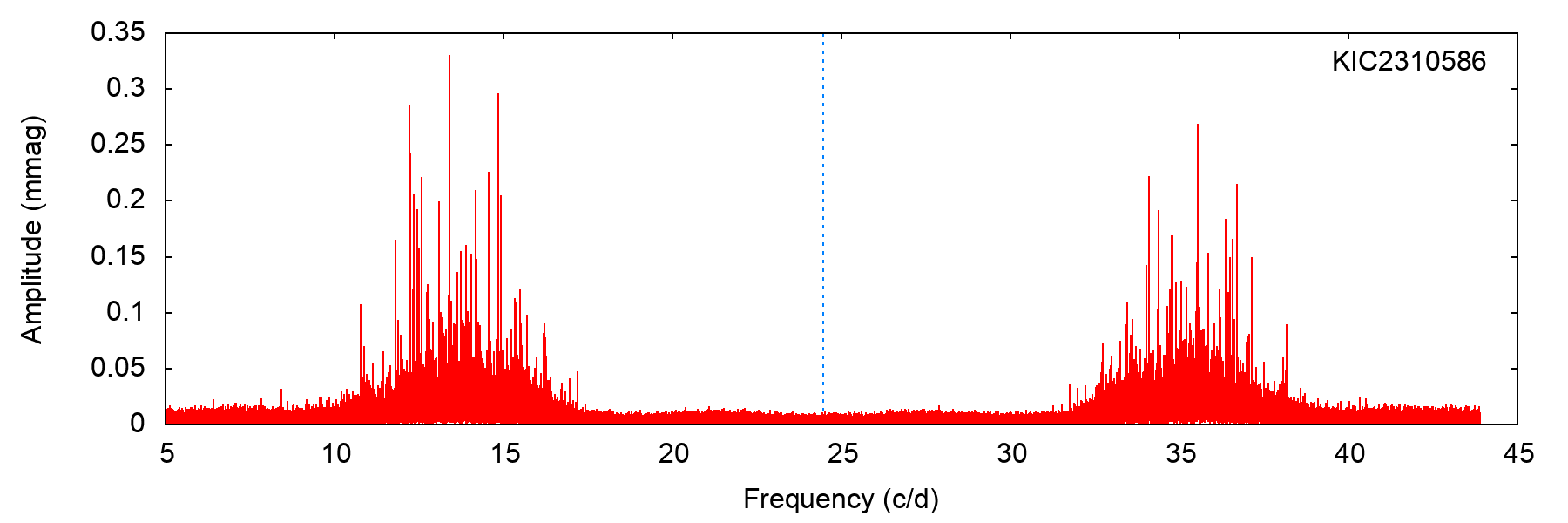}
\includegraphics[width=0.48\textwidth]{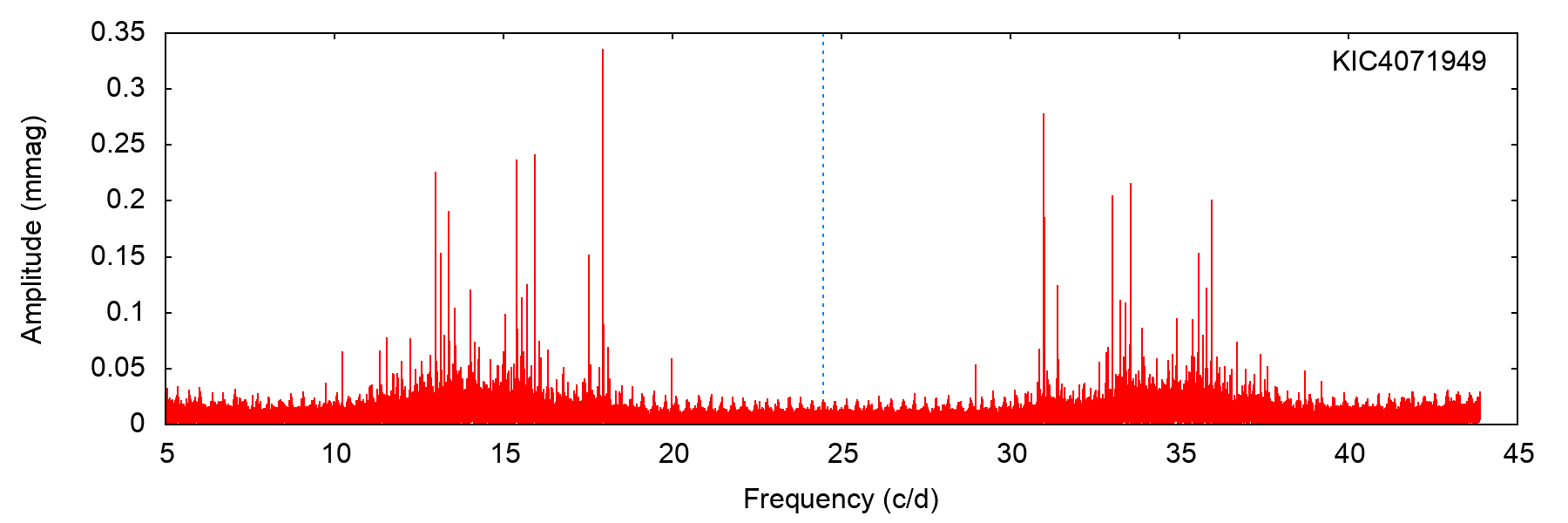}
\includegraphics[width=0.48\textwidth]{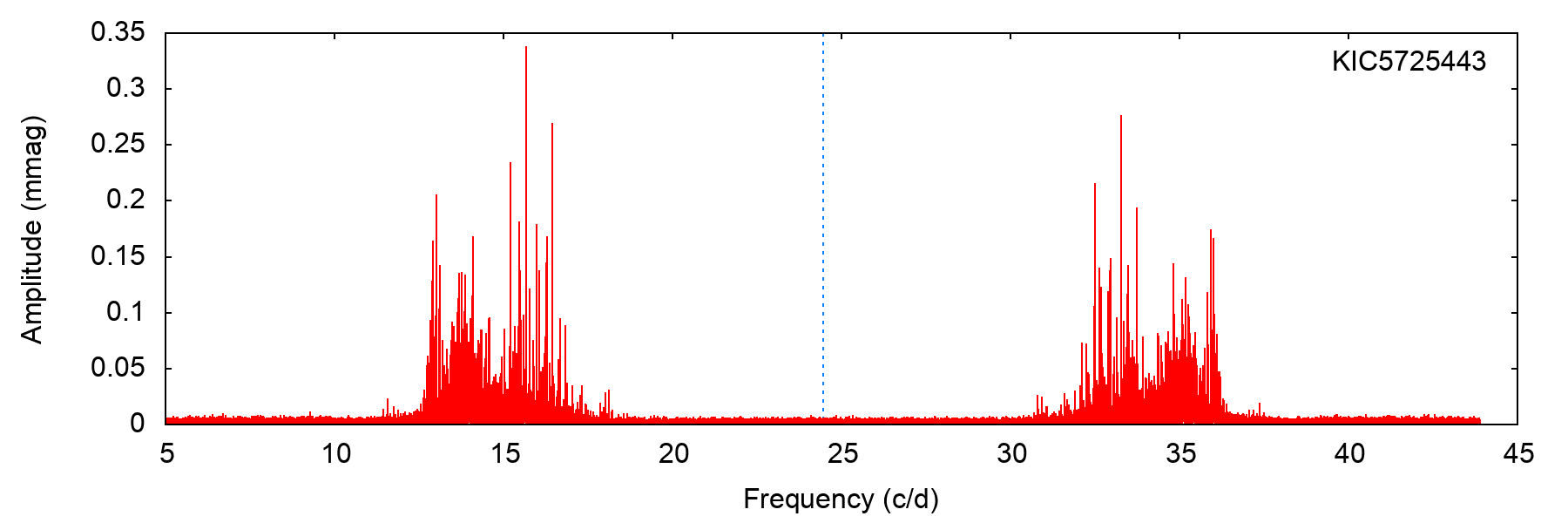}
\includegraphics[width=0.48\textwidth]{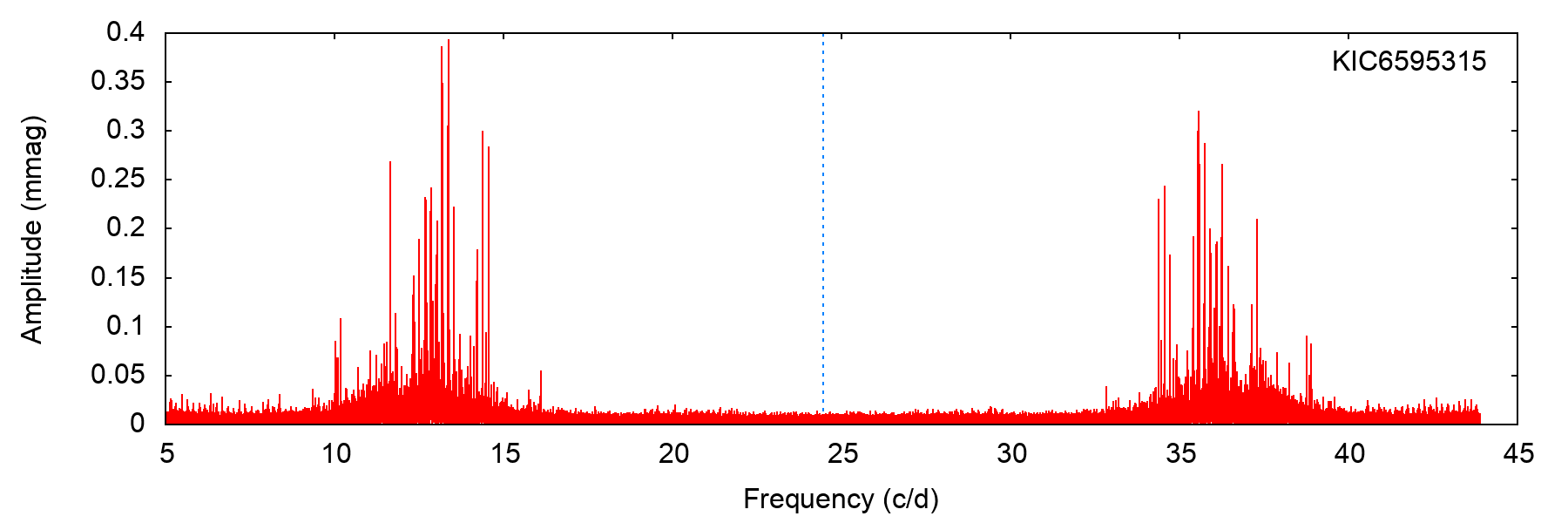}
\includegraphics[width=0.48\textwidth]{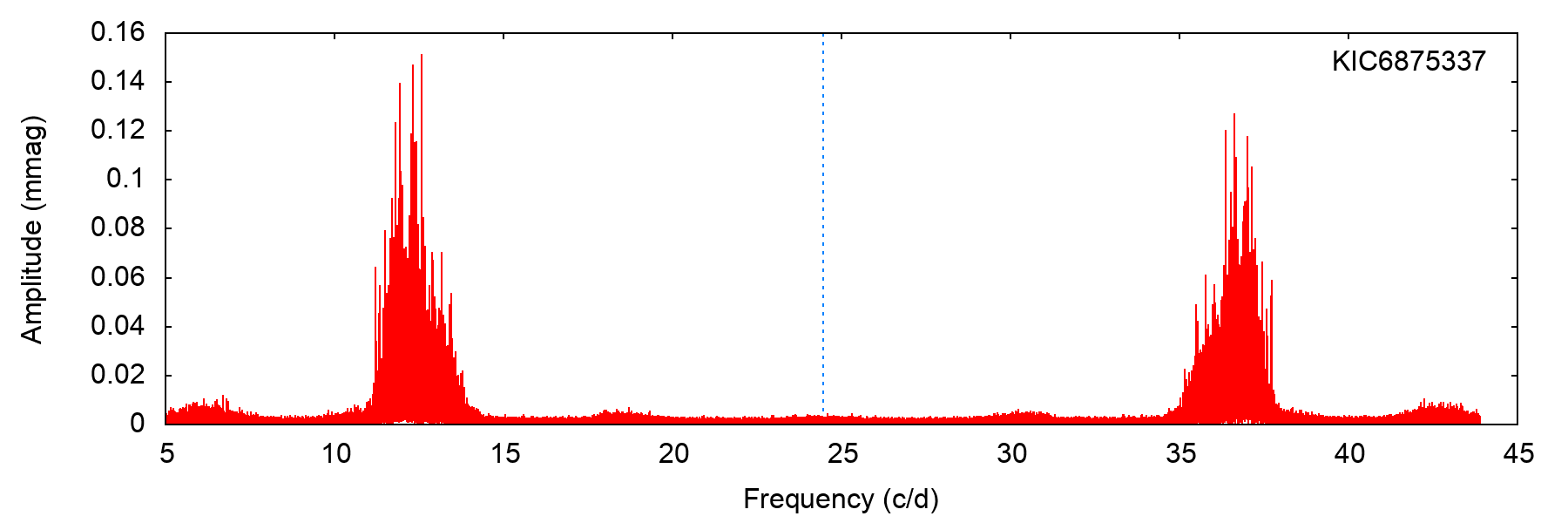}
\caption{Some typical examples of stars in this new class of puslators (see Sec.\,\ref{ssec:GroupNew}).}
\label{fig:newpulsators}
\vspace{-3mm}
\end{center}
\end{figure}

Around 30 targets have pulsation properties not described by existing groups, and probably belonging to a new pulsation class. Figure~\ref{fig:newpulsators} shows that some of these are hybrid $\delta$\,Sct pulsators (e.g. KIC\,10034489, KIC\,12055770, KIC\,12073683), while others are more `pure' (KIC\,2310586, KIC\,6595315, KIC\,6875337). The latter object, in the lower-right corner, shows smaller power excesses at approximately 7\,d$^{-1}$ either side of the main excess.

\bsp	
\end{document}